\documentclass[11pt,openany]{book} 

\usepackage[top=3cm,bottom=3cm,left=3.2cm,right=3.2cm,headsep=10pt,letterpaper]{geometry} 

\usepackage{imakeidx}
\usepackage[utf8]{inputenc}
\makeindex[program=makeindex,options=-StyleInd.ist]

\usepackage{xcolor} 
\definecolor{ocre}{RGB}{102,177,201}

\usepackage{avant}  
\usepackage{multicol}
\usepackage{orcidlink}
\usepackage{newtxtext,amsfonts}  
\usepackage{amsmath,amssymb}

\usepackage{microtype} 
\usepackage[utf8]{inputenc} 
\usepackage[T1]{fontenc} 

\usepackage{titlesec} 

\usepackage{graphicx}
\graphicspath{{Pictures/}} 

\usepackage{lipsum} 

\usepackage{tikz} 

\usepackage[english]{babel} 

\usepackage{enumitem} 
\setlist{nolistsep} 

\usepackage{booktabs} 

\usepackage{eso-pic} 

\usepackage{titletoc} 

\contentsmargin{0cm} 
\titlecontents{chapter}[0.75cm] 
{\addvspace{10pt}\large\sffamily\bfseries} 
{\color{ocre!60}\contentslabel[\Large\thecontentslabel]{0.8cm}\color{ocre}} 
{}  
{\color{ocre!60}\normalsize\sffamily\bfseries\;\titlerule*[.5pc]{.}\;\thecontentspage} 

\titlecontents{section}[2.05cm]
{\addvspace{1.2pt}\sffamily\footnotesize}
{\contentslabel[\thecontentslabel]{1.25cm}}
{}
{\sffamily\color{black}\titlerule*[0.5pc]{.}\;\thecontentspage}
[]

\titlecontents{starsection}[0cm]
{\addvspace{-10pt}\sffamily\bfseries}
{\contentslabel[\thecontentslabel]{1.25cm}}
{}
{\sffamily\color{black}\titlerule*[0.5pc]{.}\;\thecontentspage}
[]

\titlecontents{endsection}[0cm]
{\addvspace{10pt}\sffamily\bfseries}
{\contentslabel[\thecontentslabel]{1.25cm}}
{}
{\sffamily\color{black}\titlerule*[0.5pc]{.}\;\thecontentspage}
[]

\titlecontents{endsection2}[0cm]
{\addvspace{10pt}\sffamily\bfseries}
{\contentslabel[\thecontentslabel]{1.25cm}}
{}
{\sffamily\color{black}\titlerule*[0.5pc]{.}\;\thecontentspage}
[]

\titlecontents{subsection}[3.35cm] 
{\addvspace{1pt}\sffamily\scriptsize} 
{\contentslabel[\thecontentslabel]{1.25cm}} 
{}
{\sffamily\;\titlerule*[.5pc]{.}\;\thecontentspage} 
[] 

\titlecontents{lsection}[0em] 
{\footnotesize\sffamily} 
{}
{}
{}

\titlecontents{lsubsection}[.5em] 
{\normalfont\footnotesize\sffamily} 
{}
{}
{}
 
\usepackage{fancyhdr} 

\fancypagestyle{plain}{\fancyhead{}} 

\makeatletter
\renewcommand{\cleardoublepage}{
\clearpage\ifodd\c@page\else
\hbox{}
\vspace*{\fill}
\thispagestyle{empty}
\newpage
\fi}

\usepackage{amsmath,amsfonts,amssymb,amsthm} 

\newtheoremstyle{ocrenumbox}
{0pt}
{0pt}
{\normalfont}
{}
{\small\bf\sffamily\color{ocre}}
{\;}
{0.25em}
{\small\sffamily\color{ocre}\thmname{#1}\nobreakspace\thmnumber{\@ifnotempty{#1}{}\@upn{#2}}
\thmnote{\nobreakspace\the\thm@notefont\sffamily\bfseries\color{black}---\nobreakspace#3.}} 

\newtheoremstyle{blacknumex}
{5pt}
{5pt}
{\normalfont}
{} 
{\small\bf\sffamily}
{\;}
{0.25em}
{\small\sffamily{\tiny\ensuremath{\blacksquare}}\nobreakspace\thmname{#1}\nobreakspace\thmnumber{\@ifnotempty{#1}{}\@upn{#2}}
\thmnote{\nobreakspace\the\thm@notefont\sffamily\bfseries---\nobreakspace#3.}}

\newtheoremstyle{blacknumbox} 
{0pt}
{0pt}
{\normalfont}
{}
{\small\bf\sffamily}
{\;}
{0.25em}
{\small\sffamily\thmname{#1}\nobreakspace\thmnumber{\@ifnotempty{#1}{}\@upn{#2}}
\thmnote{\nobreakspace\the\thm@notefont\sffamily\bfseries---\nobreakspace#3.}}

\newtheoremstyle{ocrenum}
{5pt}
{5pt}
{\normalfont}
{}
{\small\bf\sffamily\color{ocre}}
{\;}
{0.25em}
{\small\sffamily\color{ocre}\thmname{#1}\nobreakspace\thmnumber{\@ifnotempty{#1}{}\@upn{#2}}
\thmnote{\nobreakspace\the\thm@notefont\sffamily\bfseries\color{black}---\nobreakspace#3.}} 
\makeatother

\newcounter{dummy} 
\numberwithin{dummy}{section}
\theoremstyle{ocrenumbox}
\newtheorem{theoremeT}[dummy]{Theorem}

\newtheorem{exerciseT}{Exercise}[chapter]
\theoremstyle{blacknumex}
\newtheorem{exampleT}{Example}[chapter]
\theoremstyle{blacknumbox}

\newtheorem{definitionT}{Definition}[section]
\newtheorem{corollaryT}[dummy]{Corollary}
\theoremstyle{ocrenum}

\RequirePackage[framemethod=default]{mdframed} 

\newmdenv[skipabove=7pt,
skipbelow=7pt,
backgroundcolor=black!5,
linecolor=ocre,
innerleftmargin=5pt,
innerrightmargin=5pt,
innertopmargin=5pt,
leftmargin=0cm,
rightmargin=0cm,
innerbottommargin=5pt]{tBox}

\newmdenv[skipabove=7pt,
skipbelow=7pt,
rightline=false,
leftline=true,
topline=false,
bottomline=false,
backgroundcolor=ocre!10,
linecolor=ocre,
innerleftmargin=5pt,
innerrightmargin=5pt,
innertopmargin=5pt,
innerbottommargin=5pt,
leftmargin=0cm,
rightmargin=0cm,
linewidth=4pt]{eBox}	

\newmdenv[skipabove=7pt,
skipbelow=7pt,
rightline=false,
leftline=true,
topline=false,
bottomline=false,
linecolor=ocre,
innerleftmargin=5pt,
innerrightmargin=5pt,
innertopmargin=0pt,
leftmargin=0cm,
rightmargin=0cm,
linewidth=4pt,
innerbottommargin=0pt]{dBox}	

\newmdenv[skipabove=7pt,
skipbelow=7pt,
rightline=false,
leftline=true,
topline=false,
bottomline=false,
linecolor=gray,
backgroundcolor=black!5,
innerleftmargin=5pt,
innerrightmargin=5pt,
innertopmargin=5pt,
leftmargin=0cm,
rightmargin=0cm,
linewidth=4pt,
innerbottommargin=5pt]{cBox}

\makeatletter
\renewcommand{\@seccntformat}[1]{\llap{\textcolor{ocre}{\csname the#1\endcsname}\hspace{1em}}}                    
\renewcommand{\section}{\@startsection{section}{1}{\z@}
{-4ex \@plus -1ex \@minus -.4ex}
{1ex \@plus.2ex }
{\normalfont\large\sffamily\bfseries}}
\renewcommand{\subsection}{\@startsection {subsection}{2}{\z@}
{-3ex \@plus -0.1ex \@minus -.4ex}
{0.5ex \@plus.2ex }
{\normalfont\sffamily\bfseries}}
\renewcommand{\subsubsection}{\@startsection {subsubsection}{3}{\z@}
{-2ex \@plus -0.1ex \@minus -.2ex}
{.2ex \@plus.2ex }
{\normalfont\small\sffamily\bfseries}}                        
\renewcommand\paragraph{\@startsection{paragraph}{4}{\z@}
{-2ex \@plus-.2ex \@minus .2ex}
{.1ex}
{\normalfont\small\sffamily\bfseries}}

\usepackage{hyperref}
\hypersetup{hidelinks,backref=true,pagebackref=true,hyperindex=true,colorlinks=false,breaklinks=true,urlcolor= ocre,bookmarks=true,bookmarksopen=false,pdftitle={Title},pdfauthor={Author}}

\newcommand{\thechapterimage}{}
\newcommand{\chapterimage}[1]{\renewcommand{\thechapterimage}{#1}}

\def\thechapter{\arabic{chapter}}
\def\@makechapterhead#1{
\thispagestyle{empty}
{\centering \normalfont\sffamily
\ifnum \c@secnumdepth >\m@ne
\if@mainmatter
\startcontents
\begin{tikzpicture}[remember picture,overlay]
\node at (current page.north west)
{\begin{tikzpicture}[remember picture,overlay]
\node[anchor=north west,inner sep=0pt] at (0,0) {\includegraphics[width=\paperwidth]{\thechapterimage}};
\draw[anchor=west] (5cm,-9cm) node [rounded corners=20pt,fill=ocre!10!white,text opacity=1,draw=ocre,draw opacity=1,line width=1.5pt,fill opacity=.6,inner sep=12pt]{\huge\sffamily\bfseries\textcolor{black}{\thechapter. #1\strut\makebox[22cm]{}}};
\end{tikzpicture}};
\end{tikzpicture}}
\par\vspace*{230\p@}
\fi
\fi}

\def\@makeschapterhead#1{
\thispagestyle{empty}
{\centering \normalfont\sffamily
\ifnum \c@secnumdepth >\m@ne
\if@mainmatter
\begin{tikzpicture}[remember picture,overlay]
\node at (current page.north west)
{\begin{tikzpicture}[remember picture,overlay]
\node[anchor=north west,inner sep=0pt] at (0,0) {\includegraphics[width=\paperwidth]{\thechapterimage}};
\draw[anchor=west] (5cm,-9cm) node [rounded corners=20pt,fill=ocre!10!white,fill opacity=.6,inner sep=12pt,text opacity=1,draw=ocre,draw opacity=1,line width=1.5pt]{\huge\sffamily\bfseries\textcolor{black}{#1\strut\makebox[22cm]{}}};
\end{tikzpicture}};
\end{tikzpicture}}
\par\vspace*{230\p@}
\fi
\fi
}
\makeatother 

\begin{document}
\title{Zoo of Centralities}


\begingroup
\thispagestyle{empty}
\AddToShipoutPicture*{\put(0,0){\includegraphics[scale=0.4385]{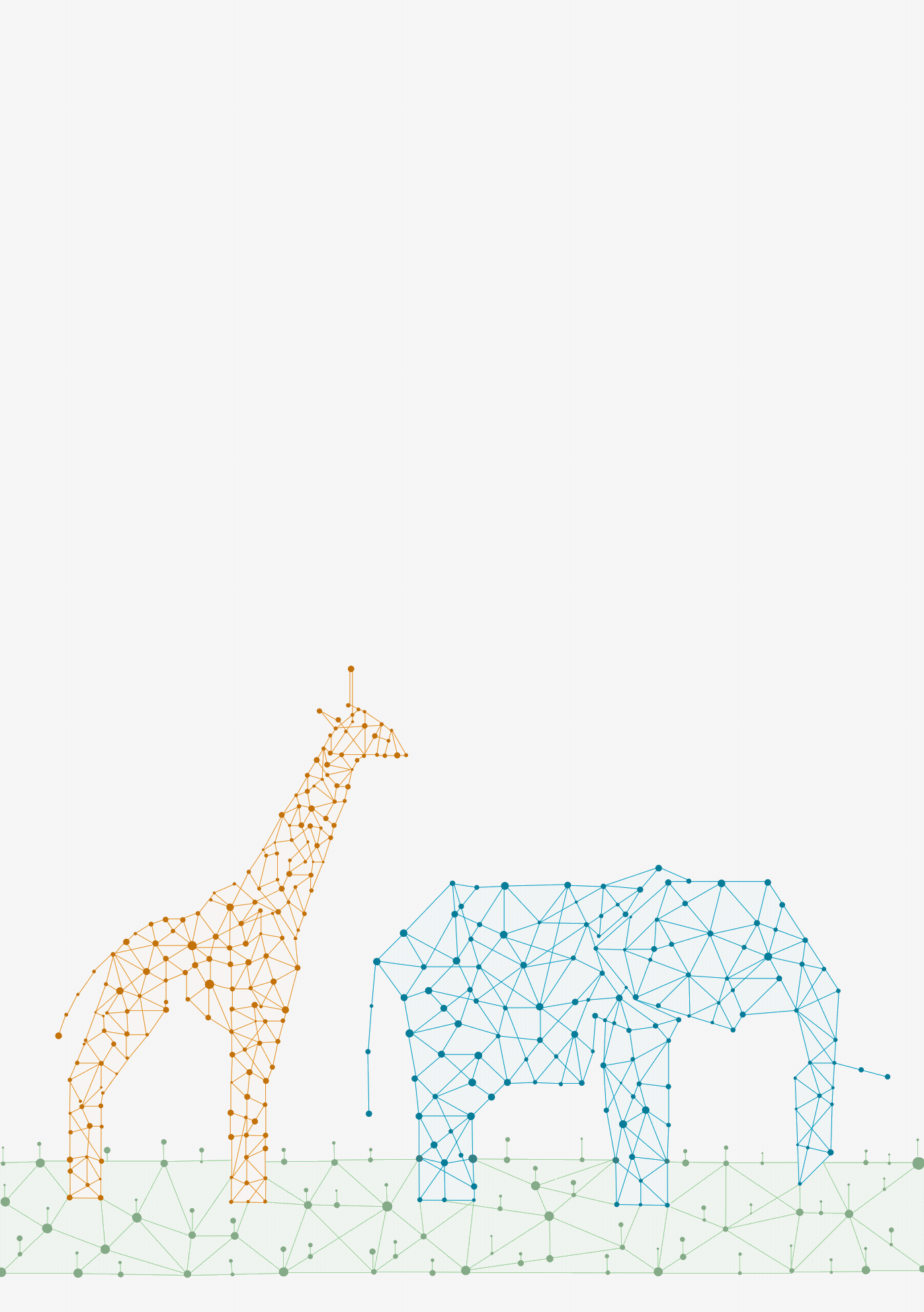}}}
\centering
\vspace*{3.0cm}
\par\normalfont\fontsize{35}{35}\sffamily\selectfont
\textbf{Zoo of Centralities:}\\
\vspace{-0.2cm}\hspace{-1cm}{\LARGE Encyclopedia of Node Metrics in Complex Networks}\par 
\vspace*{0.4cm}
{\huge by Sergey Shvydun}\par 
\endgroup

\newpage
\thispagestyle{empty}

\noindent \textsc{Zoo of Centralities: Encyclopedia of Node Metrics in Complex Networks}\\

\noindent \textsc{by Sergey Shvydun}  \orcidlink{0000-0002-6031-8614}\\

\noindent \href{https://centralityzoo.github.io/}{\textsc{https://centralityzoo.github.io/}} \\

\noindent \textsc{Contact: s.shvydun@tudelft.nl} \\

\noindent \textit{First published in November 2025} 

~\vfill

\chapterimage{head}

\pagestyle{empty} 

\tableofcontents 

\newpage
\chapterimage{head}
\textbf{$ $}
\vspace{1cm}
\section*{\Huge Preface } 
\addcontentsline{toc}{starsection}{\large Preface}

\vspace{0.5cm}

Network science provides analytical tools for studying complex systems such as social, transportation, biological, and financial networks by representing them as interconnected structures. The concept of \emph{centrality} is used to identify the most important or influential elements in a network. These elements may be considered central because they possess many connections, occupy structurally critical positions or exert significant influence over the flow of information, resources or dynamics within the system. For example, centrality can highlight the most connected individuals in a social network, critical hubs in a transportation network, essential proteins in a biological network, institutions whose failure could trigger systemic risks in a financial network or key individuals who accelerate the spread of infectious diseases.

\begin{figure}[h]
    \centering
    \includegraphics[width=0.9\textwidth]{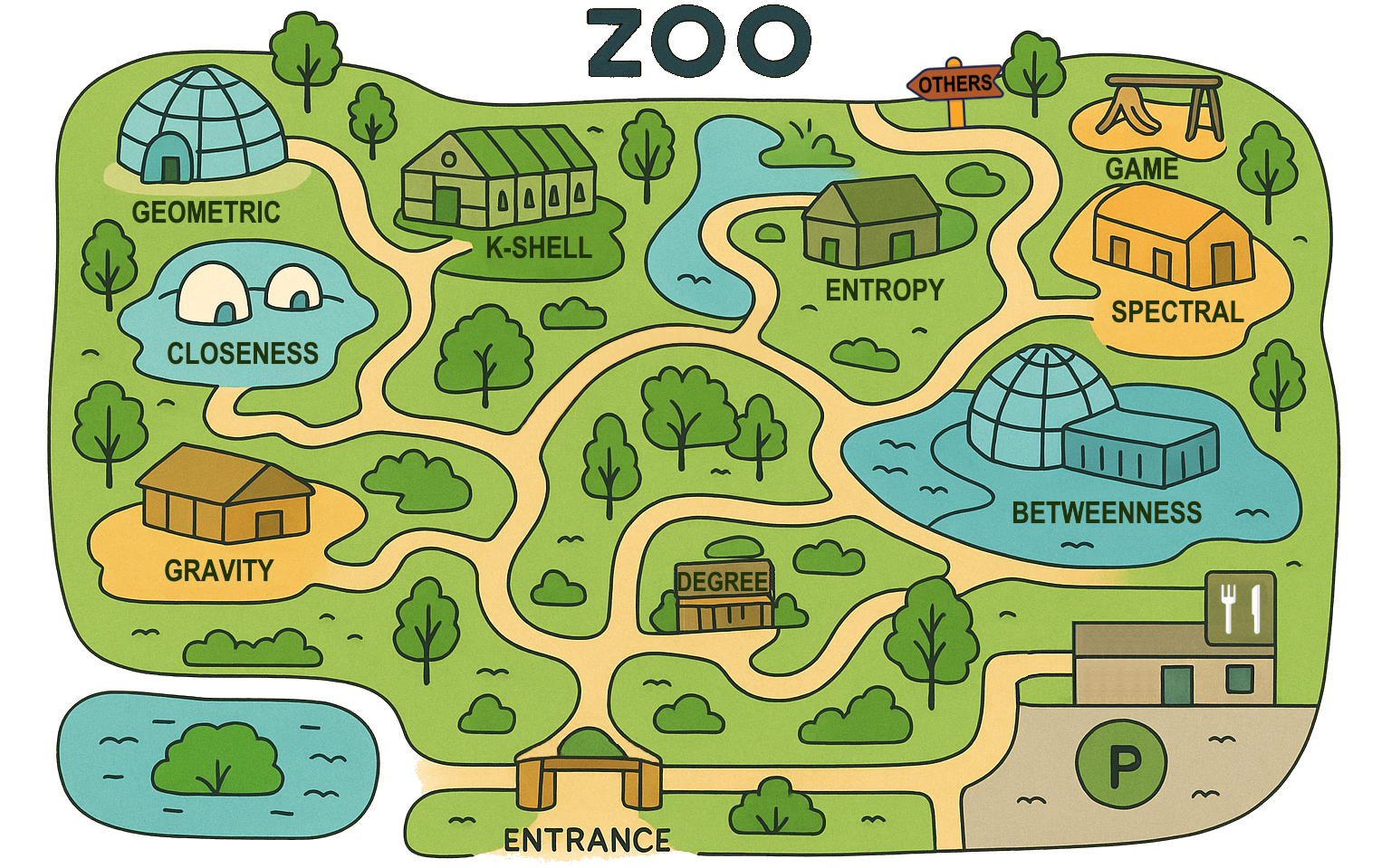}
\vspace{-0.3cm}\caption{Zoo of Centralities.}
    \label{fig:zoo_picture}
\end{figure}

Despite its wide-ranging applications, there is no universally accepted definition of node centrality. Over the years, researchers have introduced \textbf{hundreds or even thousands} of distinct centrality measures and other node metrics. Although centrality has been studied for decades, the field remains active, with numerous new measures published each year and frequently discussed at scientific conferences. Today, centrality measures are so numerous and diverse that they resemble a \emph{zoo}, with each representing a distinct "species" exhibiting its own unique characteristics and behaviors (see Figure~\ref{fig:zoo_picture}). For this reason, this work takes the title \textbf{Zoo of Centralities}, reflecting the diversity and uniqueness of the measures it presents.

\noindent However, the rapid growth in the number of proposed measures has given rise to several \textbf{challenges}:

\begin{itemize}
    \item\emph{Discoverability:} many measures are difficult to find in the literature or are not widely known outside specialized communities. Comprehensive surveys usually cover only 50-70 measures, leaving most of them largely overlooked.

\item\emph{Redundancy:} some centrality measures are identical, introduced independently under different names, while others are mathematically equivalent and yield the same node rankings (often up to a monotonic transformation), leading to unnecessary duplication.

\item\emph{Naming conflicts:} different measures sometimes share the same name, leading to confusion.

\item\emph{Validation issues:} new measures, frequently developed for specific applications (e.g., influence maximization or network dismantling), are commonly evaluated only against traditional metrics, with limited systematic benchmarking on real-world networks.

\item\emph{Accessibility:}  many measures are not readily available in open-source libraries, which makes their practical use difficult.
\end{itemize}

The aim of this work is to bring structure to the complex landscape of centrality, thereby addressing the challenges arising from the rapidly growing number of measures. The book presents over 400 centrality measures and node metrics, each accompanied by a concise description and references to the original sources.  Although the list of measures is not exhaustive, it represents the most comprehensive compilation of distinct centrality measures to date, covering a substantial portion of those most frequently employed in network science research.

This book may be valuable to several groups of readers:  

\begin{itemize}
    \item \textbf{Practitioners and applied researchers}, who can identify centrality measures relevant to their applications, where the interpretation of the most central nodes in a network corresponds to the notion of importance in their domain.

    \item \textbf{Researchers developing new node metrics}, who can use this compilation to verify that their model is novel, ensure that the measure name is not already in use and benchmark their approach against the state-of-the-art or the most relevant existing metrics.

    \item \textbf{Other researchers}, who can draw inspiration from this diverse collection of measures, applying original ideas from existing metrics to their own domain and using them to guide future research.

\end{itemize}

We invite collaborators and readers to contribute: if you are aware of, or have published, a new centrality measure that you would like to see included in this list, please let us know. The list will be \textbf{regularly updated} to reflect ongoing developments. We also encourage readers to explore and contribute to the \href{https://centralityzoo.github.io/}{\textbf{Centrality Zoo website}} at \href{https://centralityzoo.github.io/}{\textsc{https://centralityzoo.github.io/}}, which provides an \textbf{interactive platform} for discovering, comparing and implementing centrality measures.

I am deeply grateful to my supervisors, Professor Dr. Fuad Aleskerov and Professor Dr. Ir. Piet Van Mieghem, for their guidance, support and encouragement throughout my academic journey. I would like to sincerely thank Natalia Meshcheryakova for her support of this work and for her helpful suggestions, as well as my colleagues and collaborators for their valuable feedback during personal discussions and previous collaborations on related work. Finally, I am grateful to the colleagues with whom I had the pleasure of discussing this work at scientific events, whose insights and encouragement have been invaluable in shaping this resource and sustaining me throughout.

\vspace{1cm}

\noindent
November 2025 \hfill Sergey Shvydun

\pagestyle{fancy}

\chapter{Centrality in Complex Networks}

\section{Introduction}

Many real-world systems can be naturally described as networks, with nodes representing entities and links capturing their interactions. Examples span diverse domains: social networks represent individuals and their social ties; transportation networks model airports, stations or roads and their connections; biological networks capture proteins, genes or metabolites and their interactions; computer networks represent devices and the connections between them; information networks describe web pages linked through hyperlinks; financial networks represent banks or financial institutions
connected through lending, borrowing or other financial relationships. Representing complex systems as networks provides a unified framework for studying their structural organization and the dynamics of processes unfolding on them, enabling both qualitative insights and quantitative analysis.

\begin{figure}[h]
    \centering    \includegraphics[width=0.8\textwidth]{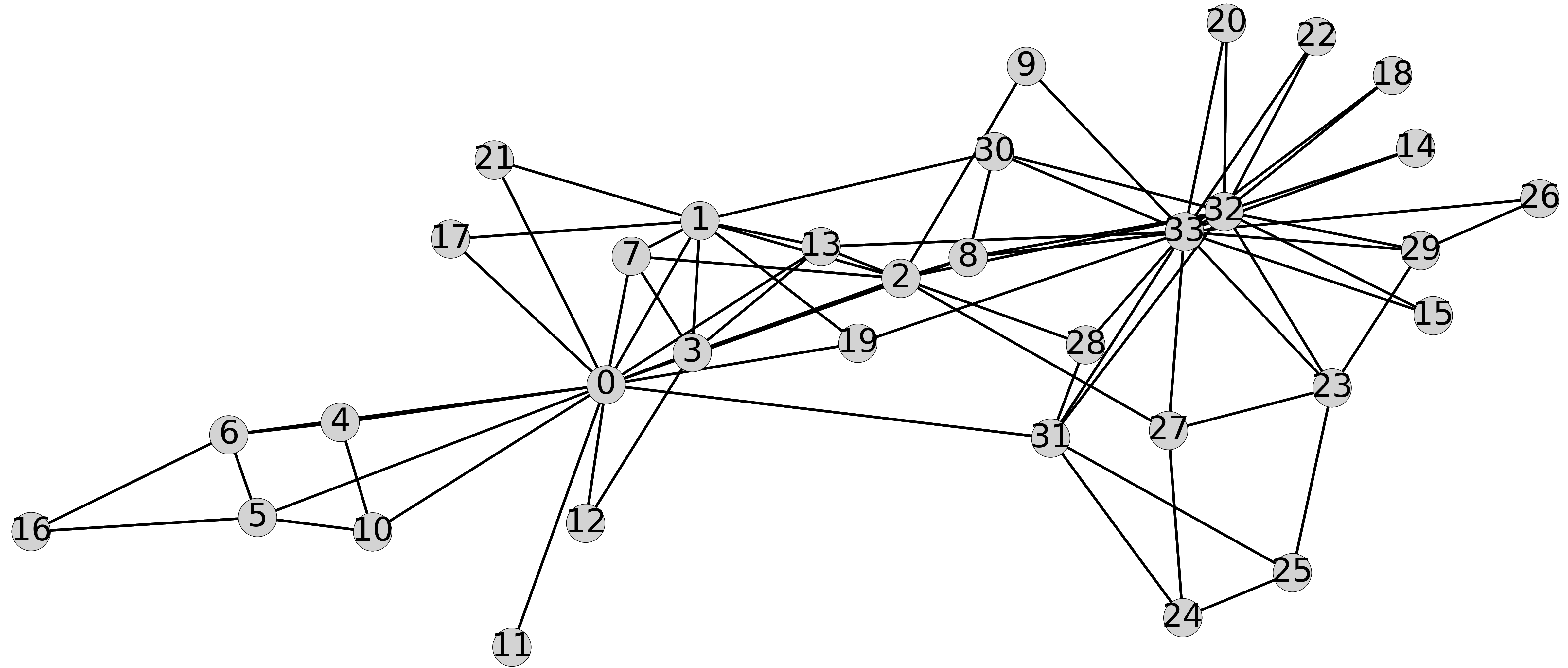}
    \vspace{-0.3cm}\caption{Zachary’s karate club network \cite{Zachary1977}.}
    \label{fig:karate}
\end{figure}

A fundamental challenge in network science is characterizing the roles of nodes to identify those that are most \emph{important} or \emph{influential} within a network. Figure \ref{fig:karate} illustrates one of the most well-known examples, the Zachary karate club network, where nodes represent members and edges represent social interactions between the club members outside the club. The concept of \emph{centrality}\index{centrality} addresses this problem by assigning a numerical value to each node that quantifies its importance or influence based on the network’s topology. Centrality has a wide range of practical applications. In epidemiology, nodes with high centrality may
correspond to super-spreaders, whose vaccination or isolation could control the spread of disease. In infrastructure and transportation networks, highly central nodes often correspond to critical components whose failure can fragment the system or create bottlenecks that disrupt the flow of traffic or resources. In social networks, centrality can highlight key influencers whose opinions propagate widely. In financial networks, highly central banks or institutions may pose systemic risks, as their distress can propagate contagiously, potentially destabilizing the entire financial system. Beyond these applications, centrality measures are increasingly utilized as node features in machine learning tasks on graphs, such as node or graph classification or link prediction.

\begin{figure}[h]
    \centering
    \includegraphics[width=0.9\textwidth]{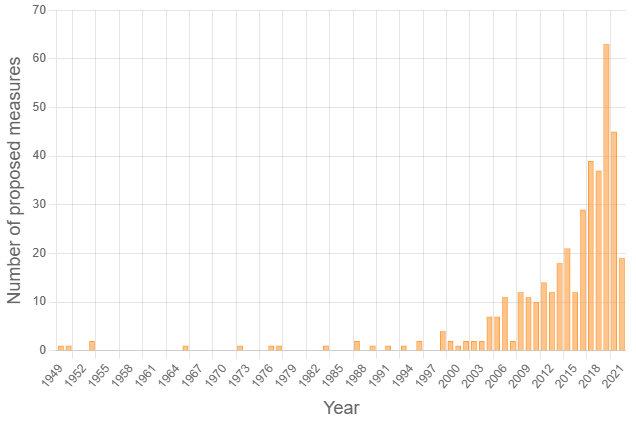}
\vspace{-0.3cm}\caption{Number of centrality measures proposed per year, showing an increasing trend over time. Source: Centiserver \cite{Mahdi2024} (data not updated after 2021).}
    \label{fig:evolution_metrics}
\end{figure}

The concept of node centrality has a rich history. Early studies in the 1950s by Bavelas \cite{Bavelas1948,Bavelas1950} and Leavitt \cite{Leavitt1951} introduced centrality to analyse communication networks, highlighting the strategic position of certain nodes in information exchange. Freeman’s work in the 1970s formalized several widely used centrality measures, including degree, closeness, and betweenness centralities \cite{Freeman1978}, while Bonacich later introduced eigenvector centrality to quantify a node’s influence in relation to the importance of its neighbors \cite{Bonacich1987}. Over time, the field has experienced a surge of new metrics, motivated by both theoretical advances and practical applications. Today, hundreds of centrality measures exist, with more than forty new measures introduced each year, reflecting the increasing interest and complexity of the field, as shown in Fig.~\ref{fig:evolution_metrics}.

Despite these advances, defining centrality remains an inherently ill-posed problem. There is no single measure that universally captures node importance, as the \emph{importance} of a node depends critically on both the network’s structure and the specific process under consideration. Consequently, each centrality measure can be seen as a distinct model that emphasizes a particular aspect of importance, reflecting the diversity of dynamical processes and structural aspects of networks that researchers may wish to capture. However, the proliferation of centrality measures has introduced several challenges. First, \emph{discoverability}: the large number of measures makes it difficult to identify all relevant ones. Second, \emph{redundancy}: many measures are reinvented under different names or highly correlated, providing overlapping information. Third, \emph{naming conflicts}: different measures are often labelled with the same or similar names. Fourth, \emph{validation issues}: only a limited number of centrality measures have been systematically validated across diverse network types and against rigorous benchmarks. In particular, new measures are often compared with only a small subset of existing metrics, typically a few classical measures, leaving their broader applicability and robustness untested. Finally, \emph{accessibility}: implementations of many measures are scattered or unavailable.

Over the years, there have been multiple attempts to organize and systematize the field of centrality, which we must acknowledge in this work. Foundational studies established the core conceptual principles of the field, while subsequent research expanded its scope through critical, process-oriented and systematic comparative analyses \cite{Freeman1978,Borgatti2005,Landherr2010,Bringmann2019}. Another line of work consists of comprehensive reviews and book chapters that summarize and classify the most widely used centrality measures \cite{Brandes2005,Das2018,Hernández2011,Lü2016,Maji2020c,Newman2018,Oldham2019,Rusinowska2011,Saxena2020,Wan2021}. Although these surveys provide valuable insights into specific measures, they generally analyze only 15-50 metrics, limiting their ability to represent the full diversity and complexity of existing models. The next line of research focuses on \emph{axiomatic analysis}, aiming to define formal properties that centrality measures should satisfy \cite{Sabidussi1966, Landherr2010, Myerson, Myerson2, Boldi, Skibski, Skibski2}. This research direction encounters two main challenges: (i) the set of axioms is not standardized and differs across studies, and (ii) testing many measures against these axioms can be computationally intensive or analytically intractable, leaving most measures largely unexplored. Other studies assess the practical performance of centrality measures in specific tasks, such as network dismantling or influence maximization \cite{Artime2024,Banerjee2020,Wan2021,Wandelt2018}, while some investigate their robustness to noise and structural perturbations, examining how sensitive rankings are to changes in the network \cite{Costenbader2003,Borgatti2006b,Frantz2009,MS2023,MS2024,Shvydun2025}. These analyses are often limited to a subset of measures due to discoverability and accessibility constraints. To facilitate exploration, several online resources, including \emph{Centiserver} \cite{Jalili2015,Mahdi2024}, the \emph{Periodic Table of Network Centrality} \cite{Schoch2024} and \emph{Axioms4Centralities} \cite{Skibski3}, were developed to catalog centrality measures and provide links to original publications or brief model descriptions. However, these resources are no longer actively maintained or updated.

In this work, we address the challenges of centrality research by providing a comprehensive overview of centrality measures, presenting detailed descriptions of a wide range of existing models in a single, consolidated manuscript to allow readers efficient access to this information. As part of this effort, we propose a unified taxonomy that standardizes the names of measures, helping to reduce naming conflicts and redundancy. To complement this, we offer the \href{https://centralityzoo.github.io/}{\textbf{Centrality Zoo website}} that lists the measures and presents comparative insights across different models, facilitating discoverability, reducing redundancy and supporting the selection of appropriate metrics in practical applications.

The remainder of this work is organized as follows. Section \ref{sec:notation} introduces the notation and provides a list of symbols used throughout the paper. In Section \ref{sec:classification}, centrality measures are grouped according to their fundamental principles, offering a structured way to compare and understand different approaches. Section \ref{sec:encyclopedia} provides detailed descriptions of each measure, including their definitions, theoretical foundations and key properties. For the reader’s convenience, a comprehensive index is provided at the end of the book.

\section{Notation}\label{sec:notation}

A complex network can be represented by a \emph{graph}\index{graph}, denoted by $G = (\mathcal{N}, \mathcal{L})$, which consists of a set of $N$ nodes (or vertices) connected by a set of $L$ links (or edges). Here, $\mathcal{N}$ denotes the set of nodes, and $\mathcal{L}$ denotes the set of links. Without loss of generality, we assume that the node set is indexed as
\[
\mathcal{N} = \{1,2,\dots,N\}.
\]

The graph \(G\) is described by an \(N {\times} N\) \emph{adjacency matrix}\index{adjacency matrix} \(A\) whose elements \(a_{ij}\) are either one or zero depending on whether node $i$ is connected to node $j$ or not, i.e.,
\[
\begin{aligned}
a_{ij} =
\begin{cases}
  1 & \text{if there is a link from $i$ to $j$},\\
  0 & \text{otherwise.}
\end{cases}
\end{aligned}
\]

If the graph is undirected, then the adjacency matrix\index{matrix!adjacency} \(A = A^{T}\) is a real symmetric matrix. Additionally, the graph $G$ can be described by a non-negative weight matrix $W$, where each element $w_{ij}$ represents the intensity of a link between nodes $i$ and $j$. Unless stated otherwise, we assume that the graph is undirected and unweighted. We also assume that the graph has no self-edges, that is, $(i,i) \notin \mathcal{L}$, or equivalently, $a_{ii} = 0$ for all $i \in \mathcal{N}$.

For a node $i \in \mathcal{N}$, we define the set of its \emph{neighbors}\index{neighbors!direct (1-hop)} as
\[
\mathcal{N}(i) = \{ j \in \mathcal{N} \mid (i,j) \in \mathcal{L} \} = \{ j \in \mathcal{N} \mid a_{ij} = a_{ji} = 1 \}.
\]

For directed networks, the set of out-neighbors of a node \(i\) is
\[
\mathcal{N}^{out}(i) = \{ j \in \mathcal{N} \mid a_{ij} = 1 \},
\]
and the set of in-neighbors of a node \(i\) is
\[
\mathcal{N}^{in}(i) = \{ j \in \mathcal{N} \mid a_{ji} = 1 \}.
\]

The \emph{degree}\index{degree} of node \(i\) is defined as the number of its neighbors:
\[
d_i = |\mathcal{N}(i)| = \sum_{j \in \mathcal{N}} a_{ij}.
\]

In directed networks, one distinguishes between the \emph{in-degree} of node $i$,
\[
d_i^{in} = \sum_{j \in \mathcal{N}} a_{ji},
\]
which counts the number of links directed \emph{to} $i$, and the \emph{out-degree} of node $i$,
\[
d_i^{out} = \sum_{j \in \mathcal{N}} a_{ij},
\]
which counts the number of links directed \emph{from} $i$ to other nodes.

A \emph{walk}\index{walk} of length $k \geq 0$ from node $i$ to node $j$ is a succession of $k$ links
\[
(i_0, i_1), (i_1,i_2), \dots, (i_{k-1}, i_k),
\]
where $i_0 = i$ and $i_k = j$. The number of walks between nodes can also be expressed using the adjacency matrix. 
Specifically, the $(i,j)$-th element of the $k$-th power $A^k$ of the adjacency matrix counts the number of walks of length $k$ from $i$ to $j$.

A \emph{path}\index{path} is a walk in which all nodes are distinct. In some literature, walks are referred to as paths, while paths are referred to as \emph{simple paths}\index{path!simple} or \emph{self-avoiding walks}\index{walk!self-avoiding}. 

A \emph{shortest path}\index{path!shortest} between two nodes $i$ and $j$ is a path connecting them that has minimal length among all paths between $i$ and $j$. The distance of the shortest path between nodes $i$ and $j$, also called the \emph{geodesic distance}\index{distance!shortest-path}, is denoted by $d_{ij}$, with the convention $d_{ij} = \infty$ if no path exists between $i$ and $j$. The \emph{diameter}\index{diameter} of the network is defined as the length of the longest shortest-path between any pair of nodes in $G$.

For \(k \ge 1\), we define the set of \emph{\(k\)-hop neighbors}\index{neighbors!\textit{k}-hop} of node \(i\) as
\[
\mathcal{N}^{(k)}(i) = \{ j \in \mathcal{N} : d_{ij} = k \},
\]
so that \(\mathcal{N}^{(1)}(i) = \mathcal{N}(i)\). In terms of the adjacency matrix, $j \in \mathcal{N}^{(k)}(i) $ if and only if $(A^k)_{ij} > 0$ and $(A^r)_{ij} = 0$ for all positive integers $r<k$.

The set of nodes within the $k$-hop neighborhood\index{neighbors!${\leq}k$-hop} of node $i$ is defined as
\[
N^{(\le k)}(i) = \{ j \in \mathcal{N} : d_{ij} \le k \} =N^{(1)}(i) \cup N^{(2)}(i)\cup ... \cup N^{(k)}(i),
\]
i.e., all nodes that can be reached from $i$ in at most $k$ steps. In terms of the adjacency matrix, $j \in \mathcal{N}^{(k)}(i) $ if and only if $\sum_{r=1}^k(A^r)_{ij}>0$.

Finally, a \emph{centrality}\index{centrality} is any mapping
\[
C: \mathcal{N} \to \mathbb{R}
\]
assigning to each node a real number representing its structural importance. In many contexts, it is assumed that \(C(i) \ge 0\) and that larger values correspond to more central or influential nodes.

Only when explicitly mentioned, will we deviate from the standard notation and symbols outlined in Table \ref{tab:notions}.

\begin{table}[h!]
  \centering
    \begin{tabular}{ l l }
    \hline
    
    \textbf{Name} & \textbf{Description}\\
    \hline
    
    \(G\) & graph \\
    \(G_i\) & subgraph of \(G\) obtained by removing node \(i\) from graph $G$\\
    \(\mathcal{N}\) & set of nodes in graph $G$\\
    \(N\) & number of nodes in graph $G$, \(N = |\mathcal{N|}\) \\
    \(\mathcal{L}\) & set of links in graph \(G\) \\
    \(L\) & number of links in graph \(G\), \(L = |\mathcal{L}|\) \\
    \(A\) & adjacency matrix of graph $G$ \\
    \(a_{ij}\) & element in the $i$-th row and $j$-th column of the adjacency matrix $A$\\
    $d_i $ & degree of node $i$\\
    $d_{ij} $ & shortest-path distance between nodes $i$ and $j$\\
    $\mathcal{N}(i)$ & neighbours of node $i$\\
    $\mathcal{N}^{(k)}(i)$ & \textit{k}-hop of node \(i\)\\
    $\mathcal{N}^{(\leq k)}(i)$ &  set of nodes within the $k$-hop neighbourhood of node \(i\)\\
    \hline
  \end{tabular}
  \caption{List of symbols and their definitions.}
  \label{tab:notions}
\end{table}

\section{Classification of centrality measures}\label{sec:classification}

Centrality measures quantify the structural importance of nodes within a network, capturing different aspects of what it means for a node to be \emph{central}. Some measures rely on the number of connections of each node, reflecting immediate influence, while others assess the position of a node within the network through distance to other nodes or involvement in paths connecting node pairs. Many measures integrate concepts from diverse disciplines, including information theory, cooperative game theory, voting theory, Dempster-Shafer evidence theory, multi-criteria decision-making, signal processing, physics, biology and geometry. As individual measures often combine multiple perspectives, no universal classification exists and the same measure may be categorized differently depending on the criteria applied.

In the literature, various approaches have been proposed to classify centrality measures. Freeman \cite{Freeman1978} classified centrality measures into three groups. \emph{Degree centralities} quantify the number of direct connections a node has and reflect its potential for communication activity. \emph{Betweenness centralities} measure how frequently a node lies on shortest or geodesic paths connecting other node pairs, capturing its potential to control information flow within the network. \emph{Closeness centralities} assess the distance of a node to all other nodes, indicating node independence or efficiency in interacting with the network and avoiding control by others.

Borgatti and Everett \cite{Borgatti2006} classify centrality measures along two dimensions: the \emph{type of nodal involvement} (radial versus medial) and the \emph{property of walks assessed} (walk versus length). Measures based on the number of connections or distance to other nodes, such as degree-like and closeness-like measures, are termed \emph{radial} measures. In contrast, measures that quantify how a node links other nodes, such as betweenness-like measures, are referred to as \emph{medial} measures. The second dimension consists of \emph{volume} measures, which count walks involving a node and \emph{length} measures, which evaluate the lengths of those walks. Cross-classifying centrality measures along these two dimensions yields four categories: radial-volume, radial-length, medial-volume and medial-length measures.

Boldi and Vigna\cite{Boldi} classify centrality measures into three categories: \emph{geometric measures} (e.g., degree and closeness centralities), \emph{spectral measures} (e.g., eigenvector centrality) and \emph{path-based measures} (e.g., betweenness centrality). Boldi and Vigna note that spectral measures admit a walk-based interpretation, since summations or powers of the adjacency matrix $A$ aggregate contributions from walks of varying lengths in the network.

Koschützki \textit{et al.} \cite{Koschützki2005} introduces a four-dimensional framework for characterizing centrality measures. The \emph{basic term} defines the function on which centrality is based and can take one of four forms: (1) \emph{reachability}, where a node is central if it can reach many others; (2) \emph{flow}, based on the amount of flow passing through a node or edge; (3) \emph{vitality}, which measures the impact of removing a node on a network function; (4) \emph{feedback}, where a node's centrality depends recursively on the centrality of other nodes. The \emph{term operator} specifies how the basic term is aggregated, \emph{personalization} allows weighting of nodes, edges or subsets, and \emph{normalization} scales the resulting values for comparability. By specifying choices along these four dimensions, any centrality measure can be systematically defined or decomposed into its functional components.

Bloch \textit{et al.} \cite{Bloch2023} classify centrality measures along two key dimensions: (i) the type of information used about nodes’ positions, referred to as \emph{nodal statistics}, and (ii) the manner in which this information is weighted as a function of distance from the focal node, referred to as \emph{weighting}. They identify three principal types of nodal statistics: the neighborhood (path) statistic, the walk statistic, and the intermediary (geodesic) statistic. Each of these can be combined with different weighting schemes, including immediate, extended or infinite weighting.

Saxena and Iyengar \cite{Saxena2020} categorize centrality measures as \emph{local} or \emph{global}. Local centrality measures, such as degree and semi-local centralities, can be computed using information from the immediate neighborhood of a node. Global centrality measures, including closeness, betweenness, eigenvector and PageRank centralities, require knowledge of the entire network structure and typically involve higher computational complexity. Alternative classification schemes extend the local-global distinction by introducing additional categories. One scheme in \cite{Ahajjam2018} divides centrality measures into \emph{local}, which rely on the immediate neighborhood of a node, \emph{global}, which require information about the entire network, and \emph{random-walk} metrics, which evaluate node influence based on random-walk dynamics. Another scheme in \cite{Chen2012} further subdivides local measures into \emph{local}, which use only direct neighbors of a node, and \emph{semi-local}, which incorporate information from nodes within an $r$-hop neighborhood (typically $r{=}2$).

In the following subsections, we classify centrality measures according to the fundamental ideas underlying these models. This classification aims to simplify the selection of appropriate measures and to illustrate the different perspectives on centrality. However, these categories are not mutually exclusive: a single measure may reflect multiple underlying concepts and, therefore, could reasonably be placed in more than one category. Moreover, alternative classification schemes can be developed depending on the theoretical perspective or criteria applied.

\subsection{Betweenness centrality measures}

Betweenness centrality measures quantify the extent to which a node lies on paths connecting other nodes in a network. Conceptually, these measures capture the \emph{mediating role} of a node and its \emph{control} or \emph{brokerage} potential: a node is considered more central if it frequently lies on paths between other nodes, potentially influencing the flow of information, resources or traffic through the network.

The mediating role of a node can be quantified in different ways depending on the underlying dynamical process, particularly on how information flows through the network. Depending on the process considered, information may propagate along \emph{shortest paths}, \emph{approximate shortest paths}, \emph{effective distances}, \emph{current-flow paths} (analogous to electrical networks) or \emph{all acyclic paths} between nodes. The contribution of a node can also be weighted by the amount of flow passing through it, reflecting how frequently it is traversed under a given flow process between pairs of nodes. Additionally, the contribution of a node may depend on the distance to the source or target node. In some formulations, contributions from long paths are excluded, as such paths may not represent realistic routes for the process being modelled.

The examples of the betweenness centrality measures discussed in this work are presented below:
{\footnotesize
\vspace{-0.8cm}
\setlength{\columnsep}{0pt} 
\begin{multicols}{2} 
\begin{itemize}
    \item All cycle betweenness (ACC) centrality;
    \item Attentive betweenness centrality (ABC);
    \item Betweenness centrality;
    \item BottleNeck centrality;
    \item Bridging centrality;
    \item Communicability betweenness centrality;
    \item Counting betweenness centrality (CBET);
    \item Current-flow betweenness centrality;
    \item $\delta$-betweenness centrality;
    \item Egocentric betweenness centrality;
    \item Endpoint betweenness centrality;
    \item \(\epsilon\)-betweenness centrality;
    \item Flow betweenness centrality;
    \item Flow coefficient;
    \item \textit{k}-betweenness centrality;
    \item Length-scaled betweenness centrality (LSBC);
    \item Linearly scaled betweenness centrality;
    \item Load centrality;
    \item Percolation centrality;
    \item Proximal betweenness centrality;
    \item Randomized shortest paths betweenness centrality;
    \item Randomized shortest paths net betweenness centrality;
    \item Ranking-betweenness centrality;
    \item Resolvent betweenness (RB) centrality;
    \item Routing betweenness centrality (RBC);
    \item Stress centrality;
    \item Transportation centrality (TC);
    \item Two-way random walk betweenness (2RW).
\end{itemize}
\end{multicols}
}
\vspace{-0.3cm}A detailed description of these betweenness measures is provided in Section \ref{sec:encyclopedia}.

\subsection{Closeness centrality measures}

Closeness centrality measures aim to quantify the extent to which a node is \emph{close} to or effectively reachable from other nodes in a network. Conceptually, these measures reflect the \emph{accessibility} and \emph{reach} of a node: a node is considered more central if its position allows it to interact with, influence or efficiently reach other nodes. 

The notion of \emph{closeness}, usually assessed in terms of the distance from a node to other nodes, can be defined in multiple ways. For instance, it can be based on \emph{shortest paths}, which measure the minimum number of steps between nodes, \emph{random walks}, based on the expected number of steps required to reach one node from another, \emph{effective distance}, reflecting how easily information or influence spreads through the network, \emph{effective resistance}, which treats the network as an electrical circuit and accounts for all parallel paths, or \emph{simple paths}, in which each acyclic path contributes to the centrality of a node.  In some formulations, the contribution of each path to the centrality of a node depends on the distance to the target node, with more distant nodes contributing less. Some closeness-like measures further focus on a \emph{restricted neighborhood}, effectively ignoring nodes beyond a certain radius or assigning them negligible influence. In addition, some measures define closeness in terms of the \emph{maximum distance} to other nodes, rather than aggregating all distances, so that the centrality reflects the farthest distance from a node in the network.

The examples of closeness centrality measures discussed in this work are presented below:
\noindent{\footnotesize
\vspace{-0.3cm}
\setlength{\columnsep}{0pt}
\begin{multicols}{2} 
\begin{itemize}
    \item Access information;
    \item Adjusted Index of Centrality (AIC);
    \item Borda centrality;
    \item Centroid centrality;
    \item Closeness centrality;
    \item Closeness vitality;
    \item Copeland centrality;
    \item Correlation centrality (CoC);
    \item Current-flow closeness centrality;
    \item Decay centrality;
    \item Decaying degree centrality (DDC);
    \item $\delta$-closeness centrality;
    \item Eccentricity centrality;
    \item Effective distance closeness centrality (EDCC);
    \item Electrical closeness centrality;
    \item Gil-Schmidt power index;
    \item Gromov centrality;
    \item Harmonic centrality;
    \item Heatmap centrality;
    \item Hide information;
    \item Immediate Effects Centrality (IEC);
    \item Improved closeness centrality (ICC);
    \item Index of CENTRality (Icentr);
    \item Information distance index (IDI);
    \item Integration centrality;
    \item Lin's index;
    \item Markov centrality;
    \item Nieminen's closeness centrality;
    \item \textit{p}-means centrality;
    \item PathRank;
    \item Radiality centrality;
    \item Residual closeness centrality;
    \item Resistance curvature;
    \item Shortest cycle closeness (SCC) centrality;
    \item Weighted \textit{k}-short node-disjoint paths (WKPaths) centrality;
    \item Zeta vector centrality.
\end{itemize}
\end{multicols}
}

A detailed description of these closeness measures is provided in Section \ref{sec:encyclopedia}.

\vfill

\subsection{Path-based centrality measures}

Path-based centrality measures evaluate the importance of a node based on the number of paths connecting it to other nodes and the relative contribution of paths of different lengths. These measures usually consider multiple paths in a network, often incorporating all paths or all walks rather than restricting attention to a single geodesic path.The contribution of each path typically decreases with increasing length, with decay factors applied to mitigate the rapid growth in the number of long walks. In other formulations, only paths of particular lengths (such as fixed-length, odd or even paths) are included. Some measures instead focus on the number of nodes reachable within a given radius $k$, considering only which nodes can be reached while disregarding both the number and the lengths of the connecting paths.
The examples of the path-based centrality measures discussed in this work are presented below:

{\footnotesize
\vspace{-0.1cm}
\setlength{\columnsep}{0pt} 
\begin{multicols}{2} 
\begin{itemize}
    \item Bipartivity index; 
    \item Edge-disjoint \textit{k}-path centrality;
    \item Even subgraph centrality;
    \item Functional centrality;
    \item Generalized subgraph centrality (GSC);
    \item Geodesic \textit{k}-path centrality;
    \item Graph-theoretic power index (GPI);
    \item \textit{k}-path centrality;
    \item Katz centrality;
    \item Local reaching centrality;
    \item \textit{m}-reach centrality;
    \item Odd subgraph centrality;
    \item Path-transfer centrality;
    \item Subgraph centrality;
    \item Total communicability centrality (TCC);
    \item Total effects centrality (TEC);
    \item Vertex-disjoint \textit{k}-path.
\end{itemize}
\end{multicols}
}

A detailed description of these path-based measures is provided in Section \ref{sec:encyclopedia}.

\subsection{Spectral centrality measures}
Spectral centrality measures quantify node importance using the eigenvalues and eigenvectors of matrices representing the network, capturing global structural properties that reflect the entire network topology. The spectral measures discussed in this subsection primarily fall into two categories. The first category, eigenvector-based centralities, assigns centrality to a node based on the centrality of neighboring nodes, formalized through an eigenvector equation in which a node score is proportional to the sum of scores of connected nodes. The second category includes measures that rely on spectral features of network matrices, such as the adjacency matrix, the Laplacian or other matrices derived from the network structure, where centrality is defined directly in terms of the eigenvalues and eigenvectors of the matrix.

The examples of spectral centrality measures discussed in this work are presented below:
{\footnotesize
\vspace{-0.2cm}
\setlength{\columnsep}{0pt} 
\begin{multicols}{2} 
\begin{itemize}
    \item Contribution centrality;
    \item Eigenvector centrality;
    \item Fitness centrality;
    \item Fractional graph Fourier transform centrality;
    \item Generalized degree centrality;
    \item Graph Fourier Transform Centrality (GFT-C);
    \item HITS (Hubs and Authorities);
    \item Hubbel centrality;
    \item Modularity centrality;
    \item Modularity density centrality;
    \item Non-backtracking centrality;
    \item Normalized wide network ranking (NWRank);
    \item Principal component centrality (PCC);
    \item Semi-local iterative algorithm (semi-IA);
    \item Silent node rank (SNR);
    \item SpectralRank (SR);
    \item Topological centrality (TC);
    \item Trophic level centrality;
    \item Wide ranking (WRank);
    \item \textit{X}-non-backtracking (X-NB) centrality.
\end{itemize}
\end{multicols}
}

A detailed description of these measures is provided in Section \ref{sec:encyclopedia}.

\vfill

\subsection{Random-walk-based centrality measures}

Random-walk-based centrality measures quantify the importance of a node by analyzing the behavior of random walks as they traverse the network. The measures, which are considered in this subsection, can be grouped into two categories. The first category transforms the adjacency matrix into a stochastic (Markov) matrix, with centrality derived from the steady-state distribution, reflecting the long-term probability of visiting each node. The second category is based on finite-length random walks, where a walker moves from node to node according to transition probabilities, and centrality is computed using statistics such as the expected number of visits or first-passage times within a given number of steps. Both approaches formalize the idea that a node is central if it is frequently traversed under network random-walk dynamics. Variations of these measures incorporate features such as absorbing states, truncated walks, or decay factors that reduce the influence of longer walks. These modifications enable the centrality measure to capture specific dynamics or constraints on the flow of information or resources.

The examples of random-walk-based centrality measures discussed in this work are as follows:
{\footnotesize
\vspace{-0.25cm}
\setlength{\columnsep}{3pt} 
\begin{multicols}{2} 
\begin{itemize}[leftmargin=*, labelindent=1pt, itemsep=0.0em]
    \item Absorbing random-walk (ARW) centrality;
    \item Adaptive LeaderRank;
    \item ArticleRank;
    \item Biased random walk (BRW) centrality;
    \item DirichletRank;
    \item Eigentrust centrality;
    \item Improved iterative resource allocation (IIRA) method;
    \item Inward accessibility;
    \item Iterative resource allocation (IRA) method;
    \item LeaderRank;
    \item LineRank;
    \item Markov centrality;
    \item Markov entropy centrality;
    \item Outward accessibility;
    \item PageRank;
    \item Probabilistic-jumping random walk centrality;
    \item Random walk accessibility (RWA);
    \item Random walk decay (RWD) centrality;
    \item Random walk centrality (RWC);
    \item Rank centrality;
    \item Second-order centrality;
    \item Seeley’s index;
    \item Semi-local ranking centrality (SLC);
    \item Similarity-based PageRank;
    \item Stochastic Approach for Link Structure Analysis (SALSA);
    \item Trust-PageRank;
    \item ViralRank;
    \item Weighted LeaderRank.
\end{itemize}
\end{multicols}
}

A detailed description of these measures is provided in Section \ref{sec:encyclopedia}.

\subsection{Local and semi-local centrality measures}

Local and semi-local centrality measures evaluate node importance by focusing on a limited neighborhood rather than the entire network. Typically, these measures consider information from direct neighbors (local measures) or from nodes up to a fixed radius \(r\) (semi-local measures), capturing connectivity within a limited neighborhood. The centrality of a node can be defined in terms of its number of neighbors, the interconnectivity among them, or other structural properties of its local neighborhood. By focusing on a local region, these measures provide a balance between computational efficiency and the ability to capture meaningful structural information, without requiring full knowledge of the network. This makes local and semi-local measures particularly suitable for large networks where global computations may be impractical.

The examples of local/semi-local centrality measures discussed in this work are presented below:
{\footnotesize
\vspace{-0.7cm}
\setlength{\columnsep}{3pt} 
\begin{multicols}{2} 
\begin{itemize}
    \item BG-index;
    \item Borgatti's effective size;
    \item Bundle index;
    \item Burt's constraint;
    \item Clustered local-degree (CLD);
    \item ClusterRank centrality;
    \item Coleman-Theil disorder index;
    \item Collective influence;
    \item Common out-neighbor (CON) score;
    \item Cumulative Contact Probability (CCP);
    \item Degree and clustering coefficient (DCC) centrality;
    \item Degree and clustering coefficient and location (DCL) centrality;
    \item Degree and Importance of Lines (DIL);
    \item Degree and structural hole count (DSHC) method;
    \item Degree centrality;
    \item Degree mass;
    \item Diffusion centrality;
    \item Diffusion degree;
    \item Disassortativity of node (DoN);
    \item Dynamics-sensitive (DS) centrality;
    \item Effective Size (ES);
    \item Extended local bridging centrality (ELBC);
    \item Extended local dimension (ELD);
    \item Extended weight degree centrality (EWDC);
    \item Fuzzy local dimension (FLD);
    \item Global and local structure (GLS) centrality;
    \item Godfather index;
    \item INF centrality;
    \item Interdependence centrality;
    \item Isolating centrality (ISC);
    \item Leverage centrality;
    \item Lobby index;
    \item Local clustering coefficient;
    \item Local entropy (LE) centrality;
    \item Local neighbor contribution (LNC) centrality;
    \item Local structural centrality (LSC);
    \item LocalRank centrality;
    \item Long-Range Interaction Centrality (LRIC);
    \item $\boldsymbol{\mu}$-Power Community Index ($\boldsymbol{\mu}$-PCI);
    \item Malatya centrality;
    \item Modified local centrality (MLC);
    \item Mutual information centrality;
    \item Neighbor distance centrality;
    \item Neighborhood connectivity;
    \item Neighborhood density (ND);
    \item NL centrality;
    \item Normalized local centrality (NLC);
    \item Pivotal index;
    \item ProfitLeader;
    \item Quantum Jensen-Shannon Divergence centrality;
    \item Rapid identifying method (RIM);
    \item Redundancy measure;
    \item Semi-global triangular centrality;
    \item Semi-local degree and clustering coefficient;
    \item Short-Range Interaction Centrality (SRIC);
    \item Spreading probability (SP) centrality;
    \item Spreading strength;
    \item Support;
    \item Topological coefficient;
    \item Volume centrality;
    \item Weight degree centrality (WDC, Liu);
    \item Weight neighborhood centrality;
    \item Weighted formal concept analysis (WFCA);
    \item Weighted \textit{h}-index;
    \item Weighted \textit{k}-shell degree neighborhood (WKSDN);
    \item Weighted volume centrality;
    \item \textit{X}-degree centrality.
\end{itemize}
\end{multicols}
}

A detailed description of these measures is provided in Section \ref{sec:encyclopedia}.

\subsection{Community-based centrality measures}

Community-based centrality measures evaluate the importance of a node with respect to the modular structure of the network. These measures consider not only the connectivity of a node within its own community but also its role in linking different communities. Formally, a node may be assigned a higher centrality if it connects densely interconnected groups, acts as a bridge between communities or facilitates interactions across communities. These centrality measures typically rely on prior community detection. Communities can be identified using methods that optimize a modularity function, spectral clustering based on network eigenvectors or the Infomap algorithm, which employs an information-theoretic metric. Node centrality is then computed based on both intra- and inter-community connectivity patterns.

The examples of community-based centrality measures discussed in this work are as follows:
{\footnotesize
\vspace{-0.4cm}
\setlength{\columnsep}{3pt} 
\begin{multicols}{2} 
\begin{itemize}
    \item BridgeRank;
    \item Comm Centrality;
    \item Community centrality;
    \item Community hub-bridge (CHB) measure;
    \item Community-based centrality (CbC); 
    \item Community-based mediator (CbM);
    \item Disassortativity and community structure centrality;
    \item Entropy-based influence disseminator (EbID);
    \item Gateway coefficient;
    \item Generalized network dismantling (GND) method;
    \item Intra-module degree (IMD);
    \item Modularity centrality;
    \item Modularity density centrality;
    \item Modularity vitality;
    \item Participation coefficient;
    \item Partition-based spreaders identification method;
    \item Weighted community betweenness (WCB) centrality.
\end{itemize}
\end{multicols}
}

A detailed description of these measures is provided in Section \ref{sec:encyclopedia}.

\subsection{Graphlet-based centrality measures}

Graphlet-based centrality measures quantify node importance by counting its participation in small subgraph patterns or graphlets, such as triangles, cliques or other motifs. Nodes involved in many such local structures are considered more central, reflecting both their participation in local connectivity patterns. Certain graphlet-based centrality measures also take into account the connectivity among a node's neighbors.

The examples of the graphlet-based centrality measures discussed in this work are as follows:
{\footnotesize
\vspace{-0.1cm}
\setlength{\columnsep}{0pt} 
\begin{multicols}{2} 
\begin{itemize}
    \item All-subgraph centrality
    \item Clique level;
    \item Cross-clique connectivity;
    \item Curvature index;
    \item Density of the Maximum Neighborhood Component;
    \item Edge clustering coefficient (NC);
    \item Edge Percolated Component (EPC);
    \item Graphlet degree centrality (GDC);
    \item Local clustering coefficient;
    \item Maximal Clique Centrality (MCC);
    \item Maximum Neighborhood Component (MNC);
    \item Truncated curvature index.
\end{itemize}
\end{multicols}
}

A detailed description of these measures is provided in Section \ref{sec:encyclopedia}.

\subsection{Gravity-based centrality measures}

Gravity-based centrality measures draw inspiration from Newton's law of gravitation, quantifying the importance of a node based on its interactions with other nodes weighted by distance. In general, the centrality of node $i$ can be expressed as
\[
c(i) = \sum_{j \neq i} \frac{m_i m_j}{d_{ij}^\alpha},
\]
where $m_i$ and $m_j$ are attributes representing the "mass" of nodes $i$ and $j$, $d_{ij}$ is a measure of distance between nodes $i$ and $j$, while $\alpha$ is a decay exponent controlling the influence of distant nodes (e.g. $\alpha=2$). Variations of this framework include restricting the sum to nodes within a truncated radius, using masses computed from other centrality measures, assigning different masses to nodes $i$ and $j$ or employing alternative distance metrics such as effective distance instead of shortest-path distance. These modifications combine node mass and network position, so that nodes with high mass and short distances to other important nodes attain higher centrality scores.

The examples of vitality-based centrality measures discussed in this work are presented below:
{\footnotesize
\vspace{-0.2cm}
\setlength{\columnsep}{3pt} 
\begin{multicols}{2} 
\begin{itemize}
    \item Adaptive omni-channel gravity centrality (AOGC);
    \item Density centrality;
    \item DK-based gravity model (DKGM);
    \item Effective distance gravity model (EffG);
    \item Effective gravity model (EGM);
    \item Entropy-based gravity model (EGM);
    \item Fusion gravity model (FGM);
    \item Gravity centrality;
    \item Gravity model;
    \item Generalized gravity centrality (GGC);
    \item Hybrid degree and \textit{k}-shell method;
    \item Hybrid degree and MDD method;
    \item \textit{k}-shell based on gravity centrality (KSGC);
    \item KDEC method;
    \item Laplacian gravity centrality;
    \item Local gravity model;
    \item Mixed gravitational centrality;
    \item Multi-characteristics gravity model (MCGM);
    \item Random walk-based gravity (RWG) centrality;
    \item Return Random Walk Gravity (RRWG) centrality;
    \item Weighted gravity model (WGravity).
\end{itemize}
\end{multicols}
}

A detailed description of these measures is provided in Section \ref{sec:encyclopedia}.

\vfill

\subsection{Vitality-based centrality measures}

Vitality-based centrality measures, also referred to as graph-induced centralities, quantify the importance of a node or edge by evaluating its impact on a network-level function. Formally, the centrality of node $i$ can be defined as
\[
C(i) = f(G) - f(G_i),
\]
where $G$ denotes the original network, $G_i$ is the network obtained after removing node $i$ (or a set of nodes or edges incident to $i$) and $f(G)$ is a function that captures a property of interest on graph $G$. The function $f(G)$ may quantify structural aspects such as connectivity, efficiency or robustness, or it may describe outcomes of dynamical processes running on the network, including flow, diffusion or synchronization. Variations of these measures include the removal of edges instead of nodes, as well as formulations in which the centrality of a node is computed by evaluating its impact across all possible subsets of nodes, as in approaches based on the Shapley value.

The examples of vitality-based centrality measures discussed in this work are presented below:

{\footnotesize
\vspace{-0.2cm}
\setlength{\columnsep}{3pt} 
\begin{multicols}{2}
\begin{itemize}
    \item Algebraic centrality;
    \item Average shortest path centrality (AC);
    \item Bridging capital;
    \item Collective network entanglement (CNE);
    \item Communicability betweenness centrality;
    \item ControlRank;
    \item Distance-weighted fragmentation (DF);
    \item Dynamical importance;
    \item Dynamical spanning tree (DST) centrality;
    \item Entropy centrality;
    \item Entropy variation (EnV);
    \item Exogenous centrality;
    \item Laplacian centrality;
    \item Link influence entropy (LInE) centrality;
    \item Local relative change of average shortest path (LRASP) centrality;
    \item Map equation centrality;
    \item Mediative effects centrality (MEC);
    \item Modularity vitality;
    \item Node contraction (IMC) centrality;
    \item Pairwise disconnectivity centrality;
    \item Quasi-Laplacian centrality (QC);
    \item Resilience centrality;
    \item Resolvent betweenness (RB) centrality;
    \item Shapley value;
    \item Shapley value based information delimiters (SVID);
    \item Spanning tree centrality (STC);
    \item Super mediator degree (SMD);
    \item Third Laplacian energy centrality (LC);
    \item Total centrality;
    \item Vertex entanglement (VE).
\end{itemize}
\end{multicols}
}

A detailed description of these measures is provided in Section \ref{sec:encyclopedia}.

\subsection{Local dimension models}

Local dimension centrality measures evaluate node importance by examining how the number of nodes or other structural properties expand at successive distance levels from a given node.  Empirical observations indicate that this growth often follows an approximately exponential pattern, motivating the use of a logarithmic transformation. For each node, these quantities are recorded at successive distance layers and a model is fitted to the logarithm of the measurements, typically using linear regression. The slope of this log-log plot represents the local dimension of the node, capturing how quickly its neighborhood expands. Nodes with faster-growing neighborhoods are considered more structurally central, as they tend to bridge clusters or connect multiple regions of the network.

The examples of the local dimension models discussed in this work are as follows:

{\footnotesize
\vspace{-0.05cm}
\setlength{\columnsep}{0pt} 
\begin{multicols}{2} 
\begin{itemize}
    \item Local degree dimension (LDD);
    \item Local dimension (LD);
    \item Local volume dimension (LVD);
    \item Local information dimensionality (LID);
    \item Multi-local dimension (MLD) centrality;
    \item Node information dimension (NID).

\end{itemize}
\end{multicols}
}

A detailed description of these measures is provided in Section \ref{sec:encyclopedia}.

\subsection{Core decomposition and hierarchical centrality measures}

Core decomposition and hierarchical centrality measures evaluate node importance based on their placement within the network's core-periphery structure. Core decomposition and hierarchical centrality measures evaluate node importance based on their placement within the network’s core-periphery structure. Two main types of methods exist: (1) methods that assign nodes to discrete groups or hierarchical levels (e.g., coloring methods), and (2) methods that decompose the network into cores of different levels (e.g., $k$-shell decomposition methods). Nodes with higher shell indices, corresponding to positions closer to the network core, are considered more central. Variants of these measures extend the basic $k$-shell approach by incorporating node-specific attributes, weighted edges, or alternative criteria for shell assignment, enabling centrality to capture both topological significance and additional relevant properties.

Many of these methods assign identical values to large groups of nodes, resulting in low discriminative power. To address this limitation, they are often extended using approaches described in the group of hybrid methods in Section \ref{subsection:hybrid}. 

The examples of the core decomposition and hierarchical centrality measures discussed in this work are as follows:
{\footnotesize
\vspace{-0.3cm}
\setlength{\columnsep}{3pt} 
\begin{multicols}{2} 
\begin{itemize}
    \item Classified neighbors centrality;
    \item CoreHD;
    \item Improved \textit{k}-shell decomposition (IKSD) method;
    \item \textit{k}-shell centrality;
    \item \textit{k}-truss index;
    \item Improved distance-based coloring method (IIS) 
    \item Independent set (IS) method
    \item Mixed degree decomposition (MDD);
    \item Onion decomposition (OD);
    \item Renewed coreness centrality;
    \item RMD-weighted degree (WD) centrality;
    \item \textit{s}-shell index;
    \item Weighted \textit{k}-shell decomposition (Wks) centrality.
\end{itemize}
\end{multicols}
}

A detailed description of the core decomposition and hierarchical centrality measures is provided in Section \ref{sec:encyclopedia}.

\subsection{Voting models}

Voting centrality measures evaluate node importance based on votes cast by nodes in the network. In one class of methods, inspired by social choice theory, each node effectively votes for other nodes according to predefined criteria, and these votes are aggregated to determine node rankings. The second class uses an iterative procedure: at each step, nodes vote for their neighbors, then the node with the highest score is selected and the scores of its neighbors are discounted to reflect the influence of the chosen node. This process continues until all nodes are ranked or a stopping criterion is reached. Variations of iterative voting models differ in how initial scores are assigned, how neighbor discounts are applied, and the rules used to update scores at each iteration.

The examples of the voting models discussed in this work are as follows:
{\footnotesize
\vspace{-0.2cm}
\setlength{\columnsep}{0pt} 
\begin{multicols}{2} 
\begin{itemize}
    \item Borda centrality;
    \item Copeland centrality;
    \item DegreeDiscountIC;
    \item DegreeDistance;
    \item DegreePunishment;
    \item EnRenew;
    \item Generalized degree discount (GDD);
    \item Improved WVoteRank centrality;
    \item NCVoteRank;
    \item SingleDiscount;
    \item Top candidate method;
    \item VMM algorithm;
    \item VoteRank centrality;
    \item VoteRank$^{++}$ centrality;
    \item Weighted top candidate method;
    \item WVoteRank.
\end{itemize}
\end{multicols}
}

A detailed description of the voting models is provided in Section \ref{sec:encyclopedia}.

\vfill

\subsection{Hybrid centrality measures}\label{subsection:hybrid}

Hybrid centrality measures integrate multiple approaches to evaluate node importance, either by combining existing centrality measures or by employing multi-criteria decision-making (MCDM) frameworks. In the first group, different measures such as degree, betweenness, closeness or $k$-shell centralities are aggregated using weighted sums or other combination rules to capture complementary aspects of centrality. The second group employs formal multi-criteria decision-making (MCDM) methods, such as the Analytic Hierarchy Process (AHP), Technique for Order Preference by Similarity to Ideal Solution (TOPSIS), multi-attribute ranking based on information entropy (MABIE) and evidence theory, to systematically integrate multiple criteria and rank nodes by their overall importance.

The examples of the hybrid centrality measures discussed in this work are as follows:
{\footnotesize
\vspace{-0.2cm}
\setlength{\columnsep}{0pt} 
\begin{multicols}{2} 
\begin{itemize}
    \item All-around score;
    \item Analytic Hierarchy Process (AHP) centrality;
    \item Balanced centrality (SWIPD);
    \item Cc-Burt centrality;
    \item ClusterRank centrality;
    \item Degree and clustering coefficient (DCC) centrality;
    \item Entropy-Burt method (E-Burt);
    \item Extended local bridging centrality (ELBC);
    \item Global and local information (GLI) method;
    \item Global importance of nodes (GIN);
    \item Global Structure Model (GSM);
    \item Hierarchical \textit{k}-shell (HKS) centrality;
    \item Hybrid characteristic centrality (HCC);
    \item Hybrid centrality (HC);
    \item Hybrid centrality measure (X);
    \item Hybrid centrality measure (Y);
    \item Hybrid centrality measure (XpY);
    \item Hybrid centrality measure (XmY);
    \item Hybrid degree centrality;
    \item Hybrid global structure model (H-GSM);
    \item Hybrid median centrality (HMC);
    \item HybridRank;
    \item Improved entropy-based centrality;
    \item Improved Global Structure Model (IGSM);
    \item Improved \textit{k}-shell algorithm (IKS);
    \item Improved \textit{k}-shell hybrid method (IKH);
    \item Influence capability (IC);
    \item Integral \textit{k}-shell centrality;
    \item \textit{k}-shell hybrid method (ksh);
    \item \textit{k}-shell iteration factor (KS-IF);
    \item \textit{k}-shell Physarum centrality;
    \item KED method;
    \item Lhc index;
    \item Local and global centrality (LGC);
    \item Local fuzzy information centrality (LFIC);
    \item Localized bridging centrality (LBC);
    \item M-centrality;
    \item Meta-centrality;
    \item Mixed core, degree and entropy (MCDE) method;
    \item Mixed core, degree and weighted entropy (MCDWE) method;
    \item Mixed core, semi-local degree and entropy (MCSDE) method;
    \item Mixed core, semi-local degree and weighted entropy (MCSDWE) method;
    \item Modified Expected Force (ExF$^M$);
    \item Multi-attribute ranking method based on information entropy (MABIE);
    \item Multi-criteria influence maximization (MCIM) method;
    \item Multiple local attributes weighted centrality (LWC);
    \item Neighborhood core diversity centrality (Cncd);
    \item New Evidential Centrality (NEC);
    \item Node importance contribution correlation matrix (NICCM) method;
    \item Node importance contribution matrix (NICM) method;
    \item Node importance evaluation matrix (NIEM) method;
    \item Node local centrality (NLC);
    \item Relative entropy;
    \item Relative local-global importance (RLGI) measure;
    \item Shell clustering coefficient (SCC);
    \item Synthesize centrality (SC);
    \item \(\boldsymbol{\theta}\)-Centrality;
    \item TOPSIS centrality;
    \item TOPSIS-RE centrality;
    \item Two-step framework (IF) centrality;
    \item Weighted \textit{k}-shell degree neighborhood (Wksd);
    \item Weighted TOPSIS (w-TOPSIS) centrality.
\end{itemize}
\end{multicols}
}

A detailed description of the hybrind measures is provided in Section \ref{sec:encyclopedia}.

\vfill

\subsection{Diffusion centrality measures}

Diffusion-based centrality measures evaluate node importance by considering the role of nodes in a network diffusion process. These methods assume a specific spreading mechanism, such as the propagation of information, influence or resources, and quantify how effectively a node can initiate or facilitate the process. Variations differ in the type of diffusion model employed, the time horizon considered, and whether probabilistic or deterministic spreading dynamics are used. 

The examples of the diffusion centrality measures discussed in this work are as follows:
{\footnotesize
\vspace{-0.1cm}
\setlength{\columnsep}{0pt} 
\begin{multicols}{2} 
\begin{itemize}
    \item Diffusion centrality;
    \item DomiRank centrality;
    \item Dynamical influence;
    \item Dynamics-sensitive (DS) centrality;
    \item Epidemic centrality;
    \item Expected Force (ExF);
    \item Game centrality (GC);
    \item Improved iterative resource allocation (IIRA) method;
    \item Independent cascade rank (ICR);
    \item Infection number;
    \item Interdependence centrality;
    \item Linear threshold centrality (LTC);
    \item Message-passing approach;
    \item Percolation centrality;
    \item Physarum centrality;
    \item PhysarumSpreader;
    \item Rumor centrality;
    \item Short-Range Interaction Centrality (SRIC);
    \item Simulations-based LRIC (LRIC-sim) index;
    \item Graph regularization centrality (GRC);
    \item Long-Range Interaction Centrality (LRIC);
    \item ModuLand centrality.
\end{itemize}
\end{multicols}
}

A detailed description of the diffusion centrality models is provided in Section \ref{sec:encyclopedia}.

\subsection{Entropy-based centrality measures}

Entropy-based centrality measures quantify the importance of a node by evaluating the uncertainty or diversity of its interactions within the network. These methods first compute relevant statistics for each node, such as the distribution of connections, paths or types of neighbors, and then apply an entropy function to these distributions. Nodes associated with higher entropy are considered more central, as they participate in more diverse or less predictable interactions, indicating a greater potential to influence or access different parts of the network. Variations of these measures differ in the choice of underlying statistics, the type of entropy used and whether local or global network information is incorporated.

The examples of the local dimension models discussed in this work are as follows:
{\footnotesize
\vspace{-0.3cm}
\setlength{\columnsep}{0pt} 
\begin{multicols}{2} 
\begin{itemize}
    \item Access information;
    \item Distance entropy (DE);
    \item Diversity coefficient;
    \item Diversity-strength centrality (DSC);
    \item Diversity-strength ranking (DSR);
    \item Entropy and mutual information-based centrality (EMI);
    \item Entropy-based ranking measure (ERM);
    \item Entropy-Burt method (E-Burt);
    \item Expected Force (ExF);
    \item Hide information;
    \item Information distance index (IDI);
    \item Local entropy (LE) centrality;
    \item Local information dimensionality (LID);
    \item Mapping entropy betweenness (MEB) centrality;
    \item Mapping entropy (ME) centrality;
    \item Markov entropy centrality;
    \item Path-transfer centrality;
    \item Random walk accessibility (RWA).
\end{itemize}
\end{multicols}
}

A detailed description of the entropy-based models is provided in Section \ref{sec:encyclopedia}.

\vfill

\subsection{Neighborhood centrality measures}

Neighborhood-based centrality measures evaluate node importance by incorporating the centrality of neighboring nodes. In a basic formulation, a node's centrality can be computed as a weighted combination of its own base centrality $f(i)$ and the sum of the centralities of its immediate neighbors:
\[
c_{\text{improved}}(i) = \alpha f(i) + \beta \sum_{j \in \mathcal{N}(i)} f(j),
\]
where $f(i)$ is an initial centrality measure of node $i$ (for example, degree, closeness or another local or global metric), while $\alpha$ and $\beta$ control the relative importance of the node's own value and the influence of its neighbors. In the special case $\alpha = 0$, the centrality of a node depends entirely on the centrality of its neighbors. 

Extended formulations propagate centrality further, so that a node's score reflects not only the centrality of its direct neighbors but also the aggregated influence of neighbors' neighbors or more distant nodes:
\[
c_{\text{extended}}(i) = \sum_{j \in \mathcal{N}(i)} c_{\text{improved}}(j).
\]

The examples of the neighborhood models discussed in this work are as follows:
{\footnotesize
\vspace{-0.3cm}
\setlength{\columnsep}{0pt} 
\begin{multicols}{2} 
\begin{itemize}
    \item Extended cluster coefficient ranking measure (ECRM);
    \item Extended diversity-strength ranking (EDSR);
    \item Extended gravity centrality (EGC);
    \item Extended \textit{h}-index centrality (EHC);
    \item Extended \textit{k}-shell hybrid method;
    \item Extended hybrid characteristic centrality (EHCC);
    \item Extended hybrid degree and \textit{k}-shell method;
    \item Extended hybrid degree and MDD method;
    \item Extended improved \textit{k}-shell hybrid method;
    \item Extended mixed gravitational centrality (EMGC);
    \item Extended neighborhood coreness;
    \item Extended RMD-weighted degree (EWD) centrality;
    \item h-index strength;
    \item Hierarchical reduction by betweenness;
    \item Improved neighbors’ \textit{k}-core (INK);
    \item Improved node contraction (IIMC) centrality;
    \item Local H-index;
    \item Neighborhood centrality;
    \item Neighborhood structure-based centrality (NSC);
    \item Network global structure-based centrality (NGSC);
    \item Node and neighbor layer information centrality (NINL).
\end{itemize}
\end{multicols}
}

A detailed description of the neighborhood centrality models is provided in Section \ref{sec:encyclopedia}.


\chapter{Encyclopedia of Models}\label{sec:encyclopedia}

\section{Absorbing random-walk (ARW) centrality}

\emph{Absorbing random-walk (ARW) centrality} \index{absorbing random-walk (ARW) centrality}\cite{Mavroforakis2015} measures the importance of nodes as \emph{absorbing states} with respect to a set of query nodes \(Q \subseteq \mathcal{N}\). The goal is to select \(C \subseteq \mathcal{N}\) of size \(k\) that minimizes the expected number of steps for random walks starting from \(Q\) to reach any node in \(C\).

Let \(D \subseteq \mathcal{N}\) be candidate nodes, and \(s\colon Q \to [0,1]\) a probability distribution over \(Q\) (\(\sum_{q \in Q} s(q)=1\)). At each step, a random walk from a transient node \(i \in T = \mathcal{N} \setminus C\) either moves uniformly to a neighbor with probability \(1-\alpha\) or restarts at a query node sampled from \(s\) with probability \(\alpha\), where \(\alpha \in [0,1]\). Nodes in \(C\) are absorbing: the walk terminates upon reaching them. For candidate set \(C \subseteq D\), let the transition matrix be
\[
P = \begin{pmatrix} P_{TT} & P_{TC} \\ 0 & I \end{pmatrix},
\]
with \(P_{TT}\) for transient-to-transient and \(P_{TC}\) for transient-to-absorbing transitions. The fundamental matrix is \(F = (I - P_{TT})^{-1}\), and the expected absorption times are
\[
L_C = \begin{pmatrix} F \\ 0 \end{pmatrix} \mathbf{1}_T.
\]
The ARW centrality of nodes in \(C\) is
\[
c_{ARW}(C) = s^\top L_C = \sum_{q \in Q} s(q) (L_C)_q.
\]

Selecting the optimal \(k\) nodes is NP-hard~\cite{Mavroforakis2015}. A greedy algorithm iteratively adds the node that maximally decreases \(c_{ARW}(C)\). This naturally favors diverse selections, as absorbing nodes placed in different graph regions intercept walks from multiple query nodes efficiently.

\section{Access information}

\emph{Access information}\index{access information} (also known as search information) quantifies how easily a node can reach other nodes in the network~\cite{Rosvall2005}. Let \(\{p(i,j)\}\) denote the set of shortest paths from node \(i\) to node \(j\), and let \(d_i\) be the degree of node \(i\). The access information of node \(i\) is defined as
\begin{equation*}
    c_{\mathcal{A}}(i) = \frac{1}{N} \sum_{j=1}^{N} S(i,j),
\end{equation*}
where 
\[
S(i,j) = - \log_2 \sum_{\{p(i,j)\}} P[p(i,j)], \quad 
P[p(i,j)] = \frac{1}{d_i} \prod_{l \neq i \neq j \in p(i,j)} \frac{1}{d_l - 1}.
\]
Here, \(S(i,j)\) represents the amount of information needed to locate node \(j\) starting from \(i\) along the shortest paths, and \(P[p(i,j)]\) is the probability of following path \(p(i,j)\) when choosing uniformly at each step. Intuitively, \(c_{\mathcal{A}}(i)\) gives the average number of “questions” required to reach any node from \(i\). For example, in a star graph, the central hub has low access information: starting at the hub it is harder to reach a specific neighbor~\cite{Rosvall2005}.

\section{Adaptive LeaderRank}

\emph{Adaptive LeaderRank}\index{LeaderRank!adaptive} (ALR) is an extension of the original LeaderRank algorithm in which transition probabilities are weighted according to each node’s H-index (also known as the lobby index) \cite{Xu2017}. Similar to LeaderRank, ALR is based on a biased random-walk process, where nodes with higher H-index are more likely to receive and transmit resources.

Specifically, the ALR model first computes the H-index \(h_i\) of each node \(i\) and then introduces a ground node \(N+1\) with \(h_{N+1} = 1\), which connects bidirectionally to all nodes in the network \(G\). This construction is equivalent to a random walk on an augmented network where the resource amount \(s_i[t]\) at node \(i\) evolves according to
\begin{equation*}
    s_i[t+1] = \sum_{j=1}^{N+1} 
    \frac{a_{ji} h_i}{\sum_{k=1}^{N+1} a_{jk} h_k} \, s_j[t],
\end{equation*}
where \(a_{ji}\) denotes the element of the adjacency matrix $A$ representing the directed edge from node \(j\) to node \(i\).  The steady-state vector \(\tilde{s} = \lim_{t \to \infty} s[t]\) quantifies the relative influence (or importance) of nodes within the network. Xu and Wang \cite{Xu2017} demonstrated that ALR exhibits improved adaptability to structural changes or local perturbations in the network topology compared with the standard LeaderRank algorithm.

\section{Adaptive omni-channel gravity centrality (AOGC)}

The \emph{adaptive omni-channel gravity centrality}\index{gravity centrality!adaptive omni-channel (AOGC)} (AOGC), originally termed the Gravity Centrality method based on an Adaptive Truncation radius and Omni-channel paths, is a gravity-based centrality measure that combines an adaptive truncation radius with omni-channel path analysis to identify influential nodes in complex networks \cite{Yang2023}. For a node \(i \in \mathcal{N}\), the AOGC score is defined as
\[
c_{AOGC}(i) = \sum_{j \in \mathcal{N}^{(\leq r)}(i)} c_{ij} \frac{m_i m_j}{LD_{ij}^2},
\]
where 
\[
c_{ij} = \exp\Bigg(\frac{k_s(i) - k_s(j)}{\max(k_s) - \min(k_s)}\Bigg)
\] 
is the attraction coefficient based on the \(k\)-shell centrality of nodes \(i\) and \(j\),  \(\mathcal{N}^{(\leq r)}(i)\) denote the set of nodes within \(r\)-hop neighborhood of node \(i\), \(m_i\) is the mass of node \(i\) defined by the neighborhood structure-based centrality (NSC) \cite{Yang2023} and \(LD_{ij}\) is the \textit{looseness distance} between nodes \(i\) and \(j\):
\[
LD_{ij} = \frac{1}{\sum_{l=1}^{r} \sigma^l (A^l)_{ij}},
\]
where \(\sigma \in (0,1)\) is a free parameter controlling the weight of paths of different lengths, \(A\) is the adjacency matrix, and \(r\) is the truncation radius.  

The adaptive truncation radius \(r\) is determined based on the average shortest path length \(\langle d \rangle\) of the network as
\[
r =
\begin{cases}
    3, & \text{if } \langle d \rangle \leq 3, \\
    \lfloor \langle d \rangle \rfloor, & \text{if } 3 < \langle d \rangle \leq \Theta, \\
    \Theta + \lfloor \ln(\langle d \rangle - \Theta + 1) \rfloor, & \text{if } \langle d \rangle > \Theta,
\end{cases}
\]
where \(\Theta\) is a threshold parameter, set to 6.  

Nodes with high AOGC values are those that have large mass (high NSC), are well-connected to other influential nodes, and are reachable via multiple strong omni-channel paths, making them particularly important for spreading processes and maintaining network cohesion.

\section{Adjusted Index of Centrality (AIC)}

The \emph{Adjusted Index of Centrality}\index{adjusted index of centrality (AIC)} (AIC) is a closeness-based centrality measure introduced by Moxley \cite{Moxley1974}. For a node \(i\), its AIC centrality, denoted \(c_{AIC}(i)\), is defined as
\begin{equation*} 
c_{AIC}(i) = \frac{\sum_{j=1}^N \sum_{k=1}^N (d_{jk} + p \cdot n_j)}{\sum_{j=1}^N (d_{ij} + p \cdot n_i)},
\end{equation*}
where \(d_{jk}\) is the shortest-path distance from node \(j\) to node \(k\), \(n_i\) is the number of nodes unreachable from \(i\), and \(p\) is a penalty parameter chosen such that \(p > \max_{i,j} d_{ij}\). The penalty ensures that nodes with unreachable nodes receive appropriately lower centrality scores.

\section{Algebraic centrality}

\emph{Algebraic centrality}\index{algebraic centrality} quantifies the importance of a node based on the algebraic connectivity of a graph \cite{Kirkland2010}. 
The algebraic connectivity \(\alpha(G)\), also known as the Fiedler value \cite{Fiedler1973}, is defined as the second smallest eigenvalue of the Laplacian matrix \(L(G)\).

Kirkland \cite{Kirkland2010} defines the algebraic centrality of node \(i\), denoted \(c_{\text{alg}}(i)\), in two alternative ways:
\begin{equation*}
c_{\text{alg}}(i) = \alpha(G) - \alpha(G_i),
\end{equation*}
or
\begin{equation*}
c_{\text{alg}}(i) = \frac{\alpha(G_i)}{\alpha(G)},
\end{equation*}
where \(G_i\) is the subgraph obtained by removing node \(i\) from \(G\).  
Thus, algebraic centrality measures either the absolute or relative change in the algebraic connectivity due to the removal of node \(i\), providing insight into the node's structural importance in the network.

\section{All-around score}

The \emph{all-around score}\index{all-around score}, also known as the \emph{degree-betweenness-\(k\)-shell} (DBK) index, aims to identify nodes that perform well across multiple centrality dimensions simultaneously \cite{Hou2012}. The all-around centrality of a node \(i\) is defined as the Euclidean distance in the normalized centrality space spanned by degree, betweenness, and \(k\)-shell measures:
\begin{equation*}
    c_{\mathrm{AA}}(i) = \sqrt{\overline{c}_d^2(i) + \overline{c}_b^2(i) + \overline{c}_k^2(i)},
\end{equation*}
where \(\overline{c}_d(i)\), \(\overline{c}_b(i)\), and \(\overline{c}_k(i)\) denote the normalized degree, betweenness, and \(k\)-shell centrality scores of node \(i\), respectively.  

Nodes with high all-around scores achieve strong performance across all three dimensions and are thus referred to as \emph{all-around nodes}.

\section{All-subgraph centrality}

\emph{All-subgraph centrality}\index{all-subgraph centrality} (ASC) quantifies the importance of a node based on its participation in all connected subgraphs of a network \cite{Riveros2020}. For any subset of nodes \(S \subseteq \mathcal{N}\), let \(G_S\) denote the subgraph induced by \(S\).  The set of all connected subgraphs that contain node \(i\) is  
\[
\mathcal{A}(i, G) = \{\, S \subseteq \mathcal{N} \mid i \in S \text{ and } G_S \text{ is connected}\}.
\]

The all-subgraph centrality of node \(i\) is
\[
C_{\mathrm{ASC}}(i, G) = \log \bigl| \mathcal{A}(i, G) \bigr|,
\]
i.e., the logarithm of the number of connected subgraphs that include \(i\). Intuitively, all-subgraph centrality counts only connected subgraphs, since there is no justification for increasing the centrality of a node by including subgraphs in which it is not directly connected to other nodes. Nodes gain higher centrality if they participate in a larger number of connected subgraphs, reflecting their involvement in more diverse structural arrangements.  

The exact computation of the number of connected subgraphs \(|\mathcal{A}(i, G)|\) containing node \(i\) is generally intractable for large networks, since the number of connected subgraphs grows exponentially with graph size. In practice, \(|\mathcal{A}(i, G)|\) is estimated using approximation methods, such as restricting subgraphs to a fixed radius around \(i\) or using sampling techniques.

\section{All cycle betweenness (ACC) centrality}

\emph{All cycle betweenness}\index{betweenness centrality!all cycle} (ACC) centrality is a variation of betweenness centrality that accounts for all simple cycles passing through a node, rather than only shortest paths \cite{Zhou2018}. For node \(i\), the ACC centrality \(c_{ACC}(i)\) is defined as
\[
c_{ACC}(i) = \sum_{k=3}^{N} \alpha^{\,k-2} \, s_k(i),
\]
where \(s_k(i)\) denotes the total number of simple cycles of length \(k\) that include node \(i\), and \(\alpha \in (0,1)\) is an attenuation factor that downweights longer cycles.  

To efficiently estimate \(s_k(i)\), Zhou \textit{et al.} employ a belief propagation algorithm. In their experiments, they consider \(\alpha = 0.1\) and \(\alpha = 0.5\) to evaluate the influence of the attenuation factor on centrality values.

\section{Analytic Hierarchy Process (AHP) centrality}
The \emph{Analytic Hierarchy Process (AHP) centrality}\index{analytic hierarchy process (AHP) centrality} is a hybrid measure for identifying influential nodes by integrating multiple centrality metrics through the AHP framework~\cite{Bian2017}. Let $R$ be the normalized $N \times m$ decision matrix, where each entry $r_{ij}$ represents the normalized influence of node $i$ with respect to centrality metric $j$. Bian \textit{et al.}~\cite{Bian2017} consider $m=3$ metrics: degree, betweenness and closeness centralities. The relative importance $w_j$ of each metric is determined using a weighted TOPSIS approach (w-TOPSIS) \cite{Hu2016}.

For each centrality criterion $j$, an $N \times N$ pairwise comparison matrix $B^{(j)}$ is constructed, where each entry $b^{(j)}_{ik}$ quantifies the relative importance of node $i$ compared to node $k$ with respect to criterion $j$ (e.g. $b^{(j)}_{ik} = r_{ij} / r_{kj}$). The AHP score of node $i$ for criterion $j$ is then computed as
\[
s_j(i) = \frac{1}{N} \sum_{k=1}^N \frac{b^{(j)}_{ik}}{\sum_{l=1}^N b^{(j)}_{lk}}.
\]

The AHP centrality of node $i$ is obtained by aggregating the weighted scores across all criteria:
\[
c_{\mathrm{AHP}}(i) = \sum_{j=1}^m w_j s_j(i).
\]

Hence,  AHP combines both the relative importance of each centrality metric and the comparative evaluation of nodes, providing a comprehensive assessment of node influence.

\section{ArticleRank}

\emph{ArticleRank}\index{ArticleRank} is a variation of the PageRank algorithm designed to measure the influence of journal articles \cite{Li2009}. It follows the general PageRank methodology but modifies the normalization term in the transition probability to prevent nodes with very few outgoing links from exerting disproportionately high influence on their neighbors. 

Formally, the ArticleRank \(c_{\text{AR}}(i)\) of a node \(i\) is defined as
\begin{equation*}
    c_{\text{AR}}(i) = (1 - p) 
    + p \sum_{j : a_{ij} = 1} 
    \frac{c_{\text{AR}}(j)}{\sum_{k = 1}^{N} a_{jk} + \overline{a}_{\text{out}}},
\end{equation*}
where \(p \in [0, 1]\) is a damping factor, \(a_{ij}\) is the adjacency matrix element indicating the presence of a link from node \(j\) to node \(i\), and \(\overline{a}_{\text{out}}\) denotes the average out-degree of the network \(G\). The ArticleRank values represent the relative influence or importance of articles within the citation network, where higher scores indicate articles that are cited by other influential papers.

\vfill

\section{Attentive betweenness centrality (ABC)}

\emph{Attentive betweenness centrality}\index{betweenness centrality!attentive (ABC)} (ABC) quantifies the importance of a node based on the amount of attention it allocates to the flow of information between other nodes \cite{Adali2012}. The method models information propagation through the network starting from a source node, with an initial flow value of 1, structured according to BFS levels. At each level, a node receives flow from neighbors in higher or the same BFS levels, attenuated by a factor \(\alpha\) to model decay. The node then accumulates this flow and redistributes a portion to neighbors in the same or next BFS level, again applying the attenuation factor.

Formally, let \(L_s(i,k) = \{ j \in \mathcal{N}(i) \mid d_{sj} = k \}\) denote the neighbors of node \(i\) at distance \(k\) from source \(s\). The total flow \(f(i)\) received by node \(i\), located at distance \(d_{si}\) from the source, is
\begin{align*}
f(i) &= \alpha \sum_{j \in L_s(i,d_{si}-1)} 
        \frac{w_{ji}}{\sum_{k \notin L_s(j,d_{sj}-1)} w_{jk}} f(j) \\
      &\quad + \alpha^2 \sum_{l \in L_s(i,d_{si})} 
        \frac{w_{li}}{\sum_{k \notin L_s(l,d_{sl}-1)} w_{lk}} 
        \sum_{j \in L_s(l,d_{sj}-1)} 
        \frac{w_{jl}}{\sum_{k \notin L_s(j,d_{sj}-1)} w_{jk}} f(j),
\end{align*}
where the first term accounts for flow received directly from higher-level neighbors, and the second term accounts for flow received indirectly via same-level neighbors, weighted by the attention factor \(\alpha\) and edge strengths \(w_{ij}\).

Attentive betweenness centrality assumes that information only propagates laterally or forward in BFS levels, never backward. After forward propagation, a backward credit assignment step distributes accumulated flow from downstream neighbors and same-level siblings proportionally to their contributions. This process continues upward, and when the source node is reached, the total credit accumulated across all nodes defines their centrality for that BFS traversal. Repeating the procedure with every node as the source and averaging the results yields the final ABC-centrality score for each node.

\section{Average shortest path centrality (AC)}

The \emph{average shortest path centrality}\index{average shortest path centrality} (AC), also known as the relative change in average shortest path (RASP), quantifies node influence by measuring the relative change in the network's average shortest path (ASP) after removing a node \cite{Lv2019}. 
The centrality of node $i$ is defined as
\begin{equation*}
   c_{AC}(i) = \frac{ASP(G_i) - ASP(G)}{ASP(G)},
\end{equation*}
where $G$ is the original network, and $G_i$ is the subgraph obtained by removing node $i$ and all its adjacent links. The average shortest path of a graph $G$ is
\begin{equation*}
   ASP(G) = \frac{\sum_{i \neq j} d_{ij}}{N(N-1)},
\end{equation*}
where $d_{ij}$ is the shortest path length between nodes $i$ and $j$ if they are connected; otherwise, $d_{ij}$ is set equal to the diameter of $G$.  

Hence, AC centrality quantifies the importance of a node in preserving efficient communication paths within the network.

\section{Balanced centrality (SWIPD)}

\emph{Balanced centrality}\index{balanced centrality (SWIPD)}, also known as SWIPD, is a hybrid centrality measure that combines multiple classical centrality indices into a single vector \cite{Debono2014}. The centrality of node $i$, denoted $c_{\text{SWIPD}}(i)$, is defined as
\[
c_{\text{SWIPD}}(i) = 
\gamma_1 A^2 D^{-1} A u \; + \;
\gamma_2 (D - \alpha D A)^{-1} A u \; + \;
\gamma_3 A D^{-2} u \; + \;
\gamma_4 A D^{-1} A u,
\]
where the four terms correspond, respectively, to \emph{square centrality}, \emph{walk centrality}, \emph{power-like centrality}, and \emph{degree-weighted centrality}.  Specifically, $D$ is the degree diagonal matrix, $u$ is the $N \times 1$ vector of ones, and the coefficients $\gamma_i$ control the relative contribution of each component. The attenuation factor in the walk centrality is $\alpha = \frac{1}{d_{\max}+1}$, where $d_{\max}$ is the maximum node degree in the network. Each of these terms captures a distinct aspect of node importance, which can be summarized as follows:
\begin{itemize}
    \item \textit{Square centrality:} emphasizes the influence of two-hop neighbors.
    \item \textit{Walk centrality:} incorporates attenuated walks starting from each node.
    \item \textit{Power-like centrality:} highlights nodes with strong direct connections while penalizing high-degree nodes.
    \item \textit{Degree-weighted centrality:} captures the influence of a node based on the degree-weighted connectivity of its neighbors.
\end{itemize}

Balanced centrality combines several centrality measures, allowing the relative contributions of different aspects of node importance to be analyzed within a single measure.

\section{Betweenness centrality}
\emph{Betweenness centrality}\index{betweenness centrality}, also known as Freeman's betweenness or sociometric betweenness, quantifies the extent to which a node lies on the communication paths connecting pairs of other nodes within a network, reflecting its potential to mediate or control the flow of information \cite{Anthonisse1971,Freeman1977}. It is formally defined as
\begin{equation*}
    c_{betw}(i) = \sum_{j\neq k \neq i}{\frac{\sigma_{jk}(i)}{\sigma_{jk}}},
\end{equation*}
where \(\sigma_{jk}\) denotes the number of shortest paths from node \(j\) to node \(k\), and \(\sigma_{jk}(i)\) represents the number of shortest paths that pass through node \(i\). A high value of betweenness centrality indicates nodes that serve as crucial hubs or bridges connecting otherwise disparate clusters within the network. Nodes with high betweenness centrality occupy strategic positions that influence the structure and dynamics of a network by shaping how information or resources flow between its parts. In weighted networks, the interpretation of the shortest paths requires careful consideration: when edge weights represent the strength of the tie rather than the cost, these values should be inverted before applying algorithms such as Dijkstra’s, thus ensuring that stronger connections correspond to shorter effective paths \cite{Opsahl2010}.

Some centrality measures are equivalent to the betwenness centrality because they provide the same ranking of nodes. For instance, Caporossi \textit{et al.} \cite{Caporossi2012} propose the \emph{adjusted betweenness centrality}\index{betweenness centrality!adjusted} \(c_{ABC}\) where \(c_{ABC}(i)=2c_{betw}(i) + N - 1\).

\vfill

\section{BG-index}

The \emph{BG-index}\index{BG-index}, also known as the \(\beta\)-measure \cite{VanDenBrink2008,Boldi}, is a social power index that quantifies a node's centrality based on its position within a network  \cite{vdBrink1994,VanDenBrink2008}. Van den Brink and Gilles assume that the network represents a social structure, in which each node may dominate some nodes while being dominated by others.

The BG-index of node \(i\) measures the expected number of times it will be selected as a predecessor by its neighbors, assuming that each neighbor chooses one of its predecessors uniformly at random. Formally,  
\begin{equation*}
    c_{\mathrm{BG{-}index}}(i) = \sum_{j \in \mathcal{N}(i)} \frac{1}{|\{k : j \in \mathcal{N}(k)\}|} = \sum_{j \in \mathcal{N}(i)} \frac{1}{|\mathcal{N}(i)|} = \sum_{j \in \mathcal{N}(i)} \frac{1}{d_j} = \sum_{j=1}^{N} \frac{a_{ij}}{\sum_{k=1}^{N} a_{kj}},
\end{equation*}
where \(\mathcal{N}(i)\) denotes the set of neighbors of node \(i\), \(d_j\) is the degree of node \(j\) and \(a_{ij}\) are entries of the adjacency matrix \(A\).  

For directed graphs, the BG-index has two versions: the positive \(\beta\)-measure (computed on the original graph \(G\)) and the negative \(\beta\)-measure (computed on the reverse graph of $G$) \cite{Boldi}.

\section{Biased random walk (BRW) centrality}

The \emph{biased random walk centrality}\index{random walk centrality!biased (BRW)} is a variation of the PageRank algorithm that incorporates both degree centrality and neighborhood density into the transition probabilities \cite{Takes2011}. Starting from a node \(i\), the walker either jumps to a random node with probability \(p\), or moves to a neighbor \(j \in \mathcal{N}(i)\) with probability
\begin{equation*}
    p_{ij} = \frac{\alpha \left(1 - 1/|\mathcal{N}(j)|\right) + (1-\alpha) c_{ND}(j)}
    {\sum_{k \in \mathcal{N}(i)} \alpha \left(1 - 1/|\mathcal{N}(k)|\right) + (1-\alpha) c_{ND}(k)},
\end{equation*}
where \(c_{ND}(j)=1 - \sum_{k \in \mathcal{N}(j)} \frac{|\mathcal{N}(j) \cap \mathcal{N}(k)|}{(|\mathcal{N}(k)| - 1) |\mathcal{N}(j)|}\) is the neighborhood density of node \(j\), and \(\alpha \in [0,1]\) balances the contributions of degree and neighborhood density. Takes and Kosters suggest \(\alpha = 0.5\) and \(p = 0.15\). This centrality measures the likelihood that a random walk biased by local connectivity features visits a node, identifying nodes that are both highly connected and embedded in dense neighborhoods.

\section{Bipartivity index}

The \emph{bipartivity index}\index{bipartivity index} quantifies the extent to which individual nodes contribute to the global bipartivity of a network \cite{Estrada2005}. Intuitively, in a perfectly bipartite network, there are no closed walks of odd length.  

Mathematically, the network bipartivity $\beta(G)$ is defined as the proportion of even-length closed walks to the total number of closed walks:
\[
\beta(G) = \frac{\langle SC \rangle_{\mathrm{even}}}{\langle SC \rangle} 
= \frac{\langle SC \rangle_{\mathrm{even}}}{\langle SC \rangle_{\mathrm{even}} + \langle SC \rangle_{\mathrm{odd}}} 
= \frac{\sum_{k=1}^N \cosh(\lambda_k)}{\sum_{k=1}^N e^{\lambda_k}} \in \left( \frac{1}{2}, 1 \right],
\]
where $\langle SC \rangle$ denotes the subgraph centralization, which can be decomposed into contributions from even- and odd-length closed walks, and $\lambda_k$ is the $k$-th eigenvalue of the adjacency matrix $A$. Note that $\beta(G) = 1$ if and only if the network $G$ is bipartite.  

The contribution of an individual node $i$ to the network bipartivity, denoted $c_\beta(i)$, can be computed using the node-level subgraph centrality:
\[
c_\beta(i) = \frac{\sum_{k=1}^N v_k(i)^2 \cosh(\lambda_k)}{\sum_{k=1}^N v_k(i)^2 e^{\lambda_k}},
\]
where $v_k(i)$ is the $i$-th component of the eigenvector $v_k$ corresponding to eigenvalue $\lambda_k$ of $A$. Thus, bipartivity index captures the extent to which node $i$ participates in even-length closed walks relative to all closed walks passing through it.

\section{Borgatti's effective size}

\emph{Borgatti’s effective size}\index{effective size!Borgatti’s}, introduced by \cite{Borgatti1997}, is a simplified reformulation of Burt’s effective size measure \cite{Burt1992}. It defines the effective size \( c_{\mathrm{BES}}(i) \) of ego \( i \)’s network as
\begin{equation*}
    c_{\mathrm{BES}}(i) = d_i - c_{r}(i),
\end{equation*}
where \( d_i \) denotes the degree of node \( i \), and \( c_{r}(i) \) represents the redundancy measure also proposed by \cite{Borgatti1997}. This formulation expresses the number of non-redundant contacts in ego \( i \)’s network, emphasizing that relationships linking \( i \) to otherwise unconnected individuals provide access to diverse and independent sources of information or resources.

\section{Borda centrality}

\emph{Borda centrality}\index{Borda centrality} \cite{Brandes2022} is a centrality measure inspired by the Borda count voting mechanism from social choice theory, which aggregates the preferences of the voters over a given set of alternatives \cite{Arrow2010,Shvydun2016}. In networks, the preferences of the nodes can be defined over a set of other nodes in the network based on their shortest-path distances. Specifically, for a node $i \in \mathcal{N}$, the \emph{distance-based preference} relation is defined as
\[
  j \succ_i k \quad \text{if and only if} \quad d_{ij} < d_{ik},
\]
that is, node $j$ is preferred to node $k$ by node $i$ if $j$ is strictly closer to $i$ than $k$ is. Thus, the distance-based preference relation of node~$i$ constitutes a weak order (irreflexive, transitive and negatively transitive binary relation) over the set $\mathcal{N} \setminus \{i\}$, where each layer~$k$ corresponds to the indifference class of nodes located at distance $k \in \{1, \ldots, \max_j d_{ij}\}$ from node~$i$.

The \emph{Borda score} of a node $i$ is then given by
\[
  c_{\mathrm{Borda}}(i) = \sum_{j \in \mathcal{N} \setminus \{i\}} 
  \left(
    \big|\{ k \in \mathcal{N} \setminus \{i\} : i \succ_j k \}\big|
    - 
    \big|\{ k \in \mathcal{N} \setminus \{i\} : k \succ_j i \}\big|
  \right).
\]

Hence, the Borda score of node $i$ is obtained by summing, over all other nodes $j$, the difference between (i) the number of nodes that are farther from $j$ than $i$ is, and (ii) the number of nodes that are closer to $j$ than $i$ is. A node receives a higher Borda score if it is, on average, relatively close to many other nodes in the network.

\section{BottleNeck centrality}

The \emph{BottleNeck} centrality identifies nodes that act as bottlenecks in a network by inspecting shortest-path trees \cite{Przulj2004}. For each source node $s$ let $T_s$ denote a shortest-path tree rooted at $s$ and let $V(T_s)$ denote the set of nodes in $T_s$ reachable from $s$, excluding the root $s$. The BottleNeck centrality of node $i$ is defined as
\begin{equation*}
    c_{\text{BottleNeck}}(i) = \sum_{s=1}^{N} P_s(i),
\end{equation*}
where
\begin{equation*}
    P_s(i) =
    \begin{cases}
        1, & \text{if more than } |V(T_s)|/4 \text{ shortest paths from } s \text{ to other nodes in } T_s \text{ pass through } i, \\
        0, & \text{otherwise}.
    \end{cases}
\end{equation*}

The total score $c_{\text{BottleNeck}}(i)$ thus counts how often node $i$ serves as a major intermediary across all sources. Intuitively, nodes with high BottleNeck centrality are critical intermediaries through which a significant fraction of shortest paths pass, making them important for the connectivity and flow within the network.

\section{BridgeRank}

\emph{BridgeRank}\index{BridgeRank} is a community-aware centrality measure proposed by Salavati \textit{et al.} \cite{Salavati2018}, aimed at identifying nodes that play a critical role in maintaining connectivity between communities. The method assumes that the graph $G$ exhibits a community structure. BridgeRank begins by partitioning the network into communities using a community detection algorithm, specifically the Louvain algorithm. Within each community, nodes are subsequently ranked based on their betweenness centrality, computed using shortest paths restricted to members of that community. From each community, the most influential node is selected to form the set of critical nodes, denoted $S$. 

The BridgeRank score of node $i$, denoted $c_{\mathrm{BridgeRank}}(i)$, is then defined as the inverse of the sum of its shortest-path distances to the critical nodes:
\[
c_{\mathrm{BridgeRank}}(i) = \frac{1}{\sum_{j \in S} d_{ij}}.
\]

Salavati \textit{et al.} \cite{Salavati2018} also proposed a modified version of BridgeRank that weights the original score and selects multiple nodes from each community based on the community’s density.

\section{Bridging capital}

\emph{Bridging capital}\index{bridging capital} identifies nodes that act as unique or vital connectors between otherwise disparate groups, with the ability to acquire and control the flow of valuable knowledge \cite{Jackson2020}. The bridging capital of node \(i\), denoted \(c_{\mathrm{brid}}(i)\), measures the total decrease in information flow over \(T\) periods after an incident edge from node \(i\) is removed from the network \(G\):
\[
    c_{\mathrm{brid}}(i) = \sum_{j \in \mathcal{N}(i)} \sum_{t=1}^{T} \sum_{k_1, k_2} v_{k_1 k_2} \left( (A^t)_{k_1 k_2} - (A_{(i,j)}^t)_{k_1 k_2} \right),
\]
where \(v_{k_1 k_2}\) represents the value of information flowing from node \(k_1\) to node \(k_2\) that is potentially affected by the removal of edge \((i,j)\), and \(A_{(i,j)}\) is the adjacency matrix of the graph \(G\) with edge \((i,j)\) removed.  Thus, \(c_{\mathrm{brid}}(i)\) captures the total contribution of node \(i\) to bridging information across the network, reflecting its role in connecting otherwise disconnected or weakly connected regions.

\section{Bridging centrality}

\emph{Bridging centrality}\index{bridging centrality} is a betweenness-based measure proposed in \cite{Hwang2006} to identify bridging nodes, which are the nodes that connect densely connected components in a network. The bridging centrality of node \(i\), denoted \(c_{\mathrm{bridging}}(i)\), is defined as the product of its betweenness centrality \(c_{\mathrm{betw}}(i)\) and its bridging coefficient \(\beta_c(i)\), capturing both global and local features of the node:
\begin{equation*}
   c_{\mathrm{bridging}}(i) = c_{\mathrm{betw}}(i) \cdot \beta_c(i) 
   = \sum_{j=1}^{N} \sum_{k=1}^{N} \frac{\sigma_{jk}(i)}{\sigma_{jk}} \cdot \frac{d_i^{-1}}{\sum_{j \in \mathcal{N}(i)} d_j^{-1}},
\end{equation*}
where  \(d_i\) is the degree of node \(i\), \(\sigma_{jk}\) is the number of shortest paths between nodes \(j\) and \(k\), and \(\sigma_{jk}(i)\) is the number of those paths that pass through node \(i\).  

A higher \(c_{\mathrm{bridging}}(i)\) indicates that more information flows through node \(i\) (high betweenness) and that the node serves as a critical connector between densely connected regions (high bridging coefficient).

\section{Bridging coefficient}
The \emph{bridging coefficient}\index{bridging coefficient} \(\beta_c(i)\) of a node determines the extent how well the node is located between high degree nodes \cite{Hwang2006}. Intuitively, there should be more congestion on the smaller degree nodes if an unit electrical current arrives on a node since the smaller degree nodes have lesser number of outlets than the bigger degree nodes have. So, if we consider the reciprocal of the degree of a node as the “resistance” of the node, the bridging coefficient \(\beta_c(i)\) of node \(i\) can be viewed as the ratio of the resistance of a node \(i\) to the sum of the resistance of the neighbors. 
\begin{equation*}
   \beta_c(i) = \frac{d_i^{-1}}{\sum_{j \in \mathcal{N}(i)}{\frac{1}{d_j}}},
\end{equation*}
where \(d_i=\sum_{j=1}^{N}{a_{ij}}\) is the degree of node \(i\) and \(\mathcal{N}(i)\) is the set of \(i\)'s neighbors. Critical bridging nodes, typically representing rate limiting points in the network and because they connect its densely connected regions, have high “resistance.”

\section{Bundle index}

The \emph{bundle index}\index{bundle index} is a power index that quantifies the individual and group influence of nodes in a network \cite{YAleskerov2020}. Each node $i$ is assigned an individual threshold of influence $q_i$, representing the level at which the node becomes affected (e.g., $q_i = 3$). A group of nodes $\Omega(i) \subset \mathcal{N}$ is called \emph{critical} for node $i$ if their combined influence exceeds the threshold:
\[
\sum_{k \in \Omega(i)} w_{ki} \geq q_i,
\]
where $w_{ki}$ denotes the weight of the link from node $k$ to node $i$.  

The bundle influence index of node $i$, denoted $c_{BI}(i)$, is defined as the number of critical groups for that node:
\[
c_{BI}(i) = \left| \left\{ \Omega(i) \subseteq \mathcal{N}(i) \;|\; \sum_{k \in \Omega(i)} w_{ki} \geq q_i \right\} \right|.
\]

Since the number of potential critical groups can grow exponentially, a variant of the bundle index considers only subsets of size up to $k$. Aleskerov and Yakuba \cite{YAleskerov2020} also propose an extension that accounts for indirect influence: in the order-$k$ version, influence is assessed over $(k+1)$-hop neighborhoods, with link strength defined as the maximum bottleneck capacity among all paths of length $(k+1)$.  

For unweighted networks, the bundle influence index reduces to
\[
c_{BI}(i) = \sum_{l=\lceil q_i \rceil}^{d_i} \binom{d_i}{l},
\]
where \(d_i\) denotes the degree of node \(i\), and \(\lceil \cdot \rceil\) denotes the ceiling function. 

The bundle index has been applied in diverse contexts, including the analysis of influential countries in food trade networks \cite{Aleskerov2022} and oil trade networks \cite{Aleskerov2023}, trade between economic sectors of different countries \cite{Aleskerov2024a} and bibliometric analysis of publications on Parkinson’s disease \cite{Aleskerov2024b}.

\section{Burt's constraint}

\emph{Burt's constraint}\index{Burt's constraint}, also called network constraint, was proposed in \cite{Burt1992,Burt2004} to quantify the extent to which a node's connections constrain its brokerage opportunities in a network. The constraint of a node \(i\) is high if its neighbors \(\mathcal{N}(i)\) communicate heavily with one another (dense network) or if they share information indirectly through a central contact (hierarchical network).  

Mathematically, the strength \(p_{ij}\) of a link \((i,j)\) is defined as the proportion of \(i\)'s time or energy invested in contact \(j\):
\begin{equation*} \label{eq_burt_energy}
   p_{ij} = \frac{a_{ij}+a_{ji}}{\sum_{k \in \mathcal{N}(i)} (a_{ik}+a_{ki})}.
\end{equation*}

Burt's constraint of node \(i\) is then
\begin{equation*}
   c_{\mathrm{Burt}}(i) = \sum_{j \in \mathcal{N}(i)} \left( p_{ij} + \sum_{k \in \mathcal{N}(i) \setminus \{j\}} p_{ik} p_{kj} \right)^2.
\end{equation*}

For unweighted undirected networks, this simplifies to
\begin{equation*}
   c_{\mathrm{Burt}}(i) = \sum_{j \in \mathcal{N}(i)} \left( \frac{1}{|\mathcal{N}(i)|} + \sum_{k \in \mathcal{N}(i) \setminus \{j\}} \frac{1}{|\mathcal{N}(i)||\mathcal{N}(k)|} \right)^2,
\end{equation*}
where \(|\mathcal{N}(i)|\) denotes the degree of node \(i\).

\section{Cc-Burt centrality}

\emph{Cc-Burt} centrality\index{Cc-Burt centrality} is a hybrid centrality measure that combines the concept of \emph{structural holes} with closeness centrality to identify key nodes in a network \cite{Zhu2017}. The structural hole theory accounts for both a node's degree and the topological relationships among its neighbors, while closeness centrality reflects the node's global position in the network.

For a node \(i\), the \textsc{Cc-Burt} centrality \(c_{\textsc{Cc-Burt}}(i)\) is defined as
\[
c_{\textsc{Cc-Burt}}(i) = \frac{1}{N} \left( \frac{1}{c_{cl}(i)} + \sum_{j \in \mathcal{N}(i)} \frac{c_{\mathrm{Burt}}(j)}{c_{cl}(j)} \right),
\]
where \(c_{cl}(i)\) is the closeness centrality of node \(i\) and \(c_{\mathrm{Burt}}(i)\) is the Burt's constraint of node \(i\).

The Cc-Burt centrality integrates local structural constraints and global network position to provide a more comprehensive evaluation of node influence.

\section{Centroid centrality}

The \emph{centroid centrality}\index{centroid centrality} evaluates how close a node is to all other nodes in a network from a game-theoretical perspective \cite{Brandes2005,Scardoni2009}. It considers pairwise comparisons between nodes to identify those that are, on average, more centrally located. 

For a pair of nodes $i$ and $j$, let $\gamma_i(j)$ denote the number of nodes that are closer (in terms of shortest-path distance) to $i$ than to $j$:
\[
\gamma_i(j) = \left| \left\{ v \in \mathcal{N} : d_{iv} < d_{jv} \right\} \right|,
\]
where $d_{iv}$ is the shortest-path distance between nodes $i$ and $v$.

The centroid centrality of node $i$ is then defined as
\begin{equation*}
    c_{\text{centroid}}(i) = \min_{j \in \mathcal{N} \setminus \{i\}} f(i,j),
\end{equation*}
where
\begin{equation*}
    f(i,j) = \gamma_i(j) - \gamma_j(i).
\end{equation*}

Intuitively, centroid centrality quantifies the positional advantage of node $i$ within the network. Nodes with high centroid values are closer, on average, to a larger portion of the network than competing nodes, making them strategically well-positioned or ``centrally dominant'' within the network structure.

\section{Classified neighbors centrality}

\emph{Classified neighbors centrality}\index{classified neighbors centrality} is an algorithm for identifying influential spreaders that classifies the neighbors of a node according to their relative order of removal during the $k$-shell decomposition \cite{CLi2018}. Each neighbor $j$ of node $i$ is assigned to one of four groups based on the $k$-shell decomposition:

\begin{enumerate}
    \item \textit{Upper neighbors ($U_i$)}: $k$-shell value of node $j$ is greater than that of node $i$.
    \item \textit{Equal-upper neighbors ($EU_i$)}: nodes $j$ and $i$ have the same $k$-shell value, but node $j$ is removed later than node $i$.
    \item \textit{Equal-lower neighbors ($EL_i$)}: nodes $j$ and $i$ have the same $k$-shell value, but node $j$ is removed earlier than node $i$.
    \item \textit{Lower neighbors ($L_i$)}: $k$-shell value of node $j$ is less than that of node $i$.
\end{enumerate}

The centrality of node $i$, denoted $c_{CN}(i)$, is defined as
\[
c_{CN}(i) = \alpha |U_i| + \beta |EU_i| + \gamma |EL_i| + \mu |L_i|,
\]
where $\alpha, \beta, \gamma, \mu \in [0,1]$ are tunable parameters. Li \textit{et al.} \cite{CLi2018} suggest $\alpha = 0.4$, $\beta = 0.35$, $\gamma = 0.25$, and $\mu = 0.1$, which correspond to the normalized values $\alpha = 0.364$, $\beta = 0.318$, $\gamma = 0.227$, and $\mu = 0.091$.

\section{Clique level}

\emph{Clique level}\index{clique level} is a centrality measure designed to identify essential nodes in protein-protein interaction networks (PINs) by quantifying how strongly a node is embedded in tightly connected subgraphs, which often indicates functional importance \cite{Hwang2009}. The clique level \(c_{\mathrm{CL}}(i)\) of node \(i\) is defined as the size of the largest clique containing \(i\):
\[
c_{\mathrm{CL}}(i) =
\begin{cases}
0, & \text{if node } i \text{ does not belong to any clique},\\
k(i), & \text{otherwise,}
\end{cases}
\]
where \(k(i)\) denotes the size of the largest clique that includes node \(i\). Hence, clique level highlights nodes that are part of highly interconnected groups, which are often critical for network stability and biological function.

\section{Closeness centrality}

\emph{Closeness centrality}\index{closeness centrality} is a measure of how central a node is within a network, based on its shortest-path distances to all other nodes \cite{Bavelas1950,Sabidussi1966,Freeman1978}.  Intuitively, a node is central in terms of closeness if it can efficiently reach all other nodes in the network, reflecting its potential to access and disseminate information, as well as to exert influence across the network. The closeness centrality \(c_{cl}(i)\) of a node \(i\) is defined as the inverse of the average shortest-path distance from \(i\) to all other nodes in the network:

\begin{equation*} \label{eq_closeness}
    c_{cl}(i) = \frac{N-1}{\sum_{j \neq i} d_{ij}},
\end{equation*}
where \(d_{ij}\) is the length of the shortest path from node \(i\) to node \(j\). Closeness centrality is typically interpreted as an indicator of either access efficiency or independence from intermediaries \cite{Brandes2016}. Nodes with shorter average distances to others can exchange information with fewer transmissions, in less time, and at lower cost \cite{Freeman1978}.

Closeness centrality is defined only for connected graphs, since shortest-path distances between nodes in different components are undefined. Extensions of closeness centrality to graphs with multiple connected components are discussed in \cite{WassermanFaust1994}. The closeness centrality without \(N-1\) in the numerator is also known as the \textit{barycenter}\index{barycenter centrality} centrality \cite{Viswanath2009}. The inverse of the barycenter centrality is also known as the \textit{Wiener index}\index{Wiener index} centrality \cite{Rocco2022}.

\section{Closeness vitality}

The \emph{closeness vitality}\index{closeness vitality} measures how the overall efficiency of a network changes when a given node is removed \cite{Brandes2005}. Specifically, it quantifies the variation in the total sum of shortest-path distances between all pairs of nodes after excluding node $i$ from the graph $G$. Let $G_i$ denote the subgraph obtained by removing node $i$ and its incident edges. The closeness vitality of node $i$ is defined as
\begin{equation*}
    c_{\text{vitality}}(i) = W(G) - W(G_i)
    = \sum_{j=1}^{N} d_G(i,j)
      + \sum_{j \neq i} \sum_{k \neq i} \big( d_G(j,k) - d_{G_i}(j,k) \big),
\end{equation*}
where $W(G)$ is the \emph{Wiener index}\index{Wiener index} of $G$, defined as the total sum of shortest-path distances between all pairs of nodes in the network \cite{Wiener1947}. Here, $d_G(i,j)$ denotes the shortest-path distance between nodes $i$ and $j$ in $G$. A lower closeness vitality value indicates a more central node, since its removal causes a smaller increase in the total pairwise distances. However, if node $i$ is a \emph{cut-vertex} (or bridge endpoint), its removal disconnects the network, resulting in $c_{\text{vitality}}(i) = -\infty$.

\section{Clustered local-degree (CLD)}

The \emph{clustered local-degree (CLD) centrality}\index{clustered local-degree (CLD) centrality} integrates both the degree of neighboring nodes and the local topological structure surrounding each node. In this measure, the local clustering coefficient of a node quantifies the connectivity among its neighbors \cite{Li2018}. 
The centrality value \( c_{\mathrm{CLD}}(i) \) for node \( i \) is defined as
\begin{equation*}
   c_{\mathrm{CLD}}(i) = (1 + c_i) \sum_{j \in \mathcal{N}(i)} d_j,
\end{equation*}
where \( c_i \) denotes the clustering coefficient of node \( i \), 
\( \mathcal{N}(i) \) is the set of neighbors of node \( i \), 
and \( d_j \) represents the degree of neighbor \( j \). Hence, the CLD centrality accounts for both the number of neighbors and the connectivity among them.

\section{Clustering degree algorithm (CDA)}

The \emph{clustering degree algorithm}\index{clustering degree algorithm (CDA)} (CDA) is designed to identify influential spreaders in weighted networks \cite{QWang2018}. CDA combines the degree and strength of a node with the network topology and the differentiated contribution of its neighbors.

The CDA score of node $i$, denoted $c_{CDA}(i)$, is defined as
\[
c_{CDA}(i) = CD(i) + \sum_{j \in \mathcal{N}(i)} \frac{w_{ij}}{\max_{ij} w_{ij}} \, CD(j),
\]
where $CD(i)$ is the clustering degree of node $i$, given by
\[
CD(i) = \frac{\alpha \sum_{j=1}^N a_{ij} + (1-\alpha) \sum_{j=1}^N w_{ij}}{1 + \exp\Bigg[-\frac{\sum_{j,k} (w_{ij} + w_{ik}) a_{ij} a_{jk} a_{ki}}{2 (\sum_{j=1}^N w_{ij}) (\sum_{j=1}^N a_{ij} - 1)}\Bigg]},
\]
with $\alpha$ as a tunable parameter (set to 0.5). 

For unweighted networks, $CD$ is independent of $\alpha$, and the CDA score simplifies to
\[
c_{CDA}(i) = \frac{d_i}{1 + e^{-cl(i)}} + \sum_{j \in \mathcal{N}(i)} \frac{d_j}{1 + e^{-cl(j)}},
\]
where $d_i$ and $cl(i)$ denote the degree and clustering coefficient of node $i$, respectively.

\section{ClusterRank centrality}

The \emph{ClusterRank centrality}\index{ClusterRank} integrates both the degree information of a node and its neighbours, as well as the local clustering structure of the network \cite{Chen2013}. The ClusterRank score of a node $i$ is defined as
\begin{equation*}
    c_{\text{ClusterRank}}(i) = f(c_i) \sum_{j \in \mathcal{N}(i)} \left( d_j^{out} + 1 \right),
\end{equation*}
where $c_i$ denotes the clustering coefficient of node $i$, $\mathcal{N}(i)$ is the set of its neighbours (or followers), and $d_j^{out}$ is the out-degree of node $j$. The function $f(c_i)$ accounts for the influence of $i$’s local clustering on its centrality and is commonly defined as $f(c_i) = 10^{-c_i}$, which downweights nodes embedded in highly clustered regions. In this way, ClusterRank emphasizes nodes that connect different local clusters while still considering the importance of their neighbors.

\vfill

\section{Coleman-Theil disorder index}

The \emph{Coleman-Theil disorder index}\index{Coleman-Theil disorder index}, also known as the hierarchy index, quantifies the extent to which a node's aggregate Burt's constraint is concentrated on a single contact \cite{Burt1992}. High values indicate that most of the constraint arises from a single relationship, whereas low values reflect a more even distribution of constraint across multiple contacts.  

For node \(i\), the Coleman-Theil disorder index \(c_{\mathrm{CTDI}}(i)\) is defined as
\begin{equation*}
    c_{\mathrm{CTDI}}(i) = \frac{\sum_{j \in \mathcal{N}(i)} \tilde{c}_{ij} \ln(\tilde{c}_{ij})}{N_i \ln(N_i)},
\end{equation*}
where \(\mathcal{N}(i)\) is the set of neighbors of node \(i\), \(N_i = |\mathcal{N}(i)|\) is the number of neighbors and \(\tilde{c}_{ij}\) represents the relative contribution of contact \(j\) to the total constraint of node \(i\), defined as
\[
\tilde{c}_{ij} = \frac{c_{ij}}{\frac{1}{N_i} \sum_{k \in \mathcal{N}(i)} c_{ik}}.
\]

The term \(c_{ij}\) denotes the Burt's constraint imposed by contact \(j\) and is given by
\[
c_{ij} = \left( p_{ij} + \sum_{k \in \mathcal{N}(i) \setminus \{j\}} p_{ik} p_{kj} \right)^2,
\]
where \(p_{ij} = \frac{w_{ij}}{\sum_{k \in \mathcal{N}(i)} w_{ik}}\) is the proportion of node \(i\)'s connections invested in contact \(j\), and \(w_{ij}\) represents the weight of the edge between nodes \(i\) and \(j\) (equal to 1 for unweighted networks).

The Coleman-Theil disorder index attains its minimum value of 0 when constraint is equally distributed among all neighbors, and reaches its maximum of 1 when all constraint is concentrated on a single neighbor.

\section{Collective influence}

The \emph{collective influence}\index{collective influence} (CI) quantifies the centrality of a node by considering not only its degree but also the degrees of nodes in its surrounding neighbourhood at a given distance $l$ \cite{Morone2015}. Formally, the CI of node $i$ is defined as
\begin{equation*}
    c_{CI}(i) = (d_i - 1) \sum_{j \in \mathcal{N}^{(l)}(i) } (d_j - 1),
\end{equation*}
where $k_i$ is the degree of node $i$, and $\mathcal{N}^{(l)}(i) $ denotes the frontier of the ball of radius $l$ centered at $i$, i.e., the set of nodes at distance exactly $l$ from $i$. The parameter $l$ is typically chosen such that it does not exceed the diameter of the network. 

Morone and Makse \cite{Morone2015} also proposed an iterative version of the CI algorithm, which allows the identification of an optimal set of influential nodes in the network.

\vfill

\section{Collective network entanglement (CNE)}

\emph{Collective network entanglement}\index{collective network entanglement (CNE)} (CNE) is an induced, entropy-based centrality measure that quantifies the role of individual nodes in preserving the functional diversity of a network \cite{Ghavasieh2021}. The CNE score \(c_{CNE}(i)\) of node \(i \in \mathcal{N}\) is defined as the change in von Neumann entropy caused by the detachment of the node and its incident edges:
\[
c_{CNE}(i) = \left( S_\beta(G^*_i) + S_\beta(G_i)\right) - S_\beta(G),
\]
where \(G_i\) denotes the subgraph obtained by removing node \(i\), \(G^*_i\) is the star graph of node \(i\), and \(S_\beta(G)\) is the von Neumann entropy of \(G\). The entropy is computed from a density matrix \(\rho\) derived from the network Laplacian \(L(G)\):
\[
\rho = \frac{e^{-\beta L(G)}}{\mathrm{tr}(e^{-\beta L(G)})}, \qquad
S_\beta(G) = -\mathrm{tr}(\rho \log \rho),
\]
with \(\beta > 0\) as a diffusion time parameter controlling the scale of information propagation.  

At very short diffusion times (\(\beta \to 0\)), CNE reduces to degree centrality, reflecting a node's immediate neighborhood. At very long times (\(\beta \to \infty\)), it captures a node's contribution to the overall network connectivity. Ghavasieh \textit{et al.} \cite{Ghavasieh2021} demonstrate that \(\beta = \beta_c\) provides an appropriate timescale for network disintegration, as confirmed in experiments on social, biological, and transportation networks.

\section{Comm Centrality}

\emph{Comm centrality}\index{Comm centrality} is a community-based centrality measure that quantifies a node’s importance by combining its intra- and inter-community connectivity through a weighted formulation \cite{Gupta2016}.  
Assume that the network \(G\) has a community structure consisting of \(K > 1\) communities.  
For a node \(i\) belonging to community \(C\), the Comm centrality \(c_{comm}(i)\) is defined as
\begin{equation*}
    c_{comm}(i) = (1+\mu_C) \left( \frac{d_i^{in}}{\max_{j \in C} d_j^{in}} \cdot R \right)
    + (1-\mu_C) \left( \frac{d_i^{out}}{\max_{j \in C} d_j^{out}} \cdot R \right)^2,
\end{equation*}
where \(d_i^{in}\) is the number of links connecting node \(i\) to other nodes within the same community, \(d_i^{out}\) is the number of links from node \(i\) to nodes in other communities, and \(R\) is a scaling parameter.  
The community mixing parameter \(\mu_C\) is defined as
\begin{equation*}
    \mu_C = \frac{1}{|C|} \sum_{j \in C} \frac{d_j^{out}}{d_j},
\end{equation*}
representing the average proportion of inter-community links within community \(C\).  
Gupta \textit{et al.} \cite{Gupta2016} suggest setting \(R = \max_{j \in C} d_j^{in}\).

Nodes with high Comm centrality values either have strong intra-community connectivity or act as key bridges between communities, depending on the structural mixing parameter \(\mu_C\).

\section{Common out-neighbor (CON) score}

The \emph{common out-neighbor (CON)}\index{common out-neighbor (CON) score} score quantifies node similarity by counting multiplicities based on the minimum number of interactions \cite{Bonato2020}. Let \(CON(i,j)\) denote the number of common out-neighbors shared by nodes \(i\) and \(j\), defined as
\begin{equation*}
    CON(i,j) = \sum_{k=1}^{N} \min(a_{ik}, a_{jk})=\sum_{k=1}^{N} a_{ik}a_{jk} = |\mathcal{N}^{out}(i) \cap \mathcal{N}^{out}(j) |.
\end{equation*}
Then, the CON score of node \(i\), denoted by \(c_{CON}(i)\), is given by
\begin{equation*}
    c_{CON}(i) = \sum_{j=1}^{N} CON(i,j) = \sum_{j=1}^{N} \sum_{k=1}^{N} a_{ik}a_{jk} = \sum_{k=1}^{N}a_{ik} \sum_{j=1}^{N} a_{jk} = \sum_{k \in \mathcal{N}^{out}(i)} d_k^{in}.
\end{equation*}

For undirected networks, the common out-neighbor (CON) score of node $i$ reduces to the sum of the degrees of its neighbors.

\section{Communicability betweenness centrality}

\emph{Communicability betweenness centrality}\index{betweenness centrality!communicability} extends traditional betweenness centrality by considering information flow along all possible walks in the network, with longer walks weighted less heavily \cite{Estrada2009}. The communicability betweenness centrality of node $i$ is defined as
\begin{equation*}
    c_{\text{comm-betw}}(i) = \frac{1}{(N-1)(N-2)} \sum_{j \neq i} \sum_{k \neq i} \frac{G_{jik}}{G_{jk}},
\end{equation*}
where $G_{jk} = (e^{A})_{jk}$ is the \emph{communicability} between nodes $j$ and $k$, obtained from the matrix exponential of the adjacency matrix $A$. It represents the total contribution of all possible walks between $j$ and $k$, with longer walks weighted less due to the factorial scaling in the series expansion of $e^{A}$. The term
\[
G_{jik} = (e^{A})_{jk} - (e^{A + E(i)})_{jk}
\]
quantifies the reduction in communicability between $j$ and $k$ when node $i$ is removed from the network. Here, $E(i)$ is an $N \times N$ matrix whose nonzero entries appear only in row and column $i$, taking the value $-1$ wherever the corresponding element of $A$ equals $+1$. Consequently, the matrix $A + E(i)$ corresponds to the adjacency matrix of the graph in which all edges incident to node $i$ have been deleted. Thus, $G_{jik}$ captures the portion of walks between $j$ and $k$ that rely on node $i$. A node with a high communicability betweenness plays a crucial intermediary role in facilitating indirect information flow across the network.

Intuitively, communicability betweenness quantifies how much a node contributes to the overall flow of information across the network, accounting for both direct and indirect paths while penalizing longer walks.

\vfill

\section{Community centrality}

\emph{Community centrality}\index{community centrality} is a community-based measure of node centrality, proposed in \cite{Kalinka2011}. It assumes that the graph \(G\) has a community structure and that nodes may belong to multiple communities.  

The community centrality of node \(i\) quantifies the number of communities it belongs to while taking into account the similarity between these communities:
\begin{equation*}
   c_{\mathrm{community}}(i) = \sum_{j \in C_i} \left( 1 - \frac{1}{|C_i|} \sum_{k \in C_i} S(j,k) \right),
\end{equation*}
where \(C_i\) is the set of communities containing node \(i\), and \(S(j,k)\) is the similarity between communities \(j\) and \(k\), computed using the Jaccard coefficient based on the number of shared nodes.  

A node achieves the highest community centrality if it belongs to many communities that are largely distinct from one another. In \cite{Kalinka2011}, community membership is determined using the link community detection algorithm proposed in \cite{Ahn2010}.

\section{Community hub-bridge (CHB) measure}

\emph{Community hub-bridge (CHB) measure}\index{community hub-bridge (CHB) measure} is a community-based centrality metric that evaluates the importance of a node by considering both its intra- and inter-community connections \cite{Ghalmane2019}.  
Assume that the network \(G\) has a community structure consisting of \(K > 1\) communities.  
For a node \(i\) belonging to community \(C_k\), the CHB centrality \(c_{CHB}(i)\) is defined as
\begin{equation*}
    c_{CHB}(i) = |C_k|\, d_i^{in} + d_i^{out} \sum_{l \neq k} \left( \lor_{j \in C_l} a_{ij} \right),
\end{equation*}
where \(d_i^{in}\) is the number of intra-community links of node \(i\) (connections within \(C_k\)), \(d_i^{out}\) is the number of inter-community links (connections to nodes in other communities), and \(\lor\) denotes the logical OR operator.  
The expression \(\left( \lor_{j \in C_l} a_{ij} \right) = 1\) if and only if node \(i\) is connected to at least one node \(j\) in community \(C_l\); otherwise, it equals zero.  

Nodes with high CHB values act as both \emph{hubs} (densely connected within their own community) and \emph{bridges} (linking multiple communities), making them crucial for information flow and inter-community connectivity.

\section{Community-based centrality (CbC)}

\emph{Community-based centrality}\index{community-based centrality} (CbC) is a measure designed to identify influential spreaders in complex networks \cite{Zhao2015}. The method distinguishes two types of links for each node: \emph{strong links}, which connect nodes within the same community, and \emph{weak links}, which connect nodes across different communities. The importance of a node is determined by both its link characteristics and the sizes of the communities to which it is connected. Mathematically, the CbC of node \(i\) is defined as
\begin{equation*}
    c_{\mathrm{CbC}}(i) = \sum_{s=1}^{K} k_{is} \frac{|C_s|}{N} = \sum_{s=1}^{K} \sum_{j \in C_s \setminus \{i\}} a_{ij} \frac{|C_s|}{N},
\end{equation*}
where \(k_{is}\) is the number of links from node \(i\) to nodes in community \(C_s\), \(K\) is the total number of communities \(C_1, \dots, C_K\), \(|C_s|\) denotes the size of community \(C_s\), and \(a_{ij}\) are elements of the adjacency matrix. In \cite{Zhao2015}, communities are identified using the CNM algorithm, which employs the Clauset-Newman-Moore greedy modularity maximization method.

CbC generalizes classical degree centrality by incorporating community structure. In the limiting case where the entire network forms a single community, CbC reduces to the standard degree centrality. Conversely, when each node constitutes its own community, CbC corresponds to the degree of the node normalized by the total number of nodes in the network.

\section{Community-based mediator (CbM)}

The \emph{community-based mediator}\index{community-based mediator (CbM)} (CbM) is a community-aware centrality measure that evaluates a node’s influence based on the entropy of its connections across different communities \cite{Tulu2018}.  
Let the network \(G\) consist of \(K\) communities \(C_1, C_2, \dots, C_K\).  
The CbM centrality \(c_{CbM}(i)\) of node \(i\) is defined as
\begin{equation*}
    c_{CbM}(i) = H_i \frac{d_i}{2L},
\end{equation*}
where \(L\) is the total number of edges in \(G\), \(d_i\) is the degree of node \(i\), and \(H_i\) is the entropy associated with the distribution of its connections among the communities:
\begin{equation*}
    H_i = -\sum_{k=1}^{K} \frac{d_i(C_k)}{d_i} \log \frac{d_i(C_k)}{d_i}.
\end{equation*}
Here, \(d_i(C_k) = \sum_{j \in C_k} a_{ij}\) denotes the number of links from node \(i\) to nodes in community \(C_k\).  

Nodes that distribute their links more evenly across different communities have higher entropy \(H_i\) and thus higher \(c_{CbM}(i)\), reflecting their stronger mediating role in inter-community connectivity.

\section{Contribution centrality}

\emph{Contribution centrality}\index{contribution centrality} is a spectral centrality measure based on structural dissimilarity \cite{Alvarez-Socorro2015}. The contribution centrality of node \(i\), denoted as \(c_{\mathrm{Contr}}(i)\), is proportional to the sum of the centralities of its neighboring nodes, weighted by their topological contributions. Formally,
\begin{equation*}
    c_{\mathrm{Contr}}(i) = \frac{1}{\lambda} \sum_{j=1}^{N} a_{ij} D_{ij} c_{\mathrm{Contr}}(j),
\end{equation*}
where \(a_{ij}\) is the adjacency matrix element and \(D_{ij}\) is a structural dissimilarity measure defined as
\begin{equation*}
    D_{ij} = 1 - \frac{|\mathcal{N}(i) \cap \mathcal{N}(j)|}{|\mathcal{N}(i) \cup \mathcal{N}(j)|},
\end{equation*}
with \(\mathcal{N}(i)\) denoting the set of neighbors of node \(i\).  

The centrality equation can be expressed as an eigenvalue problem. Defining \(A_D = A \circ D\), where \(\circ\) denotes the Hadamard (element-wise) product, the contribution centrality vector \(c_{\mathrm{Contr}}\) satisfies
\begin{equation*}
    A_D \, c_{\mathrm{Contr}} = \lambda \, c_{\mathrm{Contr}},
\end{equation*}
with \(\lambda = \lambda_{\max}\) being the dominant eigenvalue of \(A_D\). The corresponding principal eigenvector \(c_{\mathrm{Contr}}\) gives the contribution centralities of the nodes, where larger values indicate a greater structural influence within the network.

\section{ControlRank}

The \emph{ControlRank}\index{ControlRank} index is a spectral measure designed to quantify node importance in both static and dynamic networks under control settings \cite{JZhou2018}. ControlRank is based on linear control theory and evaluates the influence of a node by analyzing the minimum eigenvalues of matrices derived from the network's Laplacian. A larger minimum eigenvalue indicates that the node contributes more to the network’s overall stability and accelerates convergence dynamics.

The ControlRank index of node $i$ is defined as
\[
c_{ControlRank}(i) = \lambda_i(L^s),
\]
where $\lambda_i(L^s)$ is the minimum eigenvalue of the principal minor of $(L + L^T)/2$ obtained by removing the $i$-th row and column from the Laplacian matrix $L$.

\section{Copeland centrality}

\emph{Copeland centrality}\index{Copeland centrality} \cite{Brandes2022} is a centrality measure inspired by the Copeland voting rule from social choice theory, which aggregates the preferences of voters over a given set of alternatives based on the majority relation \cite{Arrow2010,Shvydun2016}. In networks, the preferences of the nodes can be defined over a set of other nodes based on their shortest-path distances. Specifically, for a node $i \in \mathcal{N}$, the \emph{distance-based preference} relation is defined as
\[
  j \succ_i k \quad \text{if and only if} \quad d_{ij} < d_{ik},
\]
that is, node $j$ is preferred to node $k$ by node $i$ if $j$ is strictly closer to $i$ than $k$ is. Thus, the distance-based preference relation of node~$i$ constitutes a weak order (irreflexive, transitive and negatively transitive binary relation) over the set $\mathcal{N} \setminus \{i\}$, where each layer~$k$ corresponds to the indifference class of nodes located at distance $k \in \{1, \ldots, \max_j d_{ij}\}$ from node~$i$.

The \emph{majority relation} $\mu$ between two nodes $j$ and $k$ is then defined as
\[
  j \mu k \quad \text{if and only if} \quad 
  \big|\{ i \in \mathcal{N} \setminus \{j,k\} : j \succ_i k \}\big| 
  > 
  \big|\{ i \in \mathcal{N} \setminus \{j,k\} : k \succ_i j \}\big|.
\]
In other words, node $j$ is said to dominate node $k$ if a strict majority of nodes in the network prefer $j$ to $k$, that is, if $j$ is closer than $k$ to more nodes.

The \emph{Copeland score} of a node $i$ is then defined as
\[
  c_{\mathrm{Copeland}}(i) 
  = 
  \big|\{ k \in \mathcal{N} \setminus \{i\} : i \mu k \}\big|
  - 
  \big|\{ k \in \mathcal{N} \setminus \{i\} : k \mu i \}\big|.
\]

The Copeland score of node $i$ is computed as the difference between the number of nodes it dominates and the number of nodes that dominate it. A node receives a higher Copeland score if it is preferred by more nodes, which corresponds to being relatively close to many nodes in the network.

\vfill

\section{CoreHD}

\emph{CoreHD}\index{CoreHD} is a heuristic algorithm for network decycling and dismantling, designed to identify a minimal set of nodes whose removal either eliminates all cycles or breaks the network into small disconnected components \cite{Zdeborová2016}. The algorithm generates a sequence of node removals that reflects the importance of each node in sustaining network connectivity. CoreHD operates as follows:

\begin{enumerate}
    \item Extract the 2-core of the network, consisting of nodes with degree at least 2.
    \item Remove the highest-degree node within the 2-core.
    \item Recompute the 2-core after removal. If the 2-core becomes empty, perform tree-breaking by removing the node whose removal causes the largest decrease in the size of the largest connected component.
    \item Repeat the above steps until the network is either acyclic or sufficiently fragmented.
\end{enumerate}

By focusing on the 2-core, CoreHD avoids removing peripheral nodes that do not contribute to cycles, thereby achieving near-optimal decycling and dismantling performance. The CoreHD approach has been shown to be effective for enhancing network robustness against failures and attacks.

\section{Correlation centrality (CoC)}

The \emph{correlation centrality}\index{correlation centrality (CoC)} (CoC) quantifies the influence of a node based on its correlation with all other nodes in the network \cite{Wenli2013}. Specifically, the influence of node \(i\) depends on the harmonic centrality of other nodes and the distances from node \(i\) to these nodes. The centrality of node \(i\) is defined as
\begin{equation*}
    c_{CoC}(i) = \frac{1}{N^2} \sum_{j=1}^N \left( \frac{\sigma_{ij}}{d_{ij}^\alpha} \sum_{k=1}^N \frac{1}{d_{jk}} \right),
\end{equation*}
where \(\sigma_{ij}\) is the number of shortest paths from node \(i\) to node \(j\), \(d_{ij}\) is the shortest distance from node \(i\) to node \(j\), and \(\alpha\) is an impact factor. Wenli \textit{et al.} \cite{Wenli2013} use \(\alpha = 3\). 

Correlation centrality assigns higher values to nodes that are close to highly central nodes, reflecting both the number of paths a node participates in and the overall accessibility of the network. We remark that the term “correlation” emphasizes that a node’s centrality depends on its relationship with other nodes’ centrality, rather than on a statistical correlation coefficient.

\section{Counting betweenness centrality (CBET)}

\emph{Counting betweenness centrality}\index{betweenness centrality!counting (CBET)} (CBET) extends the concept of current-flow betweenness (or Newman’s random-walk betweenness) \cite{Newman2005,Brandes2005} to directed networks with self-loops \cite{Blöchl2011}. The counting betweenness \( c_{\text{CBET}}(i) \) of a node \( i \) quantifies how frequently the node is visited during first-passage random walks, averaged over all source-target pairs.
\begin{equation*}
    c_{\text{CBET}}(i) = \frac{\sum_{s \in \mathcal{N}} \sum_{t \in \mathcal{N} \setminus \{s\}} N^{st}(i)}{N (N - 1)},
\end{equation*}
where \( N^{st}(i) \) denotes the expected number of times a random walker visits node \( i \) when traveling from source \( s \) to target \( t \), including visits due to self-loops.  If \( i \notin \{s, t\} \), then
\begin{equation*}
    N^{st}(i) = \sum_{j \neq t} \frac{N^{st}_{ij} + N^{st}_{ji}}{2},
\end{equation*}
where \( N^{st}_{ij} \) represents the expected number of times the walker uses the link \( (i, j) \).  
If \( i = s \), the walker visits the source node once more at the start of the walk, and thus
\begin{equation*}
    N^{st}(s) = \sum_{j \neq t} \frac{N^{st}_{sj} + N^{st}_{js}}{2} + 1.
\end{equation*}
If \( i = t \), the target node is visited exactly once:
\begin{equation*}
    N^{st}[t] = 1.
\end{equation*}

Blöchl \textit{et al.} \cite{Blöchl2011} emphasize that counting betweenness emphasizes self-loops more strongly than random-walk centrality. An implementation of CBET in \texttt{R} is available in \cite{DePaolis2022}.

\section{Cross-clique connectivity}

The \emph{cross-clique connectivity}\index{cross-clique connectivity} (CCC) quantifies how extensively a node participates in multiple cohesive groups within a network \cite{Faghani2013}. It is based on the idea that nodes belonging to several cliques (fully connected subgraphs) play a crucial role in linking otherwise separate dense regions of the network. Formally, the cross-clique connectivity of a node $i$ is defined as
\begin{equation*}
    c_{\text{ccc}}(i) = 
    \left| \left\{ C \subseteq \mathcal{N}: i \in C,\, C \text{ is a clique and } |C| \ge 3 \right\} \right|,
\end{equation*}
that is, the number of cliques of size at least three that include node $i$. Nodes with high $c_{\text{ccc}}(i)$ values are referred to as \emph{highly cross-connected} nodes. Such nodes serve as structural bridges between tightly connected groups, facilitating interactions and information flow across different cohesive communities within the network.

\section{Cumulative Contact Probability (CCP)}

\emph{Cumulative Contact Probability}\index{cumulative contact probability (CCP)} (CCP) was introduced by Gao \textit{et al.} \cite{Gao2009} as a centrality-based heuristic for the Single-Data Multicast (SDM) problem, which seeks to determine how to select the minimum number of relays required to achieve a target delivery ratio \(p\) within a time constraint \(T\) when delivering a data item to a set \(D\) of destinations. The CCP measure is derived from a Poisson model of contact processes in social networks. The centrality of node \(i\), denoted by \(c_{\text{CCP}}(i)\), is defined as
\begin{equation*}
c_{\text{CCP}}(i) = 1 - \frac{1}{N-1} \sum_{j=1}^{N} e^{-\lambda_{ij} T},
\end{equation*}
where \(\lambda_{ij}\) represents the contact rate between nodes \(i\) and \(j\). In unweighted networks, the contact rate \(\lambda_{ij}\) reduces to a binary indicator of link existence, that is, \(\lambda_{ij} = a_{ij}\), where \(A = [a_{ij}]\) denotes the adjacency matrix of the network. Hence, the CCP index quantifies the average probability that a randomly chosen node in the network is contacted by node \(i\) within the time interval \(T\).

\vfill

\section{Current-flow betweenness centrality}
The \emph{current-flow betweenness centrality}\index{betweenness centrality!current-flow}, also known as random-walk betweenness centrality, is discussed in \cite{Brandes2005,Newman2005}. In contrast to traditional betweenness centrality, which assumes that information spreads only along shortest paths, this measure relaxes that assumption by including contributions from essentially all paths between nodes. More precisely, information originating from a source node \(s\) can pass through randomly selected intermediate nodes before reaching a target node \(t\). Current-flow betweenness centrality captures this process by modeling the spread of information as if it moves through the network like an electrical current. The centrality of node \(i\), denoted \(c_{cfb}(i)\), is defined as
\begin{equation*}
    c_{cfb}(i) = \frac{\sum_{s,t\in \mathcal{N}} I_i^{st}}{N_B},
\end{equation*}
where \(I_i^{st}\) represents the current flowing through node \(i\) between a source node \(s\) and a target node \(t\) and \(N_B = N(N-1)\) is a normalizing constant. The resized approximation of current-flow betweenness (RCFB)\index{RCFB centrality}, a computationally efficient approximation of current-flow betweenness centrality, was proposed in \cite{Agryzkov2019}.

\section{Current-flow closeness centrality}
The \emph{current-flow closeness centrality}\index{closeness centrality!current-flow}, also known as information centrality, is a variant of closeness centrality that utilizes the concept of electrical current in a network \cite{Brandes2005,Stephenson1989}. It measures the importance of a node by considering not only the shortest paths but all possible paths through which information can flow. The centrality \(c_{cc}(i)\) of a node \(i\) is defined as
\begin{equation*}
c_{cc}(i) = \frac{N-1}{\sum_{j \neq i} \bigl( p_{ij}(i) - p_{ij}(j) \bigr)},
\end{equation*}
where \(p_{ij}(i)\) is the absolute electrical potential of node \(i\) when a unit current is injected at node \(i\) and extracted at node \(j\). The difference \(p_{ij}(i) - p_{ij}(j)\)  represents the effective resistance between nodes \(i\) and \(j\), which captures how “difficult” it is for current to flow between nodes $i$ and $j$. This definition generalizes the concept of closeness centrality by taking into account contributions from all paths, rather than only the shortest ones. A more detailed description of the electrical potential \(p_{ij}(i)\) and its computation can be found in \cite{Brandes2005}.

\section{Curvature index}

The \emph{curvature index}\index{curvature index} quantifies the local curvature properties of nodes in a network \cite{Knill2012}. It is defined based on the Euler characteristic $\chi(G)$ of a graph $G$, given by
\[
\chi(G) = \sum_{k=0}^{N-1} (-1)^k v_k,
\]
where $v_k$ denotes the number of $(k{+}1)$-cliques in $G$. The curvature $K(i)$ at node $i$ is then defined as
\[
K(i) = \sum_{k=0}^{N-1} (-1)^k \frac{V_{k-1}(i)}{k+1},
\]
where $V_k(i)$ represents the number of $(k{+}1)$-cliques incident to node $i$.

According to the discrete Gauss-Bonnet theorem \cite{Knill2011}, the Euler characteristic of a graph equals the sum of the curvatures of all its nodes, i.e.,
\[
\chi(G) = \sum_{i=1}^{N} K(i).
\]

A truncated version of the curvature index, where the summation is limited to simplices of dimension $d \leq 2$ (i.e., cliques of size up to three), was proposed by Wu \textit{et al.} \cite{Wu2015} to reduce computational complexity while retaining essential geometric information.

\section{Decay centrality}

\emph{Decay centrality}\index{decay centrality} is a centrality measure that accounts for path lengths by assigning a weight to each path that decreases exponentially with its length \cite{Jackson2008}. Specifically, the contribution of a path is given by a decay parameter \(\delta \in (0,1)\) raised to the power of the path length, summarizing the diminishing influence of distant nodes, i.e.,
 \begin{equation*}
    c_{decay}(i) = \sum_{j \neq i}{\delta^{d_{ij}}}.
\end{equation*}
The decay centrality can be interpreted as the expected number of nodes that can reach \(i\) via shortest paths, where the probability of a successful move is defined by \(\delta\). The value of \(\delta\) depends on the network, but it is commonly assumed to be \(\delta = 0.5\).

\section{Decaying degree centrality (DDC)}

The \emph{decaying degree centrality}\index{degree centrality!decaying (DDC)} (DDC) is a generalization of the classical degree centrality that accounts for the influence of all nodes in a network, with contributions decaying exponentially with distance \cite{Bandyopadhyay2018}. For a node \(i \in \mathcal{N}\), the DDC score is defined as
\[
c_{DDC}(i) = \sum_{j=1}^N \frac{d_j}{N^{2 d_{ij}}},
\]
where \(d_j\) is the degree of node \(j\), \(d_{ij}\) is the shortest-path distance between nodes \(i\) and \(j\).  

Nodes with high DDC values are not only well-connected themselves but are also close to other highly connected nodes, reflecting both local and quasi-global influence in the network. The axiomatic properties of DDC are analyzed in \cite{Bandyopadhyay2018}.

\section{Degree and clustering coefficient (DCC) centrality}

The \emph{degree and clustering coefficient}\index{degree and clustering coefficient (DCC) centrality} (DCC) centrality quantifies node importance by combining information about a node’s degree and clustering coefficient with those of its neighbors \cite{YangWang2020}. The DCC centrality of node \(i\) is defined as
\begin{equation*}
    c_{DCC}(i) = \alpha I_d(i) + (1-\alpha) I_c(i),
\end{equation*}
where the degree-based term
\begin{equation*}
    I_d(i) = d_i + \sum_{j \in \mathcal{N}(i)} d_j
\end{equation*}
captures the contribution of node \(i\) and its immediate neighbors, and the clustering-based term
\begin{equation*}
    I_c(i) = e^{-c_i} \sum_{j \in \mathcal{N}^{(2)}(i)} c_j
\end{equation*}
accounts for the clustering coefficient of node \(i\) and its second-hop neighbors. Here, \(d_i\) is the degree of node \(i\), \(c_i\) is its clustering coefficient, \(\mathcal{N}(i)\) is the set of immediate neighbors, and \(\mathcal{N}^{(2)}(i)\) denotes the set of neighbors exactly two hops away.

The parameter \(\alpha \in [0,1]\) balances the relative importance of degree and clustering effects. Yang \textit{et al.} \cite{YangWang2020} suggest determining \(\alpha\) using an entropy-based approach:
\begin{equation*}
    \alpha = \frac{1 - E_1}{2 - E_1 - E_2},
\end{equation*}
where \(E_1\) and \(E_2\) are the entropies of the distributions of \(I_d(i)\) and \(I_c(i)\), respectively. 

Nodes with high DCC centrality thus have a combination of high connectivity and tightly clustered neighborhoods, reflecting both local and semi-local structural influence.

\section{Degree and clustering coefficient and location (DCL) centrality}

The \emph{degree and clustering coefficient and location}\index{degree and clustering coefficient and location (DCL) centrality} (DCL) measure is a local centrality metric that combines a node's degree, inverse clustering coefficient, and the connectivity among its neighbors to quantify its influence in the network \cite{Berahmand2019}. The centrality of node $i$ is defined as
\[
c_{DCL}(i) = \frac{d_i}{c_i + (1/d_i)} + \frac{\sum_{j \in \mathcal{N}(i)} d_j}{|E(\mathcal{N}(i)| + 1},
\]
where $d_i$ and $c_i$ are the degree and clustering coefficient of node $i$, $\mathcal{N}(i)$ is the set of neighbors of $i$ and $|E(\mathcal{N}(i))|$ denotes the number of links among the neighbors of node $i$.

The DCL centrality captures three aspects: the node’s individual connectivity (degree), its local sparsity (inverse clustering coefficient), and the density of connections among its neighbors, which reflects its position within the local network structure. Nodes with high DCL values are those that have high degree, relatively low clustering coefficient and well-connected neighbors, indicating that they occupy structurally important positions with strong local influence and access to densely connected parts of the network.

\section{Degree and Importance of Lines (DIL)}

The \emph{Degree and Importance of Lines (DIL) centrality}\index{degree and importance of lines (DIL) centrality} evaluates the importance of a node based on its degree and the significance of its adjacent links \cite{JLiu2016}. The DIL measure reflects that the influence of a node depends on both its degree and the structural roles of its adjacent links. The centrality \( c_{\mathrm{DIL}}(i) \) of node \( i \) is defined as
\begin{equation*}
   c_{\mathrm{DIL}}(i) = d_i + \sum_{j \in \mathcal{N}(i)} 
   \left( \frac{(d_i - \Delta_{ij} - 1)(d_j - \Delta_{ij} - 1)}{\Delta_{ij}/2 + 1} \right)
   \left( \frac{d_i - 1}{d_i + d_j - 2} \right),
\end{equation*}
where \( d_i \) is the degree of node \( i \), 
\( \mathcal{N}(i) \) is the set of neighbors of \( i \), and 
\( \Delta_{ij} \) is the number of triangles that include the link \((i,j)\).

\section{Degree and structural hole count (DSHC) method}
The \emph{degree and structural hole count}\index{degree and structural hole count (DSHC)} (DSHC) method is a local centrality measure that combines information about a node's degree with the structural holes in its neighborhood~\cite{Yang2020}. The DSHC of node $i$ is defined as
\[
c_{\mathrm{DSHC}}(i) = \sum_{j \in \mathcal{N}(i)} \left( \left( \frac{1}{d_i} + \frac{1}{d_j} \right) \frac{1}{1 + \Delta_{ij}} \right)^2,
\]
where \(\mathcal{N}(i)\) is the set of neighbors of node \(i\), $d_i$ and $d_j$ denote the degrees of nodes $i$ and $j$, respectively, and 
\[
\Delta_{ij} = |\mathcal{N}(j) \setminus \mathcal{N}(i)|
\]
represents the number of structural holes between nodes $i$ and $j$, with node $i$ acting as the intermediary. This measure captures both the local connectivity and the bridging role of a node within its neighborhood.

\section{Degree centrality}

Shaw (1954) was among the first to propose using the number of direct links of a node as an indicator of its importance in a network \cite{Shaw1954}. This concept, now formalized as \textit{degree centrality}\index{degree}, measures the size of the one-hop neighborhood of a node \cite{Bonacich1972,Freeman1978}. For an undirected graph, the degree centrality of node $i$ is defined as
\begin{equation*}
    c_{degree}(i) = \sum_{j=1}^{N}{a_{ij}} = \sum_{j=1}^{N}{a_{ji}}=d_i.
\end{equation*}

For directed graphs, four variants of degree-based centrality can be considered: \textit{in-degree} centrality (number of incoming edges), \textit{out-degree} centrality (number of outgoing edges), \textit{total degree}\index{degree!total} centrality (sum of in-degree and out-degree) and the difference between in-degree and out-degree, sometimes called \textit{degree difference}. The total degree centrality reflects the overall activity of a node in the network, capturing both its incoming and outgoing connections.  The degree difference\index{degree!difference} indicates whether a node tends to be more of a receiver or a sender of connections: positive values correspond to nodes with more incoming than outgoing links, while negative values correspond to nodes with more outgoing than incoming links. Degree centrality can also be extended to weighted networks, where the adjacency matrix entry $a_{ij}$ is replaced by the link weight $w_{ij}$, reflecting the intensity of the connection \cite{Opsahl2010}.

\section{Degree mass}

\emph{Degree mass}\index{degree mass} is a family of centrality measures that generalize degree centrality \cite{PVM2015}. The \(m\)-th order degree mass of a node \(i\) is defined as the sum of the weighted degrees of nodes within its \(m\)-hop neighborhood:

\begin{equation*}
c_{dm}(i) = \sum_{k=1}^{m+1} \left( A^k u \right)_i
= \sum_{j=1}^{N} \sum_{k=1}^{m+1} \left( A^k \right)_{ij} d_j,
\end{equation*}
where \(A\) is the adjacency matrix of the graph, \(u\) is the all-ones vector and \(d_j\) is the degree of node \(j\). Here, \(m \ge 0\) specifies the order of the neighborhood considered. When \(m = 0\), the degree mass reduces to the standard degree centrality. For \(m = 1\), the degree mass of node \(i\) equals the sum of its own degree and the degrees of its immediate neighbors. As \(m\) increases, the measure incorporates the influence of nodes farther away, and for sufficiently large \(m\), it becomes proportional to the eigenvector centrality \cite{PVM2015}.

\section{DegreeDiscountIC}
\emph{DegreeDiscountIC}\index{DegreeDiscountIC} is a degree-discount heuristic for identifying influential nodes in a network~\cite{Chen2009}. The main idea is that if a node $i$ is selected as a seed, the edges connecting $i$ to its neighbors should not be fully counted when evaluating the degrees of those neighbors for further seed selection. In other words, the degree of each neighbor of a selected seed is discounted to account for the presence of the seed node.  

The algorithm proceeds iteratively as follows:
\begin{enumerate}
    \item Compute the degree $d_i$ of each node $i$, and initialize $t_i = 0$ and $dd_i = d_i$, where $t_i$ is the number of selected neighbors of $i$ and $dd_i$ is the discounted degree.
    \item Select the node $i$ with the largest $dd_i$ and add it to the seed set. Then, for each neighbor $j$ of node $i$, update
    \[
    t_j \gets t_j + 1, \quad
    dd_j = d_j - 2 t_j - (d_j - t_j) t_j p,
    \]
    where $p$ is the propagation probability (e.g., $p=0.01$). The value $dd_j$ approximates the expected number of additional nodes influenced by selecting node $j$, considering the presence of already selected neighbors.
    \item Repeat step 2 until $k$ nodes are selected for the seed set.
\end{enumerate}

\section{DegreeDistance}
\emph{DegreeDistance}\index{DegreeDistance} is a degree-based centrality measure designed to identify $k$ influential nodes in a network~\cite{Sheikhahmadi2015}. The method proceeds iteratively as follows:

\begin{enumerate}
    \item Initialize the seed set $S$ as empty. Select the node with the highest degree and add it to $S$.
    \item Consider the next highest-degree node $j$. If the distance $d_{ij}$ between $j$ and any node $i \in S$ is less than a threshold $d$, do not add $j$ to the seed set. Otherwise, include $j$ in $S$.
    \item Repeat step 2 until $|S| = k$.
\end{enumerate}

Sheikhahmadi \textit{et al.}~\cite{Sheikhahmadi2015} also proposed extensions to the DegreeDistance measure. Note that in networks with small diameter, the resulting seed set $S$ may be significantly smaller than $k$ due to the distance constraint.

\section{DegreePunishment}

\emph{DegreePunishment}\index{DegreePunishment} is a degree-based heuristic for identifying influential nodes in a network \cite{XWang2016}. The central idea is that once a node $i$ is selected as a seed, it should “punish” or restrict the selection of its neighbors in the subsequent steps. DegreePunishment iteratively performs the following steps:

\begin{enumerate}
    \item Compute the degree $d_i$ of each node $i$ and initialize $dp_i = d_i$.
    \item Select the node \(i\) with the largest $dp_i$ and add it to the seed set. The punishment that node $i$ imposes on node $j$ is calculated as
    \[
    p_{ij} = d_i \sum_{h=1}^{r} (A^h)_{ij} \, \omega^h,
    \]
    where $r$ defines the radius of influence (maximum path length) and $\omega$ is a weakening factor. The value of $dp_j$ is then updated for each node $j$ as
    \[
    dp_j = d_j - \sum_{i \neq j} p_{ij} \, \sigma_i,
    \]
    where $\sigma_i$ is an indicator function such that $\sigma_i = 1$ if node $i$ has already been selected as a seed, and $\sigma_i = 0$ otherwise.
    \item Repeat step 2 until $k$ nodes have been selected for the seed set.
\end{enumerate}

Due to computational complexity, Wang \textit{et al.} \cite{XWang2016} restrict the punishment to paths of length $r \le 2$ and suggest using $\omega = \beta_c$, where $\beta_c$ is the spreading threshold of the graph $G$.

\section{\texorpdfstring{\boldmath$\delta$-}{delta-}betweenness centrality}

Agneessens \textit{et al.}~\cite{Agneessens2017} propose a generalized version of betweenness centrality referred to as \emph{\(\delta\)-betweenness}\index{betweenness centrality!\(\delta\)-}, which incorporates a tuning parameter \(\delta \in \mathbb{R}\), reflecting the relative importance of geodesic distances in the network. The $\delta$-betweenness of a node \(i\) can be expressed as
\begin{equation*}
    c_{\delta-betw}(i) = \sum_{j=1}^{N}{\sum_{k=1}^{N}{\frac{\sigma_{jk}(i)}{\sigma_{jk}}(d_{jk}-1)^{-\delta}}},
\end{equation*}
where \(\sigma_{jk}\) is the number of shortest paths between nodes \(j\) and \(k\), \(\sigma_{jk}(i)\) is the number of paths that pass through node \(i\), and \(d_{jk}\) is the length of the shortest path from \(j\) to \(k\). Note that for \(\delta = 0\), the \(\delta\)-betweenness centrality reduces to the standard betweenness centrality.

\section{\texorpdfstring{\boldmath$\delta$-}{delta-}closeness centrality}
Agneessens \textit{et al.}~\cite{Agneessens2017} propose a generalized version of closeness centrality referred to as \emph{\(\delta\)-closeness}\index{closeness centrality!\(\delta\)-}, which incorporates a tuning parameter \(\delta \in \mathbb{R}\), reflecting the importance of geodesic distances in the network. The $\delta$-closeness of a node \(i\) can be expressed as
\begin{equation*}
    c_{\delta-cl}(i) = \frac{\sum_{j \neq i}d_{ij}^{-\delta}}{N-1},
\end{equation*}
where \(d_{ij}\) is the length of the shortest path from node \(i\) to node \(j\). Agneessens \textit{et al.}~\cite{Agneessens2017} demonstrated that, by varying the parameter \(\delta\), degree centrality and harmonic centrality can be viewed as specific instances of the generalized \(\delta\)-closeness centrality: for \(\delta \rightarrow \infty\), the index is proportional to degree centrality, whereas for \(\delta = 1\), \(\delta\)-closeness coincides with harmonic centrality.

\section{Density centrality}

\emph{Density centrality}\index{density centrality} is a semi-local measure inspired by Newton’s gravity formula \cite{Ibnoulouafi2018}. The method draws an analogy to area density in physics, measuring how much “mass” (node degree) is distributed within a fixed spatial region. 

The centrality \(c_{\textsc{Density}}(i)\) of node \(i\) is defined as
\begin{equation*}
   c_{\textsc{Density}}(i) = \sum_{j \in \mathcal{N}^{(\leq l)}(i)} \frac{d_i}{\pi\, d_{ij}^2},
\end{equation*}
where \(\mathcal{N}^{(\leq l)}(i)\) denote the set of nodes within \(l\)-hop neighborhood of node \(i\), \(d_{ij}\) is the shortest-path distance between nodes \(i\) and \(j\), and \(d_i\) is the degree of node \(i\). Ibnoulouafi and El Haziti \cite{Ibnoulouafi2018} set \(l = 3\) as the truncated radius.

\section{Density of the Maximum Neighborhood Component (DMNC)}

The \emph{Density of the Maximum Neighborhood Component}\index{density of the maximum neighbourhood component (DMNC)} (DMNC) centrality extends the maximum neighborhood component (MNC) measure by incorporating the internal link density of the largest connected component within a node’s neighborhood~\cite{Lin2008}. 
While MNC centrality considers only the size of the largest connected subgraph among the neighbors of a node, DMNC additionally evaluates how densely those neighbors are connected to each other. 

Formally, for a given node \( i \), let \( C_{\max}(G_{\mathcal{N}(i)}) \) denote the largest connected component of the induced subgraph \( G_{\mathcal{N}(i)} \). 
Then the DMNC centrality \( c_{\mathrm{dmnc}}(i) \) is defined as
\begin{equation*}
    c_{dmnc}(i) = \frac{|E(MNC(G_{\mathcal{N}(i)}))|}{|V(MNC(G_{\mathcal{N}(i)}))|^\epsilon},
\end{equation*}
where \(|E(MNC(G_{\mathcal{N}(i)}))|\) and \(|V(MNC(G_{\mathcal{N}(i)}))|\) denote the number of edges and vertices, respectively, within the largest connected component, and 
\( \epsilon \) is a tunable scaling parameter such that \( 1 \leq \epsilon \leq 2 \) (typically \( \epsilon = 1.67 \)). Thus, the DMNC centrality measures how large and how tightly connected a node’s neighbourhood is, giving higher scores to nodes whose neighbours form large, well-connected groups.

\section{Diffusion centrality}

\emph{Diffusion centrality}\index{diffusion centrality} quantifies the influence of a node in a dynamic diffusion process that starts from node \(i\) \cite{Banerjee2013}. Initially, node \(i\) passes a piece of information to each of its neighbors with probability \(\delta\). At each subsequent time step \(t > 1\), nodes that received information at time \(t-1\) pass each piece of information to their neighbors with the same probability \(\delta\).

The diffusion centrality of node \(i\), denoted \(c_{\mathrm{dif}}(i)\), is defined as the expected number of times nodes in the network have been contacted over \(T\) periods:
\[
    c_{\mathrm{dif}}(i) = \sum_{t=1}^T \sum_{j=1}^N \delta^t (A^t)_{ij},
\]
where \(A\) is the adjacency matrix of the network. For \(T = 1\), diffusion centrality is proportional to the degree centrality of node \(i\). As \(T \rightarrow \infty\), it converges to either Katz centrality or eigenvector centrality, depending on whether \(\delta\) is smaller than or greater than \(1/\lambda_{\max}\), where \(\lambda_{\max}\) is the largest eigenvalue of \(A\). In \cite{Banerjee2013}, the authors set \(\delta = 1/\lambda_{\max}\).

\section{Diffusion degree}

The \emph{diffusion degree}\index{degree!diffusion} quantifies the potential influence of a node in a network diffusion process by considering both its own propagation capability and that of its neighbors \cite{Kundu2011}. The model assumes that each node $i$ is associated with a propagation probability $x_i$, where $0 \leq x_i \leq 1$, representing its ability to transmit information or influence to adjacent nodes. The diffusion degree of node $i$ is then defined as
\begin{equation*}
    c_{\text{diffusion}}(i) = x_i d_i + \sum_{j=1}^{N} a_{ij} x_j d_j,
\end{equation*}
where $d_i$ denotes the degree of node $i$, and $a_{ij}$ is the $(i,j)$-th element of the adjacency matrix $A$. 

The first term, $x_i d_i$, captures the intrinsic contribution of node $i$, reflecting its degree and individual propagation probability, while the second term accounts for the influence of its neighbors, weighted by their respective propagation probabilities and degrees. Nodes with higher diffusion degree values are expected to play a more prominent role in spreading processes, such as information diffusion or epidemic propagation. When all nodes have the same propagation probability, i.e. $x_i = 1$ for all $i$, the diffusion degree reduces to a measure of combined connectivity within a node's immediate neighborhood, highlighting nodes that are both well-connected themselves and connected to other highly connected nodes.

\section{DirichletRank}

\emph{DirichletRank}\index{DirichletRank} is a variant of PageRank designed to address the “zero-one gap” problem\index{zero-one gap problem} inherent in the classical PageRank algorithm \cite{Wang2008}. In PageRank, a random surfer moves to one of a node’s outgoing links with probability $\alpha$, or jumps to a random node with probability $1-\alpha$. For nodes with no outgoing links (sink nodes), the surfer cannot follow a link, so they must jump to a random node with probability 1. This creates a large difference in transition behavior between a sink node and a node with even a single outgoing link, leading to the so-called “zero-one gap” in PageRank probabilities. DirichletRank overcomes this issue by using a Bayesian estimation with a Dirichlet prior to compute smoother and more realistic transition probabilities. The DirichletRank score of the nodes, denoted $c_{DR}$, is obtained by solving the eigenvector equation
\[
c_{DR} = \tilde{M} \, c_{DR},
\]
where
\[
\tilde{M} = \operatorname{diag}(1-\omega_1, \dots, 1-\omega_N) \, D^{-1}A + \frac{\operatorname{diag}(\omega_1, \dots, \omega_N)}{N} \, u u^T,
\]
with
\[
\omega_i = \frac{\mu}{\mu + \sum_{j=1}^{N} a_{ij}},
\quad \mu = 20,
\quad u \text{ is an $N \times 1$ all-one vector}.
\]

Here, $\omega_i$ represents the random jumping probability for node $i$. As defined, the more outgoing links a node has, the less likely a surfer is to jump randomly, and the more likely they are to follow one of its outgoing links.

\section{Disassortativity and community structure (mDC) centrality}

The \emph{disassortativity and community structure}\index{disassortativity and community structure (mDC) centrality} (mDC) centrality is a community-based hybrid measure that combines a node’s disassortativity (DoN) with its influence at the community boundary \cite{Wang2024}. It assumes that the network \(G\) has an identifiable community structure (e.g., Wang \textit{et al.}~\cite{Wang2024} apply the Louvain algorithm for community detection).  

For a node \(i \in C\), the mDC score \(c_{mDC}(i)\) is defined as
\[
c_{mDC}(i) = (1 - \alpha_i) \, c_{DoN}(i) + \alpha_i \, f_c(i),
\]
where \(c_{DoN}(i)\) is the node’s disassortativity score \cite{Wang2024}, and \(\alpha_i\) quantifies how isolated the community \(C\) containing node \(i\) is from the rest of the network:
\[
\alpha_i = \frac{|E^{\text{in}}_C|}{|E^{\text{in}}_C| + |E^{\text{out}}_C|}.
\]
Here, \(E^{\text{in}}_C\) and \(E^{\text{out}}_C\) denote the sets of edges within community \(C\) and connecting \(C\) to other communities, respectively. A larger \(\alpha_i\) corresponds to a more isolated community, increasing the relative weight of boundary nodes in the mDC score.  

The community boundary popularity \(f_c(i)\) captures the influence of node \(i\) at the interface between communities:
\[
f_c(i) =
\begin{cases}
\displaystyle \alpha_i \sum_{C' \in \mathcal{C}_i} \left( 1 + \frac{|C| + |C'|}{2 |C_{\max}|} \right), & d_i \neq d_i^{\text{in}}, \\[1em]
0, & d_i = d_i^{\text{in}},
\end{cases}
\]
where \(d_i\) is the degree of node \(i\), \(d_i^{\text{in}}\) is the number of neighbors within the same community \(C\), \(\mathcal{C}_i\) is the set of other communities connected to \(i\), \(|C|\) and \(|C'|\) are the sizes of communities \(C\) and \(C'\), and \(|C_{\max}|\) is the size of the largest community in the network.  

Nodes with high mDC scores are influential both locally (high DoN) and at the boundaries between communities, bridging communities and enhancing connectivity. The effectiveness of mDC has been validated on synthetic and real-world networks through analyses of network robustness and disease spreading simulations.

\section{Disassortativity of node (DoN)}

The \emph{disassortativity of a node}\index{disassortativity of node (DoN)} (DoN) quantifies a node’s tendency to connect to neighbors with lower degrees, capturing its local dominance and potential influence within the network \cite{Wang2024}. For a node \(i\) with degree \(d_i\) and neighbors \(\mathcal{N}(i)\), the DoN score is defined as
\[
c_{DoN}(i) = \sum_{j \in \mathcal{N}(i)} f(d_i, d_j),
\]
where
\[
f(d_i, d_j) =
\begin{cases} 
1, & d_i \ge d_j, \\ 
0, & d_i < d_j.
\end{cases}
\]

The DoN score of node \(i\) ranges from \(0\) to \(d_i\), where \(0\) indicates that all neighbors have higher degrees and \(d_i\) indicates that all neighbors have lower degrees. Nodes with high DoN dominate their local neighborhoods, bridging lower-degree nodes and exerting greater influence over the network’s structure and functionality. In contrast, nodes with low DoN are surrounded by more influential neighbors, limiting their impact. This aligns with the observation that in disassortative networks, high-degree nodes connected to low-degree nodes often serve as key drivers of information flow or control within the system. The effectiveness of DoN has been validated through extensive experiments on both synthetic and real-world networks, including analyses of network robustness and simulations of spreading dynamics.

\section{Distance entropy (DE)}

\emph{Distance entropy}\index{distance entropy centrality} (DE) quantifies node centrality based on the distribution of shortest-path lengths from a node to all other nodes in the network \cite{Stella2018}. The distance entropy of node $i$ is defined as
\begin{equation*}
    c_{\text{DE}}(i) = -\frac{1}{\log (M_i - m_i)} \sum_{k=1}^{M_i - m_i} p_k^{(i)} \log p_k^{(i)},
\end{equation*}
where $M_i = \max_j d_{ij}$ and $m_i = \min_j d_{ij}$ are the maximum and minimum distances from node $i$, and $p_k^{(i)}$ is the probability that the distance from $i$ to a node equals $k$. Specifically, in a connected graph $G$, if node $i$ is at distance $k$ from $n_k$ other nodes, then
\[
p_k^{(i)} = \frac{n_k}{N-1},
\]
where $N$ is the total number of nodes in the network.

Distance entropy captures how evenly distributed the distances from a node are: higher values indicate a more uniform distribution of distances, reflecting nodes that are well-positioned across multiple network layers.

\section{Distance-weighted fragmentation (DF)}

The \emph{distance-weighted fragmentation}\index{distance-weighted fragmentation (DF) centrality} (DF) centrality quantifies the effect of a node on the overall connectivity of a network by considering the average reciprocal distance among nodes after its removal \cite{Borgatti2003,Borgatti2006}. For a node \(i\), the DF centrality, denoted \(c_{DF}(i)\), is defined as

\begin{equation*} 
c_{DF}(i) = 1 - \frac{\sum_{j \neq i} \sum_{k \neq i} d_{jk}^{-1}(G_i)}{(N-1)(N-2)},
\end{equation*}
where \(G_i\) is the subgraph obtained by removing node \(i\) from \(G\), and \(d_{jk}(G_i)\) is the shortest-path distance between nodes \(j\) and \(k\) in \(G_i\).  

The DF centrality ranges from 0, when the network remains fully connected (as in a complete graph), to 1, when all nodes are isolated. Intermediate values indicate the extent to which the removal of a node increases distances in the network, thus reflecting its importance in maintaining overall network connectivity.

\section{Diversity coefficient}

\emph{Diversity coefficient}\index{diversity coefficient} is a variation of the participation coefficient based on Shannon entropy \cite{Rubinov2011}. It quantifies the distribution of a node's connections across different communities in a network with a community structure. Let the graph \(G\) consist of \(K\) communities \(C_1, \ldots, C_K\). The diversity coefficient of node \(i\) is defined as
\begin{equation*}
    c_{\mathrm{diversity}}(i) = - \sum_{s=1}^{K} p_s(i) \log p_s(i),
\end{equation*}
where 
\[
p_s(i) = \frac{d_{is}}{d_i}
\] 
denotes the fraction of links of node \(i\) connecting to community \(C_s\), with \(d_i\) representing the total degree of node $i$.

Nodes with high diversity coefficients have connections spread across many communities, indicating they serve as bridges between modules, whereas nodes with low diversity coefficients have connections concentrated within a single community.

\section{Diversity-strength centrality (DSC)}

\emph{Diversity-strength centrality}\index{diversity-strength centrality (DSC)} (DSC) is an entropy-based measure that quantifies node importance based on neighbor diversity, influence spread, and intensity \cite{Zareie2019}. The centrality of node $i$ is defined as
\begin{equation*}
    c_{\text{DSC}}(i) = - \sum_{j \in \mathcal{N}(i)} \frac{IKs(j)}{\sum_{l \in \mathcal{N}(i)} IKs(l)} \log \frac{IKs(j)}{\sum_{l \in \mathcal{N}(i)} IKs(l)},
\end{equation*}
where $\mathcal{N}(i)$ is the set of neighbors of node $i$ and $IKs(j)$ is the improved $k$-shell index of node $j$ as defined in \cite{ZLiu2015}. This term captures the relative influence of neighbors, and the entropy reflects the diversity of their influence.

Zareie \textit{et al.} \cite{Zareie2019} propose two extensions of DSC:  
\begin{enumerate}
    \item \emph{Diversity-strength ranking (DSR)}\index{diversity-strength ranking (DSR)}, which sums the DSC scores of a node's neighbors:
    \begin{equation*}
        c_{\text{DSR}}(i) = \sum_{j \in \mathcal{N}(i)} c_{\text{DSC}}(j),
    \end{equation*}  
    \item \emph{Extended diversity-strength ranking (EDSR)}\index{extended!diversity-strength ranking (EDSR)}, which sums the DSR scores of a node's neighbors:
    \begin{equation*}
        c_{\text{EDSR}}(i) = \sum_{j \in \mathcal{N}(i)} c_{\text{DSR}}(j).
    \end{equation*}
\end{enumerate}

Nodes with high DSC values are influential not only individually but also through the diversity and strength of their neighbors, reflecting both local and extended network influence.

\section{Diversity-strength ranking (DSR)}

\emph{Diversity-strength ranking}\index{diversity-strength ranking (DSR)} (DSR) is an extension of the diversity-strength centrality (DSC), proposed by Zareie \textit{et al.} \cite{Zareie2019}, designed to capture influence that extends beyond a node’s immediate neighborhood. The DSR value of node $i$ is defined as
\begin{equation*}
    c_{\text{DSR}}(i) = \sum_{j \in \mathcal{N}(i)} c_{\text{DSC}}(j)=  \sum_{j \in \mathcal{N}(i)} 
    \left(
        \sum_{k \in \mathcal{N}(j)} 
        \frac{IKs(k)}{\sum_{m \in \mathcal{N}(j)} IKs(m)} 
        \log \frac{IKs(k)}{\sum_{m \in \mathcal{N}(j)} IKs(m)}
    \right),
\end{equation*}
where $\mathcal{N}(i)$ denotes the set of neighbors of node $i$ and $IKs(k)$ is the improved $k$-shell index of node $k$ as defined by Liu \textit{et al.} \cite{ZLiu2015}. The inner summation represents the diversity-strength centrality of neighbor $j$, while the outer summation aggregates these values for all neighbors of node $i$. Thus, DSR extends DSC by capturing second-order effects through the influence of neighboring nodes. High DSR values indicate connections to neighbors that are both diverse and influential, reflecting enhanced potential for influence propagation.

\section{DK-based gravity model (DKGM)}

The \emph{DK-based gravity model}\index{gravity model!DK-based (DKGM)} (DKGM) is an extension of the local gravity model in which a node's \textit{mass} is represented by its DK value, a metric that combines the node's degree and the results from $k$-core decomposition \cite{Li2021}. Let $\mathcal{N}^{(\leq l)}(i)$ denote the set of neighbors of node $i$ within $l$ hops. Then the DKGM centrality of node $i$ can be written as
\begin{equation*}
    c_{DKGM}(i) = \sum_{j \in \mathcal{N}^{(\leq l)}(i)} \frac{DK(i) \, DK(j)}{d_{ij}^2},
\end{equation*}
where $d_{ij}$ is the shortest path distance between nodes $i$ and $j$. Following \cite{Li2021}, the truncated radius is typically set to $l=2$.  

The DK index of node $i$, denoted $DK(i)$, is given by
\begin{equation*}
    DK(i) = d_i + k_s(i) + \frac{p(i)}{\max_k q(k) + 1},
\end{equation*}
where $d_i$ is the degree of node $i$, $k_s(i)$ is the $k$-shell value of node $i$, $p(i)$ represents the iteration at which node $i$ is removed during the $k$-core decomposition and $q(k)$ denotes the total number of removal steps performed in that iteration. The DK index captures both local information (degree and $k$-shell) and global structural information (position within the $k$-core hierarchy) of nodes.  

DKGM evaluates a node’s influence by considering both its own importance and the contributions of nearby nodes, giving less weight to nodes that are farther away.

\section{DomiRank centrality}

\emph{DomiRank centrality}\index{DomiRank centrality} is a measure designed to quantify the dominance of nodes within their local neighborhoods and to highlight structurally fragile regions whose integrity and functionality depend on these dominant nodes \cite{Engsig2024}. The evolution of node fitness is governed by two processes: (i) \textit{natural relaxation}, where each node's fitness decays exponentially toward zero at rate \(\beta\); and (ii) \textit{competition}, where nodes compete with neighbors for limited resources. A node's fitness increases when it is surrounded by neighbors whose fitness is below a domination threshold \(\theta\), and decreases otherwise.  

Formally, let \(\Gamma(t) \in \mathbb{R}^{N}\) denote the vector of evolving dominance scores. The dynamics follow:
\[
\frac{d\Gamma(t)}{dt} = \alpha \, A \bigl(\theta \mathbf{1} - \Gamma(t)\bigr) - \beta \, \Gamma(t),
\]
where \(A\) is the adjacency matrix of the network \(G\), \(\alpha, \beta, \theta\) are parameters controlling competition and relaxation dynamics, and \(\mathbf{1}\) is the all-ones vector. The domination threshold \(\theta\) acts as a rescaling factor and is set to \(\theta = 1\) without loss of generality. The parameter ratio \(\sigma = \alpha / \beta\) determines the balance between local (nodal) and mesoscale (structural) information.  

At steady state (\(\lim_{t \to \infty}\)), the dominance vector \(\Gamma\) satisfies
\[
\Gamma = \theta \, \sigma \, (\sigma A + I)^{-1} A \, \mathbf{1},
\]
where \(I\) is the identity matrix. The convergence interval for \(\sigma\) is bounded as \(\sigma \in \left(0, -\frac{1}{\lambda_N}\right)\), where \(\lambda_N\) is the smallest (dominant negative) eigenvalue of \(A\).  

Nodes with high DomiRank scores dominate many neighbors that themselves have low influence, thereby identifying fragile neighborhoods highly dependent on these nodes. Engsig \textit{\textit{et al.}}~\cite{Engsig2024} demonstrate that DomiRank-based interventions can inflict more enduring network damage, impeding recovery and reducing overall system resilience.

\section{Dynamical importance}

The \emph{dynamical importance}\index{dynamical importance} quantifies the influence of individual nodes on dynamical processes occurring on networks \cite{Restrepo2006}. It measures how the removal of a node affects the largest eigenvalue of the network's adjacency matrix, which determines critical thresholds and stability conditions in dynamical processes such as synchronization, epidemic spreading and percolation. The centrality of node \( i \) is defined as the relative change in the largest eigenvalue upon the removal of node \( i \)
\begin{equation*}
    c_{dynImp}(i) = -\,\frac{\lambda(G_i) - \lambda(G)}{\lambda(G)},
\end{equation*}
where \( \lambda(G) \) is the largest eigenvalue of the adjacency matrix of graph \( G \), and \( G_i \) denotes the subgraph obtained by removing node \( i \) from \( G \).

A higher value of \( c_{dynImp}(i) \) indicates that removing node \( i \) leads to a greater reduction in the network’s largest eigenvalue, implying that the node plays a more critical role in sustaining the network’s dynamical properties. For undirected networks, the adjacency matrix is symmetric, and the largest eigenvalue corresponds to the spectral radius of the graph.

\section{Dynamical influence}

The \emph{dynamical influence}\index{dynamical influence} (DI) quantifies how strongly a node's dynamical state can affect the collective behavior of a networked system, explicitly accounting for the interplay between structure and dynamics \cite{Klemm2012}. Conceptually, it estimates the potential impact of a node on a spreading process before the contagion begins, given the system dynamics.

Klemm \textit{et al.} \cite{Klemm2012} consider the SIR model, where each node can be susceptible, infected, or recovered. Linearizing the dynamics around the stationary state in which all nodes are susceptible, small perturbations obey
\begin{equation*}
    \dot{x} = -x[t] + \beta A^T x[t],
\end{equation*}
where \(x_j[t]\) is the probability that node \(j\) is infected at time \(t\), \(\beta\) is the infection probability and \(A\) is the adjacency matrix. This equation can be rewritten as \(\dot{x} = M x\) with \(M = \beta A^T - I\). At the epidemic threshold, \(\beta = 1/\lambda_{\max}(A)\), the largest eigenvalue of \(M\) is zero, i.e., \(\dot{x} = 0\).

The dynamical influence of nodes, \(c_{DI}\), is given by the leading left eigenvector of \(M\). Equivalently, under the linearization and threshold assumption, \(c_{DI}\) corresponds to the right eigenvector of \(A\) associated with its largest eigenvalue, meaning that, in this case, dynamical influence reduces to eigenvector centrality.

\section{Dynamical spanning tree (DST) centrality}

\emph{Dynamical Spanning Tree}\index{Dynamical Spanning Tree (DST) centrality} (DST) centrality is a node importance measure that identifies the most critical nodes in a network based on their impact on the network's structural reliability \cite{Chen2003,Lü2016}. Specifically, the node whose removal results in the largest reduction in the number of spanning trees is considered the most vital. The relative importance of nodes can thus be quantified according to the decrease in the total number of spanning trees after their removal.

The DST centrality of a node \(i\) is defined as
\begin{equation*}
    c_{\mathrm{DST}}(i) = 1 - \frac{t_{G_i}}{t_G},
\end{equation*}
where \(t_G\) is the total number of spanning trees in the original graph \(G\), and \(t_{G_i}\) is the number of spanning trees in the subgraph \(G_i\) obtained by removing node \(i\) from \(G\). The total number of spanning trees can be computed using Kirchhoff’s Matrix-Tree Theorem:
\begin{equation*}
    t_G = \frac{1}{N} \prod_{k=2}^{N} \lambda_k,
\end{equation*}
where \(0 = \lambda_1 < \lambda_2 \leq \dots \leq \lambda_N\) are the eigenvalues of the Laplacian matrix \(L\) of the graph \(G\), and \(N\) is the number of nodes. For undirected networks, the Laplacian matrix is symmetric, and all eigenvalues are real and non-negative.

A higher value of \(c_{\mathrm{DST}}(i)\) indicates that node \(i\) plays a more significant role in maintaining the network’s structural connectivity.

\section{Dynamics-sensitive (DS) centrality}

\emph{Dynamics-sensitive (DS) centrality}\index{dynamics-sensitive (DS) centrality} quantifies the influence of a node by considering the weighted sum of walks originating from that node, where both the spreading rate \(\beta\) and the spreading time \(T\) are incorporated into the weighting scheme \cite{JGLiu2016}. Formally, the DS centrality of node \(i\) is defined as
\begin{equation*}
    c_{\mathrm{DS}}(i) = \left( \sum_{t=1}^{T} \beta^t A^t \right) u,
\end{equation*}
where \(A\) is the adjacency matrix of the network, \(u\) is an \(N \times 1\) vector of ones, and \(\beta\) represents the spreading rate. In their study, Liu \textit{et al.} \cite{JGLiu2016} set the time horizon to \(T=5\) and consider a spreading rate \(\beta \leq 0.1\). This definition of DS centrality effectively captures the dynamics of spreading processes on networks, as it assigns higher centrality to nodes that are reachable through multiple weighted paths within the given time frame.

\section{Eccentricity centrality}

The \emph{eccentricity centrality}\index{eccentricity centrality} \cite{Hage1995}, also referred to as Harary graph centrality \cite{Brandes2005}, measures how close a node is to the farthest node in a connected graph. For a node \(i\) in a connected graph \(G\), the eccentricity centrality, denoted by \(c_{Eccentricity}(i)\), is defined as the reciprocal of the maximum shortest-path distance from \(i\) to any other node:
\begin{equation*}
    c_{Eccentricity}(i) = \frac{1}{\max_{j \in \mathcal{N}} d_{ij}},
\end{equation*}
where \(d_{ij}\) represents the shortest-path distance between nodes \(i\) and \(j\). Nodes with the highest eccentricity centrality are considered the most central, as they are closest, on average, to the farthest nodes in the network.

\section{Edge clustering coefficient (NC)}

\emph{Edge clustering coefficient centrality}\index{clustering coefficient!edge centrality} (NC), also known as the sum of ECC (SoECC) \cite{Wang2010}, is used to identify essential proteins in networks based on the clustering of edges \cite{Wang2012}. The NC centrality of a node \(i\), denoted \(c_{\mathrm{NC}}(i)\), is defined as
\[
    c_{\mathrm{NC}}(i) = \sum_{j \in \mathcal{N}(i)} \mathrm{ECC}(i,j),
\]
where \(\mathcal{N}(i)\) is the set of neighbors of node \(i\), and \(\mathrm{ECC}(i,j)\) is the edge clustering coefficient of edge \((i,j)\), given by
\[
    \mathrm{ECC}(i,j) = \frac{z_{i,j}}{\min(d_i-1, d_j-1)}.
\]
Here, \(z_{i,j}\) denotes the number of triangles that include the edge \((i,j)\) and \(d_i\) is the degree of node \(i\). The denominator \(\min(d_i - 1, d_j - 1)\) represents the \emph{maximum number of triangles} in which the edge \((i,j)\) can potentially participate.
  
Thus, the NC centrality accounts for both the \emph{degree of the node} \(d_i\) (i.e., the number of edges incident to node \(i\)) and the \emph{clustering coefficients} of its edges, capturing the node’s involvement in tightly connected regions of the network.

\section{Edge Percolated Component (EPC)}

\emph{Edge Percolated Component}\index{edge percolated component (EPC)} (EPC) quantifies the robustness of a node’s connectivity under random edge failures \cite{Chin2003}. Specifically, it estimates the fraction of nodes that remain connected to node \(i\) when each edge in the graph \(G\) is independently removed with probability \(p\).  

Let \(G^{(k)}\) denote the \(k\)-th realization of \(G\) after random edge removal. The EPC centrality of node \(i\) is then given by
\begin{equation*}
c_{\mathrm{EPC}}(i) = \frac{1}{N K} \sum_{k=1}^{K} \sum_{j=1}^{N} \delta_{ij}^{(k)},
\end{equation*}
where \(N\) is the number of nodes, \(K\) is the total number of realizations and
\[
\delta_{ij}^{(k)} =
\begin{cases}
1, & \text{if nodes $i$ and $j$ are connected in } G^{(k)},\\
0, & \text{otherwise.}
\end{cases}
\]

Intuitively, a higher EPC centrality indicates that node \(i\) remains connected to a larger fraction of the network under random edge failures, reflecting its structural resilience.

\section{Edge-disjoint \textit{k}-path centrality}

The \emph{edge-disjoint \textit{k}-path centrality}\index{\textit{k}-path centrality!edge-disjoint} is a variant of the \textit{k}-path centrality \cite{Borgatti2006}. Unlike the original measure, which counts all simple paths, this centrality considers only \emph{edge-disjoint paths} of length up to \(k\) that originate or terminate at a given node. Formally, an \emph{edge-disjoint path} is a simple path between two nodes that does not share any edge with another counted path. The number of edge-disjoint paths between two nodes is equivalent to the maximum flow between them \cite{Ford1962}. Nodes with higher edge-disjoint \(k\)-path centrality are more robustly connected and, therefore, harder to isolate from the network.

\section{Effective distance closeness centrality (EDCC)}

\emph{Effective distance closeness centrality}\index{closeness centrality!effective distance (EDCC)} (EDCC) is a variant of classical closeness centrality in which the shortest-path distance between nodes is replaced by the \emph{effective distance} \cite{Du2015}. The effective distance $D_{j|i}$ from node $i$ to a directly connected node $j$ is defined as \cite{Brockmann2013}:
\[
D_{j|i} = 1 - \log_2 \left( \frac{a_{ij}}{di} \right),
\]
where $a_{ij}$ is the adjacency matrix entry for the edge $(i,j)$ and $d_i$ is the degree of node $i$.  

The effective shortest-path distance $\tilde{d}_{ij}$ between nodes $i$ and $j$ is then computed as the shortest path in a weighted graph where the weight of each edge $(i,j)$ is given by $D_{j|i}$. Finally, the EDCC of node $i$ is defined as
\[
c_{EDCC}(i) = \left( \sum_{j=1}^{N} \tilde{d}_{ij} \right)^{-1}.
\]

Hence, effective distance closeness centrality generalizes closeness centrality by incorporating edge weights that account for heterogeneous connectivity patterns, rather than treating all direct links as equivalent.

\section{Effective distance gravity model (EffG)}

The \emph{effective distance gravity model}\index{gravity model!effective distance} (EffG) is a variant of the gravity model that incorporates both static and dynamic interactions between nodes by utilizing the concept of \textit{effective distance}\index{distance!effective} \cite{Shang2021}. The effective distance \(D_{j|i}\) from node \(i\) to node \(j\), which are directly connected, was introduced by Brockmann and Helbing~\cite{Brockmann2013} and is defined as
\begin{equation*}
   D_{j|i} = 1 - \log_2\!\left(\frac{a_{ij}}{d_i}\right),
\end{equation*}
where \(a_{ij}\) is the element of the adjacency matrix representing the connection between nodes \(i\) and \(j\), and \(d_i\) denotes the degree of node \(i\). 

The effective distance is not necessarily symmetric, even in undirected networks, because nodes may have different degrees. The \textit{effective shortest path distance} \(\tilde{d}_{ij}\) between nodes \(i\) and \(j\) is computed as the length of the shortest path in a weighted graph, where each direct link \((i,j)\) is assigned a weight equal to \(D_{j|i}\).

The EffG centrality of node \(i\), denoted by \(c_{\text{EffG}}(i)\), is then given by
\begin{equation*}
   c_{\text{EffG}}(i) = \sum_{j \neq i} \frac{d_i\,d_j}{\tilde{d}_{ij}^2}.
\end{equation*}

\section{Effective gravity model (EGM)}

The \emph{effective gravity model}\index{gravity model!effective (EGM)} (EGM) is a variant of the classical gravity model that incorporates precise radius and value information for each node \cite{SLi2021}. The EGM score of node $i$, denoted $c_{\mathrm{EGM}}(i)$, is defined as
\[
c_{\mathrm{EGM}}(i) = \sum_{j:\, d_{ij} \le R_i} \frac{V_i V_j}{d_{ij}^2},
\]
where $d_{ij}$ is the shortest-path distance between nodes $i$ and $j$, and $V_i$ is the entropy-based value information of node $i$:
\[
V_i = \left(- \sum_{j \in L_i} \frac{d_j}{\sum_{u \in L_i} d_u} \log \frac{d_j}{\sum_{u \in L_i} d_u} \right) d_i,
\]
with $d_i$ denoting the degree of node $i$ and $L_i = \mathcal{N}(i) \cup \{i\}$ representing its neighborhood including itself. Li and Xiao \cite{SLi2021} assume that each node has a distinct influence radius $R_i$, which depends on the relationship between the node and its farthest neighbor. Specifically, $R_i$ is defined as
\[
R_i = \frac{\max_j d_{ij}}{1 + \sqrt{\frac{d_{\mathrm{max}}(i)}{d_i}}},
\]
where $d_{\mathrm{max}}(i)$ is the average degree of nodes located at the maximum distance from node $i$.

The EGM index accounts for both the local connectivity (through $V_i$) and the effective spatial reach of each node (through $R_i$), providing a nuanced measure of node influence in the network.

\section{Effective Size (ES)}

The \emph{effective size}\index{effective size} (ES) of node \(i\)'s egocentric network was introduced by Burt \cite{Burt1992} to quantify the number of nonredundant contacts in a node’s local network. The effective size \(c_{\mathrm{ES}}(i)\) of node \(i\) is defined as
\begin{equation*}
   c_{\mathrm{ES}}(i) = \sum_{j \in \mathcal{N}(i)} \left( 1 - \sum_{k \in \mathcal{N}(i) \setminus \{j\}} p_{ik} \, m_{jk} \right),
\end{equation*}
where \(p_{ik}\) denotes the proportion of \(i\)’s time or energy invested in the relationship with node \(k\), computed as
\[
p_{ik} = \frac{a_{ik} + a_{ki}}{\sum_{q \in \mathcal{N}(i)} (a_{iq} + a_{qi})},
\]
and \(m_{jk}\) represents the marginal strength of contact \(j\)’s relation with contact \(k\), given by
\begin{equation*}
   m_{jk} = \frac{a_{jk} + a_{kj}}{\max_{q} (a_{jq} + a_{qj})}.
\end{equation*}

According to Burt \cite{Burt1992}, the inner term \(\sum_{k \in \mathcal{N}(i) \setminus \{j\}} p_{ik} m_{jk}\) quantifies the \emph{redundancy} of node \(j\), that is, the extent to which \(i\)’s connection to \(j\) is duplicated by other ties in \(i\)’s network. If node \(j\) is completely disconnected from all other neighbors of \(i\), this term equals zero, indicating that \(j\) provides a fully nonredundant contact.  Thus, the effective size \(c_{\mathrm{ES}}(i)\) measures the total number of nonredundant contacts of node $i$.

\section{Efficiency centrality (EffC)}

The \emph{efficiency centrality}\index{efficiency centrality (EffC)} (EffC), also known as information centrality, quantifies the contribution of each node to the overall efficiency of a network \cite{Latora2007,SWang2017}. The centrality of node $i$ is defined as the relative decrease in network efficiency resulting from its removal:
\[
c_{EffC}(i) = \frac{E(G) - E(G_i)}{E(G)},
\]
where $G_i$ is the subgraph obtained by removing node $i$, and $E(G)$ denotes the global efficiency of graph $G$, calculated as
\[
E(G) = \frac{1}{N(N-1)} \sum_{i \neq j} \frac{1}{d_{ij}},
\]
with $d_{ij}$ being the length of the shortest path between nodes $i$ and $j$. If no path exists between $i$ and $j$, it is assumed that $d_{ij} = \infty$.  

Thus, the global efficiency $E(G)$ can be interpreted as the sum of the harmonic centralities of all nodes in $G$, linking efficiency centrality directly to the nodes’ ability to facilitate information flow across the network.

\section{Egocentric betweenness centrality}

\emph{Egocentric betweenness centrality}\index{betweenness centrality!egocentric} measures the betweenness role of a node within its egocentric network \cite{Freeman1982}. The egocentric (or centered) network of node \(i\) is defined as the subgraph of \(G\) consisting of \(i\) and its 1-hop neighbors.  Formally, the egocentric betweenness centrality of node \(i\) is defined as the betweenness centrality of \(i\) within its egocentric network.

For computational purposes, the egocentric betweenness of node \(i\) can be expressed directly using the adjacency matrix \(A\) of the full graph \(G\) \cite{Marsden2002,Everett2005}:
\begin{equation*}
   c_{\mathrm{ego}}(i) = \sum_{j \neq i} \sum_{k \neq i} \frac{1}{\left( A^2 \cdot (1-A) \right)_{jk}},
\end{equation*}
where \(\left( A^2 \cdot (1-A) \right)_{jk}\) counts the number of 2-step paths between neighbors \(j\) and \(k\) that pass through \(i\) but exclude direct connections between \(j\) and \(k\).  

The egocentric betweenness centrality captures how much node \(i\) mediates interactions among its neighbors, reflecting its local brokerage role.

\section{Eigentrust centrality}

\emph{Eigentrust centrality}\index{eigentrust centrality} is designed to quantify trust in a network and to reduce the impact of malicious peers in a peer-to-peer (P2P) system \cite{Kamvar2003}. The eigentrust values \(t\) of the nodes are defined as the limit
\begin{equation*} 
t = \lim_{n \rightarrow \infty} \left(C^T \right)^n \cdot c,
\end{equation*}
where \(c_i = 1/N\) for all nodes, and the elements \(c_{ij}\) of the matrix \(C\) are the normalized trust values:

\begin{equation*} 
c_{ij} = \frac{\max(s_{ij},0)}{\sum_{k=1}^{N} \max(s_{ik},0)}.
\end{equation*}
Here, \(s_{ij}\) is the local trust value, defined as the sum of ratings for transactions that peer \(i\) has received from peer \(j\).  

Eigentrust centrality corresponds to the left principal eigenvector of the matrix \(C\), which is equivalent to the stationary distribution of the Markov chain defined by \(C\). In particular, if the local trust values are derived from the adjacency matrix \(A\), then \(C\) can be expressed as the row-normalized matrix \(C = D^{-1} A\), where \(D\) is a diagonal \(N \times N\) matrix with the degree (number of neighbors) of each node on the diagonal. In this case, eigentrust centrality reduces to the left principal eigenvector of $C$, equivalent to PageRank with a damping factor of 1.

\section{Eigenvector centrality}

\emph{Eigenvector centrality}\index{eigenvector centrality} (also known as the principal eigenvector or left dominant eigenvector) was initially introduced by Landau \cite{Landau1895} in the context of chess tournaments. The concept was later independently rediscovered by Wei \cite{Wei1952} and subsequently popularized by Kendall \cite{Kendall1955} for sports ranking. Berge \cite{Berge1958} extended the idea by proposing a general definition of eigenvector centrality for graphs based on social connections. Later, Bonacich \cite{Bonacich1972b} reintroduced and further popularized the measure, particularly in the context of link analysis.

Eigenvector centrality generalizes degree centrality by accounting not only for the number of connections of a node, but also for the centrality of its neighbours \cite{Bonacich1972,Newman2018}. Formally, the importance $c_{ev}(i)$ of a node $i$ is proportional to the sum of the importances of its neighbours, which themselves depend on the importances of their neighbours, and so on, i.e.,
\begin{equation*}
    c_{ev}(i) = \frac{1}{\lambda_{max}} \sum_{(i,j)\in \mathcal{L}}{c_{ev}(j)}=\frac{1}{\lambda_{max}} \sum_{j=1}^{N}{a_{ij} \cdot c_{ev}(j)}.
\end{equation*}

The calculation of eigenvector centrality can be formulated as an eigenvalue problem, where $\lambda_{\text{max}}$ is the largest eigenvalue of the adjacency matrix $A$, and $c_{ev}$ is the corresponding eigenvector. Eigenvector centrality is typically applied to undirected networks; however, it can, in theory, also be computed for directed networks, although certain complications arise in the directed case \cite{Newman2018}.

\section{Electrical closeness centrality}
\emph{Electrical closeness}\index{closeness centrality!electrical (EleClose)} (EleClose) centrality is a variant of the classical closeness centrality that uses the effective resistance to measure the distance between pairs of nodes~\cite{Wang2010,Ellens2011}. The EleClose centrality of node $i$ is defined as
\[
c_{\mathrm{EleClose}}(i) = \frac{1}{\sum_{j=1}^N \Omega_{ij}},
\]
where 
\[
\Omega_{ij} = Q^{\dagger}_{ii} + Q^{\dagger}_{jj} - 2 Q^{\dagger}_{ij}
\]
denotes the effective resistance between nodes $i$ and $j$, and $Q^{\dagger}$ is the Moore-Penrose pseudo-inverse of the Laplacian matrix of the graph $G$. This measure accounts for all paths in the network weighted by their effective resistance, capturing both direct and indirect connections between nodes and providing a centrality value that reflects the node's influence on the overall network connectivity.

\section{Endpoint betweenness centrality}

\emph{Endpoint betweenness}\index{betweenness centrality!endpoint (EPBC)} centrality (EPBC) is a variant of standard betweenness centrality that considers not only a node's role as an intermediary on shortest paths but also as a source or target \cite{Brandes2008}. This extension is particularly relevant in networks such as information exchange systems, where the origin or destination of information can be as influential as the nodes that relay it. It is formally defined as
\begin{equation*}
    c_{\mathrm{EPBC}}(i) = \sum_{j \neq k} \frac{\sigma_{jk}(i)}{\sigma_{jk}},
\end{equation*}
where $\sigma_{jk}$ denotes the total number of shortest paths from node $j$ to node $k$, and $\sigma_{jk}(i)$ counts the number of those paths that pass through node $i$, including the cases where $i$ acts as the source ($j=i$) or the target ($k=i$).

In fully connected directed graphs (excluding trivial self-paths), including endpoints results in a uniform increase of $2(n-1)$ in the centrality of each node relative to standard betweenness, leaving the relative rankings unchanged. However, in networks where some nodes cannot reach all others, this increase becomes non-uniform. In such cases, endpoint betweenness centrality provides a more accurate measure of node importance by capturing how frequently a node participates as a source or target in shortest-path connections.

\section{EnRenew}

The \emph{EnRenew}\index{EnRenew} algorithm is a variant of the VoteRank algorithm that incorporates information entropy to evaluate node influence \cite{Guo2020}. Initially, the seed set \(S\) is empty. Each node \(i\) votes for its neighbor \(j\) with a vote weight
\begin{equation*}
    h_{ij} = -\frac{d_i}{\sum_{l \in \mathcal{N}(j)} d_l} 
              \log \left(\frac{d_i}{\sum_{l \in \mathcal{N}(j)} d_l} \right),
\end{equation*}
where \(d_i\) is the degree of node \(i\) and \(\mathcal{N}(j)\) denotes the neighbors of node \(j\).  

The voting procedure iteratively executes the following steps:

\begin{enumerate}
    \item Each node \(i\) votes for its neighbors \(j\) using the weights \(h_{ij}\).
    \item Select the node \(k \notin S\) with the highest total votes 
    \begin{equation*}
        s_k = \sum_{i=1}^{N} a_{ik} h_{ik}
    \end{equation*}
    and add it to the seed set \(S\).
    \item Update the voting weights \(h_{ij}\) for nodes within the \(l\)-hop neighborhood of node \(k\):
    \begin{equation*}
        h^{\mathrm{new}}_{ij} = \left( 1 - \frac{1}{2^{l-1} E_{\langle d \rangle}} \right) h_{ij},
    \end{equation*}
    where
    \begin{equation*}
        E_{\langle d \rangle} = -\log \frac{1}{\langle d \rangle}
    \end{equation*}
    represents the information entropy of a node in a \(\langle d \rangle\)-regular graph, with \(\langle d \rangle\) being the average degree of the network. Guo \textit{et al.} \cite{Guo2020} suggest \(l = 2\).
\end{enumerate}

\section{Entropy and mutual information-based centrality (EMI)}

\emph{Entropy and mutual information-based centrality}\index{entropy!mutual information-based centrality (EMI)} (EMI) quantifies the importance of a node by combining its structural entropy with the mutual information shared with its neighbors, which can highlight nodes that are overvalued in terms of connectivity \cite{YLi2019}. For an unweighted, undirected network, the centrality of node \(i\) is defined as
\begin{equation*}
    c_{\mathrm{EMI}}(i) = S(i) + MI(i),
\end{equation*}
where \(S(i)\) is the \emph{structural entropy}\index{entropy!structural} of node \(i\):
\begin{equation*}
    S(i) = - \sum_{j \in \mathcal{N}(i)} \frac{d_j}{\sum_{l \in \mathcal{N}(i)} d_l} 
           \log \frac{d_j}{\sum_{l \in \mathcal{N}(i)} d_l},
\end{equation*}
with \(d_j\) being the degree of neighbor \(j\). The term \(MI(i)\) captures the mutual information between node \(i\) and its neighbors:
\begin{equation*}
    MI(i) =\sum_{j \in \mathcal{N}(i)} \frac{|\mathcal{N}(i) \cap \mathcal{N}(j)|}{|\mathcal{N}(i)| |\mathcal{N}(j)|} \log \frac{|\mathcal{N}(i)|+|\mathcal{N}(j)|-|\mathcal{N}(i) \cap \mathcal{N}(j)|}{|\mathcal{N}(i)| |\mathcal{N}(j)|}
\end{equation*}

\section{Entropy centrality}
\emph{Entropy centrality}\index{entropy centrality} measures the structural importance of a node based on the concept of information entropy in a network \cite{OrtizArroyo2008}. It quantifies how much the overall uncertainty or information diversity of the network decreases when node \( i \) is removed. Formally,
\[
c_{\mathrm{Entropy}}(i) = H_{ce}(G) - H_{ce}(G_i),
\]
where \( G_i \) is the graph obtained by deleting node \( i \) (and its associated edges) from \( G \). The term \( H_{ce}(G) \) denotes the \emph{centrality entropy} of the graph, defined as
\[
H_{ce}(G) = -\sum_{i=1}^{N} \gamma(i) \log_2 \gamma(i),
\]
with
\[
\gamma(i) = \frac{\sum_{j=1}^N\sigma_{ij}}{\sum_{k=1}^{N}\sum_{j=1}^{N}\sigma_{kj}}.
\]
where \(\sigma_{ij}\) denotes the number of shortest paths from node \(j\) to node \(k\). Thus, \( \gamma(i) \) denotes the normalized contribution of node \( i \) to the network’s connectivity structure, computed as the fraction of all geodesic paths that originate from it.

The underlying intuition of entropy centrality is that the structural configuration of a network can be regarded as an information system, where entropy quantifies the heterogeneity of connections among nodes. The removal of a structurally important node reduces this heterogeneity and, consequently, the network’s overall entropy. Therefore, nodes whose removal results in a larger decrease in entropy are considered more central or influential.

\section{Entropy variation (EnV)}

\emph{Entropy variation}\index{entropy variation (EnV)} (EnV) is a vitality-based centrality measure that quantifies the change in graph entropy caused by the removal of a node \cite{Ai2017}. Let $G_i$ denote the subgraph obtained by removing node $i$ from $G$. The entropy variation of node $i$ is defined as
\begin{equation*}
    c_{\text{EnV}}(i) = I(G) - I(G_i),
\end{equation*}
where $I(G)$ is the entropy of the graph with respect to a chosen centrality measure $f$:
\begin{equation*}
    I(G) = - \sum_{i=1}^N \frac{f(i)}{\sum_{j=1}^N f(j)} \log \frac{f(i)}{\sum_{j=1}^N f(j)}.
\end{equation*}

Ai \cite{Ai2017} considers four choices for $f$: in-degree centrality, out-degree centrality, degree centrality, and betweenness centrality. A higher EnV indicates that the removal of the node causes a larger redistribution of centrality values, highlighting nodes that are critical for maintaining the overall structural balance and information flow in the network.

\section{Entropy-based gravity model (EGM)}

The \emph{entropy-based gravity model}\index{gravity model!entropy-based} (EGM) is a variant of the local gravity model in which a node’s mass is determined by an entropy-based hybrid centrality measure \cite{Yan2020}. Specifically, the mass \( m(i) \) of node \( i \) is defined as a weighted linear combination of \( m \) normalized centrality measures:
\begin{equation*}
    m(i) = \sum_{j=1}^m w_j c_j(i),
\end{equation*}
where \( c_j(i) \) denotes the normalized value of the \( j\)-th centrality index, and \( w_j \) represents its corresponding weight. The weights \( w_j \) are determined using the entropy weight method, which quantifies the amount of information each centrality measure contributes. Specifically,
\begin{equation*}
    w_j = \frac{1 - S_j}{m - \sum_{k=1}^m S_k},
\end{equation*}
where
\begin{equation*}
    S_j = -\frac{1}{\ln N} \sum_{i=1}^N 
    \frac{c_j(i)}{\sum_{k=1}^N c_j(k)} 
    \ln \left( \frac{c_j(i)}{\sum_{k=1}^N c_j(k)} \right),
\end{equation*}
and \( S_j \) represents the entropy value of the \( j\)-th centrality measure across all \( N \) nodes.

Using the entropy-weighted mass, the EGM centrality of node \( i \) is given by
\begin{equation*}
    c_{\text{egm}}(i) = \sum_{j \in \mathcal{N}^{(\leq l)}(i)} \frac{m(i)\, m(j)}{d_{ij}^2},
\end{equation*}
where \( d_{ij} \) is the shortest path distance between nodes \( i \) and \( j \), and \( \mathcal{N}^{(\leq l)}(i) \) denotes the set of nodes whose shortest-path distance from $i$ is less than or equal to $l$ (typically \( l = 2 \)).

The performance of the entropy-based gravity model depends on the selection of centrality measures included in the hybrid formulation. Yan \textit{et al.} \cite{Yan2020} compared nine different combinations of centrality measures and demonstrated that the combination incorporating the h-index (Lobby index), closeness centrality, betweenness centrality and PageRank yields the best performance. The effectiveness of this approach was validated on six real-world networks through simulations of the Susceptible-Infected-Recovered (SIR) spreading process.

\section{Entropy-based influence disseminator (EbID)}

Entropy-Based Influence Disseminator (EbID) quantifies node influence by combining an entropy-based node quality index with the community structure of its neighbors \cite{Saxena2020b}. The centrality of node $i$ is defined as
\begin{equation*}
    c_{\text{EbID}}(i) = \sum_{j \in \mathcal{N}(i)} \frac{1}{d_i (d_j - 1)} \log \frac{1}{d_i (d_j - 1)} 
    + \sum_{j \in \mathcal{N}(i)} v(C_j),
\end{equation*}
where $\mathcal{N}(i)$ is the set of neighbors of node $i$, $d_i$ is the degree of node $i$. The first term captures the entropy of the probability distribution of reaching a node in two hops, and the second term, $v(C_j)$, measures the relative edge density of the community to which node $j$ belongs. Saxena \textit{et al.} \cite{Saxena2020b} employ the Louvain method to detect communities.  

Nodes with higher $c_{\text{EbID}}(i)$ are those that not only provide efficient two-hop reachability but also are connected to well-connected communities, making them effective disseminators of influence in the network.

\section{Entropy-based ranking measure (ERM)}

The \emph{Entropy-Based Ranking Measure}\index{entropy-based ranking measure (ERM)} (ERM) is a centrality metric that quantifies the influence of a node based on the degrees of its first- and second-order neighbors \cite{Zareie2017}. Let \(d_i^{(1)}\) denote the total degree of the neighbors of node \(i\), defined as
\begin{equation*}
    d_i^{(1)} = \sum_{j \in \mathcal{N}(i)} d_j,
\end{equation*}
where \(d_j\) is the degree of neighbor \(j\). Similarly, let \(d_i^{(2)}\) be the total degree of the neighbors of node \(i\)'s neighbors:
\begin{equation*}
    d_i^{(2)} = \sum_{j \in \mathcal{N}(i)} d_j^{(1)}.
\end{equation*}

The ERM centrality of node \(i\) is defined as
\begin{equation*}
    c_{\mathrm{ERM}}(i) = \sum_{j \in \mathcal{N}(i)} \sum_{k \in \mathcal{N}(j)} EC(i),
\end{equation*}
where \(EC(i)\) represents \emph{the entropy centrality}\index{entropy centrality} of node \(i\), given by
\begin{equation*}
    EC(i) = E_1(i) + \lambda_i E_2(i) 
    = -\sum_{j \in \mathcal{N}(i)} \frac{d_j}{d_i^{(1)}} \log{\frac{d_j}{d_i^{(1)}}} 
      + \lambda_i \left(-\sum_{j \in \mathcal{N}(i)} \frac{d_j^{(1)}}{d_i^{(2)}} \log{\frac{d_j^{(1)}}{d_i^{(2)}}} \right).
\end{equation*}
Here, \(E_1(i)\) and \(E_2(i)\) denote the entropy of the degrees of the first- and second-order neighbors of node \(i\), respectively, and \(\lambda_i \in [0,1]\) is a tunable parameter that balances their contributions. Following \cite{Zareie2017}, \(\lambda_i\) can be set as
\begin{equation*}
    \lambda_i = \frac{d_i^{(2)}}{\max_k d_k^{(2)}},
\end{equation*}
so that nodes with larger second-order neighborhoods give proportionally more weight to \(E_2(i)\).

\section{Entropy-Burt method (E-Burt)}

The \emph{Entropy-Burt method}\index{entropy-Burt method (E-Burt)} (E-Burt) is an entropy-based extension of Burt's constraint that accounts for both the weights of connections and the distribution of a node’s total connection strength across its edges \cite{Hu2018}. The centrality of node \(i\) is defined as
\begin{equation*}
    c_{E-Burt}(i) = \sum_{j \in \mathcal{N}(i)} \left( p_{ij} + \sum_{k \in \mathcal{N}(i) \setminus \{j\}} p_{ik} p_{ki} \right)^2,
\end{equation*}
where
\begin{equation*}
    p_{ij} = \frac{h_i}{\sum_{k \in \mathcal{N}(i)} h_k}.
\end{equation*}

The term \(h_i\) represents the effective connection strength of node \(i\) and is defined for weighted networks as
\begin{equation*}
    h_i = \left(1 - \sum_{j \in \mathcal{N}(i)} \frac{w_{ij}}{\sum_{k \in \mathcal{N}(i)} w_{ik}} \ln \frac{w_{ij}}{\sum_{k \in \mathcal{N}(i)} w_{ik}} \right) \sum_{j \in \mathcal{N}(i)} w_{ij},
\end{equation*}
where \(w_{ij}\) is the edge weight. For unweighted networks, Hu and Mei \cite{Hu2018} consider \(w_{ij} = d_i \cdot d_j\), with \(d_i\) denoting the degree of node \(i\).

E-Burt centrality assigns higher values to nodes that are constrained yet connected to diverse neighbors, reflecting both uneven distribution of connection strengths and redundancy in the local neighborhood.

\section{Epidemic centrality}

\emph{Epidemic centrality}\index{epidemic centrality} quantifies the expected influence of a node in SIR (Susceptible–Infected–Recovered) processes by averaging the epidemic impact of an outbreak originating from that node over the full range of infection and recovery probabilities \cite{Sikic2013}. In the SIR model, each node can be in one of three states: susceptible (S), infected (I) or recovered (R). An infected node transmits the infection to each susceptible neighbor with probability \(\beta\) and recovers independently with probability \(\mu\). The process continues until no infected nodes remain, producing a final outbreak size that depends on both the initially infected node and the epidemic parameters \((\beta, \mu)\).  

The epidemic centrality \(c_{EC}(i)\) of node \(i\) is defined as
\begin{equation*}
    c_{EC}(i) = \int_0^1 \int_0^1 w(p, q) \, X_i(p,q) \, dp \, dq,
\end{equation*}
where \(X_i(p,q)\) is the expected fraction of nodes infected in an SIR process starting from node \(i\) with infection probability \(p = \beta\) and recovery probability \(q = \mu\), while \(w(p,q)\) is a nonuniform weight function encoding the relative importance of different epidemic regimes \((p, q)\).  

Šikić \textit{et al.} \cite{Sikic2013} propose using a product of beta distributions for the weight:
\begin{equation*}
    w(p,q) = f_{\alpha,\alpha}(p) \, f_{\alpha,\alpha}(q),
\end{equation*}
with
\begin{equation*}
    f_{\alpha,\beta}(x) = \frac{\Gamma(\alpha + \beta)}{\Gamma(\alpha) \Gamma(\beta)} x^{\alpha-1} (1-x)^{\beta-1}, \quad 0 < x < 1,
\end{equation*}
where \(\Gamma(\cdot)\) is the gamma function. For \(\alpha = \beta\), the distribution \(f_{\alpha,\alpha}(x)\) is symmetric around its mean \(x = 1/2\). As a special case, \(\alpha = \beta = 1\) yields a uniform distribution over \([0,1]\).

Epidemic centrality thus quantifies the average epidemic impact of node \(i\), under the assumption that the epidemic starts from \(i\), across all considered epidemic regimes. Nodes with higher epidemic centrality are expected to generate larger outbreaks on average, across the range of infection and recovery probabilities considered. The epidemic centrality captures not only central nodes but also structurally peripheral nodes that may nonetheless exert significant influence on epidemic dynamics.

\section{\(\epsilon\)-betweenness centrality}

The \emph{\(\epsilon\)-betweenness centrality}\index{betweenness centrality!$\epsilon$-} was introduced by Carpenter \textit{et al.} \cite{Carpenter2002} to make betweenness centrality more robust in the presence of uncertain or noisy network data. In many real-world networks, small changes in edge weights or connectivity can dramatically alter shortest paths. To address this, the authors define an \(\epsilon\)-shortest path as a path \(P_{i\rightarrow j}\) from node \(i\) to node \(j\) whose length satisfies
\[
\text{length}(P_{i\rightarrow j}) \leq (1 + \epsilon) \, d_{ij},
\]
where \(d_{ij}\) is the shortest-path distance between \(i\) and \(j\). Therefore, \(\epsilon\)-betweenness considers all paths with lengths close to the shortest path, not only the exact shortest paths.

The \(\epsilon\)-betweenness centrality of node \(i\), denoted \(c_{e\text{-}betw}(i)\), is then defined as
\begin{equation*}
c_{e\text{-}betw}(i) = \sum_{j=1}^{N} \sum_{k=1}^{N} \frac{\sigma_{jk}^{\epsilon}(i)}{\sigma_{jk}^{\epsilon}},
\end{equation*}
where \(\sigma_{jk}^{\epsilon}\) is the total number of \(\epsilon\)-shortest paths between nodes \(j\) and \(k\), and \(\sigma_{jk}^{\epsilon}(i)\) is the number of such paths that pass through node \(i\). This definition generalizes standard betweenness, reducing sensitivity to minor changes in the network structure.

\section{Even subgraph centrality}

\emph{Even subgraph centrality}\index{subgraph centrality!even} counts the number of closed walks of \emph{even} length in a network \cite{Rodriguez2007}. Even-length walks include contributions from both cyclic and acyclic structures, reflecting back-and-forth movements that capture redundancy, potential signal propagation, and indirect interactions. The even subgraph centrality of node $i$, denoted $c_{even}(i)$, is defined as
\[
c_{even}(i) = \sum_{k=0}^{\infty} \frac{(A^{2k})_{ii}}{(2k)!} 
= \sum_{j=1}^{N} \left( v_j(i) \right)^2 \cosh(\lambda_j),
\]
where $A$ is the adjacency matrix of the network, and $v_j(i)$ is the $i$-th component of the eigenvector $v_j$ corresponding to eigenvalue $\lambda_j$.

\section{Exogenous centrality}
\emph{Exogenous centrality}\index{exogenous centrality} quantifies the effect of a node on the centrality of other nodes in a graph \cite{Everett2010}. Formally, the exogenous centrality of node \(i\) is defined as
\begin{equation*}
c_{E}(i) = \sum_{j \in \mathcal{N} \setminus \{i\}} \bigl( c(j,G) - c(j,G_i) \bigr),
\end{equation*}
where \(G_i\) is the subgraph obtained by removing node \(i\) from \(G\), and \(c(j,G)\) is a standard centrality measure of node \(j\) (e.g., degree, closeness, or betweenness).  

Exogenous centrality quantifies the contribution of node \(i\) to the centrality of all other nodes in the network. In other words, it measures how the presence of \(i\) enhances the centrality of the rest of the graph. For example, if the underlying centrality \(c(i,G)\) is the degree \(d_i\), the exogenous centrality of \(i\) equals \(d_i\). Similarly, if \(c(i,G)\) is closeness centrality, the exogenous centrality of \(i\) reflects how the removal of \(i\) would increase the average shortest-path distances among the remaining nodes, thereby reducing their closeness.

\section{Expected Force (ExF)}

The \emph{expected force}\index{expected force (ExF)} (ExF) is a semi-local, entropy-based measure that quantifies the spreading power of nodes in a network \cite{Lawyer2015}. In a continuous-time epidemiological framework, a node's spreading potential can be estimated by summarizing the distribution of infected-susceptible edges after a small number of transmission events originating from that node in an otherwise fully susceptible network.

Consider a network where a single node $i$ is initially infected, and all other nodes are susceptible. Let $1, \dots, J$ denote all possible clusters of infected nodes after $x=2$ transmission events, assuming no recovery. Each cluster $j$ represents either (i) node $i$ plus two neighbors at distance one, or (ii) node $i$ plus one neighbor at distance one and another at distance two. The expected force of node $i$ is defined as
\begin{equation*}
    c_{\text{ExF}}(i) = - \sum_{j=1}^J \frac{D_j}{\sum_{k=1}^J D_k} \log \frac{D_j}{\sum_{k=1}^J D_k},
\end{equation*}
where $D_j$ is the degree of cluster $j$, i.e., the total number of neighbors of nodes in the cluster.

\section{Expected rank}

\emph{Expected rank}\index{expected rank} is a centrality measure based on \emph{neighborhood inclusion}, which induces a partial ranking among nodes and has been shown to be preserved by many existing centrality indices \cite{Schoch2018}. Schoch and Brandes \cite{Schoch2016} formalized a simple principle: if a node has the same or a superset of the connections of another node, it cannot be less central. Formally,
\begin{equation*}
\mathcal{N}(u) \subseteq \mathcal{N}(v) \cup \{v\} \quad \Rightarrow \quad c(u) \leq c(v),
\end{equation*}
where $\mathcal{N}(u)$ denotes the set of neighbors of node $u$.

Neighborhood inclusion defines a \emph{pre-order} (a reflexive and transitive binary relation) on the nodes of a graph. From this partial order, full rankings (linear extensions) can be generated by arranging the nodes into a total order that preserves all pairwise relations implied by the partial order. The expected rank of node $i$, denoted $c_{ER}(i)$, is then defined as
\[
c_{ER}(i) = \sum_{k=1}^{N} k \, \Pr[\mathrm{rk}(i) = k],
\]
where $\Pr[\mathrm{rk}(i) = k]$ is the probability that node $i$ occupies rank $k$ in a full ranking.

A limitation of this approach is that enumerating all possible linear extensions is computationally prohibitive for large networks. Schoch and Brandes \cite{Schoch2018} suggest using approximation techniques, such as sampling or heuristic methods, to estimate expected ranks efficiently.

\section{Extended cluster coefficient ranking measure (ECRM)}

The \emph{extended cluster coefficient ranking measure}\index{extended!cluster coefficient ranking measure (ECRM)} (ECRM) is an extension of the shell clustering coefficient (SCC), where the centrality of a node depends not only on its own SCC but also on the SCC values of its neighbors. Specifically, the ECRM score \(c_{ECRM}(i)\) is defined as
\[
c_{ECRM}(i) = \sum_{j \in \mathcal{N}(i)} \sum_{l \in \mathcal{N}(j)} SCC(l),
\]
where the shell clustering coefficient of node \(i\) is given by
\[
SCC(i) = \sum_{j \in \mathcal{N}(i)} \left[ 2 - \mathrm{corr}\bigl[ sv(i), sv(j) \bigr] + \left(\frac{2 d_j}{\max_l d_l} + 1 \right) \right].
\]
Here, \(d_j\) denotes the degree of node \(j\), and \(\mathrm{corr}[sv(i), sv(j)]\) is the Pearson correlation between the shell vectors of nodes \(i\) and \(j\):
\[
sv(i) = \bigl(|N_{ks}^{(1)}(i)|, \dots, |N_{ks}^{(f)}(i)| \bigr),
\]
where \(|N_{ks}^{(k)}(i)|\) represents the number of neighbors of node \(i\) belonging to the \(k\)-th hierarchy in the $k$-shell decomposition, and \(f\) is the maximum hierarchy level in the network.  

Nodes with high ECRM values are those that are connected to neighbors with diverse shell hierarchies and locally distinct structures, indicating that they occupy positions that bridge multiple structural layers and potentially influence the network across different $k$-shell levels.

\section{Extended diversity-strength ranking (EDSR)}

\emph{Extended diversity-strength ranking}\index{diversity-strength ranking (DSR)!extended (EDSR)} (EDSR) further extends the diversity-strength ranking (DSR) measure by incorporating the influence potential of nodes over a wider network range, as proposed by Zareie \textit{et al.} \cite{Zareie2019}. The EDSR value of node $i$ is given by
\begin{equation*}
    c_{\text{EDSR}}(i) =\sum_{j \in \mathcal{N}(i)}  c_{\text{DSR}}(i) = \sum_{j \in \mathcal{N}(i)} 
    \left[
        \sum_{k \in \mathcal{N}(j)} 
        \left(
            \sum_{p \in \mathcal{N}(k)} 
            \frac{IKs(p)}{\sum_{q \in \mathcal{N}(k)} IKs(q)} 
            \log \frac{IKs(p)}{\sum_{q \in \mathcal{N}(k)} IKs(q)}
        \right)
    \right],
\end{equation*}
where $\mathcal{N}(i)$ denotes the set of neighbors of node $i$, and $IKs(p)$ is the improved $k$-shell index of node $p$ as defined by Liu \textit{et al.} \cite{ZLiu2015}. The innermost term represents the diversity-strength centrality of node $k$, the middle summation yields the DSR of node $j$, and the outermost summation aggregates these across the neighbors of node $i$. EDSR thus captures multi-level influence by integrating local, second-order, and higher-order structural information. Nodes with high EDSR values lie within regions of strong and diverse influence, indicating their importance in diffusion and spreading processes across the network.

\section{Extended gravity centrality (EGC)}

Inspired by the LocalRank centrality \cite{Chen2012} and the extended neighborhood coreness \cite{Bae2014}, Ma \textit{et al.}~\cite{Ma2015} proposed the \emph{extended gravity centrality}\index{gravity centrality!extended} (EGC) of node \(i\), denoted by \(c_{\text{EGC}}(i)\), as
\begin{equation*}
   c_{\text{EGC}}(i) = \sum_{j \in \mathcal{N}(i)} c_{\text{Gravity}}(j)
   = \sum_{j \in \mathcal{N}(i)} \sum_{l \in \mathcal{N}^{(\leq l)}(j)} \frac{k_s(j)\,k_s(l)}{d_{jl}^2},
\end{equation*}
where \(\mathcal{N}(i)\) denotes the set of neighbors of node \(i\), \(d_{jl}\) is the shortest path distance between nodes \(j\) and \(l\), \(k_s(j)\) represents the $k$-shell value of node \(j\) and \(\mathcal{N}^{(\leq l)}(j)\) denotes the set of nodes within the $l$-hop neighborhood of node \(j\).

The EGC thus integrates the gravity centralities of a node’s immediate neighbors, capturing both local and higher-order topological influences in the network.

\section{Extended \textit{h}-index centrality (EHC)}

\emph{Extended \textit{h}-index centrality}\index{\textit{h}-index!extended (EHC)} (EHC) is a node centrality measure that ranks nodes based on the degrees of their neighbors \cite{Zareie2019b}. For each node $i$, a cumulative function $CMC(i)$ is first computed as
\[
CMC(i) = \sum_{k=1}^{h} p^{\,1 + k \frac{p}{r}} \, s_k(i),
\]
where $h = \max_j d_j$ is the maximum degree in the network, $s_k(i)$ is the number of neighbors of node $i$ with degree greater than or equal to $k$, and $p$ and $r$ are tunable parameters. Experimentally, Zareie and Sheikhahmadi \cite{Zareie2019b} suggest $p = 0.8$ and $r = 100$.

The extended H-index centrality of node $i$, denoted $c_{EHC}(i)$, is then defined as the sum of the cumulative functions of its neighbors:
\[
c_{EHC}(i) = \sum_{j \in \mathcal{N}(i)} CMC(j).
\]

\section{Extended \textit{k}-shell hybrid method}

The \emph{extended $k$-shell hybrid method}\index{\textit{k}-shell hybrid method!extended (ESKH)} (ESKH), proposed by Namtirtha \textit{et al.} \cite{Namtirtha2018}, is an extension of the $k$-shell hybrid method (ksh) in which the centrality of a node depends on the hybrid centralities of its neighbors. The centrality \( c_{\mathrm{ESKH}}(i) \) of node \( i \) is defined as
\begin{equation*}
   c_{\mathrm{ESKH}}(i) 
   = \sum_{j \in \mathcal{N}(i)} c_{\mathrm{ksh}}(j) 
   = \sum_{j \in \mathcal{N}(i)} \sum_{t \in \mathcal{N}^{(\leq l)}(j)} 
    \frac{\sqrt{k_s(j) + k_s(t)} + \mu\,k_t}{d_{jt}^2},
\end{equation*}
where \( \mathcal{N}(i) \) denotes the set of neighbors of node \( i \), 
\( \mathcal{N}^{(\leq l)}(j) \) represents the set of nodes within the \( l \)-hop neighborhood of node \( j \),
\( d_{jt} \) is the shortest path distance between nodes \( j \) and \( t \),
\( k_s(j) \) and \( k_s(t) \) are the $k$-shell indices of nodes \( j \) and \( t \), respectively,
\( k_t \) is the degree of node \( t \), 
and \( \mu \in (0,1) \) is a tunable parameter that balances the relative influence of the two components.

\section{Extended hybrid characteristic centrality (EHCC)}

The \emph{Extended Hybrid Characteristic Centrality}\index{hybrid characteristic centrality (HCC)!extended (EHCC)} (EHCC) extends the hybrid characteristic centrality (HCC) by incorporating the contributions of a node's neighbors \cite{LiuJ2023}. The EHCC of node $i$ is defined as
\begin{equation*}
    c_{EHCC}(i) = c_{HCC}(i) + \sum_{j \in \mathcal{N}(i)} c_{HCC}(j),
\end{equation*}
where $\mathcal{N}(i)$ denotes the set of neighbors of node $i$ and $c_{HCC}(i)$ is the HCC score of node $i$, given by
\begin{equation*}
    c_{HCC}(i) = \frac{d^{ex}(i)}{\max_j d^{ex}(j)} + \frac{pos(i)}{\max_j pos(j)},
\end{equation*}
with $d^{ex}(i)$ representing the extended degree of node $i$, and $pos(i)$ denoting the iteration at which node $i$ is removed during the E-shell decomposition.    

Nodes with high EHCC values are influential both due to their own structural position in the network and the importance of their immediate neighbors.

\section{Extended hybrid degree and \textit{k}-shell method}

The \textit{extended hybrid degree and \(k\)-shell method}\index{hybrid degree and \textit{k}-shell method!extended} ranks nodes in complex networks by considering both the \(k\)-shell index and the degree of each node and its neighbors \cite{Maji2021}. It extends the hybrid degree and \(k\)-shell centrality (\(c_{x\text{-}ks}\)) by defining the centrality of node \(i\) as the sum of the \(c_{x\text{-}ks}\) scores of its immediate neighbors:
\[
c_{\text{ex-ks}}(i) = \sum_{j \in \mathcal{N}(i)} c_{x\text{-}ks}(j)
= \sum_{j \in \mathcal{N}(i)} \left( k_s(j) \sum_{k \in \mathcal{N}^{(\leq r)}(j)} \frac{d_k}{d_{jk}^2} \right),
\]
where \(\mathcal{N}(i)\) denotes the set of neighbors of node \(i\), \(\mathcal{N}^{(\leq r)}(j)\) is the set of nodes within distance \(r\) from node \(j\) (excluding \(j\) itself), \(k_s(j)\) and \(d_j\) are the \(k\)-shell index and degree of node \(j\), and \(d_{jk}\) is the shortest path distance between nodes \(j\) and \(k\).  Maji \textit{et al.}~\cite{Maji2021} consider a three-hop neighborhood, i.e., \(r = 3\).

The extended hybrid degree and \(k\)-shell method has been evaluated using the susceptible-infected-recovered (SIR) model and metrics such as spreadability, monotonicity, and Kendall’s tau, demonstrating superior performance compared to several existing centrality measures in identifying influential seed nodes on real networks.

\section{Extended hybrid degree and MDD method}

The \textit{extended hybrid degree and MDD method}\index{hybrid degree and MDD method!extended} ranks nodes in complex networks by considering both the mixed degree decomposition (MDD) index and the degree of each node and its neighbors \cite{Maji2021}. It extends the hybrid degree and MDD centrality (\(c_{x\text{-}MDD}\)) by defining the centrality of node \(i\) as the sum of the \(c_{x\text{-}MDD}\) scores of its immediate neighbors:
\[
c_{\text{ex-MDD}}(i) = \sum_{j \in \mathcal{N}(i)} c_{x\text{-}MDD}(j)
= \sum_{j \in \mathcal{N}(i)} \left( MDD(j) \sum_{k \in \mathcal{N}^{(\leq r)}(j)} \frac{d_k}{d_{jk}^2} \right),
\]
where \(\mathcal{N}(i)\) denotes the set of neighbors of node \(i\), \(\mathcal{N}^{(\leq r)}(j)\) is the set of nodes within distance \(r\) from node \(j\) (excluding \(j\) itself), \(MDD(j)\) and \(d_j\) are the MDD index and degree of node \(j\), and \(d_{jk}\) is the shortest path distance between nodes \(j\) and \(k\).  Maji \textit{et al.}~\cite{Maji2021} consider a three-hop neighborhood, i.e., \(r = 3\).

The extended hybrid degree and MDD method has been evaluated using the susceptible-infected-recovered (SIR) model and metrics such as spreadability, monotonicity, and Kendall’s tau, demonstrating superior performance compared to several existing centrality measures in identifying influential seed nodes on real networks.

\section{Extended improved \textit{k}-shell hybrid method}

Maji \textit{et al.} \cite{Maji2020b} introduced the \emph{extended improved $k$-shell hybrid}\index{\textit{k}-shell hybrid method!extended improved (EIKH)} (EIKH) centrality, in which the improved $k$-shell hybrid (IKH) centrality ($c_{\mathrm{IKH}}$) replaces the standard $k$-shell hybrid measure ($c_{\mathrm{ksh}}$) to enhance sensitivity to structural variations within the network. Specifically, the centrality \( c_{\mathrm{EIKH}}(i) \) of node \( i \) is defined as
\begin{equation*}
   c_{\mathrm{ESKH}}(i) 
   = \sum_{j \in \mathcal{N}(i)} c_{\mathrm{IKH}}(j) 
   = \sum_{j \in \mathcal{N}(i)} \sum_{t \in \mathcal{N}^{(\leq l)}(j)} 
    \frac{\sqrt{k_s(j) + k_s(t)} + \mu(d_{jt}) \cdot k_t}{d_{jt}^2},
\end{equation*}
where \( \mathcal{N}(i) \) denotes the set of neighbors of node \( i \), 
\( \mathcal{N}^{(\leq l)}(j) \) represents the set of nodes within the \( l \)-hop neighborhood of node \( j \),
\( d_{jt} \) is the shortest path distance between nodes \( j \) and \( t \),
\( k_s(j) \) and \( k_s(t) \) are the $k$-shell indices of nodes \( j \) and \( t \), respectively,
\( k_t \) is the degree of node \( t \), 
and \( \mu\) is defined as
\begin{equation*}
    \mu (d_{jt}) = \frac{2(l - d_{jt} + 1)}{l(l + 1)}.
\end{equation*}

\section{Extended local bridging centrality (ELBC)}

The \emph{extended local bridging centrality}\index{bridging centrality!extended local} (ELBC), also known as the 2-hop local bridging centrality (LBC2) or Extended LBC, extends the localized bridging centrality by incorporating information from a broader, two-hop neighborhood around each node \cite{Macker2016}. It is defined as
\begin{equation*}
   c_{\mathrm{LBC2}}(i) = c_{\mathrm{ego2}}(i) \cdot \beta_c(i),
\end{equation*}
where \(c_{\mathrm{ego2}}(i)\) denotes the betweenness centrality of node \(i\) computed within its 2-hop egocentric network, and \(\beta_c(i)\) is the bridging coefficient of node \(i\).  

This extension captures not only the immediate (1-hop) brokerage of a node but also its mediating role across the next layer of neighbors, offering a more comprehensive view of local connectivity.

\section{Extended local dimension (ELD)}

\emph{Extended local dimension (ELD)}\index{local dimension!extended (ELD)} was proposed by Pu \textit{et al.}~\cite{Pu2014} as an extension of the \emph{local dimension} (LD)~\cite{Silva2012} to account for variations in topological distance across nodes in a network. While the original LD by Silva and Costa \cite{Silva2012} measures the scaling of neighborhood volume $B_i(r)$ at a fixed distance $r$, the extended LD allows the distance parameter to vary for each node, capturing heterogeneity in local network structure.  

For a given node $i$, let $B_i(r)$ denote the number of nodes within a topological distance $r$ from $i$, and let $n_i(r)$ denote the number of nodes exactly at distance $r$. The extended local dimension $D_i(r_i)$ is defined as
\[
D_i(r_i) \simeq r \frac{n_i(r_i)}{B_i(r_i)},
\]
where $r_i$ denotes the maximum value of shortest distances
between the central node $i$ and all others. That means the
maximum $r$ is different between nodes.

ELD is particularly useful in irregular or spatially embedded networks, as it reflects the effective dimensionality around each node and captures local deviations from uniform scaling.

\section{Extended mixed gravitational centrality (EMGC)}

The \emph{extended mixed gravitational centrality}\index{gravitational centrality!extended mixed (EMGC)} (EMGC), proposed by Wang \textit{et al.} \cite{Wang2018}, is an extension of the mixed gravitational centrality (MGC) measure. In this formulation, the centrality \( c_{\mathrm{EMGC}}(i) \) of node \( i \) is determined not only by its immediate structural characteristics but also by the mixed gravitational centralities of its neighboring nodes. The EMGC is defined as
\begin{equation*}
   c_{\mathrm{EMGC}}(i) = \sum_{j \in \mathcal{N}(i)} c_{\mathrm{MGC}}(j)
   = \sum_{j \in \mathcal{N}(i)} \sum_{l \in \mathcal{N}(j)} \frac{k_s(j)\,d_l}{d_{jl}^2},
\end{equation*}
where \( \mathcal{N}(i) \) denotes the set of neighbors of node \( i \), \( d_{jl} \) is the shortest path distance between nodes \( j \) and \( l \), \( k_s(j) \) represents the $k$-shell index of node \( j \), and \( d_l \) is the degree of node \( l \).

The extended mixed gravitational centrality incorporates higher-order neighborhood effects by aggregating the influence of neighboring nodes’ MGC values, thereby providing a more comprehensive assessment of node importance within the network’s multi-level structure.

\section{Extended neighborhood coreness}

The \emph{extended neighborhood coreness}\index{neighborhood coreness!extended} is an extension of the INK score (neighborhood coreness) proposed by Bae \textit{\textit{et al.}}~\cite{Bae2014}. The centrality of node \(i\), denoted by \(c_{nc+}(i)\), is defined as
\begin{equation*}
   c_{nc+}(i) = \sum_{j \in \mathcal{N}(i)} c_{INK}(j) = \sum_{j \in \mathcal{N}(i)} \sum_{l \in \mathcal{N}(j)} k_s(l),
\end{equation*}
where \(c_{INK}(i)\) is the INK score of node \(i\), \(k_s(l)\) is the \(k\)-core value of node \(l\), and \(\mathcal{N}(i)\) denotes the set of neighbors of node \(i\).

\section{Extended RMD-weighted degree (EWD) centrality}

The \emph{extended RMD-weighted degree}\index{RMD-weighted degree centrality!extended (EWD)} (EWD) centrality, originally called extended weighted degree, is an extension of the RMD-weighted degree (WD) centrality \cite{Yang2018}. Motivated by the idea that incorporating neighbor information over a broader range can improve the accuracy of node influence ranking, the EWD centrality of node $i$ is defined as
\[
c_{EWD}(i) = \sum_{j \in \mathcal{N}(i)} c_{RMD}(j)= \sum_{j \in \mathcal{N}(i)}\sum_{k \in \mathcal{N}(j)} \frac{Iter(k)}{MaxIter},
\]
where $\mathcal{N}(i)$ denotes the set of neighbors of node $i$, $Iter(j)$ is the iteration at which node $j$ is removed during the RMD decomposition, and $MaxIter$ is the total number of iterations. The RMD decomposition iteratively removes the node with the smallest degree, ranking nodes according to their structural importance in the network.

Nodes with high EWD values are those whose neighbors are themselves influential according to the RMD-weighted degree. This means the node is not only locally well-connected but also strategically positioned near other structurally important nodes, enhancing its potential impact on the network.

\section{Extended weight degree centrality (EWDC)}

The \emph{extended weight degree centrality}\index{degree centrality!extended weight (EWDC)} (EWDC) is an extension of the weight degree centrality (WDC) \cite{Liu2016}, which incorporates the centrality of a node's neighbors to improve its discriminative power.

The EWDC centrality of node \(i\) is defined as
\begin{equation*}
    c_{EWdc}(i) = \left( \sum_{j \in \mathcal{N}(i)} c_{Wdc}(j) - c_{Wdc}(i) \right) c_{Wdc}(i)^{|r|},
\end{equation*}
where \(c_{Wdc}(i)\) is the weight degree centrality of node \(i\), given by
\begin{equation*}
    c_{Wdc}(i) = \left( \sum_{j \in \mathcal{N}(i)} d_j - d_i \right) d_i^{\alpha}.
\end{equation*}
Here, \(d_i\) is the degree of node \(i\), \(\mathcal{N}(i)\) denotes the set of neighbors of node \(i\), and \(\alpha\) is a tunable parameter controlling the contribution of the node's own degree. Following Liu \textit{et al.}~\cite{Liu2016}, \(\alpha\) can be set to \(|r|\), where \(r\) is the network's degree assortativity coefficient. This allows the centrality measure to adapt to assortative, disassortative, and neutral networks. EWDC further incorporates the WDC values of neighbors, enhancing its ability to distinguish influential nodes across different network structures.

\section{Fitness centrality}

\emph{Fitness centrality}\index{fitness centrality} is a network measure designed to assess node vulnerability within complex networks \cite{Servedio2025}. It incorporates the concept of node fitness, capturing the criticality of nodes that sustain network functionality by supporting many nodes with low connectivity.  

Formally, the fitness centrality \(c_{\mathrm{fitness}}(i,t)\) of node \(i\) at time \(t\) is defined as
\[
c_{\mathrm{fitness}}(i,t) = \delta + \sum_{j \in \mathcal{N}(i)} \frac{1}{c_{\mathrm{fitness}}(j,t-1)},
\]
or, equivalently, in a more compact matrix form:
\[
c_{\mathrm{fitness}}(t) = \delta \mathbf{1} + A \left(c_{\mathrm{fitness}}(t-1)\right)^{-1},
\]
where \(\mathcal{N}(i)\) denotes the set of neighbors of node \(i\), \(A\) is the adjacency matrix of the network, and \(\delta > 0\) is a small constant (typically \(\delta = 0.01\)) to ensure convergence.  

The initial scores are set as \(c_{\mathrm{fitness}}(i,0) = 1\) for all nodes \(i\). This recursive definition reflects the intuition that a node’s fitness centrality is higher when it is connected to many nodes with low fitness centrality, thus identifying nodes that are crucial for maintaining network integrity. The fitness centrality of node \(i\) is taken as the steady-state value \(c_{\mathrm{fitness}}(i,t^\infty)\).  

Servedio \textit{et al.} \cite{Servedio2025} demonstrate that fitness centrality effectively identifies these essential nodes across diverse network topologies, providing a robust tool for vulnerability assessment and targeted interventions in complex systems.

\section{Flow betweenness centrality}
\emph{Flow betweenness}\index{betweenness centrality!flow} is a variant of betweenness centrality where shortest paths are substituted with edge-independent paths \cite{Freeman1991,Newman2018}, i.e.,
\begin{equation*}
    c_{flow-betw}(i) = \sum_{j=1}^{N}{\sum_{k=1}^{N}{\frac{m_{jk}(i)}{m_{jk}}}},
\end{equation*}
where \(m_{jk}\) denotes the maximum flow from node \(j\) to node \(k\) and \(m_{jk}(i)\) denotes the maximum flow from node \(j\) to node \(k\) that passes through node \(i\). A node with high flow betweenness plays a critical role in the network, acting as a bridge or bottleneck that supports the transfer of resources between many other nodes. This measure captures the influence of nodes that may not lie on shortest paths but are nevertheless essential for maintaining network connectivity and facilitating flow.

\section{Flow coefficient}

The \emph{flow coefficient}\index{flow coefficient} is a measure of local centrality introduced by Honey \textit{et al.} \cite{Honey2007} to quantify the capacity of a node to mediate information flow among its immediate neighbors. For a node \(i\), the flow coefficient \(c_{\mathrm{fc}}(i)\) is defined as the fraction of all possible two-step paths between pairs of its neighbors that actually pass through node \(i\):
\begin{equation*}
    c_{\mathrm{fc}}(i) = 
    \frac{\sum_{j \neq k \in \mathcal{N}(i)} (A^2)_{jk}}
    {|\mathcal{N}(i)| \, (|\mathcal{N}(i)| - 1)},
\end{equation*}
where \(\mathcal{N}(i)\) denotes the set of neighbors of node \(i\) and \(A\) is the adjacency matrix of the network. The term \((A^2)_{jk}\) counts the number of paths of length two between nodes \(j\) and \(k\).  

High values of \(c_{\mathrm{fc}}(i)\) indicate that node \(i\) plays an important role in facilitating information flow among its neighbors, whereas low values suggest that the node’s neighbors are more directly connected to one another.

\section{Fractional graph Fourier transform centrality (FrGFTC)}

\emph{Fractional graph Fourier transform centrality}\index{graph Fourier transform centrality (GFT-C)!fractional (FrGFTC)} (FrGFTC) is an extension of the graph Fourier transform centrality (GFT-C) that incorporates fractional powers of the Laplacian eigenvectors to capture more nuanced structural variations in a network \cite{Tseng2024}.  

Let \(U = [u_1, \dots, u_N]\) be the eigenvector matrix of the graph Laplacian \(L\). In FrGFTC, the fractional power \(U^\alpha\) is used instead of \(U\), where \(\alpha \in (0,1]\) is the fractional order controlling the influence of higher-frequency components in the network.  

For node \(i\), the FrGFTC centrality \(c_{\mathrm{FrGFTC}}(i)\) is defined as
\[
c_{\mathrm{FrGFTC}}(i) = \sum_{l=1}^N e^{k \lambda_l} \left| \sum_{j=1}^N f_i(j) \, (U^\alpha)_{jl} \right|,
\]
where \(\lambda_l\) is the \(l\)-th eigenvalue of \(L\), \(k\) is a scaling parameter (e.g., \(k=0.1\)), and \(f_i(j)\) is the node importance signal defined by
\[
f_i(j) =
\begin{cases}
1, & i = j, \\
\dfrac{1/d_{ij}}{\sum_{k \neq i} 1/d_{ik}}, & i \neq j,
\end{cases}
\]
with \(d_{ij}\) denoting the shortest-path distance between nodes \(i\) and \(j\).  

Intuitively, FrGFTC generalizes GFT-C by allowing partial contributions from higher-frequency modes, which can better highlight nodes that are influential in both local and global network structures. Nodes with high FrGFTC values are those that strongly affect the propagation of signals across the network, capturing both immediate neighborhood and long-range connectivity patterns.

\section{Functional centrality}

\emph{Functional centrality}\index{functional centrality} quantifies the importance of nodes according to their participation in closed walks of lengths up to $k$, as proposed in \cite{Rodriguez2007}. The functional centrality of node $i$ is defined as
\[
c_f(i) = \sum_{l=0}^{k} a_l (A^l)_{ii} 
= \sum_{j=1}^{N} \left( v_j(i) \right)^2 \left( \sum_{l=0}^{k} a_l \lambda_j^l \right),
\]
where $A$ is the adjacency matrix of the network, $v_j(i)$ is the $i$-th component of the eigenvector $v_j$ corresponding to eigenvalue $\lambda_j$ of $A$, and the coefficients $a_l$ are defined as $a_0 = 1$ and $a_l = 1/l$ for any integer $l>0$.  

In functional centrality, the contribution of a node to walks of length $l$ is weighted inversely by $l$, so that shorter walks are given higher importance, while longer walks contribute progressively less. Functional centrality thus provides a way to quantify node importance with an emphasis on shorter, more localized interactions in the network.

\section{Fusion gravity model (FGM)}

The \emph{fusion gravity model}\index{gravity model!fusion (FGM)} (FGM) is a variant of the traditional gravity model that integrates multiple node attributes to evaluate the influence of nodes in complex networks \cite{Guo2024}. Specifically, the FGM score \(c_{FGM}(i)\) of node \(i\) is defined as
\begin{equation*}
c_{FGM}(i) = e_i \sum_{j \in \mathcal{N}^{(\leq r)}(i)} \frac{v_i v_j}{d_{ij}^2},
\end{equation*}
where \(\mathcal{N}^{(\leq r)}(i)\) denotes the set of nodes whose shortest path distance from node \(i\) is less than or equal to \(r\), \(d_{ij}\) is the shortest path distance between nodes \(i\) and \(j\), \(e_i\) represents the eigenvector centrality of node \(i\), and \(r\) is the radius of influence. Guo \textit{\textit{et al.}}~\cite{Guo2024} suggest setting \(r = 0.5 \langle d \rangle\), where \(\langle d \rangle\) is the average shortest path length in the network \(G\).  

The \textit{mass} \(v_i\) of node \(i\) is given by
\[
v_i = \frac{d_i}{\max_j d_j} + \frac{k_s(i)}{\max_j k_s(j)},
\]
where \(d_i\) and \(k_s(i)\) denote the degree and the \(k\)-shell index of node \(i\), respectively.  

The FGM index is designed to identify influential nodes in a network by combining structural and positional attributes. Its effectiveness is typically evaluated using the Susceptible-Infected (SI) propagation model.

\section{Fuzzy local dimension (FLD)}

\emph{Fuzzy local dimension}\index{local dimension!fuzzy (FLD)} (FLD) is a measure designed to identify influential nodes by combining fuzzy set theory with local dimension concepts \cite{Wen2019}. In the original local dimension (LD) measure \cite{Silva2012}, all nodes within a given radius (or box-size) contribute equally to the central node. FLD extends this approach by assigning different contributions to nodes based on their distance from the center node $i$: the closer a node $j$ is to $i$, the greater its contribution.

Specifically, Wen and Jiang \cite{Wen2019} use a fuzzy membership function to weight each node within the radius. The weighted number of nodes within radius $r$ of node $i$ is given by
\[
B_i(r) = \frac{\sum_{j \in N^{(\leq r)}(i)} e^{-d_{ij}^2 / r^2}}{|N^{(\leq r)}(i)|},
\]
where $d_{ij}$ is the shortest distance between nodes $i$ and $j$, and $N^{(\leq r)}(i)$ denotes the set of nodes satisfying $d_{ij} \le r$. 

The fuzzy local dimension $c_{FLD}(i)$ of node $i$ is then obtained as the slope of the line fitting $\log_2(B_i(r))$ versus $\log_2(r)$ on a double logarithmic scale using linear regression. Nodes with higher fuzzy local dimension values are those whose neighborhoods are both dense and close to the node, indicating strong local influence and a greater potential to affect the surrounding network.

\section{Game centrality (GC)}

\emph{Game centrality}\index{game centrality (GC)} (GC) is a game-theoretical measure that quantifies the ability of individual nodes to influence others to adopt their strategy \cite{Simko2013}. Simko and Csermely consider the two-player canonical Prisoner's Dilemma game with strategies \(\{C,D\}\), where the payoff matrix is given by
\[
\begin{array}{c|c|c}
 & \text{Cooperate (C)} & \text{Defect (D)} \\ \hline
\text{Cooperate (C)} & (3, 3) & (0, 6) \\ \hline
\text{Defect (D)} & (6, 0) & (1, 1)
\end{array}
\]

In the GC framework, all nodes initially cooperate except for node \(i\), which defects. In each round, nodes play the Prisoner's Dilemma with their neighbors in the underlying contact network and subsequently update their strategies according to the \emph{best-takes-over} rule (also called the imitation of the best strategy): each node adopts the strategy of the neighbor (or itself) with the highest total payoff in the previous round, with updates occurring synchronously across the network.

The game centrality of node \(i\), \(c_{\mathrm{GC}}(i)\), is defined as the proportion of defectors in the network, averaged over the last 50 simulation steps. Intuitively, GC measures the influence of node \(i\) in converting cooperating nodes to defectors.

\section{Gateway coefficient}

The \emph{gateway coefficient}\index{gateway coefficient} quantifies a node’s role in connecting different network modules by combining information about community structure and nodal centrality \cite{Vargas2014}. It extends the participation coefficient by introducing a weight that reflects the “importance” of the node’s intermodular connections.  

Assume that the network \(G\) has a community structure consisting of \(K\) communities \(C_1, \ldots, C_K\). The gateway coefficient of node \(i\) is defined as
\begin{equation*}
    c_{\mathrm{gateway}}(i) = 1 - \sum_{s=1}^{K} 
    \left( \frac{d_{is}}{d_i} \right)^2 
    \left( 1 - \overline{d}_{iS} \cdot  \overline{c}_{iS} \right)^2,
\end{equation*}
where \(d_{is}\) is the number of links from node \(i\) to nodes in community \(C_s\), and \(d_i\) is the total degree of node \(i\). The term \(\overline{d}_{iS} = \frac{d_{iS}}{\sum_{j \in C_s} d_{jS}}\) represents the fraction of all connections in community \(C_s\) that belong to node \(i\), and  
\[
\overline{c}_{iS} = \frac{\sum_{j \in \mathcal{N}(i,s)} c(j)}{\max_s \sum_{j \in C_s} c(j)}
\]
accounts for the average centrality \(c(j)\) of the neighbors \(\mathcal{N}(i,s)\) of node \(i\) within community \(C_s\).  Ruiz Vargas and Wahl \cite{Vargas2014} define \(c(j)\) as either the degree or betweenness centrality of node \(j\), depending on the application.

\vfill

\section{Generalized degree centrality}

The \emph{generalized degree centrality}\index{degree centrality!generalized} extends standard degree centrality by incorporating the influence of a node's neighbors \cite{Csató2017}. Rather than counting only direct connections, it redistributes the total sum of degrees across the network, assigning greater importance to nodes connected to highly central neighbors. A positive parameter \(\varepsilon > 0\) controls the strength of neighbor influence, interpolating between standard degree centrality (\(\varepsilon \to 0\)) and equal centrality for all nodes within a connected component (\(\varepsilon \to \infty\)).  

Formally, the generalized degree centrality vector \(\mathbf{x}(\varepsilon)\) is defined as
\begin{equation*}
(I + \varepsilon L) \, \mathbf{x}(\varepsilon) = d,
\end{equation*}
where \(I\) is the identity matrix, \(L\) is the graph Laplacian, and \(d\) is the vector of node degrees. For an individual node \(i\), the generalized degree centrality can also be expressed as
\begin{equation*}
d_i = x_i(\varepsilon) + \varepsilon \sum_{j \in \mathcal{N}(i)} a_{ij} \big(x_i(\varepsilon) - x_j(\varepsilon)\big),
\end{equation*}
illustrating that a node's centrality increases if it is connected to less central neighbors and decreases if it is connected to more central neighbors.  

Nodes with high generalized degree centrality are not only well-connected themselves but are also linked to other influential nodes, providing a more nuanced reflection of their global importance in the network. In practice, small values of \(\varepsilon\) are recommended to preserve the properties of standard degree centrality while enhancing differentiation among nodes.

\section{Generalized degree discount (GDD)}

The \emph{Generalized Degree Discount (GDD)}\index{generalized degree discount (GDD)} is a degree‐discount heuristic for identifying a set of influential spreaders in a network~\cite{Wang2016b}. It builds on the earlier DegreeDiscount method by introducing for each node a “status” (probability of not being influenced yet) and computing a \emph{generalized discounted degree} that accounts not only for the number of non‐seed neighbours, but also for how covered those neighbours already are by selected seeds. In each step a node with the highest generalized discounted degree is selected as a spreader, and thereafter the effective degrees of its neighbors (and their neighbors) are updated to reflect the diminished marginal influence.  

The algorithm proceeds iteratively as follows:
\begin{enumerate}
    \item Compute the degree $d_i$ of each node $i$, and initialize $t_i = 0$ and $dd_i = d_i$, where $t_i$ is the number of selected neighbors of $i$ and $dd_i$ is the discounted degree.
    \item \textit{Selection and update:} select the node $i$ with the largest $dd_i$ and add it to the seed set $S$. Then, for each 1-hop and 2-hop neighbor $j$ of node $i$, update
    \[
    t_j \gets t_j + 1, \quad
    dd_j = (d_j - 2 t_j - (d_j - t_j) t_j p) 
           + \frac{t_j (t_j - 1) p}{2} 
           - \sum_{k \in \mathcal{N}(j) \setminus S} t_k p,
    \]
    where $p$ is the propagation probability (e.g., $p=0.01$). The value $dd_j$ approximates the expected number of additional nodes influenced by selecting node $j$, accounting for both 1-hop and 2-hop neighbor effects.

    \item Repeat step 2 until $k$ nodes are selected for the seed set.
\end{enumerate}

\section{Generalized gravity centrality (GGC)}

The \emph{Generalized gravity centrality}\index{gravity centrality!generalized (GGC)} (GGC), also known as the clustering gravity model \cite{Zhao2022}, is a variant of the gravity model that incorporates both degree and clustering coefficient to evaluate the “mass” or influence of a node \cite{HLi2021}. The centrality of node $i$ is defined as
\[
c_{GGC}(i) = \sum_{j \in \mathcal{N}^{(\leq r)}(i)} \frac{sp_i \, sp_j}{d_{ij}^2},
\]
where $d_{ij}$ is the shortest distance between nodes $i$ and $j$, and $\mathcal{N}^{(\leq r)}$ denotes the set of nodes within radius $r$ of node $i$.  

The spreading ability of a node is given by
\[
sp_i = d_i \, e^{\alpha c_i},
\]
where $d_i$ and $c_i$ are the degree and clustering coefficient of node $i$, and $\alpha$ is a tunable parameter controlling the influence of the clustering coefficient. Li \textit{et al.} \cite{HLi2021} suggest $\alpha = 2$ in their experiments.

\section{Generalized network dismantling (GND) method}

The \textit{generalized network dismantling}\index{generalized network dismantling (GND) method} (GND) method is designed to identify a minimal-cost set of nodes whose removal fragments a network into small, disconnected components \cite{Ren2019}. The key idea of GND is that the network is assumed to possess a community structure and is recursively split into two groups, with the most effective nodes for disconnecting these groups identified and removed at each step. The order in which nodes are removed can be interpreted as a form of node centrality, reflecting each node’s importance in maintaining network connectivity.  

The GND method proceeds iteratively on the largest connected component of the network as follows:
\begin{enumerate}
  \item \textit{Cost-weighted Laplacian construction:} construct the \emph{cost-weighted Laplacian matrix}\index{matrix!cost-weighted Laplacian}
  \[
  L_w = D_B - B,
  \]
  where \(B = AW + WA - A\) is the cost-weighted matrix, \(W = \mathrm{diag}(w_1, w_2, \dots, w_n)\) is a diagonal matrix of node removal costs \(w_i > 0\) (if \(W = I\), then \(B = A\)), and \(D_B\) is the degree diagonal matrix of \(B\).

  \item \textit{Spectral bipartition:} compute the second smallest eigenvector \(\mathbf{v}^{(2)}\) of \(L_w\) and use the signs of its components to bipartition the network nodes into two subgraphs. Ren \textit{\textit{et al.}}~\cite{Ren2019} propose an approximation algorithm based on the power-iteration method.

  \item \textit{Weighted vertex cover refinement:} identify edges crossing between the two subgraphs and apply a weighted vertex cover algorithm to determine the minimal-cost set of nodes whose removal covers all crossing edges, effectively disconnecting the subgraphs.

  \item \textit{Node removal and update:} remove the selected nodes from the network, recompute connected components, and repeat the process on the largest remaining connected component.
\end{enumerate}

The GND method has been applied to the network dismantling problem and tested on social, criminal, corruption and power infrastructure networks.

\section{Generalized subgraph centrality (GSC)}

The \emph{generalized subgraph centrality}\index{subgraph centrality!generalized (GSC)} (GSC), also referred to as $t$-subgraph centrality, extends the classical subgraph centrality by allowing a flexible weighting of closed walks based on their length, capturing both local and global aspects of a node's influence \cite{Estrada2010c}.  

The subgraph centrality of node $i$, denoted $c_s(i)$, is defined as
\[
c_s(i) = \sum_{k=0}^{\infty} \frac{(A^k)_{ii}}{k!} = [e^A]_{ii},
\]
where $A$ is the adjacency matrix and $(A^k)_{ii}$ counts the number of closed walks of length $k$ starting and ending at node $i$.  In the generalized form, the factorial weighting is rescaled by a parameter $t$, allowing the emphasis on short or long walks to be adjusted:

\begin{itemize}
    \item \textit{Positive rescaling} ($t \geq 0$): longer walks are increasingly penalized, yielding a more localized centrality measure
    \[
    c_{GSC}(i) = \sum_{k=0}^{\infty} \frac{(A^k)_{ii}}{(t+k)!}.
    \]
    \item \textit{Negative rescaling} ($t < 0$): the emphasis shifts toward longer walks, capturing a node's global environment
    \[
    c_{GSC}(i) = \sum_{k=0}^{|t|-1} (A^k)_{ii} + \sum_{k=0}^{\infty} \frac{(A^{|t|+k})_{ii}}{k!} 
    = \sum_{k=0}^{|t|-1} (A^k)_{ii} + [A^{|t|} e^A]_{ii}.
    \]
\end{itemize}

By varying $t$, one can tune the centrality to emphasize either local structure (large positive $t$) or global network connectivity (negative $t$). The generalized subgraph centrality has been applied to protein-protein interaction (PPI) networks, with $t = 7$ shown to outperform subgraph centrality and other measures in identifying essential proteins in the yeast network.

\section{Geodesic \textit{k}-path centrality}

The \emph{geodesic \(k\)-path centrality}\index{geodesic \textit{k}-path centrality} measures the influence of a node by counting the number of shortest (geodesic) paths of length at most \(k\) that originate from it~\cite{Borgatti2006}.  
Formally, for a node \(i\), it is defined as the total number of geodesic paths of length \(1 \le \ell \le k\) starting from \(i\).

A notable special case occurs when \(k = 1\), in which case the geodesic \(k\)-path centrality reduces to the degree centrality. By counting shortest paths of length greater than one, this measure extends degree centrality to capture the ability to reach and potentially influence nodes that are not directly adjacent. Importantly, unlike \(m\)-reach centrality, which considers only the number of nodes reachable within \(m\) steps, the geodesic \(k\)-path centrality accounts for the number of shortest paths originating from a node.

\section{Gil-Schmidt power index}

The \emph{Gil-Schmidt power centrality index}\index{power index!Gil-Schmidt} is a normalized variant of harmonic centrality \cite{Sinclair2009}, which quantifies the influence of a node in a network while explicitly accounting for the set of nodes it can reach. Let \( R(i,G) \) denote the set of nodes reachable from node \( i \) in the graph \( G \). The Gil-Schmidt power index \( c_{GS}(i) \) of node \( i \) is defined as
\begin{equation*}
    c_{GS}(i) = \frac{1}{|R(i,G)|} \sum_{j \in R(i,G)} \frac{1}{d_{ij}} 
    = \frac{1}{|R(i,G)|} c_{harmonic}(i),
\end{equation*}
where \( d_{ij} \) is the length of the shortest path from \( i \) to \( j \). Intuitively, the index computes the average of the inverse shortest-path distances from node \( i \) to all nodes it can reach, so that nodes located closer to many others receive higher scores. The normalization by \( |R(i,G)| \) ensures that the measure is comparable across nodes with differing numbers of reachable nodes. In connected graphs, where every node can reach all others, the Gil-Schmidt power index coincides with the harmonic centrality.

\section{Global and local information (GLI) method}

The \emph{global and local information} (GLI) method is a hybrid centrality measure that combines global and local node information using degree and $k$-shell decomposition \cite{YangY2020}. 

First, the \emph{improved $k$-shell score} of node \(i\) is defined as
\[
Iks(i) = k_s(i) + nit(i),
\]
where \(k_s(i)\) is the standard $k$-shell index of node \(i\) and \(nit(i)\) denotes the iteration number at which node \(i\) is removed during the $k$-shell decomposition, corresponding to the layer of node $i$ in the onion decomposition of the graph.

The \textsc{GLI} centrality of node \(i\) is then given by
\[
c_{\textsc{GLI}}(i) = \exp \left( \frac{Iks(i) + d_i}{\sum_{j=1}^N (Iks(j) + d_j)} \right) 
\left( \sum_{j \in \mathcal{N}^{(\leq r)}(i)} \frac{Iks(j) + d_j}{d_{ij}} \right),
\]
where \(d_i\) is the degree of node \(i\), \(d_{ij}\) is the shortest-path distance between nodes \(i\) and \(j\), and \(\mathcal{N}^{(\leq r)}(i)\) denotes the set of nodes within a truncated radius \(r\) from node \(i\).  

Yang \textit{et al.} \cite{YangY2020} set \(r = 3\) to reduce computational complexity. This measure integrates a node’s global hierarchical position (via improved $k$-shell) with the local connectivity of its neighborhood to better capture its influence in the network.

\section{Global and local structure (GLS) centrality}

The \emph{global and local structure (GLS) centrality}\index{global and local structure (GLS) centrality} evaluates node importance by integrating both local and global network structures \cite{Sheng2020}. The global influence of node \(i\) is defined as
\begin{equation*}
    f_g(i) = d_i \sum_{j \in \mathcal{N}(i)} \alpha^{|\mathcal{N}(i) \cap \mathcal{N}(j)|},
\end{equation*}
where \(d_i\) is the degree of node \(i\), \(\mathcal{N}(i)\) is the set of neighbors of node \(i\), and \(\alpha = 1.1\) is a constant.  

The local influence of node \(i\) accounts for the normalized degree of its neighbors and the inverse average degree of each neighbor's neighbors:
\begin{equation*}
    f_l(i) = \sum_{j \in \mathcal{N}(i)} \frac{d_j}{N-1} \cdot \frac{d_j}{\sum_{l \in \mathcal{N}(j)} d_l}.
\end{equation*}

The GLS centrality of node \(i\) combines its global and local influence as
\begin{equation*}
    c_{GLS}(i) = f_g(i) \cdot f_l(i) 
    = \frac{d_i}{N-1} \sum_{j \in \mathcal{N}(i)} \frac{d_j^2 \alpha^{|\mathcal{N}(i) \cap \mathcal{N}(j)|}}{\sum_{l \in \mathcal{N}(j)} d_l}.
\end{equation*}

Hence, the GLS centrality combines the weighted contributions of immediate neighbors, accounting for their degrees, with the overlap between the neighborhoods of connected nodes, thereby integrating local connectivity and semi-global structural information.

\section{Global importance of nodes (GIN)}

The \emph{global importance of nodes}\index{global importance of nodes (GIN)} (GIN) quantifies the influence of a node by combining its degree, representing its direct connectivity, with its potential impact on other nodes, weighted by their degrees and distances within the network \cite{Zhao2020}. The GIN centrality of node \(i\) is defined as
\begin{equation*}
    c_{GIN}(i) = 
    \begin{cases}
        e^{\frac{\alpha d_i}{N}} \cdot \sum_{j \neq i} \frac{\beta d_j}{d_{ij}}, & \text{if } d_i \neq 0, \\
        0, & \text{if } d_i = 0,
    \end{cases}
\end{equation*}
where \(d_i\) is the degree of node \(i\), \(d_{ij}\) is the shortest-path distance between nodes \(i\) and \(j\), and \(\alpha\) and \(\beta\) are tunable parameters. Zhao \textit{et al.} \cite{Zhao2020} set \(\alpha = \beta = 1\) in their analysis. Hence, the GIN measure captures both the direct connectivity of a node, measured by its degree, and the potential influence on other nodes, weighted by their degrees and the shortest-path distances within the network.

\section{Global Structure Model (GSM)}

The \emph{Global Structure Model}\index{global structure model (GSM)} (GSM) centrality accounts for self-influence (SI) and global influence (GI) \cite{Ullah2021}. For node \(i\), the GSM score is defined as
\[
c_{GSM}(i) = SI(i) \cdot GI(i) = e^{k_s(i)/N} \cdot \sum_{j \neq i} \frac{k_s(j)}{d_{ij}},
\]
where \(k_s(i)\) is the $k$-shell value of \(i\), \(d_{ij}\) is the shortest-path distance between \(i\) and \(j\) and \(N\) is the total number of nodes in the network. The term \(SI(i)\) reflects intrinsic influence based on the core-periphery position, while \(GI(i)\) captures contributions from all other nodes weighted by distance.

\section{Godfather index}

The \emph{Godfather index}\index{Godfather index} measures a node's brokerage and coordination capital by counting the number of pairs of a node’s neighbors who are not connected to each other \cite{Jackson2020}. The centrality of node \(i\), denoted \(c_{\mathrm{GF}}(i)\), can be expressed as
\[
    c_{\mathrm{GF}}(i) = \sum_{k>j} a_{ik} a_{ij} (1 - a_{kj}) = \sum_{k > j \in \mathcal{N}(i)} (1 - a_{kj}),
\]
where \(a_{ij}\) is the adjacency matrix of the network and \(\mathcal{N}(i)\) is the set of neighbors of node \(i\). 

Jackson \cite{Jackson2020} shows that the Godfather Index is inversely related to the clustering coefficient \(c_{\mathrm{cl}}(i)\), weighted by the number of neighbor pairs:
\[
    c_{\mathrm{GF}}(i) = (1 - c_{\mathrm{cl}}(i)) \frac{d_i (d_i - 1)}{2},
\]
where \(d_i\) is the degree of node \(i\).  

The Godfather Index is also related to the redundancy coefficient \(c_r(i)\) as
\[
    c_{\mathrm{GF}}(i) = \frac{d_i (d_i - 1)}{2} - \frac{d_i}{2} c_r(i).
\]
This formulation highlights that nodes with many unconnected neighbor pairs (low redundancy) have higher Godfather centrality, reflecting their potential as brokers or coordinators in the network.

\section{Graph Fourier Transform Centrality (GFT-C)}

\emph{Graph Fourier Transform Centrality}\index{graph Fourier transform centrality (GFT-C)} (GFT-C) is a spectral measure that evaluates node importance based on the Laplacian eigendecomposition of a graph \cite{Singh2017}. The eigenvalues of the Laplacian \(L\) capture the global variation of a graph signal across nodes: eigenvectors associated with large eigenvalues vary rapidly across adjacent nodes, whereas those corresponding to small eigenvalues vary slowly.  

For node \(i\), the GFT-C centrality \(c_{GFT-C}(i)\) is defined as
\[
c_{GFT-C}(i) = \sum_{l=1}^N e^{k \lambda_l} \left| \sum_{j=1}^N f_i(j) u_l(j) \right|,
\]
where \(\lambda_l\) and \(u_l\) are the \(l\)-th eigenvalue and eigenvector of \(L\), \(k\) is a parameter (Singh \textit{et al.} suggest \(k=0.1\)), and \(f_i(j)\) quantifies the relative importance of node \(j\) with respect to node \(i\):
\[
f_i(j) =
\begin{cases}
1, & i = j, \\
\dfrac{1/d_{ij}}{\sum_{k \neq i} 1/d_{ik}}, & i \neq j,
\end{cases}
\]
with \(d_{ij}\) denoting the shortest-path distance between nodes \(i\) and \(j\).  

GFT-C of node \(i\) can be interpreted as the weighted sum of the graph Fourier transform coefficients of the importance signal \(f_i\). GFT-C is extended to the \emph{fractional graph Fourier transform centrality}\index{graph Fourier transform centrality (GFT-C)!fractional (FrGFTC)} (FrGFTC) in \cite{Tseng2024}, where the eigenvector matrix \(U = [u_1, \dots, u_N]\) is replaced by \(U^\alpha\), with \(\alpha\) denoting the fractional order.

\section{Graph regularization centrality (GRC)}

\emph{Graph regularization centrality}\index{graph regularization centrality (GRC)} (GRC) is a centrality measure derived from graph signal processing theory \cite{DalCol2023}. It evaluates node importance by examining how a delta signal centered on each node spreads across the graph under a regularization constraint.

For node \(i\), the GRC centrality \(c_{GRC}(i)\) is defined as
\[
c_{GRC}(i) = \frac{1}{s_i(i)},
\]
where \(s_i\) is obtained by solving the regularized optimization problem
\[
s_i = \arg\min_g \left( \|g - \delta_i\|^2 + \gamma g^T L g \right).
\]
Here, \(g \in \mathbb{R}^{N \times 1}\) is an \(N\)-dimensional graph signal over all nodes, \(\delta_i\) is an \(N \times 1\) delta signal with \(\delta_i(k) = 1\) if \(k = i\) and \(\delta_i(k) = 0\) otherwise, \(L\) is the Laplacian matrix of the graph \(G\), and \(\gamma\) is a regularization parameter controlling the spread of the signal across the network. Larger values of \(\gamma\) allow the delta signals to diffuse further along the graph. The optimization balances two objectives: the signal should remain close to the original delta (locality) while spreading smoothly over the network according to the Laplacian (global influence). Nodes whose signals diffuse more widely have smaller \(s_i(i)\) values and thus higher GRC centrality, capturing both local and global network characteristics. 

When \(\gamma = 0\), \(s_i = \delta_i\), so \(s_i(i) = 1\) and all nodes have centrality equal to one. According to Dal Col and Petronetto \cite{DalCol2023}, setting \(\gamma = 1\) produces centrality values that balance local and global network characteristics without additional parameter tuning.

\section{Graph-theoretic power index (GPI)}

The \emph{graph-theoretic power index}\index{power index!graph-theoretic} (GPI) was proposed by Markovsky \textit{et al.}~\cite{Markovsky1988} to measure power in exchange networks. Here, power is conceived as an unobservable, structurally determined potential for obtaining relatively favorable resource levels. Let \(m_{ik}\) denote the number of nonintersecting paths (i.e., paths that share only the source node) of length \(k\) originating from node \(i\). 

Markovsky \textit{et al.} \cite{Markovsky1988} observed that odd-length nonintersecting paths are advantageous, while even-length paths are disadvantageous. Advantageous paths provide direct exchange alternatives or mitigate the effects of disadvantageous paths. The GPI of node \(i\) is defined as the difference between the number of advantageous and disadvantageous paths:
\begin{equation*}
    c_{GPI}(i) = \sum_{k=1}^8 (-1)^{k-1} m_{ik}.
\end{equation*}

\section{Graphlet degree centrality (GDC)}

\emph{Graphlet degree centrality}\index{degree centrality!Graphlet (GDC)} (GDC) was introduced in \cite{Milenković2011} for biological networks to quantify the density and complexity of a node's extended neighborhood by counting the number of different graphlets that the node participates in. GDC is based on 2- to 5-node graphlets, which are small, connected, induced, non-isomorphic graphs where nodes correspond to particular symmetry groups (automorphism orbits). There are a total of 73 orbits across all 2-5-node graphlets. 

The graphlet degree centrality \(c_{GDC}(i)\) of a node \(i\) is defined as
\begin{equation*}
    c_{GDC}(i) = \sum_{j=0}^{72} w_j \cdot \log(v_j(i) + 1),
\end{equation*}
where \(v_j(i)\) denotes the number of times node \(i\) touches orbit \(j\), and \(w_j \in [0,1]\) is the weight of orbit \(j\), accounting for dependencies between orbits \cite{Milenković2008}:
\begin{equation*}
    w_j = 1 - \frac{\log(o_j)}{\log(73)},
\end{equation*}
where \(o_j\) is the number of orbits that influence orbit \(j\). The weighting scheme assigns higher weights to “important” orbits that are minimally affected by other orbits and lower weights to “less important” orbits that are highly dependent on others. 

Nodes located in dense extended network neighborhoods will have higher GDC values than nodes in sparser neighborhoods.

\section{Gravity centrality}

The \emph{gravity centrality}\index{gravity centrality} (also known as the \textit{gravity $k$-shell} metric) is inspired by Newton’s law of gravitation. In this measure, the $k$-shell value of a node is regarded as its mass, while the shortest path length between two nodes represents their distance \cite{Ma2015}. 

Let \(\mathcal{N}^{(\leq l)}(i)\) denote the set of nodes within the $l$-hop neighborhood of node \(i\). The gravity centrality \(c_{\text{Gravity}}(i)\) of node \(i\) is defined as
\begin{equation*}
   c_{\text{Gravity}}(i) = \sum_{j \in \mathcal{N}^{(\leq l)}(i)} \frac{k_s(i)\,k_s(j)}{d_{ij}^2},
\end{equation*}
where \(d_{ij}\) is the shortest path distance between nodes \(i\) and \(j\) and \(k_s(i)\) denotes the $k$-shell value of node \(i\). To reduce computational complexity, Ma \textit{et al.}~\cite{Ma2015} set \(l = 3\), meaning that only the nearest neighbors, next-nearest neighbors, and third-order neighbors of node \(i\) are considered.

\section{Gravity model}

The \emph{gravity model}\index{gravity model} (originally proposed in \cite{Fei2018} as the inverse-square law method), inspired by Newton’s law of gravitation, evaluates a node’s importance in spreading dynamics by incorporating both neighborhood and path information \cite{Li2019}. In this model, the degree of a node is regarded as its mass, while the shortest path length between two nodes represents their distance. 

The centrality \(c_{\text{GM}}(i)\) of node \(i\) is defined as
\begin{equation*}
   c_{\text{GM}}(i) = \sum_{j \neq i} \frac{d_i\,d_j}{d_{ij}^2},
\end{equation*}
where \(d_{ij}\) denotes the shortest path distance between nodes \(i\) and \(j\) and \(d_i\) represents the degree of \(i\). 

According to the gravity model, a node with a larger degree (reflecting stronger local connectivity) and shorter average distances to other nodes (indicating higher global accessibility) is considered more influential in the network.

\section{Gromov centrality}

\emph{Gromov centrality}\index{Gromov centrality} is a multi-scale measure of node centrality that quantifies the average triangle excess over all pairs of nodes in a given \(l\)-hop neighborhood~\cite{Babul2022}. Let \(\mathcal{N}^{(\leq l)}(i)\) denote 
The set of nodes within the $l$-hop neighborhood of node \(i\), and let \(d_{ij}\) represent the shortest-path distance between nodes \(i\) and \(j\). The Gromov centrality of node \(i\), denoted \(c_{\text{Gromov}}(i)\), is defined as
\begin{equation*}
   c_{\text{Gromov}}(i) = \frac{1}{|T(\mathcal{N}^{(\leq l)}(i))|} \sum_{(j,k) \in T(\mathcal{N}^{(\leq l)}(i))} \Delta_i(j,k),
\end{equation*}
where
\[
\Delta_i(j,k) = d_{jk} - d_{ij} - d_{ik} \le 0
\]
is the triangle inequality excess (equivalent to the Gromov product) between nodes \(j\) and \(k\) with respect to \(i\), and 
\[
T(\mathcal{N}^{(\leq l)}(i)) = \{(j,k) \mid j,k \in \mathcal{N}^{(\leq l)}(i), \ j \neq k \}
\]
is the set of all unordered pairs of nodes in the \(l\)-neighborhood of \(i\).

The triangle inequality excess \(\Delta_i(j,k)\) equals zero if and only if node \(i\) lies on a geodesic (shortest path) between nodes \(j\) and \(k\). Very negative values of \(\Delta_i(j,k)\) indicate that passing through node \(i\) induces a significant detour between \(j\) and \(k\).  

By definition, \(c_{\text{Gromov}}(i)\) is always non-positive. Gromov centrality thus characterizes the extent to which a node lies between other pairs of nodes in its \(l\)-neighborhood. When \(l = 1\), it reflects the proportion of triangles formed by a node's neighbors, and a locally central node exhibits a star-like structure. When \(l\) equals the diameter of the network, Gromov centrality becomes equivalent to closeness centrality.

\section{h-index strength}

The \emph{h-index strength}\index{\textit{h}-index strength} is an extension of the weighted h-index that incorporates the influence of a node's neighbors \cite{Gao2019}. For node $i$, the h-index strength $c_{hs}(i)$ is defined as the sum of the weighted h-indexes of all its neighbors:
\begin{equation*}
    c_{hs}(i) = \sum_{j \in \mathcal{N}(i)} c_{wh}(j),
\end{equation*}
where $\mathcal{N}(i)$ is a set of neighbors of node $i$ and $c_{wh}(j)$ denotes the weighted h-index of neighbor $j$, which is defined in Gao \textit{et al.} \cite{Gao2019}.  

Nodes with high h-index strength are connected to neighbors with high weighted h-index values, reflecting both local connectivity and the strength of neighboring nodes.

 \section{Harmonic centrality}
\emph{Harmonic centrality}\index{harmonic centrality} (also known as Latora closeness centrality \cite{Latora2001}, nodal efficiency \cite{Achard2007}, reciprocal closeness \cite{Agneessens2017} or efficiency centrality \cite{Zhou2012}) was introduced in \cite{Harris1954,Beauchamp1965} and discussed in \cite{Rochat2009}. It is an extension of closeness centrality, in which the centrality of node \(i\) is computed as the sum of the inverse distances to all other nodes, i.e.,
\begin{equation*}
    c_{harmonic}(i) = \sum_{j \neq i}{\frac{1}{d_{ij}}}.
\end{equation*}
where \(d_{ij}\) is the length of the shortest path from node \(i\) to node \(j\). Intuitively, harmonic centrality quantifies a node’s closeness to all others by summing the reciprocals of shortest-path distances, remaining well-defined even in disconnected networks: if no path exists between a pair of nodes, the shortest-path distance is considered infinite, and consequently, their contribution to the sum is taken as zero.

\section{Heatmap centrality}

\emph{Heatmap centrality}\index{heatmap centrality} captures both local and global network information by comparing the farness of a node with the average farness of its neighbors \cite{Duron2020}.  The centrality of node \(i\), denoted \(c_{\text{heatmap}}(i)\), is defined as
\begin{equation*}
c_{\text{heatmap}}(i) = \sum_{j=1}^{N} d_{ij} - \frac{\sum_{j=1}^{N} \left( a_{ij} \sum_{k=1}^{N} d_{jk} \right)}{\sum_{j=1}^{N} a_{ij}},
\end{equation*}
where \(d_{ij}\) is the shortest-path distance between nodes \(i\) and \(j\), so that \(\sum_{j=1}^{N} d_{ij}\) represents the farness of node \(i\).  
Intuitively, this measure identifies “hot spot” nodes within their local neighborhoods: a node whose farness is smaller than the average farness of its neighbors is considered more influential in the network.

\section{Hide information}

\emph{Hide information}\index{hide information} measures how easily other nodes can reach a given node~\cite{Rosvall2005}. The hide information of node \(i\) is defined as
\begin{equation*}
    c_{\mathcal{H}}(i) = \frac{1}{N} \sum_{j=1}^{N} S(j,i),
\end{equation*}
where
\[
S(j,i) = - \log_2 \sum_{\{p(j,i)\}} P[p(j,i)], \quad 
P[p(j,i)] = \frac{1}{k_j} \prod_{l \neq j \neq i \in p(j,i)} \frac{1}{k_l - 1}.
\]
Here, \(S(j,i)\) represents the amount of information required to locate node \(i\) starting from node \(j\), and \(P[p(j,i)]\) is the probability of following path \(p(j,i)\) when choosing neighbors uniformly at each step. A node with high hide information is easier to locate from other nodes. For example, in a star graph, the central hub has high hide information as it is easily reached from peripheral nodes~\cite{Rosvall2005}.

\section{Hierarchical \textit{k}-shell (HKS) centrality}

The \emph{hierarchical $k$-shell}\index{\textit{k}-shell centrality!hierarchical (HKS)} (HKS) centrality is a hybrid extension of the $k$-shell centrality \cite{Zareie2018} that combines $k$-shell decomposition with node distances. The centrality of node \(i\) is defined as
\begin{equation*}
    c_{HKS}(i) = \sum_{j \in \mathcal{N}(i)} \sum_{l \in \mathcal{N}(j)} s(l),
\end{equation*}
where 
\begin{equation*}
    s(l) = d_l (b_l + f_l).
\end{equation*}
Here, \(d_l\) is the degree of node \(l\), \(b_l\) is the iteration at which node \(l\) is removed during $k$-shell decomposition, and \(f_l\) captures the distance of node \(l\) to the nodes with the highest $k$-shell score. Specifically, let \(K\) denote the set of nodes with the highest $k$-shell index. Then
\begin{equation*}
    f_l = \max_{u \in K} (b_u - d_{lu}),
\end{equation*}
where \(d_{lu}\) is the shortest path distance between nodes \(l\) and \(u\). This formulation integrates both local connectivity and hierarchical position to more accurately identify influential nodes.

\section{Hierarchical reduction by betweenness}

\emph{Hierarchical reduction by betweenness}\index{hierarchical reduction by betweenness} is an iterative node-removal procedure that reveals the hierarchical organization of a network based on betweenness centrality \cite{Hanneman2005}. The method is conceptually analogous to the \(k\)-core decomposition, but instead of using node degree as the removal criterion, it employs betweenness centrality.

The algorithm proceeds as follows:
\begin{enumerate}
    \item Compute the betweenness centrality for all nodes in the network.  
    \item Identify and remove the nodes with the minimum betweenness centrality value.  
    \item Recalculate betweenness centrality on the resulting (reduced) graph.  
    \item Repeat the removal and recalculation steps until all nodes have been eliminated.  
\end{enumerate}

Each node is assigned a \emph{hierarchical level} corresponding to the iteration (or reduction step) at which it is removed from the network. Nodes removed in the early stages occupy peripheral positions, whereas those that persist until later iterations represent structurally more central elements of the network hierarchy.

\section{HITS (Hubs and Authorities)}

The \emph{Hyperlink-Induced Topic Search (HITS)}\index{HITS} algorithm was originally introduced in \cite{Kleinberg1999}. It assigns two scores to each node: the \emph{hub score}\index{hub} and the \emph{authority score}\index{authority}. A node \(i\) has a high hub score \(c_{\text{hub}}(i)\) if it links to nodes with high authority scores, i.e.,
\begin{equation*}
    c_{\text{hub}}(i) =  \sum_{j=1}^{N} a_{ij} \, c_{\text{authority}}(j),
\end{equation*}
where \(a_{ij}\) denotes the adjacency matrix element from node \(i\) to node \(j\). Similarly, a node \(i\) has a high authority score \(c_{\text{authority}}(i)\) if it is pointed to by nodes with high hub scores:
\begin{equation*}
    c_{\text{authority}}(i) = \sum_{j=1}^{N} a_{ji} \, c_{\text{hub}}(j).
\end{equation*}

The iterative calculation of HITS can be formulated as an eigenvalue problem. Specifically, the hub and authority vectors correspond to the principal eigenvectors of \(AA^T\) and \(A^TA\), respectively, associated with the largest eigenvalue \(\lambda_{\max}\) of \(AA^T\) (or equivalently \(A^TA\)). In practice, HITS is designed for directed networks. For undirected networks, the hub and authority scores are identical and reduce to the standard eigenvector centrality.

\section{Hybrid characteristic centrality (HCC)}

\emph{Hybrid characteristic centrality}\index{hybrid characteristic centrality (HCC)} (HCC) is a hybrid measure that combines E-shell hierarchy decomposition and extended degree to evaluate node importance in a network \cite{LiuJ2023}.  

The \emph{extended degree}\index{degree!extended} of node $i$ is defined as
\begin{equation*}
    d^{ex}(i) = \delta d_i + (1-\delta) \sum_{j \in \mathcal{N}(i)} d_j,
\end{equation*}
where $d_i$ is the degree of node $i$, $\mathcal{N}(i)$ is the set of neighbors of $i$ and $\delta \in [0,1]$ balances the contribution of $i$'s own degree and its neighbors' degrees. Liu \textit{et al.} \cite{LiuJ2023} set $\delta = 0.5$.  

The \emph{E-shell hierarchy decomposition}\index{E-shell hierarchy decomposition} is a variant of $k$-shell decomposition that uses $d^{ex}(i)$ instead of the standard degree $d_i$. The HCC of node $i$ is then defined as
\begin{equation*}
    c_{HCC}(i) = \frac{d^{ex}(i)}{\max_j d^{ex}(j)} + \frac{pos(i)}{\max_j pos(j)},
\end{equation*}
where $pos(i)$ denotes the iteration at which node $i$ is removed during the E-shell decomposition.

Nodes with high HCC values are not only structurally central according to the E-shell hierarchy but also well-connected to influential neighbors, reflecting both local and global importance.

\section{Hybrid centrality (HC)}

\emph{Hybrid centrality}\index{hybrid centrality!HC} (HC) is designed to identify core, intermediate, and peripheral nodes in a network \cite{Christensen2018}. This measure builds upon the hybrid centrality framework proposed by Pozzi \textit{et al.} \cite{Pozzi2013}, combining multiple classical centrality rankings into a single index.  

The hybrid centrality of node \(i\) is defined as
\begin{equation*}
    c_{\mathrm{HC}}(i) = 
    \frac{
        c_{BC}^u(i) + c_{BC}^w(i) + c_C^u(i) + c_C^w(i) + c_D^u(i) + c_D^w(i) + c_{EC}^u(i) + c_{EC}^w(i) - 8
    }{8(N-1)},
\end{equation*}
where \(c_{BC}\), \(c_C\), \(c_D\), and \(c_{EC}\) denote the rankings of nodes by betweenness, closeness, degree, and eigenvector centralities in unweighted (\(u\)) and weighted (\(w\)) networks.  

High HC values indicate nodes that are central across multiple measures, while low values correspond to peripheral nodes in the network structure.

\section{Hybrid centrality measure (X)}

\emph{Hybrid centrality}\index{hybrid centrality!X} (X) was proposed by Pozzi \textit{et al.}~\cite{Pozzi2013} to improve the stability and robustness of node rankings by combining classical centrality measures, which are often positively correlated. Specifically, \(X\) aggregates the rankings of degree and betweenness centralities in both unweighted and weighted networks:
\[
X = \frac{c_D^u + c_D^w + c_{BC}^u + c_{BC}^w - 4}{4(N-1)},
\]
where \(c_D\) and \(c_{BC}\) denote the rankings of nodes by degree and betweenness centralities in unweighted (\(u\)) and weighted (\(w\)) networks.  Nodes with low \(X\) values are highly central, whereas high values indicate peripheral nodes.

\section{Hybrid centrality measure (Y)}

\emph{Hybrid centrality}\index{hybrid centrality!Y} (Y) was introduced to provide a more comprehensive assessment of node importance by combining several classical centrality measures, which tend to be positively correlated \cite{Pozzi2013}. It aggregates the rankings of eccentricity, eigenvector, and closeness centralities in unweighted and weighted networks:
\[
Y = \frac{c_E^u + c_E^w + c_{EC}^u + c_{EC}^w + c_C^u + c_C^w - 6}{6(N-1)},
\]
where \(c_E\), \(c_{EC}\) and \(c_C\) denote the rankings of nodes by eccentricity, eigenvector and closeness centralities in unweighted (\(u\)) and weighted (\(w\)) networks.  Nodes with low \(Y\) values are central, while high values correspond to peripheral nodes.

\section{Hybrid centrality measure (XpY)}

\emph{Hybrid centrality}\index{hybrid centrality!XpY} (XpY) combines two hybrid indices \(X\) and \(Y\) to capture node centrality across multiple classical measures, improving robustness relative to any single measure \cite{Pozzi2013}:
\[
XpY = X + Y =  \frac{c_D^u + c_D^w + c_{BC}^u + c_{BC}^w - 4}{4(N-1)} + \frac{c_E^u + c_E^w + c_{EC}^u + c_{EC}^w + c_C^u + c_C^w - 6}{6(N-1)}.
\]
where \(c_D\), \(c_{BC}\), \(c_E\), \(c_{EC}\), and \(c_C\) denote the rankings of nodes by degree, betweenness, eccentricity, eigenvector, and closeness centralities in unweighted (\(u\)) and weighted (\(w\)) networks. Low values of \(XpY\) indicate central nodes, while high values correspond to peripheral nodes. XpY is introduced for the selection of central stocks in financial networks \cite{Pozzi2013}.

\section{Hybrid centrality measure (XmY)}

\emph{Hybrid centrality}\index{hybrid centrality!XmY} (XmY) evaluates asymmetries in node influence by taking the difference between two hybrid indices, \(X\) and \(Y\) \cite{Pozzi2013}:
\[
XmY = X - Y = \frac{c_D^u + c_D^w + c_{BC}^u + c_{BC}^w - 4}{4(N-1)} - \frac{c_E^u + c_E^w + c_{EC}^u + c_{EC}^w + c_C^u + c_C^w - 6}{6(N-1)}.
\]
where \(c_D\), \(c_{BC}\), \(c_E\), \(c_{EC}\), and \(c_C\) denote the rankings of nodes by degree, betweenness, eccentricity, eigenvector, and closeness centralities in unweighted (\(u\)) and weighted (\(w\)) networks.  High \(XmY\) values indicate nodes with relatively few important connections, whereas low values correspond to nodes with many unimportant connections. This measure highlights nodes whose influence differs across the underlying centrality dimensions. XmY is introduced for the selection of central stocks in financial networks \cite{Pozzi2013}.

\section{Hybrid degree and \textit{k}-shell method}

The \emph{hybrid degree and \textit{k}-shell method}\index{hybrid degree and \textit{k}-shell method} is a variant of the local gravity model that ranks nodes in complex networks based on the degree and \(k\)-shell index of each node and its \(r\)-hop neighbors \cite{Maji2021}. The centrality \(c_{x\text{-}ks}(i)\) of node \(i\) is defined as
\[
c_{x\text{-}ks}(i) = \sum_{j \in \mathcal{N}^{(\leq r)}(i)} \frac{k_s(i) d_j}{d_{ij}^2} = k_s(i) \sum_{j \in \mathcal{N}^{(\leq r)}(i)} \frac{d_j}{d_{ij}^2},
\]
where \(\mathcal{N}^{(\leq r)}(i)\) denotes the set of nodes within distance \(r\) from node \(i\) (excluding node \(i\)), \(k_s(i)\) and \(d_i\) are the \(k\)-shell index and degree of node \(i\), and \(d_{ij}\) is the shortest path distance between nodes \(i\) and \(j\). Maji \textit{et al.}~\cite{Maji2021} consider a three-hop neighborhood, i.e., \(r = 3\).

The hybrid degree and \(k\)-shell method was evaluated using the susceptible-infected-recovered (SIR) model and metrics including spreadability, monotonicity, and Kendall’s tau, and it outperformed seven existing centrality measures in identifying influential seed nodes on real networks.

\section{Hybrid degree and MDD method}

The \emph{hybrid degree and MDD method}\index{hybrid degree and MDD method} is a variant of the local gravity model that ranks nodes in complex networks based on the degree and mixed degree decomposition (MDD) index of each node and its \(r\)-hop neighbors \cite{Maji2021}. The centrality \(c_{x\text{-}MDD}(i)\) of node \(i\) is defined as
\[
c_{x\text{-}MDD}(i) = \sum_{j \in \mathcal{N}^{(\leq r)}(i)} \frac{MDD(i) d_j}{d_{ij}^2} = MDD(i) \sum_{j \in \mathcal{N}^{(\leq r)}(i)} \frac{d_j}{d_{ij}^2},
\]
where \(\mathcal{N}^{(\leq r)}(i)\) denotes the set of nodes within distance \(r\) from node \(i\) (excluding node \(i\)), \(MDD(i)\) and \(d_i\) are the MDD index and degree of node \(i\), and \(d_{ij}\) is the shortest path distance between nodes \(i\) and \(j\). Maji \textit{et al.}~\cite{Maji2021} consider a three-hop neighborhood, i.e., \(r = 3\).

The hybrid degree and MDD method was evaluated using the susceptible-infected-recovered (SIR) model and metrics including spreadability, monotonicity, and Kendall’s tau, and it outperformed seven existing centrality measures in identifying influential seed nodes on real networks.

\section{Hybrid degree centrality}

\emph{Hybrid degree centrality}\index{degree centrality!hybrid (HDC)} (HDC) \cite{Ma2017} quantifies a node's influence by combining contributions from both near-source and distal effects under varying spreading probabilities. 
The near-source influence of node $i$ is represented by its degree, while the distal influence is captured by the \emph{modified local centrality}\index{modified local centrality (MLC)} (MLC). Specifically, MLC adjusts the semi-local LocalRank centrality by subtracting the contribution of neighbors' direct connections:
\begin{equation*}
    c_{MLC}(i) = \sum_{j \in \mathcal{N}(i)} \sum_{k \in \mathcal{N}(j)} n(k) - 2 \sum_{j \in \mathcal{N}(i)} |\mathcal{N}(j)|,
\end{equation*}
where $\mathcal{N}(i)$ is the set of neighbors of node $i$, and $n(k)=|\mathcal{N}^{(\leq 2)}(k)|$ denotes the number of nearest and next-nearest neighbors of node $k$.

The hybrid degree centrality of node $i$ is defined as
\begin{equation*}
    c_{HDC}(i) = (\beta - p) \, \alpha \, |\mathcal{N}(i)| + p \, c_{MLC}(i),
\end{equation*}
where $p$ is the spreading probability, and $\alpha$ and $\beta$ are parameters controlling the relative contributions of near-source and distal influence. 
Ma and Ma \cite{Ma2017} suggest $\alpha = 1000$, $\beta \in [0.1, 0.2]$, and $p < 0.6$. When $\beta = 0.2$ and $p = 0.1$, the contributions from degree centrality (DC) and modified local centrality (MLC) in the hybrid centrality (HDC) are each approximately half of the total HDC value.

\section{Hybrid global structure model (H-GSM)}

The \emph{hybrid global structure model}\index{global structure model (GSM)!hybrid (H-GSM)} (H-GSM) is an extension of the GSM model that combines the local information (\(iSI\)) and global influence (\(iGI\)) of nodes in a network \cite{Mukhtar2023}. The centrality \( c_{H-GSM}(i) \) of node \( i \) is defined as
\begin{equation*}
    c_{H-GSM}(i) = iSI(i) \cdot iGI(i) 
    = e^{k_s(i) d_i / N} \cdot \sum_{j \neq i} \frac{e^{k_s(j) d_j / N}}{d_{ij}^{\lceil \log_2 \overline{iSI} \rceil}},
\end{equation*}
where \( k_s(i) \) and \( d_i \) are the $k$-shell index and degree of node \( i \), respectively,  
\( d_{ij} \) is the shortest path distance between nodes \( i \) and \( j \), \( \overline{iSI} \) is the average local information across all nodes, and \(\lceil x \rceil\) denotes the smallest integer greater than or equal to \(x\). 

H-GSM integrates both local connectivity and global network influence, providing a more comprehensive measure of node importance compared to purely local or global metrics.

\section{Hybrid median centrality (HMC)}

The \emph{hybrid median centrality}\index{hybrid median centrality (HMC)} (HMC) identifies influential nodes by aggregating rankings from multiple existing centrality measures \cite{Fei2017}. Let there be \(m\) rankings of nodes based on \(m\) different centrality measures. For example, Fei \textit{et al.} \cite{Fei2017} consider \(m = 3\) measures: degree, closeness and betweenness centralities.  

The node rankings are organized in an \(N \times m\) multi-attribute decision-making matrix:
\[
R = 
\begin{bmatrix} 
r_{11} & r_{12} & r_{13}  \\
r_{21} & r_{22} & r_{23}  \\ 
\vdots & \vdots & \vdots  \\ 
r_{N1} & r_{N2} & r_{N3}  
\end{bmatrix},
\]
where \(r_{ij}\) denotes the ranking of node \(i\) with respect to the \(j\)-th centrality measure.  

The hybrid median centrality \(c_{\mathrm{HMC}}(i)\) of node \(i\) is then defined as the median of its maximum and minimum rankings:
\[
c_{\mathrm{HMC}}(i) = \frac{\min_j r_{ij} + \max_j r_{ij}}{2}.
\]

The effectiveness of HMC is typically evaluated using the susceptible-infected (SI) propagation model on real-world networks, such as Email, US Air97, Karate Club, and Jazz Musicians.

\section{HybridRank}

\emph{HybridRank}\index{HybridRank} is a hybrid centrality measure designed to identify a set of influential spreaders in a network by combining topological properties of nodes \cite{Ahajjam2018}. The method operates in two main steps: computing a hybrid centrality score for each node and selecting a subset of influential spreaders.

The hybrid centrality \( c_{HC}(i) \) of node \( i \) is defined as
\begin{equation*}
    c_{HC}(i) = c_{EV}(i) \sum_{j \in \mathcal{N}(i)} k_s(j),
\end{equation*}
where \( c_{EV}(i) \) denotes the eigenvector centrality of node \( i \), \( k_s(j) \) is the $k$-shell index of neighbor \( j \), and \( \mathcal{N}(i) \) represents the set of neighbors of \( i \).

In the second step, nodes are ranked according to their hybrid centrality scores. The algorithm then iteratively selects the node with the highest \( c_{HC} \) value as a seed and removes it along with its immediate neighbors from consideration. This process continues until the desired number of non-adjacent influential spreaders is obtained.

\section{Hubbel centrality}

\emph{Hubbell centrality}\index{Hubbel centrality} is a generalization of Leontief’s input-output model for economic systems \cite{Hubbell1965}. The centrality \( c_{\text{Hubbell}}(i) \) of node \( i \) depends on its exogenous contribution \( e_i \) (self-contribution), the status of its neighbors, and the strength with which those neighbors influence node \(i\). It is formally defined as
\begin{equation*}
    c_{Hubbel}(i) = e_i + \sum_{j=1}^N{w_{ij} \cdot c_{Hubbel}(j)},
\end{equation*}
or, equivalently
\begin{equation*}
    c_{Hubbel} = (I - W)^{-1} \cdot E,
\end{equation*}
where \(I\) is an \(N \times N\) identity matrix, \(E = (e_1, \ldots, e_N)^{T}\) is an \(N \times 1\) vector representing the exogenous contributions (or self-contributions) and \(W\) is an \(N \times N\) weight matrix capturing the influence among nodes. This formulation can be seen as a generalization of the Katz centrality when
\[
e_i = \beta, \quad \forall i = 1, \ldots, N, 
\quad \text{and} \quad 
W = \alpha A,
\]
where \(A\) is the adjacency matrix of the network, \(\alpha\) is an attenuation (scaling) factor and \( \beta \) is a constant exogenous input. However, unlike Katz centrality, Hubbell~\cite{Hubbell1965} does not assume that $W = \alpha A$. Instead, the rows of the weight matrix \(W\) are normalized so that the total influence on each node does not exceed one:
\[
\sum_{j=1}^{N} w_{ij} \leq 1, \quad \forall i \in \mathcal{N}.
\]

As an example, \(W\) can be constructed by normalizing each row through division by \(N-1\),  which ensures that the total influence each node receives from its neighbours is proportionally scaled across the network.

\section{Immediate Effects Centrality (IEC)}

\emph{Immediate effects centrality}\index{effects centrality!immediate (IEC)} (IEC) is a variant of closeness centrality that accounts for both the lengths and strengths of sequences of interpersonal influence connecting nodes~\cite{Friedkin1991}. The IEC score of node \(i\) is defined as
\begin{equation*}
    c_{IEC}(i) = \frac{N-1}{\sum_{j \neq i} m_{ji}},
\end{equation*}
where \(m_{ji}\) is the mean length of the sequences of interpersonal influence from node \(i\) to node \(j\). The matrix \(M = [m_{ji}]\) is given by
\[
M = (I - Z + E Z_{dg}) D,
\]
with the following definitions: \(I\) is the \(N \times N\) identity matrix; \(D\) is diagonal with entries \(d_{ii} = 1/v_i\), where \(v_i\) is an element of the left eigenvector of \(W\); \(W\) is the normalized adjacency matrix with self-loops, with each nonzero row divided by its row sum; \(E\) is an \(N \times N\) matrix of ones; \(Z = (I - W + W^{\infty})^{-1}\); and \(Z_{dg}\) is obtained from \(Z\) by setting all off-diagonal entries to zero.

IEC is the reciprocal of the mean length of influence sequences from node \(i\) to all other nodes. Larger IEC values indicate that an actor’s influence spreads more rapidly through the network.

\section{Improved closeness centrality (ICC)}

The \emph{improved closeness centrality}\index{closeness centrality!improved (ICC)} (\textsc{ICC}) is a variation of standard closeness centrality that accounts for the number of shortest paths between nodes \cite{Luan2021}. 

For a node \(i\), the \textsc{ICC} centrality \(c_{\textsc{ICC}}(i)\) is defined as
\[
c_{\textsc{ICC}}(i) = \frac{N-1}{\sum_{j \neq i} d_{ij} \left( \frac{1}{n_{ij}} \right)^{\alpha}},
\]
where \(d_{ij}\) is the shortest-path distance between nodes \(i\) and \(j\), \(n_{ij}\) is the number of shortest paths connecting \(i\) and \(j\), and \(0 \leq \alpha \leq 1\) is a tunable parameter. 

Luan \textit{et al.} \cite{Luan2021} show that the \textsc{ICC} performs best for \(\alpha = 0.2\). Note that \textsc{ICC} reduces to standard closeness centrality when \(\alpha = 0\).

\section{Improved distance-based coloring method (IIS)}

The \emph{improved distance-based coloring (IIS) method}\index{improved!distance-based coloring method (IIS)} builds upon the independent set (IS) approach in \cite{ZhaoX2015} by introducing a distance constraint to further enhance the spatial dispersiveness of the selected spreaders \cite{Guo2016}. The IIS procedure can be summarized as follows:

\begin{enumerate}
    \item \textit{Node ranking:} all nodes are ranked based on a selected centrality measure, such as degree, betweenness, closeness, or coreness centrality.
    \item \textit{Distance-based coloring:} nodes are sequentially assigned colors with the rule that any two nodes sharing the same color must be separated by a network distance of at least $r$. The parameter $r$ can be set, for example, to the average shortest path length in the network. This step ensures that nodes with the same color are sufficiently dispersed.
    \item \textit{Independent set construction:} after coloring, nodes with the same color form an independent set, similar to the IS method.
    \item \textit{Spreader selection:} within each independent set, the node with the highest centrality is selected. Priority is often given to the set containing the node with the maximum centrality value in the network. This guarantees that the chosen spreaders are both highly influential and well-distributed.
\end{enumerate}

By introducing the distance constraint, the IIS method improves upon the IS method by reducing overlap between the influence regions of multiple spreaders, which can enhance overall spreading efficiency.

\section{Improved entropy-based centrality}

The \emph{improved entropy-based centrality}\index{improved!entropy-based centrality} is a semi-local measure that accounts for connection weights and the heterogeneity in neighbors’ degrees of confidence \cite{Peng2017,Ni2020,Qiao2018}. The total influence $I(i)$ of node $i$ is defined as a weighted sum of its direct influence $DI(i)$ and its average indirect influence $II(i)$ on two-hop neighbors:
\begin{equation*}
    I(i) = \phi_1 DI(i) + \phi_2 II(i),
\end{equation*}
where $\phi_1$ and $\phi_2$ are the weights of direct and indirect influence, respectively, with $\phi_1 + \phi_2 = 1$.  

The direct influence of node $i$ is given by
\begin{equation*}
    DI(i) {=} -\sum_{j \in \mathcal{N}(i)} \left(\theta_1  \frac{w_{ij}}{\sum_{l \in \mathcal{N}(i)} w_{il}} \log_{10} \frac{w_{ij}}{\sum_{l \in \mathcal{N}(i)} w_{il}} 
    {+} \theta_2  \frac{d_j^\beta}{\sum_{l \in \mathcal{N}(i)} d_l^\beta} \log_{10} \frac{d_j^\beta}{\sum_{l \in \mathcal{N}(i)} d_l^\beta}\right),
\end{equation*}
where $\theta_1$ and $\theta_2$ are weighting coefficients with $\theta_1 + \theta_2 = 1$, $d_j$ denotes the degree of node $j$, $\beta$ is a tunable parameter representing confidence strength, $w_{ij}$ is the weight of the edge between nodes $i$ and $j$, and $\mathcal{N}(i)$ denotes the set of neighbors of node $i$.  

The average indirect influence of node $i$ on its two-hop neighbors $\mathcal{N}^{(2)}(i)$ is
\begin{equation*}
    II(i) = \frac{1}{|\mathcal{N}^{(2)}(i)|} \sum_{j \in \mathcal{N}^{(2)}(i)} \frac{\sum_{k=1}^N DI(i) DI(k)}{\sum_{k=1}^N a_{ik} a_{kj}},
\end{equation*}
where $a_{ik}$ are entries of the adjacency matrix.  

Peng \textit{et al.} \cite{Peng2017,Ni2020} consider $\beta=2$ and $\theta_1 = \theta_2 = \phi_1 = \phi_2 = 0.5$, while Qiao \textit{et al.} \cite{Qiao2018} suggest $\theta_1 = 0.4$, $\theta_2 = 0.6$, $\phi_1 = 0.6$, $\phi_2 = 0.4$, and $\beta = 1$. For $\theta_1 = 0$ and $\beta = 1$, the measure reduces to another entropy-based centrality proposed in \cite{Qiao2017}.

\section{Improved Global Structure Model (IGSM)}

The \emph{Improved Global Structure Model}\index{global structure model (GSM)!improved (IGSM)} (IGSM) is a variant of the GSM model in which the $k$-shell index is replaced by node degree \cite{Zhu2022}. The centrality of node \(i\) is defined as
\begin{equation*}
    c_{IGSM}(i) = e^{d_i / N} \cdot \sum_{j \neq i} \frac{d_j}{d_{ij}^{\lceil \log_2 \langle d \rangle \rceil}},
\end{equation*}
where \(d_i\) is the degree of node \(i\), \(\langle d \rangle\) is the average degree of the network, \(d_{ij}\) is the shortest path distance between nodes \(i\) and \(j\), and \(\lceil x \rceil\) denotes the smallest integer greater than or equal to \(x\). The exponent in the denominator reduces the contribution of distant nodes while emphasizing the influence of highly connected nodes, providing a balance between local connectivity and global network structure.

\section{Improved iterative resource allocation (IIRA) method}

The \emph{improved Iterative Resource Allocation}\index{improved!iterative resource allocation (IIRA) method} (IIRA) method is an extension of the iterative resource allocation (IRA) method that incorporates both neighbor centrality and the spreading rate to evaluate node influence \cite{Zhong2015}. In IIRA, each node $i$ is initially assigned a resource $I_i(0)$, which is iteratively distributed to its neighbors according to their centrality and the spreading rate $\beta$. After a sufficient number of iterations $t$, the resource $I_i(t)$ of each node approaches a steady state, and the final resource values are used to identify influential nodes.

The diffusion process can be formalized using the $N \times N$ stochastic matrix $P$, with entries
\[
p_{ij} = \left( 1 - (1-\beta)^{d_i} \right) \frac{a_{ij} c_i}{\sum_{k=1}^N a_{ik} c_k},
\]
where $a_{ij}$ is the adjacency matrix, $c_i$ represents a chosen centrality of node $i$, and $d_i$ is its degree. Zhong \textit{et al.} \cite{Zhong2015} set the spreading rate $\beta = 0.2$ and the number of iterations $t = 50$, demonstrating that IIRA computed using closeness centrality provides more accurate identification of influential nodes than the version based on eigenvector centrality.

\section{Improved \textit{k}-shell algorithm (IKS)}

The \emph{improved $k$-shell}\index{improved!\textit{k}-shell algorithm (IKS)} (IKS) algorithm identifies influential spreaders by combining $k$-shell centrality with node information entropy \cite{Wang2020}. The entropy of node \(i\) is defined as
\begin{equation*}
    e(i) = - \sum_{j \in \mathcal{N}(i)} \frac{d_j}{2L} \ln \left( \frac{d_j}{2L} \right),
\end{equation*}
where \(d_j\) is the degree of neighbor node \(j\), \(\mathcal{N}(i)\) is the set of neighbors of node \(i\), and \(L\) is the total number of links in the network. Node information entropy quantifies the propagation potential of a node: higher entropy indicates that the node can more effectively influence its neighbors.

In the improved \(k\)-shell (IKS) method, nodes are selected iteratively based on their \(k\)-shell index and information entropy. 
In each iteration, the process starts from the highest \(k\)-shell, which represents the most central core, and then moves toward lower shells. 
From each shell, one node with the highest information entropy is selected. 
After one node has been chosen from every shell, a new iteration begins, starting again from the highest \(k\)-shell and moving inward. 
This procedure continues until all nodes are selected, ensuring that nodes chosen early are both structurally central and information-rich.

\section{Improved \textit{k}-shell decomposition (IKSD) method}

The \emph{improved \textit{k}-shell decomposition}\index{improved!\textit{k}-shell decomposition (IKSD) method} (IKSD) is a variant of the standard $k$-shell decomposition \cite{ZLiu2015}, designed to better capture the hierarchical importance of nodes in a network. Unlike the standard $k$-shell, IKSD refines node ranking by iteratively removing nodes starting from those with the lowest degree and recalculating degrees after each removal. The process proceeds as follows:

\begin{enumerate}
    \item Initialize $k = 1$ and consider the current graph $G$.
    \item Identify all nodes in $G$ with the minimum degree.
    \item Assign an IKSD value of $k$ to these nodes.
    \item Remove the identified nodes from the graph and update the degrees of remaining nodes.
    \item Increment $k$ and repeat steps 2-4 until all nodes have been assigned an IKSD value.
\end{enumerate}

Nodes with higher IKSD values are located in the core of the network and are considered structurally more influential, whereas nodes with lower IKSD values occupy the periphery.

\section{Improved \textit{k}-shell hybrid method (IKH)}

The \textit{improved $k$-shell hybrid}\index{\textit{k}-shell hybrid method!improved (IKH)} (IKH) centrality, proposed by Maji \textit{\textit{et al.}}~\cite{Maji2020b}, is a refined variant of the $k$-shell hybrid (ksh) method\index{\textit{k}-shell hybrid method}. It introduces a distance-dependent weighting parameter that adjusts the contribution of neighboring nodes based on their shortest path distance from the focal node, increasing sensitivity to both local and extended network structures.

The centrality score \( c_{\mathrm{IKH}}(i) \) of node \( i \) is defined as
\begin{equation*}
    c_{\mathrm{IKH}}(i) = 
    \sum_{j \in \mathcal{N}^{(\leq l)}(i)} 
    \frac{\sqrt{k_s(i) + k_s(j)} + \mu(d_{ij}) \cdot d_j}{d_{ij}^2},
\end{equation*}
where \( \mathcal{N}^{(\leq l)}(i) \) is the set of nodes within the \( l \)-hop neighborhood of node \( i \), 
\( d_{ij} \) is the shortest path distance between nodes \( i \) and \( j \), 
\( k_s(i) \) and \( k_s(j) \) are the $k$-shell indices of nodes \( i \) and \( j \), 
\( d_j \) is the degree of node \( j \), and 
\(\mu(d_{ij})\) is a tunable parameter defined as
\begin{equation*}
    \mu(d_{ij}) = \frac{2(l - d_{ij} + 1)}{l(l + 1)},
\end{equation*}
where \( l \) is the radius of the considered neighborhood. This formulation ensures that closer nodes exert a stronger influence than more distant ones.

\section{Improved neighbors’ \textit{k}-core (INK)}

The \emph{improved neighbors’ \textit{k}-core (INK) method}\index{improved!neighbors’ \textit{k}-core (INK) method}, also known as neighborhood coreness or INK score, is an extension of the classical $k$-shell centrality \cite{Bae2014,Lin2014}. While nodes with the largest $k$-core values may have different spreading influence, INK accounts for the $k$-core values of neighboring nodes to better distinguish their influence. 

The centrality of node \(i\) is defined as
\begin{equation*}
    c_{\mathrm{INK}}(i) = \sum_{j \in \mathcal{N}(i)} k_s(j)^{\alpha},
\end{equation*}
where \(k_s(j)\) is the $k$-core value of neighbor \(j\), \(\mathcal{N}(i)\) denotes the set of neighbors of node \(i\), and \(\alpha\) is a tunable parameter controlling the contribution of neighbors’ influence.

Nodes connected to influential neighbors attain higher INK scores. The parameter \(\alpha\) modulates this effect: for \(\alpha < 1\), the influence of neighbors with large $k$-core values is weakened; when \(\alpha = 1\), the INK score reduces to the sum of the neighbors’ $k$-core values.

\section{Improved node contraction (IIMC) centrality}

The \emph{Improved IMC}\index{node contraction (IMC) centrality!improved} (IIMC) method \cite{Jia2011} extends the node contraction (IMC) centrality \cite{Tan2006} by incorporating the importance of edges in addition to nodes. The IIMC centrality of node \(i\) is defined as
\[
    c_{\mathrm{IIMC}}(i) = \alpha \, c_{\mathrm{IMC}}(i,G) + \beta \sum_{j \in \mathcal{N}(i)} c_{\mathrm{IMC}}((i,j), G^{*}),
\]
where \(c_{\mathrm{IMC}}(i,G)\) denotes the IMC centrality of node \(i\) in the original graph \(G\), \(\mathcal{N}(i)\) is the set of neighbors of node \(i\), and \(c_{\mathrm{IMC}}((i,j), G^{*})\) represents the IMC centrality of edge \((i,j)\) computed on the line graph \(G^{*}\) of \(G\). The parameters \(\alpha\) and \(\beta\) control the relative contributions of node and edge importance. In \cite{Jia2011}, the authors recommend \(\alpha / \beta = 5\) as a suitable balance between the two contributions.

\section{Improved WVoteRank centrality}

The \emph{improved WVoteRank centrality}\index{WVoteRank!improved} is a modification of WVoteRank that incorporates both 1-hop and 2-hop neighbors in the voting process \cite{Kumar2022}. Each node $i$ is characterized by the tuple $(s_i, v_i)$, where $s_i$ is the voting score and $v_i$ is the voting ability. Initially, $(s_i, v_i) = (0,1)$ for all $i \in \mathcal{N}$. The voting procedure iteratively performs the following steps:

\begin{enumerate}
    \item \textit{Vote:} each node distributes votes to its neighbors. The voting score of node $i$ is computed as
    \begin{equation*}
        s_i = \sqrt{ d_i \sum_{j=1}^{N} w_{ij} v_j + \sum_{j \in \mathcal{N}(i)} \sum_{k \in \mathcal{N}(j) \setminus \{i\}} w_{jk} v_k },
    \end{equation*}
    where $w_{ij}$ is the weight of the edge $(i,j)$, $d_i$ is the degree of node $i$ and $\mathcal{N}(j)$ denotes the neighbors of node $j$.
    
    \item \textit{Select:} the node $k$ with the highest voting score $s_k$ is elected. This node is removed from subsequent voting rounds by setting its voting ability to zero, $v_k = 0$.
    
    \item \textit{Update:} the voting ability of nodes within the 2-hop neighborhood of the elected node is reduced. For each $j \in \mathcal{N}^{( \leq 2)}(k)$, the updated voting ability is
    \begin{equation*}
        v_j \leftarrow \max(0, v_j - f_{kj}),
    \end{equation*}
    where $f_{kj} = \frac{1}{\langle d \rangle d_{kj}}$, $\langle d \rangle$ is the average degree of the network, and $d_{kj}$ is the shortest-path distance from the elected node $k$.
\end{enumerate}

The improved WVoteRank centrality accounts for both direct and indirect neighbors, ensuring that the influence of selected nodes appropriately diminishes the voting power of surrounding nodes.

\section{Independent cascade rank (ICR)}

The \emph{Independent cascade rank}\index{independent cascade rank (ICR)} (ICR) is a centrality measure that quantifies the expected influence of a node under the independent cascade (IC) diffusion model \cite{Kempe2005,Riquelme2018}. In the IC model, when a node becomes active, it has a single chance to activate each of its inactive neighbors with a probability \(p_{ij}\) associated with edge \((i,j)\). The ICR of a node \(i\) is defined as the expected number of nodes that would be activated if \(i\) were chosen as the initial seed.  

Since the activation process is probabilistic, the nodes activated from a given seed set may differ in each realization, causing the node rankings to vary between executions. To obtain a stable estimate of ICR, the expected influence is typically computed by averaging over multiple simulation runs.

\section{Independent set (IS) method}

The \emph{independent set}\index{independent set (IS) method} (IS) method is a widely used approach for identifying multiple influential spreaders in complex networks \cite{ZhaoX2015}. The method leverages the concept of \textit{independent sets}, which are groups of nodes in which no two nodes are directly connected. The IS method consists of the following steps:

\begin{enumerate}
    \item \textit{Node ranking:} all nodes in the network $G$ are ranked according to a chosen centrality measure, typically the node degree.
    \item \textit{Network coloring:} the network is colored using the Welsh-Powell algorithm. This algorithm assigns colors to nodes such that no two adjacent nodes share the same color. The computational complexity of this step is $O(N^2)$.
    \item \textit{Independent set formation:} Nodes with the same color form an \textit{independent set}, guaranteeing that the selected nodes are not directly connected to one another.
    \item \textit{Spreader selection:} within each independent set, the node with the highest centrality is chosen as an initial spreader. This strategy ensures that multiple spreaders are both influential and spatially well-distributed across the network.
\end{enumerate}

\section{Index of CENTRality (Icentr)}

\emph{Index of CENTRality}\index{Index of CENTRality (Icentr)} (Icentr) is a centrality measure developed to evaluate the performance of transportation networks \cite{Mussone2022}. Icentr defines the centrality of a node based on the weights of its incident edges, which in turn depend on the distances between nodes.  

For unweighted networks, the centrality of node \(i\) is defined as
\[
c_{Icentr}(i) = \sum_{(j,k) \in \mathcal{L}} \frac{1}{2^{w(j,k)}},
\]
where \(\mathcal{L}\) is the set of edges in the network and
\[
w(j,k) =
\begin{cases}
\max(d_{ij}, d_{ik}), & d_{ij} \neq d_{ik}, \\
d_{ij} + 1, & d_{ij} = d_{ik},
\end{cases}
\]
with \(d_{ij}\) denoting the shortest-path distance between nodes \(i\) and \(j\).  

Icentr assigns higher centrality to nodes connected via shorter paths, reflecting their greater accessibility and potential influence within the network. Mussone \textit{et al.} \cite{Mussone2022} also extend Icentr to weighted networks, where both node and edge weights are considered in the computation.

\section{Infection number}

The \emph{infection number}\index{infection number} quantifies the expected impact of a node $i$ as the initial source of infection \cite{Bauer2012}. Under the assumption that all infection walks are independent, Bauer and Lizier \cite{Bauer2012} approximate the infection number using \emph{self-avoiding walks} (SAWs) in a graph $G$, which is also referred to as the walks-based method. 

For the SIR (susceptible/infected/removed) model, where an infected node is removed from the network with probability $\lambda = 1$ (representing either death or full recovery with immunity), the expected number of resulting infections from node $i$ is
\[
I_i^{SIR} = \sum_{k=1}^{N-1} \sum_{j=1}^{N} \Bigl( 1 - (1 - \beta^k)^{s_{ij}^{k,0}} \Bigr) 
\Bigl( 1 - \sum_{t=1}^{k-1} \bigl( 1 - (1 - \beta^t)^{(A^t)_{ij}} \bigr) \Bigr),
\]
where $(A^k)_{ij}$ is the total number of paths of length $k$ from $i$ to $j$, $s_{ij}^{k,0}$ is the number of self-avoiding walks of length $k$ from $i$ to $j$ and $\beta$ is the infection probability.

The infection number of node $i$, considering only paths up to a maximum length $K$, is denoted $I_i^{SIR}(K)$. Bauer and Lizier \cite{Bauer2012} show that $I_i^{SIR}(K)$ with $K=4$ provides a good approximation of the true infection spread across a wide range of $\beta$ values.

\section{Influence capability (IC)}
\emph{Influence capability}\index{influence capability (IC)} (IC) is a hybrid centrality measure that integrates node position information from the $k$-shell decomposition with the influence of the neighborhood~\cite{ZWang2017}. The centrality of node $i$ is defined as
\[
c_{\mathrm{IC}}(i) = w_1 \cdot IC_p(i) + (1 - w_1) \cdot IC_N(i),
\]
where $IC_p(i)$ captures the influence of node $i$ based on its positional attributes:
\[
IC_p(i) = d_r(i) + \sum_{j \in \mathcal{N}(i)} \frac{2}{\pi} \arctan \left( \bigl(Iter(j)\bigr)^{1/3} \right).
\]
Here, $Iter(j)$ is the iteration at which node $j$ is removed in the $k$-shell decomposition, and $d_r(i)$ is the residual degree of node $i$ after removing neighbors $l$ with $Iter(l) < Iter(j)$.  

The neighborhood influence is measured by
\[
IC_N(i) = \sum_{j \in \mathcal{N}(i)} \sum_{l \in \mathcal{N}(j)} d_l,
\]
which aggregates the degrees of all neighbors of node $i$ and of their neighbors, counting each node as many times as it appears in these neighbor sets. The parameter $w_1 \in [0,1]$ balances the contributions of $IC_p(i)$ and $IC_N(i)$. Wang \textit{et al.}~\cite{ZWang2017} suggest setting
\[
w_1 = \frac{1 - E_1}{2 - E_1 - E_2},
\]
where $E_1$ and $E_2$ are the entropies of the $IC_p(i)$ and $IC_N(i)$ distributions, respectively.

\section{Information distance index (IDI)}

The \emph{information distance index}\index{information distance index (IDI)} (IDI) is an entropy-based centrality measure that quantifies node importance based on its distances to all other nodes in the network \cite{Konstantinova2006}. The index of node $i$ is defined as
\begin{equation*}
    c_{\text{IDI}}(i) = - \sum_{j \neq i} \frac{d_{ij}}{\sum_{k=1}^N d_{ik}} \log_2 \frac{d_{ij}}{\sum_{k=1}^N d_{ik}},
\end{equation*}
where $d_{ij}$ is the shortest-path distance from node $i$ to node $j$, and $N$ is the total number of nodes in the network.  

The IDI measure captures how uniformly a node is positioned with respect to all others: higher IDI values indicate nodes whose distances to the rest of the network are more evenly distributed, highlighting their central role in connecting the network.

\section{Integral \textit{k}-shell centrality}

The \emph{integral $k$-shell}\index{\textit{k}-shell centrality!integral (IKS)} (IKS) centrality is an extension of the $k$-shell centrality that incorporates both 2-step neighborhood information and the historical $k$-shell values of nodes \cite{Wang2019}. The historical $k$-shell of a node $i$ is defined as the sum of its $k$-shell values across previous iterations of the decomposition.  

The integral $k$-shell value $c_{IKS}(i)$ of node $i$ is given by
\begin{equation*}
    c_{IKS}(i) = Q_2(i) + \sum_{j=1}^{m_i} k_s^{(j)},
\end{equation*}
where $Q_2(i)$ denotes the number of nodes within 2 hops of node $i$, $m_i$ is the iteration at which node $i$ is removed during the $k$-shell decomposition, and $k_s^{(j)}$ is the historical $k$-shell index assigned to the nodes removed at iteration $j$.

Nodes with high IKS values are considered influential both because of their local 2-step connectivity and their persistence in the core layers of the network during the $k$-shell decomposition.

\section{Integration centrality}

\emph{Integration centrality}\index{integration centrality} quantifies how well connected a node is within the network \cite{Guimaraes1972,Valente1998}. The integration centrality of node \(i\) is defined as
\begin{equation*}
c_{\mathrm{Integration}}(i) = \frac{\sum_{j \neq i} \left( d_G + 1 - d_{ji} \right)}{N-1},
\end{equation*}
where \(d_G\) is the diameter of \(G\) and \(d_{ji}\) is the length of the shortest path from node \(j\) to node \(i\). Integration effectively inverts distances to provide a closeness-like measure, averaged over all other nodes. High integration centrality indicates that a node can be reached efficiently from most other nodes in the network, whereas low integration centrality indicates that it is more peripheral. For undirected networks, integration centrality coincides with radiality centrality.

\section{Interdependence centrality}
\emph{Interdependence centrality}\index{interdependence centrality} considers the possibility that two nodes may influence each other through the same link \cite{AleskerovBook, Shvydun2020}. A typical example of such networks is a trade network, where a reduction in trade flow between two countries can lead to both economic and non-economic losses for the exporter and/or the importer.  

Interdependence centrality generalizes the concept of individual and group influence from the long-range interaction centrality (LRIC) \cite{AleskerovBook} by introducing two node-specific parameters, $q_i^{\mathrm{in}}$ and $q_i^{\mathrm{out}}$, which represent the threshold levels at which node $i$ becomes affected. Using these thresholds, the importance of a link $(i,j)$ can be evaluated in two complementary dimensions: \emph{influence} (the effect of $i$ on $j$) and \emph{dependence} (the effect of $j$ on $i$). The direct influence of node~$i$ on node~$j$ through link $(i,j)$ is defined as
\begin{equation*}
c_{ij}^{\mathrm{in}} = \max_{\Omega_k(j):\, i \in \Omega_k^{P}(j)} 
\frac{w_{ij}}{\sum_{h \in \Omega_k(j)} w_{hj}},
\end{equation*}
where $\Omega_k(j)$ denotes a critical group of node~$j$, and $\Omega_k^{P}(j)$ is the subset of its pivotal nodes. The value $c_{ij}^{\mathrm{in}} \in [0,1]$ measures the relative importance of node~$i$ for node~$j$: $c_{ij}^{\mathrm{in}} = 0$ indicates that $i$ has no direct influence on $j$, while $c_{ij}^{\mathrm{in}} = 1$ implies that node~$i$ alone can exceed the influence threshold of node~$j$.

Analogously, to quantify how critical the same link $(i,j)$ is for node~$i$, we consider node~$i$’s own influence threshold $q_i^{\mathrm{out}}$. The direct dependence of node~$i$ on node~$j$ is given by
\begin{equation*}
c_{ji}^{\mathrm{out}} = \max_{\Omega_k(i):\, j \in \Omega_k^{P}(i)} 
\frac{w_{ij}}{\sum_{h \in \Omega_k(i)} w_{ih}},
\end{equation*}
where $\Omega_k(i)$ and $\Omega_k^{P}(i)$ are defined analogously for node~$i$. This formulation enables representing the network as a two-layer structure, consisting of an \emph{influence layer}, capturing how nodes affect others, and a \emph{dependence layer}, capturing how nodes are affected in return.

Once the direct influence of each edge is defined, the indirect influence between nodes can be estimated along different paths that traverse both layers. Building on this framework, Shvydun \cite{Shvydun2020} proposes three models to evaluate indirect influence and subsequently aggregate these estimates into the interdependence centrality measure.

\section{Intra-module degree (IMD)}

The \emph{intra-module degree}\index{degree!intra-module (IMD)} (IMD), also known as the within-module degree, is a community-based measure that quantifies the role of nodes within their respective communities~\cite{Guimerà2005}. Consider a graph \(G\) with a community structure consisting of \(K\) communities \(C_1, \dots, C_K\). Each community may have a different internal structure, ranging from fully centralized (e.g., a star) to fully decentralized (e.g., a complete graph). Nodes with similar functional roles are expected to exhibit similar relative intra-community connectivity.

The intra-module degree of a node \(i\) belonging to community \(C_s\) is defined as the Z-score of its internal degree relative to other nodes in the same community:
\begin{equation*}
   c_{IMD}(i) = \frac{k_{is} - \mu_s}{\omega_s},
\end{equation*}
where 
\[
k_{is} = \sum_{j \in C_s \setminus \{i\}} a_{ij}
\]
is the number of links from node \(i\) to other nodes in module \(C_s\), and \(\mu_s\) and \(\omega_s\) are the mean and standard deviation, respectively, of the internal degrees of nodes in \(C_s\). A higher intra-module degree indicates that node \(i\) is more strongly connected within its community.

\section{Inward accessibility}

The \emph{inward accessibility}\index{accessibility!inward} quantifies how likely a node is to be reached from other nodes in a network after a fixed number of steps along self-avoiding walks \cite{Travençolo2008}. Let \( N \) be the total number of nodes in the network, and let \( P_h(j,i) \) denote the transition probability that an agent starting from node \( j \) reaches node \( i \) in exactly \( h \) steps along a self-avoiding walk (i.e., a simple path without revisiting nodes). The inward accessibility of node \( i \) after \( h \) steps is defined as:
\begin{equation*}
    c_{IA_h}(i)
    = \frac{1}{N - 1}
    e^{\left(
    -\sum_{j:\, P_h(j,i) \neq 0}
    \frac{P_h(j,i)}{N - 1}
    \log\!\left(\frac{P_h(j,i)}{N - 1}\right)
    \right)}.
\end{equation*}

The term inside the exponential represents the Shannon entropy of the distribution of probabilities that node \( i \) is reached from all other nodes after \( h \) steps. A higher entropy indicates that node \( i \) is reached with similar probabilities from many distinct sources, reflecting higher inward accessibility. The normalization factor \( 1/(N - 1) \) ensures comparability across networks of different sizes.

\section{Isolating centrality (ISC)}

\emph{Isolating centrality}\index{isolating centrality (ISC)} (ISC) identifies nodes that critically affect network connectivity \cite{Ugurlu2022}. The ISC of node \(i\), denoted \(c_{\text{ISC}}(i)\), is defined as
\begin{equation*}
c_{\text{ISC}}(i) = |\mathcal{N}(i) \cap d_{\delta}| \cdot |\mathcal{N}(i)|,
\end{equation*}
where \(\mathcal{N}(i)\) is the set of neighbors of node \(i\), and \(d_{\delta}\) is the set of nodes with minimal degree \(\delta = \min_j |\mathcal{N}(j)|= \min_j d_j\).  

A high ISC value indicates that the node lies between densely connected internal nodes and sparsely connected terminal nodes. Removing such central nodes can weaken or even disrupt communication paths between groups of active nodes.

\section{Iterative resource allocation (IRA) method}

The \emph{iterative resource allocation}\index{iterative resource allocation (IRA) method} (IRA) method ranks nodes based on the distribution of resources according to neighbors’ centrality \cite{Ren2014}. Initially, each node is assigned a unit resource, which is then iteratively redistributed to its neighbors proportionally to their centrality. After several iterations, the resources of the nodes converge to a steady state, and the final resource values are used to measure the spreading influence of nodes. 

Formally, the influence of nodes is given by the principal left eigenvector of the \(N \times N\) stochastic matrix \(P\), whose elements are
\begin{equation*}
    p_{ij} = \frac{a_{ij} c_j^{\alpha}}{\sum_{k=1}^{N} a_{ik} c_k^{\alpha}},
\end{equation*}
where \(c_j\) denotes a chosen centrality of node \(j\) and \(\alpha\) is a tunable parameter controlling the nonlinear weighting of the centrality. Ren \textit{et al.} \cite{Ren2014} implement the IRA process by taking \(k\)-shell centrality as \(c_j\) and choosing \(\alpha = 1\).

\section{\textit{k}-betweenness centrality}

The \emph{\(k\)-betweenness centrality}\index{betweenness centrality!\textit{k}-}, also known as bounded-distance betweenness (BDBC) \cite{Brandes2008} or range-limited betweenness centrality \cite{Ravasz2012}, is a variant of the standard betweenness centrality in which only paths of length at most \(k\) are considered \cite{Borgatti2006}. It is formally defined as
\begin{equation*}
    c_{k-betw}(i) = \sum_{\substack{j\neq l \neq i \\ d_{jl} \leq k}}{\frac{\sigma_{jl}(i)}{\sigma_{jl}}},
\end{equation*}
where \(\sigma_{jl}\) denotes the number of shortest paths from node \(j\) to node \(l\), and \(\sigma_{jl}(i)\) represents the number of paths that pass through node \(i\). The rationale behind the \(k\)-betweenness centrality is that very long paths are less likely to be used in real processes and should therefore contribute less, or not at all, to a node’s centrality. By restricting attention to shorter paths, \(k\)-betweenness provides a more localized measure of a node’s role in network flow.

\section{\textit{k}-path centrality}

The \emph{\(k\)-path centrality}\index{\textit{k}-path centrality}, introduced by Sade~\cite{Sade1989}, quantifies the influence of a node by counting all simple (cycle-free) paths of length at most \(k\) that originate from it. 
Formally, for a node \(i\), the \(k\)-path centrality \(c_{k\text{-path}}(i)\) is the number of distinct simple paths of length \(1 \le \ell \le k\) starting at \(i\).  For \(k = 1\), the \(k\)-path centrality equals the degree centrality.

The \(k\)-path centrality generalizes degree centrality by considering not only immediate neighbors but also nodes reachable within paths of length up to \(k\), capturing broader influence within the network.

\section{\textit{k}-shell based on gravity centrality (KSGC)}

The \emph{\textit{k}-shell based on gravity centrality} (KSGC) is an extension of the local gravity model that integrates both \(k\)-shell and degree centralities to identify influential nodes in networks \cite{Yang2021}. 

Let $\mathcal{N}^{(\leq k)}(i)$ denote set of nodes within the $k$-hop neighbourhood of node \(i\). The centrality \( c_{\textsc{KSGC}}(i) \) of node \( i \) is defined as
\begin{equation*}
   c_{\textsc{KSGC}}(i) = 
   \sum_{j \in \mathcal{N}^{(\leq k)}(i)}
   e^{\frac{k_s(i) - k_s(j)}{ks_{\max} - ks_{\min}}} 
   \cdot 
   \frac{d_i d_j}{d_{ij}^2},
\end{equation*}
where \( d_{ij} \) is the shortest-path distance between nodes \( i \) and \( j \), \( d_i \) is the degree of node \( i \) and \( k_s(i) \) is the \(k\)-shell index of node \( i \). The terms \( ks_{\max} \) and \( ks_{\min} \) represent the maximum and minimum \(k\)-shell values in the network, respectively. Yang and Xiao \cite{Yang2021} consider \( l = 2 \) as the truncated radius.

\vfill

\section{\textit{k}-shell centrality}

The \emph{\textit{k}-shell centrality}\index{\textit{k}-shell centrality} (also called node coreness, coreness centrality, or \textit{k}-coreness centrality \cite{PVM2014}) is based on the \textit{k}-shell (or \textit{k}-core) decomposition of a network. The \textit{k}-core of a graph $G$ is defined as the maximal subgraph in which every node has degree at least $k$ \cite{Seidman1983,Bollobas1984}.Nodes in the $k$-core may also belong to higher-order $(k{+}1)$-cores. The \textit{k}-shell is then defined as the set of nodes that belong to the $k$-core but not to any higher-order $(k{+}1)$-core. The identification of the $k$-shell can be performed iteratively as follows:

\begin{enumerate}
    \item Remove all nodes with degree less than 1 and their incident links. Repeat until no such nodes remain. The removed nodes constitute the $k_s = 1$ shell.
    \item Remove all nodes with degree less than 2 in the remaining network and their links. Repeat until no such nodes remain. These nodes form the $k_s = 2$ shell.
    \item Continue this iterative process for increasing $k$ until all nodes are assigned a shell index $k_s$.
\end{enumerate}

The \textit{k}-shell centrality of a node $i$ is the index $k_s$ of the highest-order core containing $i$ \cite{Batagelj2002}. Intuitively, nodes in higher-order shells are considered more central because they are embedded in denser, more interconnected regions of the network. Kitsak \textit{et al.}~\cite{Kitsak2010} showed that nodes with high \textit{k}-shell indices often play a more influential role in spreading processes than nodes with high degree. However, a limitation of \textit{k}-shell centrality is its relatively low resolution: since the \textit{k}-shell index takes only a small number of discrete values, it may not always distinguish effectively between nodes in large or heterogeneous networks.

\section{\textit{k}-shell hybrid method (ksh)}

The \emph{$k$-shell hybrid method}\index{\textit{k}-shell hybrid method} (ksh) is a centrality measure that combines the $k$-shell decomposition and degree information within the framework of gravitational centrality \cite{Namtirtha2018}. Specifically, the centrality score \( c_{\mathrm{ksh}}(i) \) of node \( i \) is defined as
\begin{equation*}
    c_{\mathrm{ksh}}(i) = \sum_{j \in \mathcal{N}^{(\leq l)}(i)} 
    \frac{\sqrt{k_s(i) + k_s(j)} + \mu\,d_j}{d_{ij}^2},
\end{equation*}
where \( \mathcal{N}^{(\leq l)}(i) \) denotes the set of nodes within the \( l \)-hop neighborhood of node \( i \), \( d_{ij} \) is the shortest path distance between nodes \( i \) and \( j \), \( k_s(i) \) and \( k_s(j) \) represent the $k$-shell indices of nodes \( i \) and \( j \), respectively, \( d_j \) is the degree of node \( j \), and \( \mu \in (0,1) \) is a tunable parameter that balances the contributions of the two components. 

Namtirtha \textit{et al.} \cite{Namtirtha2018} recommend using \( l = 3 \) and \( \mu = 0.4 \) for optimal performance.

\section{\textit{k}-shell iteration factor (KS-IF)}

The \emph{$k$-shell iteration factor}\index{\textit{k}-shell iteration factor (KS-IF)} (KS-IF) ranks nodes by combining their degree and $k$-shell decomposition information \cite{Wang2016}. The centrality of node \(i\) is defined as
\begin{equation*}
    c_{KS-IF}(i) = \delta(i) d_i + \sum_{j \in \mathcal{N}(i)} \delta(j) d_j,
\end{equation*}
where \(d_i\) is the degree of node \(i\) and \(\delta(i)\) is the $k$-shell iteration factor given by
\begin{equation*}
    \delta(i) = k_s(i) \left( 1 + \frac{n(i)}{m(i)} \right).
\end{equation*}
Here, \(k_s(i)\) is the $k$-shell index of node \(i\), \(n(i)\) is the iteration at which node \(i\) is removed, and \(m(i)\) is the total number of iterations in that step. KS-IF combines node degree and $k$-shell position to better identify influential nodes.

\section{\textit{k}-shell Physarum centrality}
The \emph{\textit{k}-shell Physarum centrality}\index{\textit{k}-shell Physarum centrality} is a hybrid measure that combines Physarum centrality and $k$-shell centrality to identify influential nodes in weighted networks~\cite{Gao2013}. The centrality of node $i$ is defined as
\[
c_{kp}(i) = c_{\mathrm{Physarum}}(i) \cdot \frac{k_s(i)}{\sum_{j=1}^N k_s(j)},
\]
where $c_{\mathrm{Physarum}}(i)$ is the Physarum centrality of node $i$ and $k_s(i)$ is its $k$-shell index. This formulation weights the Physarum centrality by the relative position of the node within the $k$-shell decomposition, emphasizing nodes that are both structurally central and carry significant flux.

\section{\textit{k}-truss index}

The \emph{\textit{k}-truss index}\index{\textit{k}-truss index} evaluates the influence of a node based on the $k$-truss decomposition of a graph $G$ \cite{Cohen2008}. A $k$-truss is the maximal subgraph in which every edge participates in at least $k-2$ triangles. Equivalently, a $k$-clique corresponds to a $(k-2)$-truss and a $k$-truss is contained within a $(k+1)$-core. Malliaros \textit{et al.} \cite{Malliaros2016} demonstrate that nodes belonging to the maximal $k$-truss subgraph exhibit stronger spreading influence than nodes identified by other centrality measures, including degree and $k$-core index.

\section{Katz centrality}

\emph{Katz centrality}\index{Katz centrality}, also known as Katz prestige, is a generalization of eigenvector centrality \cite{Katz1953,Newman2018}. In the original formulation, the influence of a node \(i\) is measured by a weighted sum of all powers of the adjacency matrix \(A\), with an attenuation factor \(\alpha\):
\begin{equation*}
    c_{\mathrm{Katz}}(i) = \sum_{k=1}^{\infty} \sum_{j=1}^{N} \alpha^k \, (A^k)_{ij},
\end{equation*}
or, equivalently, in vector form:
\begin{equation*}
    \mathbf{c}_{\mathrm{Katz}} = (I - \alpha A)^{-1} \mathbf{1},
\end{equation*}
where \(\mathbf{1}\) is the vector of all ones. The series converges only if the attenuation factor \(\alpha\) is smaller than the reciprocal of the largest eigenvalue of the adjacency matrix \(A\), i.e., \(\alpha < 1 / \lambda_{\max}\), where \(\lambda_{\max}\) denotes the principal eigenvalue of \(A\).

Katz centrality is closely related to Bonacich’s power centrality \(c_{\mathrm{power}}\) \cite{Bonacich1987}, which is defined as
\begin{equation*}
    c_{\mathrm{power}} = \sum_{k=0}^{\infty} \sum_{j=1}^{N} \alpha^k \, (A^{k+1})_{ij}.
\end{equation*}
The two measures are directly proportional, with Bonacich's power centrality expressed in terms of Katz centrality as
\begin{equation*}
    c_{\mathrm{power}} = \alpha \, c_{\mathrm{Katz}}.
\end{equation*}

A further generalization, often referred to as \emph{alpha centrality}\index{alpha centrality} or \emph{Bonacich alpha centrality}, introduces an exogenous factor \(\beta\) and allows \(\alpha\) to take negative values to model negative influence \cite{Bonacich1987,Bonacich2001,Poulin2000}. The centrality of node \(i\) is then defined by
\begin{equation*} 
    c_{\alpha}(i) = \alpha \sum_{j=1}^{N} a_{ij} \, c_{\alpha}(j) + \beta,
\end{equation*}
where \(\alpha\) and \(\beta\) are positive constants. This formulation balances the contribution from the eigenvector-like term with the constant \(\beta\). In most current implementations, \(\alpha\) is set to \(0.1\) and \(\beta\) to \(1.0\). Bonacich and Lloyd \cite{Bonacich2001} demonstrated that Katz and alpha centralities differ only by a constant \(\beta\) under the condition 
\(\alpha < \frac{1}{\lambda_{\max}}\), with \(\lambda_{\max}\) being the principal eigenvalue of \(A\).

\section{KDEC method}

The \emph{KDEC}\index{KDEC method} method identifies influential nodes by combining the gravity model with the concept of effective distance \cite{Zhang2019}. In this model, a node's mass is given by the product of its degree and \(k\)-shell index, while the shortest-path distance is replaced by the effective distance to better capture the network's dynamic spreading pathways.

Let $\mathcal{N}^{(\leq l)}(i)$ denote set of nodes within the $l$-hop neighbourhood of node \(i\). The centrality \( c_{\textsc{KDEC}}(i) \) of node \( i \) is defined as
\begin{equation*}
   c_{\textsc{KDEC}}(i) = 
   \sum_{j \in \mathcal{N}^{(\leq l)}(i)} 
   \frac{m(i)\, m(j)}{\tilde{d}_{ij}^2},
\end{equation*}
where \( \tilde{d}_{ij} \) is the \emph{effective shortest-path distance} proposed by Brockmann and Helbing~\cite{Brockmann2013}. The parameter \( m(i) \) represents the mass of node \( i \), given by
\begin{equation*}
   m(i) = d_i \cdot k_s(i),
\end{equation*}
where \( d_i \) is the degree of node \( i \) and \( k_s(i) \) is its \(k\)-shell index. Zhang \textit{et al.}~\cite{Zhang2019} consider \( l = 2 \) as the truncated radius.

\section{KED method}

The \emph{KED method}\index{KED method} is a local spreader ranking algorithm designed to identify influential nodes in large-scale social networks \cite{Chen2014}. KED is based on the concept of \emph{path diversity}, which quantifies the diversity of spreading paths originating from each node. 

Let \(K_i\) denote the total degree of node \(i\)'s neighbors, i.e.,
\begin{equation*}
    K_i = \sum_{j \in \mathcal{N}(i)} k_j,
\end{equation*}
where \(k_j\) is the degree of neighbor \(j\) and \(\mathcal{N}(i)\) is the set of neighbors of node \(i\). The centrality of node \(i\) is then defined as
\begin{equation*}
    c_{\mathrm{KED}}(i) = k_i \, E_i^\alpha \, D_i^\beta,
\end{equation*}
where:  
\begin{itemize}
    \item \(k_i\) is the degree of node \(i\),  
    \item \(E_i = \frac{\sum_{j \in \mathcal{N}(i)} - p_j \log(p_j)}{\log(k_i)}\) is the local path diversity, with \(p_j = k_j / K_i\),  
    \item \(D_i = \exp\!\Big(\frac{K_i}{\max_l K_l}\Big)\) captures the contribution of the degrees of neighboring nodes,  
    \item \(\alpha\) and \(\beta\) are tunable parameters that control the relative importance of path diversity \(E_i\) and neighbor influence \(D_i\).  
\end{itemize}

Chen \textit{et al.} \cite{Chen2014} suggest using \(\alpha = \beta = 1\), giving equal weight to path diversity and neighbor connectivity in the ranking.

\section{Laplacian centrality}
The concept of \emph{Laplacian centrality}\index{Laplacian centrality} quantifies the importance of a node based on the network's response to its removal \cite{Qi2012}. Formally, the Laplacian centrality $c_{\text{Laplacian}}(i)$ of node $i$ is defined as the relative drop in the \emph{Laplacian energy} of the graph $G$ upon deletion of $i$:
\begin{equation*}
c_{\text{Laplacian}}(i) = \frac{E_L(G) - E_L(G_i)}{E_L(G)},
\end{equation*}
where $E_L(G)$ denotes the Laplacian energy of $G$, given by
\begin{equation*}
E_L(G) = \sum_{k=1}^{N} \lambda_k^2 = \sum_{i=1}^{N} \left(\sum_{j=1}^{N} w_{ij} \right)^2 + 2 \sum_{i<j} w_{ij}^2,
\end{equation*}
with $\lambda_k$ representing the eigenvalues of the Laplacian matrix of $G$, and $w_{ij}$ the weight of the edge between nodes $i$ and $j$. Here, $G_i$ denotes the graph obtained by deleting node $i$ from $G$. Qi \textit{et al.}~\cite{Qi2012} also showed that the Laplacian centrality of a node is closely related to the number of 2-walks in which it participates.

\section{Laplacian gravity centrality}

The \emph{Laplacian gravity centrality}\index{gravity centrality!Laplacian (LGC)} (LGC) is an extension of the local gravity model that combines Laplacian centrality with the network structure to identify influential nodes in complex networks \cite{ZhangQ2022}. The centrality \(c_{\textsc{LGC}}(i)\) of node \(i\) is defined as
\begin{equation*}
   c_{\textsc{LGC}}(i) = \sum_{j \in \mathcal{N}^{(\leq l)}(i)} \frac{c_L(i)\, c_L(j)}{d_{ij}^2},
\end{equation*}
where \(\mathcal{N}^{(\leq l)}(i)\) denote the set of nodes within \(l\)-hop neighborhood of node \(i\), \(d_{ij}\) is the shortest-path distance between nodes \(i\) and \(j\), and \(c_L(i)\) is the Laplacian centrality of node \(i\). Zhang \textit{et al.} \cite{ZhangQ2022} set the truncated radius to \(l = \langle d \rangle / 2\), where \(\langle d \rangle\) is the average shortest-path distance in the network.

\section{LeaderRank}

\emph{LeaderRank}\index{LeaderRank} is a parameter-free counterpart of the PageRank algorithm that is based on random walks in a graph~\cite{Lü2011}. The method introduces a ground node \(N+1\) that connects bidirectionally to every node in the network \(G\), ensuring that the network is strongly connected. To initiate the ranking process, one unit of resource is assigned to each node, except for the ground node and this resource is then evenly distributed among the neighbours of each node. Mathematically, this process is equivalent to a random walk on \(G\), where the resource \(s_i[k+1]\) at node \(i\) and discrete time \(k+1\) is updated according to
\begin{equation*}
    s_i[k+1] = \sum_{j=1}^{N+1} \frac{a_{ji}}{d_j^{out}} \, s_j[k],
\end{equation*}
with \(d_j^{out}\) denoting the number of successors of node \(j\) in \(G\). 
The initial scores are set as \(s_{N+1}[0] = 0\) for the ground node and \(s_i[0] = 1\) for all other nodes in \(G\). 
At the steady state \(\lim_{k \to \infty}s[k]=\Tilde{s}\), the score of the ground node is evenly redistributed to the other nodes, yielding the final LeaderRank score
\begin{equation*}
    c_{\text{LeaderRank}}(i) = \Tilde{s}_i + \frac{\Tilde{s}_{N+1}}{N}.
\end{equation*}

\section{Length-scaled betweenness centrality (LSBC)}

\emph{Length-scaled betweenness centrality}\index{betweenness centrality!length-scaled} (LSBC), also known as distance-scaled betweenness, is a variant of betweenness centrality that weights shortest paths inversely proportional to their length~\cite{Borgatti2006,Brandes2008}. The centrality of node \(i\) is defined as
\begin{equation*}
    c_{b_{dist}}(i) = \sum_{j=1}^{N} \sum_{k=1}^{N} \frac{1}{d_{jk}} \frac{\sigma_{jk}(i)}{\sigma_{jk}},
\end{equation*}
where \(d_{jk}\) is the length of the shortest path from \(j\) to \(k\), \(\sigma_{jk}\) is the total number of shortest paths between \(j\) and \(k\), and \(\sigma_{jk}(i)\) is the number of those paths passing through \(i\). This measure reflects the intuition that longer paths contribute less to a node's centrality.

\section{Leverage centrality}

\emph{Leverage centrality}\index{leverage centrality} quantifies the relative connectivity of a node compared to its neighbors \cite{Joyce2010}. For a node \(i\) with degree \(d_i\) and neighbors \(\mathcal{N}(i)\), the leverage centrality is defined as
\begin{equation*}
c_{\mathrm{Leverage}}(i) = \frac{1}{d_i} \sum_{j \in \mathcal{N}(i)} \frac{d_i - d_j}{d_i + d_j}.
\end{equation*}

This measure identifies nodes that are more connected than their neighbors, indicating their potential to control the flow of information. Nodes with negative leverage centrality are less connected than their neighbors and are thus influenced by them. Leverage centrality can also be extended to directed graphs by computing in-leverage and out-leverage using in-degree and out-degree, respectively \cite{Joyce2010}.

\section{Lhc index}

\emph{Lhc index}\index{Lhc index} is a hybrid approach for identifying highly influential spreaders in complex networks. It integrates both neighbor information and topological connectivity information among neighboring nodes \cite{Wang2021}. The neighbor information is represented by the degree of a node, which reflects the number of its direct connections. The connectivity among a node’s neighbors is characterized by the number of triangular structures centered on that node, indicating how tightly its neighbors are interconnected.

The centrality \( c_{\textsc{lhc}}(i) \) of node \( i \) is defined as
\begin{equation*}
   c_{\textsc{lhc}}(i) = 
   \sum_{j \in \mathcal{N}(i)} 
   \sum_{l \in \mathcal{N}^{(\leq k)}(j)} 
   \frac{
      d_l \left( 1 + \frac{\Delta_l}{\Delta} \right)
   }{d_{jl}^2},
\end{equation*}
where \( \mathcal{N}^{(\leq k)}(j) \) denotes the set of nodes within the $k$-hop neighbourhood of node \(j\), \( d_{jl} \) is the shortest-path distance between nodes \( j \) and \( l \), \( d_l \) is the degree of node \( l \), \( \Delta_l \) is the number of triangles including node \( l \), and \( \Delta \) is the total number of triangular structures in the network. 

To reduce computational complexity, Wang \textit{et al.}~\cite{Wang2021} set the distance parameter to \( k = 2 \).

\section{Lin's index}

\emph{Lin's index}\index{Lin's index} is an adaptation of closeness centrality for disconnected graphs \cite{Lin1976}. For a node \(i\), it is defined as
\begin{equation*}
c_{\mathrm{Lin}}(i) = \frac{\bigl| \{ j \in \mathcal{N} \mid d_{ij} < \infty \} \bigr|^2}{\sum_{j \in \mathcal{N},\, d_{ij} < \infty} d_{ij}},
\end{equation*}
where \(d_{ij}\) is the shortest-path distance between nodes \(i\) and \(j\) and the numerator is the square of the number of nodes reachable from \(i\). Squaring the numerator gives greater weight to nodes with larger reachable sets, which is particularly important in disconnected graphs. By definition, isolated nodes are assigned a centrality of 1.

For connected graphs, Lin's index reduces to closeness centrality, as all nodes are reachable.

\section{Linear threshold centrality (LTC)}

The \emph{Linear threshold centrality}\index{linear threshold centrality (LTC)} (LTC), also called Linear Threshold Rank (LTR), quantifies the influence of nodes based on the linear threshold (LT) model \cite{Riquelme2018}. In the LT model, a node \(i\) becomes active once the weighted sum of its active neighbors exceeds its individual threshold \(\theta_i\).

The LTC centrality of node \(i\) is defined as the fraction of nodes that can be activated when \(i\) and its neighbors are taken as the seed set:
\[
c_{LTC}(i) = \frac{|F(\{i\} \cup \mathcal{N}(i))|}{N},
\]
where \(\mathcal{N}(i)\) is the set of neighbors of \(i\) and \(F(X)\) is the set of nodes eventually activated by the linear threshold process starting from the seed set \(X\). Formally,
\begin{equation*}
F(X) = \bigcup_{t=0}^{k} F_t(X) = F_0(X) \cup F_1(X) \cup \dots \cup F_k(X),
\end{equation*}
where
\begin{equation*}
k = \min \{ t \in \mathbb{N} \mid F_t(X) = F_{t+1}(X) \} \le N.
\end{equation*}

At each time step \(t > 0\), a node \(j \notin F_{t-1}(X)\) becomes active if the weighted fraction of its active neighbors exceeds its threshold, i.e.,
\begin{equation*}
j \in F_t(X) \quad \text{if} \quad 
\sum_{l \in F_{t-1}(X) \cap \mathcal{N}^{\mathrm{in}}(j)} w_{lj} \ge \theta_j.
\end{equation*}

LTC thus captures both the local influence of a node through its immediate neighborhood and its potential to trigger wider cascades in the network. Nodes with high LTC centrality are those that, together with their immediate neighbors, can activate a large fraction of the network, indicating strong local influence and high cascading potential.

\section{Linearly scaled betweenness centrality}

\emph{Linearly scaled betweenness centrality}\index{betweenness centrality!linearly scaled} is a variant of betweenness centrality that weights shortest paths according to the relative position of intermediate nodes along the path from the source~\cite{Geisberger2008,Brandes2008}. The centrality of node \(i\) is defined as
\begin{equation*}
    c_{b-lin}(i) = \sum_{j=1}^{N} \sum_{k=1}^{N} \frac{d_{ji}}{d_{jk}} \frac{\sigma_{jk}(i)}{\sigma_{jk}},
\end{equation*}
where \(d_{jk}\) is the length of the shortest path from \(j\) to \(k\), \(\sigma_{jk}\) is the total number of shortest paths between \(j\) and \(k\), and \(\sigma_{jk}(i)\) is the number passing through \(i\). Nodes farther from the source, and thus closer to the target, contribute more to centrality. In undirected graphs, however, linearly scaled betweenness reduces to standard betweenness centrality, since the relative distances along paths in opposite directions sum to one:
\[
\frac{d_{ji}}{d_{jk}} + \frac{d_{ki}}{d_{kj}} = 1.
\]

\section{LineRank}

\emph{LineRank}\index{LineRank} is a flow-based centrality measure for large-scale graphs that quantifies the “flow” through each node \cite{Kang2011}. The method first transforms the original graph \(G\) into its corresponding line graph \(L(G)\), where nodes in \(L(G)\) represent edges in \(G\). PageRank centrality is then computed on the nodes of \(L(G)\).  

The LineRank centrality of a node in the original graph is obtained by aggregating the stationary probabilities of its incident edges in the line graph. This value represents the total amount of flow passing through the node, capturing its importance in the propagation of information or connectivity within the network.

\section{Link influence entropy (LInE) centrality}
\emph{Link Influence Entropy}\index{link influence entropy (LInE) centrality} (LInE) centrality quantifies the importance of each node based on the influence of the links connected to it~\cite{Singh2017}. The centrality of node $i$ is determined by the cumulative influence of all links incident to that node:
\[
c_{\mathrm{LInE}}(i) = \sum_{j \in \mathcal{N}(i)} p_{ij},
\]
where $p_{ij}$ represents the influence of link $(i,j)$, defined according to the change in the average shortest path length after the removal of that link:
\[
p_{ij} = \frac{|\langle d_G \rangle - \langle d_{G_{ij}} \rangle|}{\sum_{i \neq j} |\langle d_G \rangle - \langle d_{G_{ij}} \rangle|}.
\]
Here, $\langle d_G \rangle$ denotes the average shortest path length in the original graph $G$, and $\langle d_{G_{ij}} \rangle$ denotes the corresponding value for the graph obtained by removing link $(i,j)$. 

If link $(i,j)$ acts as a bridge (i.e., its removal disconnects the graph), Singh \textit{\textit{et al.}}~\cite{Singh2017} compute $\langle d_{G_{ij}} \rangle$ using only the largest connected component when it contains more than 80\% of the nodes; otherwise, $\langle d_{G_{ij}} \rangle$ is averaged over the two resulting components.

If link $(i,j)$ is a bridge (i.e., its removal disconnects the graph), Singh \textit{\textit{et al.}}~\cite{Singh2017} compute $\langle d_{G_{ij}} \rangle$ using the largest connected component if it contains more than 80\% of the nodes; otherwise, $\langle d_{G_{ij}} \rangle$ is averaged over the two resulting components.

\section{Load centrality}
\emph{Load centrality}\index{load centrality}, also referred to as \emph{traffic load centrality} (TLC), was introduced by Goh \textit{\textit{et al.}}~\cite{Goh2001} and later reformulated by Brandes~\cite{Brandes2008}. 
It is a flow-based variant of betweenness centrality that quantifies the extent to which a node participates in the transport of information or resources across the network.

In this model, a source node sends a unit quantity of a commodity to a target node along all shortest paths connecting them. 
At each intermediate step, if multiple adjacent nodes are equally close to the target, the transmitted flow is divided equally among them and propagated recursively until the commodity reaches the destination.

Formally, the load centrality \( c_{\mathrm{load}}(i) \) of node \( i \) is defined as the total amount of flow passing through \( i \) during all pairwise exchanges between nodes:
\[
c_{\mathrm{load}}(i) = \sum_{s \neq t \neq i} \ell_{st}(i),
\]
where \( \ell_{st}(i) \) denotes the fraction of the unit flow sent from source \( s \) to target \( t \) that passes through node \( i \). Thus, load centrality captures the node’s contribution to network traffic, assigning higher values to nodes that frequently lie on shortest paths connecting other nodes.

\section{Lobby index}

The \emph{lobby index}\index{lobby index}, or \(l\)-index, is inspired by Hirsch’s h-index\index{Hirsch’s h-index}\index{\textit{h}-index}, which quantifies the scientific output of a researcher \cite{Korn2009,Lu2016}. For a node \(i\), the lobby centrality \(c_{\mathrm{Lobby}}(i)\) is defined as the largest integer \(k\) such that \(i\) has at least \(k\) neighbors with degree at least \(k\):
\begin{equation*}
c_{\mathrm{Lobby}}(i) = \max \{ k \mid d_{i(k)} \ge k \},
\end{equation*}
where \(i(k)\) denotes the neighbor of \(i\) with the \(k\)-th largest degree.

Several extensions of the lobby index to weighted networks have been proposed:  
\begin{enumerate}
    \item The \emph{collaboration index}\index{collaboration index} (\(c\)-index) computes the Hirsch h-index of the sequence formed by multiplying each neighbor’s strength by the weight of the connecting edge \cite{Yan2013}.
    \item The \emph{communication ability}\index{communication ability} replaces each neighbor’s degree with the product of its \emph{weighted lobby index} (also called \(h\)-degree), which is the lobby index calculated using the strengths of the incident edges, and the weight of the edge connecting it to the node \cite{Zhai2013}.
\end{enumerate}

\section{Local and global centrality (LGC)}

\emph{Local and global centrality}\index{local and global centrality (LGC)} (LGC) identifies influential nodes by integrating both local and global topological properties of a network \cite{Ullah2021b}. The centrality of a node \(i\) is defined as
\[
c_{LGC}(i) = \frac{d_i}{N} \sum_{j \neq i} \frac{\sqrt{d_j + \alpha}}{d_{ij}},
\]
where \(d_i\) is the degree of node \(i\), \(d_{ij}\) is the shortest-path distance between nodes \(i\) and \(j\), and \(\alpha \in [0,1]\) is a parameter that controls the relative influence of neighboring node degrees (e.g., $\alpha=0.4$).  

Nodes with high LGC scores are those that not only have many connections but are also closely connected to other well-connected nodes, making them critical for information spreading and network cohesion. The effectiveness of LGC has been evaluated on six real-world networks and validated using the Susceptible-Infected-Recovered (SIR) epidemic model.

\section{Local clustering coefficient}

The \emph{local clustering coefficient}\index{local clustering coefficient}\index{clustering coefficient!local}, also known simply as the clustering coefficient, of a node \(i\) measures the probability that two randomly chosen neighbors of \(i\) are connected to each other \cite{Newman2018}. Formally, the local clustering coefficient \(c_{\mathrm{cl}}(i)\) of node \(i\) is defined as the ratio between the number of actual links among its neighbors and the number of all possible links between them:
\begin{equation*} 
   c_{\mathrm{cl}}(i) =
   \begin{cases}
      \dfrac{\sum_{j \in \mathcal{N}(i)} \sum_{k \in \mathcal{N}(i),\, k \neq j} a_{jk}}{d_i (d_i - 1)}, & \text{if } d_i > 1, \\[1.0em]
      0, & \text{otherwise,}
   \end{cases}
\end{equation*}
where \(d_i\) is the degree of node \(i\). Thus, \(c_{\mathrm{cl}}(i)\) quantifies how close the neighborhood of node \(i\) is to forming a clique (complete subgraph).  The local clustering coefficient has been extended to weighted graphs in \cite{Barrat2004}.

The local clustering coefficient can be used to identify \emph{structural holes} in a network, indicating how influential a node may be in mediating or controlling information flow between its neighbors. In many networks, it is empirically observed that the local clustering coefficient of nodes depends roughly on their degree, with high-degree nodes typically exhibiting lower clustering \cite{Newman2018}.

\section{Local degree dimension (LDD)}

The \emph{local degree dimension}\index{local dimension!degree (LDD)} (\textsc{LDD}) is a centrality measure for identifying influential nodes, based on the assumption that both the number of neighbors at each topological layer and the pattern of how this number changes across layers reflect a node's importance \cite{Zhong2022}.

For each node \(i\), \textsc{LDD} first computes \(n_i(l)\), the number of nodes at shortest-path distance \(l\) from \(i\). Next, it analyzes how \(n_i(l)\) changes with \(l\), identifying the number of rising layers \(l_{i+}\) and declining layers \(l_{i-}\). The \textsc{LDD} score \(c_{\textsc{LDD}}(i)\) is then given by
\[
c_{\textsc{LDD}}(i) = d_i\, D_{i+}\, l_{i+} + D_{i-}\, l_{i-},
\]
where \(d_i\) is the degree of node \(i\), and \(D_{i+}\) and \(D_{i-}\) are the rates of increase and decrease, obtained by linear fitting of \(n_i(l)\) in the rising and declining layers, respectively. 

Hence, the local degree dimension (LDD) captures a node's degree as well as the upward and downward trends in the number of its neighbors, reflecting both the breadth and potential spreading speed of its influence.

\section{Local dimension (LD)}

The \emph{local dimension}\index{local dimension} (LD) quantifies the dimensionality of nodes in a network by examining how the volume of the neighborhood around each node scales with increasing topological distance \cite{Silva2012}. In most spatially embedded real networks, which typically lack the small-world property, the distribution of link lengths follows a power law. Consequently, the number of nodes $B_i(r)$ located within a topological distance $r$ from a node $i$ obeys the relationship
\[
B_i(r) \sim r^d,
\]
where the constant $d$ characterizes the effective dimension of the network. Silva and Costa \cite{Silva2012} refined this power-law relationship by allowing the dimensionality to vary locally, proposing that
\[
B_i(r) = \alpha r^{D_i(r)},
\]
where $D_i(r)$ represents the local dimension around node $i$. The local dimension coefficient $D_i(r)$ can be estimated from the slope of the $B_i(r)$ curve on a double-logarithmic scale and is discretized as
\[
D_i(r) \simeq r \frac{n_i(r)}{B_i(r)},
\]
where $n_i(r)$ denotes the number of nodes that are exactly at a topological distance $r$ from the reference node $i$. Pu \textit{et al.} \cite{Pu2014} further extended the local dimension measure by allowing the distance parameter $r$ to vary across different nodes in the network.

\section{Local entropy (LE) centrality}
\emph{Local entropy}\index{entropy centrality!local (LE)} (LE) is a semi-local centrality measure that accounts for the degrees of a node's neighbors~\cite{Zhang2014,Nie2016}. The LE of node $i$ is defined as
\[
c_{\mathrm{LE}}(i) = - \sum_{j \in \mathcal{N}(i)} d_j \log d_j,
\]
where $d_j$ is the degree of neighbor node $j$, and $\mathcal{N}(i)$ denotes the set of neighbors of node $i$. This measure captures the heterogeneity of the local neighborhood: higher LE values indicate that node $i$ is connected to neighbors with diverse degrees.

\section{Local fuzzy information centrality (LFIC)}

\emph{Local fuzzy information centrality}\index{local fuzzy information centrality (LFIC)} (LFIC) is a centrality measure for identifying influential nodes based on the local dimension of nodes and fuzzy theory \cite{Zhang2021}. 

The \textsc{LFIC} centrality \(c_{\textsc{LFIC}}(i)\) of node \(i\) is defined as
\[
c_{\textsc{LFIC}}(i) = \sum_{l=1}^{K} \frac{-p_i(l) \ln p_i(l)}{l^2},
\]
where \(l\) is the distance from the center node \(i\), and \(K\) is the maximal box size, defined as \(K = \lceil \max_j d_{ij} / 2 \rceil\). Here, \(p_i(l)\) is the probability associated with neighbor nodes at distance \(l\) from node \(i\):
\[
p_i(l) = \frac{1}{e} \frac{f_i(l)}{\sum_{l=1}^{K} f_i(l)},
\]
with
\[
f_i(l) = n_i(l) \, e^{-l^2 / K^2},
\]
where \(n_i(l)\) is the number of nodes whose shortest-path distance from node \(i\) equals \(l\).

Hence, LFIC combines local node structure with fuzzy weighting to capture the influence of nodes at varying distances from the center node.

\section{Local gravity model}

The \emph{local gravity model}\index{gravity model!local} is a variant of the gravity model in which a node’s centrality depends only on its \(l\)-hop neighborhood \cite{Li2019}. Let \(\mathcal{N}^{(\geq l)}(i)\) denote the set of nodes within \(l\)-hop neighborhood of node \(i\). The centrality \(c_{\text{Local-Gravity}}(i)\) of node \(i\) is defined as
\begin{equation*}
   c_{\text{Local-Gravity}}(i) = \sum_{j \in \mathcal{N}^{(\geq l)}(i)} \frac{d_i\,d_j}{d_{ij}^2},
\end{equation*}
where \(d_{ij}\) represents the shortest path distance between nodes \(i\) and \(j\), and \(d_i\) is the degree of node \(i\). 

Thus, the local gravity model incorporates only local structural information within an \(l\)-hop neighborhood. When \(l\) equals the diameter of the network, the local gravity model becomes equivalent to the original gravity model.

\section{Local H-index}

The \emph{local H-index (LH-index)}\index{Local H-index} \cite{Liu2018} is a semi-local centrality measure and an improved version of the traditional H-index, also known as the lobby index or \(l\)-index \cite{Korn2009}. The local H-index of a node \(i\), denoted by \(c_{LH}(i)\), is defined as
\begin{equation*}
    c_{LH}(i) = h(i) + \sum_{j \in \mathcal{N}(i)} h(j),
\end{equation*}
where \(h(i)\) denotes the H-index of node \(i\), and \(\mathcal{N}(i)\) represents the set of its neighbors. Thus, the local H-index accounts for both the H-index of a node and those of its neighboring nodes.

\section{Local information dimensionality (LID)}

The \emph{local information dimensionality}\index{local information dimensionality (LID)} (LID) is an entropy-based, semi-local centrality measure derived from the concept of local dimension \cite{Wen2020}. It characterizes the structural complexity of the neighborhood around each node by quantifying how local information changes with scale.

For a given node $i$, the local information dimensionality $D_{I_i}$ is defined as
\[
D_{I_i} = - \frac{dI_i(r)}{d \ln r},
\]
where the derivative is taken with respect to the logarithm of the topological distance $r$, which represents the scale of locality around node $i$. The information content $I_i(r)$ is computed based on the number of nodes $B_i(r)$ within a topological distance $r$ from node $i$ as
\[
I_i(r) = - \frac{B_i(r)}{N}\ln \frac{B_i(r)}{N},
\]
where $N$ is the total number of nodes in the network.

The local information dimensionality $D_{I_i}$ can be estimated from the slope of the relationship between $I_i(r)$ and $\ln r$ on a logarithmic scale, and discretized as
\[
D_{I_i} \simeq \frac{n_i(r)\left[1 + \ln \frac{B_i(r)}{N}\right]r}{N},
\]
where $n_i(r)$ denotes the number of nodes located exactly at distance $r$ from the reference node $i$.

The scale of locality $r$ is typically chosen as half of the maximum shortest-path distance from node $i$, defined as
\[
r = \lceil d_{\max}(i)/2 \rceil,
\]
where $d_{\max}(i)$ is the maximum geodesic distance from node $i$, and $\lceil \cdot \rceil$ denotes the ceiling function.

\section{Local neighbor contribution (LNC) centrality}

The \emph{local neighbor contribution}\index{local neighbor contribution (LNC) centrality} (LNC) centrality quantifies node importance by combining the node’s own structural influence, based on its degree and local neighborhood size, with the aggregated contributions from its nearest and next-nearest neighbors \cite{Dai2019}. The local neighbor contribution (LNC) centrality of node \(i\) is defined as
\begin{equation*}
    c_{LNC}(i) = O_c(i) \cdot N_c(i),
\end{equation*}
where \(O_c(i)\) represents the node's own structural influence, determined by its degree and the connectivity of its nearest and next-nearest neighbors:
\begin{equation*}
    O_c(i) = d_i \cdot |\mathcal{N}^{(\leq 2)}(i)| \sum_{j \in \mathcal{N}^{(\leq 2)}(i)} \frac{1}{d_j} \left(1 - \frac{1}{d_j}\right)^{|\mathcal{N}^{(\leq 2)}(i)|-1},
\end{equation*}
and \(N_c(i)\) represents the contribution from the nearest and next-nearest neighbors of \(i\):
\begin{equation*}
    N_c(i) = \frac{|\mathcal{N}^{(\leq 2)}(i)|}{N-1} \sum_{j \in \mathcal{N}(i)} d_j,
\end{equation*}
where \(|\mathcal{N}^{(\leq 2)}(i)|\) denotes the number of nearest and next-nearest neighbors of node \(i\) and \(d_j\) is the degree of node \(j\).

\section{Local reaching centrality}

The \emph{local reaching centrality}\index{local reaching centrality} of a node \(i\) in an unweighted directed graph \(G\) is equivalent to the \(m\)-reach centrality with \(m = N{-}1\), where \(N\) is the total number of nodes in the graph~\cite{Mones2012}.  
It is defined as the proportion of nodes in the graph that can be reached from \(i\) following outgoing edges:
\[
c_{\mathrm{LR}}(i) = \frac{|\{ j \in \mathcal{N}: \sum_{k=1}^{N-1}(A^k)_{ij} > 0\}|}{N-1}.
\]

Local reaching centrality quantifies the extent to which a node can reach other nodes in the network, providing a normalized measure of its local influence in terms of reachability.  
In undirected networks, all nodes within the same connected component have identical local reaching centrality, equal to the fraction of nodes in the component relative to the total number of nodes in the network.

\section{Local relative change of average shortest path (LRASP) centrality}

The \emph{local relative change of average shortest path}\index{local relative change of average shortest path (LRASP) centrality} (LRASP) centrality is a modification of the average shortest path centrality (AC or RASP) \cite{Hajarathaiah2022}. Let \(G_{N_i(l)}\) denote the subgraph containing node \(i\) and all nodes within \(l\) hops from \(i\). The LRASP centrality of node \(i\) is defined as
\begin{equation*}
    c_{LRASP}(i) = \frac{ASP(G'_{N_i(l)}) - ASP(G_{N_i(l)})}{ASP(G_{N_i(l)})},
\end{equation*}
where \(ASP(G_{N_i(l)})\) is the average shortest path length of \(G_{N_i(l)}\):
\begin{equation*}
    ASP(G_{N_i(l)}) = \frac{\sum_{j \neq k \in N_i(l)} d_{jk}}{|N_i(l)|(|N_i(l)|-1)},
\end{equation*}
with \(d_{jk}\) being the shortest path distance between nodes \(j\) and \(k\) if reachable; otherwise, \(d_{jk}\) is set to the diameter of \(G_{N_i(l)}\). The subgraph \(G'_{N_i(l)}\) is obtained by removing all links adjacent to node \(i\).

LRASP quantifies the relative change in average shortest path when the immediate connections of node \(i\) are removed, capturing its local structural importance. When \(l\) equals the network diameter \(d(G)\), LRASP reduces to the standard AC centrality. In \cite{Hajarathaiah2022}, the authors use \(l = d(G)/2\) to balance local and semi-global structural effects.

\section{Local structural centrality (LSC)}

The \emph{local structural centrality}\index{local structural centrality (LSC)} (LSC) accounts for both the number of nearest and next-nearest neighbors and their topological connections, reflecting that influence depends on local neighborhood size and connectivity \cite{Gao2014}. The centrality \( c_{\mathrm{LSC}}(i) \) of node \( i \) is defined as
\begin{equation*}
   c_{\mathrm{LSC}}(i) = \sum_{j \in \mathcal{N}(i)} 
   \left( \alpha |\mathcal{N}^{(\leq 2)}(i)| + (1 - \alpha) \sum_{k \in \mathcal{N}^{(\leq 2)}(j)} c_k \right),
\end{equation*}
where \( \mathcal{N}^{(\leq 2)}(i) \) denotes the set of nearest and next-nearest neighbors of node \( i \), 
\( c_k \) is the clustering coefficient of neighbor \( k \), 
and \( \alpha \) is a tunable parameter between 0 and 1. In \cite{Gao2014}, two values of \(\alpha\) are considered: \(\alpha = 0.2\) and \(\alpha = 0.7\).

\section{Local volume dimension (LVD)}

The \emph{local volume dimension}\index{local volume dimension (LVD)} (LVD) is a centrality measure for identifying influential nodes based on the local dimension concept \cite{HLi2021b}. For each node \(i\), the LVD centrality evaluates the total degree within a box \(V_i(l)\), defined as the sum of the degrees of all nodes at distance \(l\) from node \(i\). It is assumed that the total degree within each box follows a power-law distribution. 

The LVD centrality \(c_{\textsc{LVD}}(i)\) of node \(i\) is given by
\[
c_{\textsc{LVD}}(i) = \frac{d \ln V_i(l)}{d \ln l}.
\]

The LVD centrality of node $i$ is estimated numerically as the slope of the linear regression of \(\ln V_i(l)\) against \(\ln l\).

\section{Localized bridging centrality (LBC)}

\emph{Localized bridging centrality}\index{bridging centrality!localized} (LBC) is a variant of bridging centrality that combines a node's local brokerage and connectivity features \cite{Nanda2008}. It is defined as the product of the egocentric betweenness centrality \(c_{\mathrm{ego}}(i)\) and the bridging coefficient \(\beta_c(i)\):
\begin{equation*}
   c_{\mathrm{LBC}}(i) = c_{\mathrm{ego}}(i) \cdot \beta_c(i),
\end{equation*}
where \(c_{\mathrm{ego}}(i)\) is the egocentric betweenness centrality of node \(i\), which is computed as the betweenness of node $i$ in its egocentric network \cite{Freeman1982}.  

The LBC measure identifies nodes that serve as local brokers within their immediate neighborhood while also connecting to high-degree neighbors.

\section{LocalRank centrality}

\emph{LocalRank centrality}\index{LocalRank centrality} (also called semi-local or local centrality) quantifies a node’s importance by considering both its nearest and next-nearest neighbors \cite{Chen2012}. For a node \(i\), it is defined as
\begin{equation*}
c_{\mathrm{LR}}(i) = \sum_{j \in \mathcal{N}(i)} \sum_{k \in \mathcal{N}(j)} n(k),
\end{equation*}
where \(\mathcal{N}(i)\) is the set of neighbors of \(i\), and \(n(k)=|N^{(\le 2)}(k)|\) is the number of nearest and next-nearest neighbors of node \(k\).  

Intuitively, LocalRank captures both the direct connectivity of a node and the connectivity of its neighbors, allowing nodes connected to highly connected neighborhoods to achieve higher centrality. This makes it more discriminative than degree centrality while remaining computationally efficient.

\section{Long-Range Interaction Centrality (LRIC)}
The \emph{Long-Range Interaction Centrality (LRIC)}\index{long-range interaction centrality (LRIC)} index is an extension of SRIC that accounts for the indirect influence of nodes \cite{Aleskerov2017,AleskerovBook}. LRIC is based on the concepts of direct and group influence, as in SRIC, where each node \(i\) has an individual threshold of influence \(q_i\), while \(\Omega(i)\) and \(\Omega^{p}(i)\) denote, respectively, the sets of critical and pivotal neighbours of node \(i\). However, the direct influence \(c_{ij}\) of node \(i\) on node \(j\) is defined as
\begin{equation*}
    c_{ij} = 
    \begin{cases}
        \max_{i \in \Omega^{p}_k(j)} \frac{a_{ij}}{\sum_{h \in \Omega_k(j)} a_{hj}}, & \text{if } \exists k: i \in \Omega^{p}_k(j),\\
        0, & \text{otherwise}.
    \end{cases}
\end{equation*}

The direct influence \(c_{ij}\) can be interpreted as the maximal possible influence of node \(i\) within any group \(\Omega_k(j)\) where it is pivotal. LRIC further considers the indirect influence of nodes by examining paths of length \(\leq s\) in the network of direct influences. There are three common variations of the LRIC index:

\begin{enumerate}
    \item \textit{LRIC(max)}: the influence \(f(P_{i \rightarrow j})\) of node \(i\) on node \(j\) along a path \(P_{i \rightarrow j}\), characterized by a sequence of edges \((i,k_1),(k_1,k_2),\ldots,(k_{s-1},j)\), is defined as the \emph{joint probability} of the edges:
    \begin{equation*}
        f(P_{i \rightarrow j}) = c_{ik_1} \times c_{k_1k_2} \times \ldots \times c_{k_{s-1}j}.
    \end{equation*}

    \item \textit{LRIC(maxmin)}: the influence \(f(P_{i \rightarrow j})\) of node \(i\) on node \(j\) along a path \(P_{i \rightarrow j}\) is defined by the \emph{bottleneck capacity}:
    \begin{equation*}
        f(P_{i \rightarrow j}) = \min(c_{ik_1}, c_{k_1k_2}, \ldots, c_{k_{s-1}j}).
    \end{equation*}
    In both LRIC(max) and LRIC(maxmin), the indirect influence $\Tilde{c}_{ij}$ of node \(i\) on node \(j\) is determined by the path with the greatest strength, i.e.,
    \[
        \Tilde{c}_{ij} = \max_{P_{i \rightarrow j}}f(P_{i \rightarrow j}).
    \]

    \item \textit{LRIC(PPR)}: the influence \(f(P_{i \rightarrow j})\) of node \(i\) on node \(j\) along a path \(P_{i \rightarrow j}\) is determined by considering all paths between them \cite{Aleskerov2020b}. Specifically, the indirect influence $\tilde{c}_{ij}$ of node $i$ on node $j$ is quantified using the personalized PageRank (PPR) algorithm, which estimates the probability of reaching node $j$ starting from node $i$. This computation uses a modified graph of direct influences, where an additional link is introduced from each node \(k\) to node \(i\) with strength
    \begin{equation*}
        c_{ki} = N-1 - \sum_{j \neq i} c_{kj}.
    \end{equation*}
\end{enumerate}

The final LRIC score of node $i$ is obtained by aggregating its indirect influence on all other nodes in the network. For instance, one possible aggregation is given by
\[
c_{\mathrm{LRIC}}(i) = \sum_{j=1}^{N} \Tilde{c}_{ij}.
\]

The LRIC index has been applied in diverse domains, including the identification of influential countries in global food trade networks \cite{Aleskerov2017b}, analysis of financial \cite{Aleskerov2020,Shvydun2020b}, global arms transfer \cite{Shvydun2019}, international conflict \cite{Aleskerov2016}, and international migration networks \cite{Aleskerov2016b}, as well as the detection of key actors in terrorist networks \cite{Shvydun2019} and citation networks of economic journals \cite{Aleskerov2016c}.

\section{M-centrality}

The \emph{M-centrality}\index{M-centrality} evaluates the influence of a node by combining local information from its neighborhood with global information about its position in the network \cite{Ibnoulouafi2018}. 
The centrality of node $i$ is defined as
\begin{equation*}
   c_{M}(i) = \mu \, k_s(i) + (1-\mu) \, \Delta D(i),
\end{equation*}
where $k_s(i)$ is the $k$-shell centrality of node $i$, representing its global importance, and $\Delta D(i)$ captures local degree variation:
\begin{equation*}
   \Delta D(i) = \sum_{j \in \mathcal{N}(i)} d_i \left| \frac{d_j - d_i}{\sum_{l \in \mathcal{N}(i)} d_l} \right|,
\end{equation*}
with $\mathcal{N}(i)$ denoting the set of neighbors of node $i$, and $d_j$ the degree of neighbor $j$.  

The parameter $\mu \in [0,1]$ balances the contributions of the global ($k$-shell) and local (degree variation) measures. 
Ibnoulouafi \textit{et al.} \cite{Ibnoulouafi2018} suggest setting $\mu$ based on the relative entropies of the two distributions:
\begin{equation*}
   \mu = \frac{1 - E_1}{2 - E_1 - E_2},
\end{equation*}
where $E_1$ and $E_2$ are the entropies of the $k$-shell centrality and $\Delta D(i)$ distributions, respectively.

\section{$\boldsymbol{\mu}$-Power Community Index ($\boldsymbol{\mu}$-PCI)}

The \emph{\(\mu\)-Power Community Index}\index{\(\mu\)-Power Community Index} (\(\mu\)-PCI) was introduced by Katsaros \textit{et al.} \cite{Papadimitriou2009,Katsaros2013} to identify nodes situated in densely connected regions of a network, which are therefore likely to act as influential spreaders. This metric combines the concepts of betweenness centrality, which captures nodes that lie on numerous communication paths between pairs of nodes, and the transitive network density reflected by the coreness measure.

Let \(\mathcal{N}^{(\mu)}(i)\) denote the \(\mu\)-hop neighborhood of node \(i\). The centrality of node \(i\), denoted by \(c_{\mu\text{-}PCI}(i)\), is defined as the largest integer \(k\) such that there exist at least \(\mu k\) nodes in \(\mathcal{N}^{(\mu)}(i)\) whose degree is greater than or equal to \(k\), while all remaining nodes in \(\mathcal{N}^{(\mu)}(i)\) have a degree less than or equal to \(k\):
\begin{equation*}
c_{\mu\text{-}PCI}(i) = \max \left\{ k : \left| \{ j \in \mathcal{N}^{(\mu)}(i) \, : \, d_j \geq k \} \right| \geq \mu k \right\},
\end{equation*}
where \(d_i\) is the degree of node \(i\). This definition highlights two key components: the parameter \(\mu\), which controls the extent of the local neighborhood considered, and the degree threshold \(k\), which represents the structural strength of node \(i\)’s surroundings. A higher \(c_{\mu\text{-}PCI}(i)\) value indicates that node \(i\) is embedded within a densely interconnected region where many neighboring nodes also possess high degrees, suggesting a strong potential for local influence.

\section{\textit{m}-reach centrality}

The \emph{\(m\)-reach centrality}\index{\textit{m}-reach centrality}, also referred to as \(m\)-step reach centrality or the \(K\)-order propagation number~\cite{LiYa2019}, quantifies the extent of a node’s influence by counting the number of nodes that can be reached within \(m\) steps from it~\cite{Borgatti2003}.  Formally, for a node \(i\), the \(m\)-reach centrality \(c_{m\text{-reach}}(i)\) is defined as
\[
c_{m\text{-reach}}(i) = |\{ j \in \mathcal{N} : d_{ij} \le m \}| = |N^{(\le m)}(i)|,
\]
where \(\mathcal{N}\) is the set of all nodes in the network and \(d_{ij}\) denotes the shortest-path distance between nodes \(i\) and \(j\). The \(m\)-reach centrality generalizes several well-known measures:  
\begin{itemize}
    \item For \(m = 1\), it coincides with the degree centrality.
    \item For \(m = 2\), it is equivalent to the reachability measure~\cite{Higley1991}.
\end{itemize}

The parameter \(m\) should not exceed the diameter of the network, as values larger than the diameter would include all nodes and thus provide no further discrimination between nodes.

\section{Malatya centrality}

\emph{Malatya centrality}\index{Malatya centrality} is a local centrality measure that evaluates a node's importance relative to the degrees of its neighbors \cite{Oztemiz2024}. For a node \(i\), it is defined as
\[
c_{Malatya}(i) = \sum_{j \in \mathcal{N}(i)} \frac{|\mathcal{N}(i)|}{|\mathcal{N}(j)|}=d_i\sum_{j \in \mathcal{N}(i)} \frac{1}{d_j},
\]
where \(\mathcal{N}(i)\) denotes the set of neighbors of nodes \(i\) and $d_i$ is the degree of node $i$. 

Malatya centrality assigns higher centrality to nodes that have more connections than their neighbors, highlighting nodes that are locally more connected than their immediate surroundings.

\section{Map equation centrality}

\emph{Map equation centrality}\index{map equation centrality} is a vitality-based centrality measure derived from the information-theoretic community detection framework known as the \emph{map equation} \cite{Blöcker2022}. It quantifies the importance of a node based on how much it contributes to the overall description length of flow on the network.  

Assume that the network \(G\) has a non-overlapping community structure represented by the partition \(M\). The map equation centrality \(c_{ME}(i)\) of node \(i\) measures the reduction in the average codeword length if node \(i\) is silenced, meaning that when a random walker visits \(i\), no codeword for \(i\) is transmitted. Formally,
\begin{equation*}
    c_{ME}(i) = L(G, M) - L^{*}_i(G, M),
\end{equation*}
where \(L(G, M)\) denotes the optimal description length of flow for the original network under partition \(M\), and \(L^{*}_i(G, M)\) is the optimal description length when node \(i\) is silenced.

Hence, \(c_{ME}(i)\) represents the \emph{marginal contribution} of node \(i\) to the total information cost of describing flow on the network. Nodes with higher \(c_{ME}(i)\) values are more important for maintaining the network’s information flow structure, while nodes with lower or negative values have less or even disruptive influence on modular information dynamics.

\section{Mapping entropy betweenness (MEB) centrality}

\emph{Mapping Entropy Betweenness}\index{mapping entropy betweenness (MEB) centrality} (MEB) extends the concept of mapping entropy by weighting nodes with betweenness centrality instead of degree centrality \cite{Gialampoukidis2016}. The centrality of node $i$ is defined as
\begin{equation*}
    c_{\text{MEB}}(i) = - BC_i \sum_{j \in \mathcal{N}(i)} \log BC_j,
\end{equation*}
where $BC_i$ is the \emph{normalized betweenness centrality} of node $i$, and $\mathcal{N}(i)$ denotes its set of neighbors. The normalized betweenness centrality is computed as
\begin{equation*}
    BC_i = \frac{\sum_{j\neq k \neq i} \frac{\sigma_{jk}(i)}{\sigma_{jk}}}{N^2 - 3N + 2},
\end{equation*} 
where \(\sigma_{jk}\) denotes the number of shortest paths from node \(j\) to node \(k\), and \(\sigma_{jk}(i)\) represents the number of paths that pass through node \(i\).

The MEB centrality highlights nodes with high betweenness that are also connected to other high-betweenness neighbors, emphasizing their key role in facilitating information flow.

\section{Mapping entropy (ME) centrality}
\emph{Mapping entropy}\index{entropy centrality!mapping (ME)} (ME) is a semi-local measure that incorporates both the degree of a node and the degrees of its neighbors~\cite{Zhang2014,Nie2016}. The ME of node $i$ is defined as
\[
c_{\mathrm{ME}}(i) = - d_i \sum_{j \in \mathcal{N}(i)} \log d_j,
\]
where $d_i$ is the degree of node $i$ and $\mathcal{N}(i)$ denotes the set of neighbors of node $i$. ME emphasizes nodes with high degree that are connected to neighbors with a wide range of degrees, thus combining node-level and neighborhood-level information.

\section{Markov centrality}

\emph{Markov centrality}\index{Markov centrality}, also known as random-walk closeness \cite{Blöchl2011}, is based on the concept of mean first passage time (MFPT) in a Markov chain \cite{White2003}. The MFPT \(m_{ij}\) from node \(i\) to node \(j\) is the expected number of steps required to reach \(j\) for the first time starting from \(i\):
\begin{equation*}
m_{ij} = \sum_{n=1}^{\infty} n \cdot f_{ij}^{(n)},
\end{equation*}
where \(f_{ij}^{(n)}\) is the probability that the chain first reaches \(j\) in exactly \(n\) steps.

The Markov centrality of node \(i\) is defined as the inverse of the average MFPT to \(i\) from a set of root nodes \(R\) (e.g., \(R = \mathcal{N} \setminus \{i\}\)):
\begin{equation*}
c_{\mathrm{Markov}}(i) = \frac{1}{\frac{1}{|R|}\sum_{j \in R} m_{ji}}.
\end{equation*}

This measure applies to both directed and undirected graphs. Intuitively, \(m_{ij}\) represents an average distance from \(i\) to \(j\) under random-walk dynamics, so Markov centrality can be interpreted as an averaged random-walk closeness centrality.

\section{Markov entropy centrality}

\emph{Markov entropy centrality}\index{entropy centrality!Markov}, originally called entropy centrality, is an entropy-based node centrality metric derived from a discrete random Markovian transfer process \cite{Nikolaev2015}. In this model, an object is transferred from a given node according to the following rules: at each step, the object is either absorbed by the current node with probability $a$, terminating the process, or passed to one of its neighbors with probability $1-a$, allowing the process to continue.

The centrality of node $i$, $c_{\mathrm{MEC}}(i)$, is quantified by the entropy of the distribution of destinations reached by the object originating from $i$ after $t$ transitions:
\[
c_{\mathrm{MEC}}(i) = - \sum_{j=1}^N \left(p_{ij}^t + p_{ij'}^t\right) \log \left(p_{ij}^t + p_{ij'}^t\right),
\]
where $(p_{ij}^t + p_{ij'}^t)$ denotes the probability that the object, starting at node $i$, is held by node $j$ after $t$ steps. The original $2N \times 2N$ transition probability matrix $P$ is defined as
\[
p_{ij} = 
\begin{cases}
    a, & \text{if } j = i',\\
    1, & \text{if } i = j = i',\\
    \frac{(1-a) a_{ij}}{d_i}, & \text{otherwise,}
\end{cases}
\]
where $i'$ denotes the absorbing state corresponding to node $i$, $a_{ij}$ is the adjacency matrix entry, and $d_i$ is the degree of node $i$.

By design, Markov entropy centrality measures a node's potential for information spread: nodes with high entropy can reach a diverse set of destinations with relatively even probability, indicating a structurally influential and versatile role. Conversely, low entropy implies that walks starting from the node are concentrated on a few targets, reflecting lower reach. Experimentally, Nikolaev \textit{et al.} \cite{Nikolaev2015} suggest using $t=5$ and absorption probability $a \in [0.1, 0.2]$.

\section{Maximal Clique Centrality (MCC)}

The \emph{maximal clique centrality}\index{maximal clique centrality (MCC)} (MCC) is based on the observation that essential proteins in a yeast protein-protein interaction network tend to be highly clustered \cite{Chin2014}. A \emph{maximal clique} is a fully connected subgraph that is not contained in any larger fully connected subgraph. Let \(S_i\) denote the set of maximal cliques containing node \(i\). The MCC of node \(i\) is then defined as
\begin{equation*}
    c_{\mathrm{MCC}}(i) = \sum_{C \in S_i} (|C|-1)!,
\end{equation*}
where \(|C|\) is the size of clique \(C\).  Under this definition, the MCC of an isolated node is \(1\). For a node $i$ whose neighbors are all disconnected (i.e., there is no edge between any two neighbors of node \(i\)), the MCC reduces to the degree of node $i$:
\begin{equation*}
    c_{\mathrm{MCC}}(i) = \sum_{j=1}^{N} a_{ij} = d_i.
\end{equation*}

\section{Maximum Neighborhood Component (MNC)}

The \emph{Maximum Neighborhood Component}\index{maximum neighborhood component centrality} (MNC) centrality quantifies the importance of a node based on the internal connectivity structure among its immediate neighbors~\cite{Lin2008}. 
For a given node \( i \), let its neighborhood be denoted by \( \mathcal{N}(i) \), that is, the set of nodes directly connected to \( i \). 
The neighborhood \( \mathcal{N}(i) \) induces a subgraph \( G_{\mathcal{N}(i)} \) composed solely of these neighboring nodes and the edges between them. The MNC centrality \( c_{\mathrm{MNC}}(i) \) of node \( i \) is then defined as the size of the largest connected component within this induced subgraph, formally expressed as:
\[
c_{\mathrm{MNC}}(i)
= \max_{\substack{C \subseteq \mathcal{N}(i)\\ C\text{ is connected in } G_{\mathcal{N}(i)} }}|C|.
\]

Intuitively, the MNC centrality measures how well the neighbors of a node are connected to each other. 
Nodes whose neighbors form a densely connected group (i.e., a large maximum connected component) receive higher MNC scores, reflecting their structural significance within the network.

\section{Mediative effects centrality (MEC)}

\emph{Mediative effects centrality}\index{effects centrality!mediative (MEC)} (MEC) quantifies the extent to which a node channels or transmits the influence of other nodes, reflecting its mediating role within the network \cite{Friedkin1991}. The MEC score of node \(i\) is defined as
\begin{equation*}
    c_{MEC}(i) = \frac{\sum_{k \neq i} \tilde{t}_{(k)i}}{N-1},
\end{equation*}
where \(\tilde{t}_{(k)i}\) quantifies the contribution of node \(i\) in transmitting the impersonal effects of node \(k\):
\begin{equation*}
    \tilde{t}_{(k)i} = \frac{\sum_{j \neq i \neq k} t_{(k)ji}}{(N-2) t_{(k)ii}}.
\end{equation*}
Here, \(t_{(k)ji}\) is the \((i,j)\) entry of \(T_{(k)} = (I - W_{(k)})^{-1}\), where \(W_{(k)}\) is the \((N-1) \times (N-1)\) matrix obtained by removing the \(k\)th row and column from the \(N \times N\) row-normalized adjacency matrix \(W\).

\section{Message-passing approach}

The \emph{message-passing approach}\index{message-passing approach} identifies influential spreaders in complex networks under the susceptible-infected-recovered (SIR) model, specifically when diffusion starts from a single seed node \cite{Min2018}. It assumes the network is \textit{locally tree-like}, so that the infection events along different paths are approximately independent.  

Let \(H_{ij}\) denote the probability that node \(j\), reached by following an edge from node \(i\), \textit{does not} trigger a large-scale epidemic, given the transmissibility \(T\). These probabilities satisfy the recursive relation
\begin{equation*}
H_{ij} = 1 - T + T \prod_{k \in \mathcal{N}(j) \setminus \{i\}} H_{jk},
\end{equation*}
where \(\mathcal{N}(j)\) is the set of neighbors of node \(j\). This equation can be solved iteratively for all links in the network.  

The probability that a seed node \(i\) triggers a global epidemic is then
\begin{equation*}
P_i = 1 - \prod_{j \in \mathcal{N}(i)} H_{ij},
\end{equation*}
which represents the likelihood that infection spreads from node \(i\) to a significant fraction of the network. Under the tree-like approximation, the expected fraction of nodes infected when an epidemic occurs starting from node \(i\) can be estimated as
\begin{equation*}
S_i = \frac{1}{N} \left( 1 + \sum_{\substack{j \neq i}} P_j \right),
\end{equation*}
where the sum approximates the contribution of all other nodes.  

Finally, an influence score for node \(i\) can be defined as
\begin{equation*}
\rho_i = P_i \, S_i,
\end{equation*}
which combines the probability that an epidemic occurs with the expected fraction of nodes affected. This measure provides a ranking of nodes according to their spreading potential within the network, under the locally tree-like assumption.

\section{Meta-centrality}

\emph{Meta-centrality}\index{meta-centrality} is a hybrid centrality measure that integrates multiple centrality rankings using the Borda count from social choice theory \cite{Madotto2016}. Given $n$ rankings of nodes derived from different centrality measures, the method aims to select the most informative rankings and aggregate them into a single meta-centrality score. For example, Madotto \& Liu \cite{Madotto2016} consider $n=8$ measures for weighted networks: degree, strength, closeness, eigenvector, PageRank, $k$-shell, weighted $k$-shell, and expected force.

The Borda count aggregation-based meta-centrality proceeds in three steps:

\begin{enumerate}
    \item \textit{Slicing}: identify subsets of rankings to be used in the aggregation. Form the set $X$ as
    \[
        X = H \cup L \cup HL,
    \]
    where $H = \{h_i\}_{i=1,...,n}$, $L = \{l_i\}_{i=1,...,n}$, and $HL = \{h_i \cup l_i\}_{i=1,...,n}$ are defined based on the Spearman correlation $m_{ij}$ between rankings $i$ and $j$:
    \begin{align*}
        h_i &= \{i\} \cup \{j \mid m_{ij} \geq t_b, i \leq j\},\\
        l_i &= \{i\} \cup \{j \mid m_{ij} \leq t_s, i \leq j\},
    \end{align*}
    with thresholds $t_b = 0.8$ and $t_s = 0.3$.
    
    \item \textit{Selection}: choose the most informative subsets of rankings by selecting two sets $T_1, T_2 \subset X$ with the highest entropy. The entropy of a set $x_i \in X$ is given by
    \[
        E(x_i) = \frac{1}{|x_i|}\sum_{j \in x_i} \frac{m_{ij}}{\sum_j m_{ij}} \log \frac{m_{ij}}{\sum_j m_{ij}}.
    \]
    
    \item \textit{Aggregation}: compute the Borda count $B(i)$ of node $i$ using the selected subset $T_1$ (or $T_2$):
    \[
        B(i) = \sum_{j \in T_1} (N - \tau_j(i)),
    \]
    where $\tau_j(i)$ is the position of node $i$ in ranking $j$.
\end{enumerate}

\section{Mixed core, degree and entropy (MCDE) method}

The \emph{mixed core, degree, and entropy}\index{mixed!core, degree and entropy (MCDE)} (MCDE) method is a hybrid centrality measure that combines $k$-shell and degree centralities with a weighted entropy measure \cite{Sheikhahmadi2017}. The entropy of node \(i\) is defined as
\begin{equation*}
    H(i) = \sum_{k=0}^{k_{max}} p_k(i) \log_2 p_k(i),
\end{equation*}
where \(k_{\max}\) denotes the maximum $k$-shell index in the network and \(p_k(i)\) is the fraction of node \(i\)'s neighbors in the $k$th core,
\begin{equation*}
    p_k(i) = \frac{|\{ j \in \mathcal{N}(i) : k_s(j) = k \}|}{d_i}.
\end{equation*}

The MCDE centrality of node \(i\) is then given by
\begin{equation*}
    c_{MCDE}(i) = \alpha k_s(i) + \beta d_i + \gamma H(i),
\end{equation*}
where \(d_i\) is the degree, \(k_s(i)\) is the $k$-shell score, and \(\alpha, \beta, \gamma\) are weights controlling the contribution of each component. Sheikhahmadi and Nematbakhsh \cite{Sheikhahmadi2017} suggest \(\alpha = \beta = \gamma = 1\).

Three variations of MCDE have also been proposed:
\begin{itemize}
    \item \emph{Mixed Core, Degree, and Weighted Entropy (MCDWE)}\index{mixed!core, degree and weighted entropy (MCDWE)}: computes a weighted entropy
    \begin{equation*}
        H(i) = \sum_{k=0}^{k_{max}} \frac{p_k(i) \log_2 p_k(i)}{k_{max} - |\{r : \exists j \in \mathcal{N}(i), k_s(j) = r\}| + 1}.
    \end{equation*}
    \item \emph{Mixed Core, Semi-local Degree, and Entropy (MCSDE)}\index{mixed!core, semi-local degree and entropy (MCSDE)}: replaces degree \(k_i\) with LocalRank (semi-local) centrality.
    \item \emph{Mixed Core, Semi-local Degree, and Weighted Entropy (MCSDWE)}\index{mixed!core, semi-local degree and weighted entropy (MCSDWE)}: combines the MCDWE and MCSDE approaches.
\end{itemize}

\section{Mixed core, degree and weighted entropy (MCDWE) method}

The \emph{mixed core, degree and weighted entropy}\index{mixed!core, degree and weighted entropy (MCDWE)} (MCDWE) method extends MCDE by computing a weighted entropy for each node \cite{Sheikhahmadi2017}. For node $i$, the weighted entropy is
\[
H(i) = \sum_{k=0}^{k_{\max}} \frac{p_k(i) \log_2 p_k(i)}{k_{\max} - |\{r : \exists j \in \mathcal{N}(i), k_s(j) = r\}| + 1},
\]
where $p_k(i)$ is the fraction of neighbors of node $i$ in the $k$th shell. The centrality of node $i$ is then
\[
c_{MCDWE}(i) = \alpha k_s(i) + \beta d_i + \gamma H(i),
\]
where $\alpha, \beta, \gamma$ are weights controlling the relative contributions of $k$-shell, degree, and weighted entropy. Sheikhahmadi and Nematbakhsh \cite{Sheikhahmadi2017} suggest \(\alpha = \beta = \gamma = 1\). MCDWE emphasizes nodes whose neighbors are spread across shells, refining the ranking of influential nodes.

\section{Mixed core, semi-local degree and entropy (MCSDE) method}

The \emph{mixed core, semi-local degree, and entropy} (MCSDE) method\index{mixed!core, semi-local degree and entropy (MCSDE)} \cite{Sheikhahmadi2017} is a variant of MCDE that replaces the standard degree $d_i$ with the LocalRank (semi-local) centrality. For node $i$, the centrality is defined as
\[
c_{MCSDE}(i) = \alpha k_s(i) + \beta \ c_{\mathrm{LR}}(i) + \gamma H(i),
\]
where $k_s(i)$ is the $k$-shell index, $c_{\mathrm{LR}}(i)$ is the LocalRank centrality of node $i$, which is defined in \cite{Chen2012}, $\alpha, \beta, \gamma$ are weights controlling the contributions of each component (Sheikhahmadi and Nematbakhsh \cite{Sheikhahmadi2017} suggest \(\alpha = \beta = \gamma = 1\)), and $H(i)$ is the entropy of node $i$:
\[
H(i) = \sum_{k=0}^{k_{\max}} p_k(i) \log_2 p_k(i),
\]
with $k_{\max}$ being the maximum $k$-shell index in the network and $p_k(i)$ denoting the fraction of neighbors of node $i$ in the $k$th shell. MCSDE integrates semi-local neighborhood information with hierarchical node structure, improving the identification of nodes that are critical for spreading processes and network connectivity.

\section{Mixed core, semi-local degree and weighted entropy (MCSDWE) method}

The \emph{mixed core, semi-local degree, and weighted entropy} (MCSDWE) method\index{mixed!core, semi-local degree and weighted entropy (MCSDWE)} combines the MCDWE and MCSDE approaches to produce a fully weighted, semi-local centrality measure \cite{Sheikhahmadi2017}. The centrality of node $i$ is defined as
\[
c_{MCSDWE}(i) = \alpha k_s(i) + \beta \, c_{\mathrm{LR}}(i) + \gamma H(i),
\]
where $k_s(i)$ is the $k$-shell index, $c_{\mathrm{LR}}(i)$ is the LocalRank centrality of node $i$ \cite{Chen2012}, and $\alpha, \beta, \gamma$ are weights controlling the contributions of each component (Sheikhahmadi and Nematbakhsh \cite{Sheikhahmadi2017} suggest $\alpha = \beta = \gamma = 1$).  The weighted entropy $H(i)$ is given by
\[
H(i) = \sum_{k=0}^{k_{\max}} \frac{p_k(i) \log_2 p_k(i)}{k_{\max} - |\{r : \exists j \in \mathcal{N}(i), k_s(j) = r\}| + 1},
\]
where $p_k(i)$ is the fraction of neighbors of node $i$ in the $k$th shell and $k_{\max}$ denotes the maximum $k$-shell index in the network.  

MCSDWE integrates local, semi-local, and hierarchical neighborhood information, providing a comprehensive and nuanced assessment of node influence in complex networks.

\section{Mixed degree decomposition (MDD)}

\emph{Mixed degree decomposition (MDD)}\index{mixed!degree decomposition (MDD)}, also referred to as the $m$-shell method, is an extension of the classical $k$-shell method that introduces a tunable parameter \(\lambda\) to better rank the spreading ability of nodes in complex networks \cite{Zeng2013}. MDD decomposes the network based on both the residual and exhausted degrees of nodes.  

The \emph{residual degree}\index{degree!residual} \(d_i^{(r)}\) of node \(i\) is defined as the number of links connecting it to nodes that remain in the network, while the \emph{exhausted degree}\index{degree!exhausted} \(d_i^{(e)}\) counts the links connecting node \(i\) to nodes that have already been removed. The \emph{mixed degree}\index{degree!mixed} is then given by
\begin{equation*}
    d_i^{(m)} = d_i^{(r)} + \lambda d_i^{(e)},
\end{equation*}
where \(\lambda \in [0,1]\) controls the relative contribution of exhausted links.  

The MDD procedure performs a $k$-shell decomposition using \(d_i^{(m)}\) to iteratively remove nodes. In the limiting cases, when \(\lambda = 0\), the MDD score reduces to the standard $k$-shell centrality, and when \(\lambda = 1\), it is equivalent to the degree centrality. Zeng and Zhang \cite{Zeng2013} suggest using \(\lambda = 0.7\) for optimal performance.

\section{Mixed gravitational centrality}

The \emph{mixed gravitational centrality}\index{gravitational centrality!mixed (MGC)} (MGC), also referred to as the \textit{improved gravitational centrality}, represents an enhanced formulation of the classical gravitational centrality measure. In this variant, the mass of the focal node is determined by its $k$-shell index, while the masses of its neighboring nodes are characterized by their degrees \cite{Wang2018}. The centrality score \( c_{\mathrm{MGC}}(i) \) for node \( i \) is expressed as
\begin{equation*}
    c_{\mathrm{MGC}}(i) = \sum_{j \in \mathcal{N}(i)} \frac{k_s(i)\,d_j}{d_{ij}^2},
\end{equation*}
where \( \mathcal{N}(i) \) denotes the set of neighbors of node \( i \), \( d_{ij} \) is the shortest path distance between nodes \( i \) and \( j \), \( k_s(i) \) is the $k$-shell index of node \( i \), and \( d_j \) is the degree of node \( j \). 

The mixed gravitational centrality integrates the hierarchical structure of the network, captured by the $k$-shell decomposition, with local connectivity information, represented by node degrees, thereby providing a more comprehensive quantification of node influence within complex networks.

\section{Modified Expected Force (ExF$^M$)}

Lawyer \cite{Lawyer2015} proposed a \emph{modified version of the expected force}\index{modified expected force (ExF$^M$)}, denoted ExF$^M$\index{expected force (ExF)!modified (ExF$^M$)}, which incorporates the degree \(d_i\) of the node \(i\) as
\[
c_{\text{ExF}^M}(i) = \log(\alpha d_i) \, c_{\text{ExF}}(i),
\]
where \(\alpha > 1\) is a scaling parameter (e.g., \(\alpha = 2\)) and \(c_{\text{ExF}}(i)\) is the original expected force of node \(i\), defined by
\[
c_{\text{ExF}}(i) = - \sum_{j=1}^J \frac{D_j}{\sum_{k=1}^J D_k} \log \frac{D_j}{\sum_{k=1}^J D_k}.
\]
Here, \(D_j\) denotes the degree of cluster \(j\), i.e., the total number of neighbors of nodes in the cluster, and \(j = 1, \dots, J\) enumerate all possible clusters of infected nodes after \(x=2\) transmission events, assuming no recovery. ExF$^M$ thus adjusts the original expected force by giving additional weight to the seed node’s degree, capturing both its local connectivity and the potential spreading capacity of its early infections.

\section{Modified local centrality (MLC)}

\emph{Modified local centrality}\index{modified local centrality (MLC)} (MLC) is a semi-local centrality measure that quantifies the distal influence of a node by considering its neighbors and the neighbors of its neighbors, while adjusting for direct connections \cite{Ma2017}.  The MLC score of node $i$ is defined as
\[
c_{MLC}(i) = \sum_{j \in \mathcal{N}(i)} \sum_{k \in \mathcal{N}(j)} n(k) - 2 \sum_{j \in \mathcal{N}(i)} |\mathcal{N}(j)|,
\]
where \(\mathcal{N}(i)\) is the set of neighbors of node \(i\), and 
\[
n(k) = |\mathcal{N}^{(\leq 2)}(k)| 
\] 
denotes the number of nearest and next-nearest neighbors of node \(k\).  

Intuitively, MLC captures the broader influence of a node beyond its immediate neighbors, while removing the double-counted contributions of direct connections. Nodes with high MLC values are those that not only connect to many neighbors but are also positioned in dense local neighborhoods that can facilitate spreading processes over multiple steps in the network.

\section{ModuLand centrality}

\emph{ModuLand centrality}\index{ModuLand centrality} quantifies a node's importance based on its role within the influence-function-based community landscape constructed by the NodeLand algorithm \cite{Kovács2010}. For each node \(k\), NodeLand computes an influence function \(f_k\) by iteratively building the set \(A_k\) of nodes strongly influenced by \(k\). Starting from \(A_k = \{k\}\), neighboring nodes are added one at a time only if their inclusion increases the density of the subgraph induced by \(A_k\). This process continues until no further improvement in density is possible.

The ModuLand centrality \(c_{\mathrm{ModuLand}}(i)\) of node \(i\) is then defined as
\[
c_{\mathrm{ModuLand}}(i) = \sum_{j=1}^{N} c(i,j) = \sum_{j=1}^{N} \sum_{k=1}^{N} f_k(i,j),
\]
where
\[
f_k(i,j) =
\begin{cases}
w_{ij}, & \text{if } (i,j) \in A_k,\\
0, & \text{otherwise,}
\end{cases}
\]
and \(w_{ij}\) is the weight of edge \((i,j)\). This formulation captures how strongly node \(i\) participates in regions of high influence across the network, reflecting its centrality in the overlapping community landscape.

\section{Modularity centrality}

\emph{Modularity centrality}\index{modularity!centrality} is a spectral measure of node importance that quantifies a node’s contribution to the modular (community) structure of a network \cite{GWang2008}. It is based on the modularity matrix $M$, whose elements are defined as
\[
M_{ij} = A_{ij} - \frac{d_i d_j}{2L},
\]
where $A_{ij}$ is the adjacency matrix, $d_i$ and $d_j$ are the degrees of nodes $i$ and $j$, respectively, and $L$ is the total number of edges in the network.

The centrality of a node is determined by the corresponding component in the leading eigenvector of $M$, i.e., the eigenvector associated with the eigenvalue of largest magnitude. Nodes with larger components in this eigenvector play a stronger role in reinforcing the network’s modular structure, indicating higher importance within their communities. Hence, modularity centrality provides a measure of how central or influential a node is in maintaining the community organization of the network.

\section{Modularity density centrality}

\emph{Modularity density centrality}\index{modularity!density centrality} is a variant of modularity centrality based on the spectral optimization of modularity density \cite{Fu2010}. Unlike modularity centrality, which relies on the modularity matrix $M$, modularity density centrality is derived from the kernel matrix $K$ defined as
\[
K = \sigma I + 2A - D,
\]
where $I$ is the identity matrix, $D$ is the diagonal degree matrix, and $\sigma$ is a real number chosen sufficiently large to make $K$ positive definite. The centrality of a node is determined by the corresponding component in the leading eigenvector of $K$, i.e., the eigenvector associated with the eigenvalue of largest magnitude.

Standard modularity suffers from a resolution limit, which can prevent the detection of smaller communities and may overemphasize node degree rather than the structural role of nodes. Modularity density centrality addresses this limitation by optimizing a spectral relaxation of modularity density using $K$. The centrality score of a node is given by its component in the dominant eigenvector of $K$, reflecting how strongly the node contributes to the network's community structure under the modularity density criterion.

\section{Modularity vitality}

\emph{Modularity vitality}\index{modularity!vitality} is a community-based centrality measure that quantifies the contribution of a node to the overall modularity of a network \cite{Magelinski2021}.  
Assume that the network \(G\) has a community structure consisting of \(K > 1\) communities (Magelinski \textit{et al.} \cite{Magelinski2021} employ the Leiden community detection algorithm).  
Given a community partition \(C\), the modularity vitality \(c_{MV}(i)\) of node \(i\) is defined as
\begin{equation*}
    c_{MV}(i) = Q(G, C) - Q(G_i, C \setminus \{i\}),
\end{equation*}
where \(Q(G, C)\) denotes the modularity of the network \(G\) under partition \(C\), and \(G_i\) is the graph obtained by removing node \(i\) from \(G\).  Importantly, the community structure \(C\) is \emph{not recomputed} after the removal of node \(i\); modularity is recalculated using the same partition excluding \(i\).

A positive modularity vitality \(c_{MV}(i)\) indicates that node \(i\) enhances the modular structure of the network, functioning as a community hub.  
Conversely, a negative \(c_{MV}(i)\) suggests that the node weakens modularity, typically acting as a bridge between communities.  
Thus, modularity vitality not only measures the importance of a node but also distinguishes between hub-like and bridge-like roles within the network.

\section{Multi-attribute ranking method based on information entropy (MABIE)}
\emph{Multi-Attribute Ranking Method Based on Information Entropy}\index{multi-attribute ranking method based on information entropy (MABIE)} (MABIE) is a hybrid centrality measure that integrates both local and global information of a network using four classical centrality metrics: degree centrality (DC), harmonic centrality (HC), betweenness centrality (BC), and correlation centrality (CoC)~\cite{Wenli2013}. 

MABIE constructs an $N \times 4$ multi-attribute node-importance decision matrix
\[
R = 
\begin{bmatrix} 
r_{11} & r_{12} & r_{13} & r_{14} \\
r_{21} & r_{22} & r_{23} & r_{24} \\ 
\vdots & \vdots & \vdots & \vdots \\ 
r_{N1} & r_{N2} & r_{N3} & r_{N4} \\ 
\end{bmatrix},
\]
where $r_{ij} = \frac{c_j(i)}{\sum_{k=1}^N c_j(k)}$ and $c_j(i)$ denotes the $j$-th centrality value of node $i$, with $j \in \{\mathrm{DC}, \mathrm{HC}, \mathrm{BC}, \mathrm{CoC}\}$. Thus, matrix $R$ contains the normalized centrality values of all nodes.

The information entropy vector $E = (E_1, E_2, E_3, E_4)$ quantifies the information content of each centrality metric and is defined as
\[
E_j = -\frac{1}{\ln N} \sum_{i=1}^N r_{ij} \ln r_{ij}.
\]
Specifically, it measures the degree of differentiation among nodes with respect to each metric: higher entropy values correspond to a more uniform (and thus less informative) distribution of centrality values, whereas lower entropy values indicate greater variability and stronger discriminative power.

The MABIE centrality of node $i$ is then defined as a weighted linear combination of the normalized centrality measures:
\[
c_{\mathrm{MABIE}}(i) = \sum_{j=1}^4 w_j r_{ij},
\]
where the weight $w_j$ represents the relative importance of the $j$-th centrality measure and is computed as
\[
w_j = \frac{1 - E_j}{\sum_{k=1}^{4} (1 - E_k)}.
\]

While Wenli \textit{\textit{et al.}}~\cite{Wenli2013} originally considered four centrality measures, the MABIE framework can be extended to any number $K$ of centrality metrics. Building on this concept, Zhang \textit{et al.}~\cite{Zhang2022} proposed the \textit{Multiple Local Attributes Weighted Centrality}\index{multiple local attributes weighted centrality} (LWC). LWC extends the MABIE framework by incorporating local structural information through four metrics: degree, two-hop degree, clustering coefficient, and two-hop clustering coefficient (the sum of the clustering coefficients of a node's neighbors).

\section{Multi-characteristics gravity model (MCGM)}

The \emph{multi-characteristics gravity model}\index{gravity model!multi-characteristics (MCGM)} (MCGM) is a variant of the local gravity model designed to identify influential spreaders in complex networks. In this model, a node’s mass is determined by a combination of three structural features: degree, \(k\)-shell index and eigenvector centrality \cite{Li2022}. 

Let \(\mathcal{N}^{(\leq l)}(i)\) denote the set of nodes whose shortest-path distance from $i$ is less than or equal to $l$. The centrality \(c_{\text{MCGM}}(i)\) of node \(i\) is then defined as
\begin{equation*}
   c_{\text{MCGM}}(i) = 
   \sum_{j \in \mathcal{N}^{(\leq l)}(i)} 
   \frac{
      \left( 
         \frac{d_i}{d_{\max}} 
         + \frac{\alpha\, k_s(i)}{ks_{\max}} 
         + \frac{ev(i)}{ev_{\max}} 
      \right)
      \left( 
         \frac{d_j}{d_{\max}} 
         + \frac{\alpha\, k_s(j)}{ks_{\max}} 
         + \frac{ev(j)}{ev_{\max}} 
      \right)
   }{d_{ij}^2},
\end{equation*}
where \(d_{ij}\) is the shortest-path distance between nodes \(i\) and \(j\); \(d_i\), \(k_s(i)\), and \(ev(i)\) denote the degree, \(k\)-shell index and eigenvector centrality of node \(i\), respectively. The terms \(d_{\max}\), \(ks_{\max}\) and \(ev_{\max}\) represent the corresponding maximum values across all nodes in the network.

The coefficient \(\alpha\) adjusts the relative influence of the \(k\)-shell index and is computed as
\begin{equation*}
   \alpha = 
   \frac{
      \max \left( 
         \frac{d_{\text{mid}}}{d_{\max}}, 
         \frac{ev_{\text{mid}}}{ev_{\max}} 
      \right)
   }{
      \frac{ks_{\text{mid}}}{ks_{\max}}
   },
\end{equation*}
where \(d_{\text{mid}}\), \(ks_{\text{mid}}\), and \(ev_{\text{mid}}\) denote the median values of the degree, \(k\)-shell index and eigenvector centrality, respectively.  Li and Huang \cite{Li2022} consider \(l = 2\) as the truncated radius.

\section{Multi-criteria influence maximization (MCIM) method}

\emph{Multi-criteria influence maximization}\index{multi-criteria influence maximization (MCIM)} (MCIM) method is an iterative hybrid method that selects $k$ influential nodes using the Technique for Order of Preference by Similarity to Ideal Solution (TOPSIS) \cite{Zareie2018b}. In MCIM, an $N \times 4$ decision matrix $R$ is constructed as
\[
R = \begin{bmatrix}
    d_1 & IDS_1 & -DO_1 & -IDO_1 \\
    d_2 & IDS_2 & -DO_2 & -IDO_2 \\
    \vdots & \vdots & \vdots & \vdots \\
    d_N & IDS_N & -DO_N & -IDO_N \\
\end{bmatrix},
\]
where $d_i$ is the degree of node $i$, $IDS_i$ is the entropy-based ranking measure (ERM) \cite{Zareie2017}, and $DO_i$ and $IDO_i$ denote the direct overlap and indirect overlap of node $i$, respectively:
\[
DO_i = \sum_{j \in \mathcal{N}(i)} s_j, 
\quad 
IDO_i = \sum_{j \in \mathcal{N}(i)} s_j \, |\mathcal{N}(i) \cap \mathcal{N}(j)|,
\]
with $s_j = 1$ if node $j$ is in the seed set $S$ and $s_j = 0$ otherwise.

Initially, the seed set $S$ contains the nodes with the highest degree, which are then removed from the decision matrix $R$. At each iteration, the most important node $u$ is selected from $R$ using the TOPSIS method. The selected node $u$ is added to $S$ and removed from $R$, and the values of $DO$ and $IDO$ for the remaining nodes are updated accordingly. This process continues until either $|S| = k$, meaning $k$ nodes have been selected, or $S = \mathcal{N}$, meaning all nodes have been added to the seed set.

\section{Multi-local dimension (MLD) centrality}

The \emph{multi-local dimension (MLD) centrality}\index{local dimension!multi- (MLD)} is a variant of the local dimension measure for identifying influential spreaders in complex networks \cite{Wen2020b}. MLD evaluates the structural information around a node by considering concentric boxes of increasing radius. For a given node \(i\), the box radius \(l\) ranges from 1 to the maximum shortest-path distance from \(i\).

The proportion of nodes within a box of radius \(l\) is
\begin{equation*}
   \mu_i(l) = \frac{N_i(l)}{N},
\end{equation*}
where \(N_i(l)\) is the number of nodes covered by the box, and \(N\) is the total number of nodes in the network. Based on \(\mu_i(l)\), the generalized partition function \(Z_i(q,l)\) is defined as
\begin{equation*}
   Z_i(q,l) = 
   \begin{cases}
       \mu_i(l)^q, & q \notin \{0,1\},\\[1mm]
       1/\mu_i(l), & q = 0,\\[1mm]
       \mu_i(l)\, \log_2 \mu_i(l), & q = 1,
   \end{cases}
\end{equation*}
where \(q \in \mathbb{R}\) is a tunable parameter controlling the emphasis on different structural scales.

The multi-local dimension \(c_{\textsc{MLD}}(i,q)\) of node \(i\) is then defined as
\begin{equation*}
   c_{\textsc{MLD}}(i,q) = 
   \begin{cases}
       \displaystyle \lim_{l \to 0} \frac{\log_2 Z_i(q,l)}{(q-1)\, \log_2 l}, & q \neq 1,\\[1mm]
       \displaystyle \lim_{l \to 0} \frac{Z_i(q,l)}{\log_2 l}, & q = 1.
   \end{cases}
\end{equation*}

In practice, \(c_{\textsc{MLD}}(i,q)\) is estimated numerically as the slope of a linear regression: if \(q \neq 1\), the regression is of \(\frac{\log_2 Z_i(q,l)}{q-1}\) versus \(\log_2 l\); if \(q = 1\), it is of \(Z_i(q,l)\) versus \(\log_2 l\). Wen \textit{et al.}~\cite{Wen2020b} show that \textsc{MLD} reduces to the local information dimensionality (LID) \cite{Wen2020} when \(q=1\), and to the local dimension measure \cite{Pu2014} when \(q=0\).

\section{Multiple local attributes weighted centrality (LWC)}

The \emph{Multiple Local Attributes Weighted Centrality}\index{multiple local attributes weighted centrality} (LWC) is an extension of the MABIE framework~\cite{Wenli2013} that incorporates local structural information to evaluate node importance in complex networks \cite{Zhang2022}. LWC integrates four local metrics: the degree, two-hop degree, clustering coefficient and two-hop clustering coefficient (the sum of the clustering coefficients of a node's neighbors).  

For a network with $N$ nodes, LWC constructs an $N \times 4$ decision matrix
\[
R = 
\begin{bmatrix} 
r_{11} & r_{12} & r_{13} & r_{14} \\
r_{21} & r_{22} & r_{23} & r_{24} \\ 
\vdots & \vdots & \vdots & \vdots \\ 
r_{N1} & r_{N2} & r_{N3} & r_{N4} \\ 
\end{bmatrix},
\]
where $r_{ij} = \frac{c_j(i)}{\sum_{k=1}^N c_j(k)}$ and $c_j(i)$ denotes the $j$-th local attribute of node $i$.  

The information entropy of each attribute is computed as
\[
E_j = -\frac{1}{\ln N} \sum_{i=1}^N r_{ij} \ln r_{ij},
\]
which quantifies the discriminative power of the $j$-th metric. Lower entropy values indicate greater variability among nodes and stronger ability to distinguish influential nodes.  

The LWC centrality of node $i$ is then defined as a weighted sum of the normalized attributes:
\[
c_{\mathrm{LWC}}(i) = \sum_{j=1}^{4} w_j r_{ij}, \quad 
w_j = \frac{1 - E_j}{\sum_{k=1}^{4} (1 - E_k)},
\]
where $w_j$ is the weight of the $j$-th attribute, computed from its entropy.  

Nodes with high LWC values are those that simultaneously exhibit strong local connectivity and high structural influence in their immediate and extended neighborhoods, making them critical for spreading processes or network cohesion.

\section{Mutual information centrality}

\emph{Mutual information}\index{mutual information} is a local centrality measure that evaluates a node's importance based on the information content associated with its edges \cite{Liu2014}. Specifically, the mutual information \(I(i)\) of node \(i\) is defined as the sum of the mutual information between node \(i\) and its neighbors:
\[
I(i) = \sum_{j \in \mathcal{N}(i)} I(i,j) = \sum_{j \in \mathcal{N}(i)} (\ln d_i - \ln d_j) = \sum_{j \in \mathcal{N}(i)} \ln \frac{d_i}{d_j},
\]
where \(d_i\) and \(d_j\) are the degrees of nodes \(i\) and \(j\), respectively. 

Nodes with higher mutual information are considered more important, as they contain more structural information relative to their neighbors.

\section{NCVoteRank}

The \emph{NCVoteRank}\index{NCVoteRank}\index{VoteRank!NCVoteRank} centrality is a modification of VoteRank that incorporates the coreness values of neighbors into the voting process \cite{Kumar2020}. Each node \(i\) is represented by the tuple \((s_i, v_i)\), where \(s_i\) is the voting score and \(v_i\) is the voting ability, initialized as \((s_i, v_i) = (0, 1)\) for all \(i \in \mathcal{N}\). The voting procedure iteratively performs the following steps:

\begin{enumerate}
    \item \textit{Vote:} Each node votes for its neighbors using its voting ability. The voting score of node \(i\) is updated as
    \begin{equation*}
        s_i = \sum_{j=1}^{N} a_{ji} \, v_j \, \big(\theta + (1-\theta)c_{INK}(j)\big),
    \end{equation*}
    where \(c_{INK}(j)\) is the improved neighbors’ \textit{k}-core (INK) score of node \(j\), which is defined in \cite{Bae2014,Lin2014}, while \(\theta\) is a parameter. Kumar and Panda \cite{Kumar2020} suggest \(\theta = 0.5\).
    
    \item \textit{Select:} The node \(k\) with the highest voting score \(s_k\) is elected. Node \(k\) will not participate in subsequent voting turns, meaning its voting ability is set to zero (\(v_k = 0\)).
    
    \item \textit{Update:} The voting ability of 1-hop and 2-hop neighbors of node \(k\) is reduced to account for influence spread. Specifically, for each neighbor \(i \in \mathcal{N}(k)\), the updated voting ability is
    \begin{equation*}
        v_i \leftarrow \max\big(0, v_i - f\big),
    \end{equation*}
    where \(f = 1 / \langle d \rangle\) for 1-hop neighbors and \(f = 1 / (2 \langle d \rangle)\) for 2-hop neighbors, with \(\langle d \rangle\) denoting the average degree of the network.
\end{enumerate}

NCVoteRank identifies influential nodes by combining local coreness information with iterative voting, ensuring that nodes with high coreness and connectivity are prioritized while the influence of selected nodes propagates through their neighbors.

\section{Neighbor distance centrality}

The \emph{neighbor distance centrality}\index{neighbor distance centrality} is a specific case of neighborhood centrality \cite{YLiu2016}, obtained by using degree centrality as the benchmark measure, a decay parameter \(a = 0.2\), and considering neighbors up to two steps (\(n = 2\)) away \cite{YLiuTang2015}.

Formally, for a node \(i\), the neighbor distance centrality is defined as
\begin{equation}
c_{nd}(i) = d_i + \sum_{j \in \mathcal{N}^{(1)}(i)} (0.2) d_j + + \sum_{j \in \mathcal{N}^{(2)}(i)} (0.2)^2 d_j,
\end{equation}
where \(d_i\) is the degree centrality of node \(i\) and \(\mathcal{N}^{(k)}(i)\) denotes the set of \(k\)-hop neighbors.

The neighbor distance centrality captures the influence of a node by combining its own degree with the degrees of its immediate and secondary neighbors, with contributions decaying with distance. Liu \textit{\textit{et al.}} \cite{YLiu2016} reported that this configuration achieves high performance in identifying influential nodes in various network structures.

\section{Neighborhood centrality}

The \emph{neighborhood centrality}\index{neighborhood centrality} quantifies the influence of a node by aggregating its own centrality and that of its neighbors up to \( n \) steps away \cite{YLiu2016}. Let \( f \) denote a benchmark centrality measure. Then, the neighborhood centrality \( c_{nc}(i) \) of node \( i \) is defined as
\begin{equation*}
   c_{nc}(i) = f(i) + \sum_{k=1}^n \sum_{j \in \mathcal{N}^{(k)}(i)} a^k f(j),
\end{equation*}
where \( a \in [0,1] \) is a decay parameter and \( \mathcal{N}^{(k)}(i) \) is the set of \( k \)-hop neighbors of node \( i \).

Liu \textit{\textit{et al.}}~\cite{YLiu2016} considered degree or $k$-shell centrality as the benchmark \( f \) and reported the highest performance for neighborhood centrality with \( n = 2 \) and \( a > 0.2 \). When \( f \) is defined as the degree centrality with \( n = 2 \) and \( a = 0.2 \), the measure is referred to as the \emph{neighbor distance centrality}\index{neighbor distance centrality} \cite{YLiuTang2015}.

\section{Neighborhood connectivity}

The \emph{neighborhood connectivity}\index{neighborhood connectivity} (also referred to as the \emph{average neighborhood degree}) of a node \(i\), denoted as \(c_{NC}(i)\), is defined as the average degree of all its nearest neighbors \cite{Maslov2002}. Formally,
\begin{equation*}
    c_{NC}(i) = \frac{\sum_{j \in \mathcal{N}(i)} |\mathcal{N}(j)|}{|\mathcal{N}(i)|} = \frac{\sum_{j \in \mathcal{N}(i)} d_j}{d_i},
\end{equation*}
where \(\mathcal{N}(i)\) represents the set of neighbors of node \(i\) and $d_i$ is the degree of node $i$. For isolated nodes (i.e., nodes with no neighbors), the neighborhood connectivity is defined to be zero.  

For weighted networks, a corresponding generalization known as the weighted average nearest-neighbors degree was introduced by Barrat \textit{et al.} \cite{Barrat2004}.

\section{Neighborhood core diversity centrality (Cncd)}

The \emph{neighborhood core diversity centrality}\index{neighborhood core diversity centrality (Cncd)} (Cncd) is inspired by the extended neighborhood coreness (ENC) \cite{Bae2014} and incorporates information entropy to quantify path diversity \cite{Yang2021}. 

First, the path diversity \(p(i)\) of node \(i\) is defined as the ratio of its degree \(d_i\) to the sum of the degrees of its neighbors:
\[
p(i) = \frac{d_i}{\sum_{j \in \mathcal{N}(i)} d_j}.
\]

The \textsc{Cncd} centrality \(c_{\textsc{Cncd}}(i)\) of node \(i\) is then defined as
\[
c_{\textsc{Cncd}}(i) = c_{nc+}(i) + \delta \, \frac{\sum_{j \in \mathcal{N}(i)} p(j) \ln p(j)}{\ln(1/d_i)} \, \frac{\sum_{j \in \mathcal{N}(i)} c_{nc+}(j)}{c_{nc+}(i)},
\]
where \(c_{nc+}(i)\) is the extended neighborhood coreness of node \(i\) as defined in \cite{Bae2014} and \(\delta\) is a tunable parameter (e.g., \(\delta = 2\)).

The Cncd measure captures both the coreness of a node and the diversity of paths in its local neighborhood, providing a more nuanced evaluation of influence in complex networks.

\section{Neighborhood density (ND)}

The \emph{neighborhood density}\index{neighborhood density (ND)} quantifies the connectivity among a node's neighbors \cite{Takes2011}. Takes and Kosters observe that neighbors of prominent nodes tend to share more connections than those of regular nodes. For node \(i\), the neighborhood density \(c_{ND}(i)\) is defined as
\begin{equation*}
    c_{ND}(i) = 1 - \sum_{j \in \mathcal{N}(i)} \frac{|\mathcal{N}(i) \cap \mathcal{N}(j)|}{(|\mathcal{N}(j)| - 1) |\mathcal{N}(i)|},
\end{equation*}
where \(|\mathcal{N}(i) \cap \mathcal{N}(j)|\) counts the number of neighbors shared by nodes \(i\) and \(j\), and the denominator normalizes the measure so that it is independent of the degrees of \(i\) and \(j\). The neighborhood density is minimal when all neighbors of \(i\) are fully connected, and increases as fewer neighbor pairs are connected.

\section{Neighborhood structure-based centrality (NSC)}

The \emph{neighborhood structure-based centrality}\index{neighborhood structure-based centrality (NSC)} (NSC) is a hybrid measure that integrates the Lobby index (\(l\)-index) and the \(k\)-shell centrality of a node and its neighbors \cite{Yang2023}. For a node \(i \in \mathcal{N}\), the NSC score \(c_{NSC}(i)\) is defined as
\[
c_{NSC}(i) = \frac{c_{Lobby}(i)}{\langle c_{Lobby} \rangle} + \frac{k_s(i)}{\langle k_s \rangle} 
+ \sum_{j \in \mathcal{N}(i)} \left( \frac{c_{Lobby}(j)}{\langle c_{Lobby} \rangle} + \frac{k_s(j)}{\langle k_s \rangle} \right),
\]
where \(\mathcal{N}(i)\) denotes the set of neighbors of node \(i\), \(c_{Lobby}(i)\) is the Lobby index \cite{Korn2009} of node \(i\), \(k_s(i)\) is its \(k\)-shell centrality, and \(\langle c_{Lobby} \rangle\) and \(\langle k_s \rangle\) are the mean values of the Lobby index and \(k\)-shell centrality, respectively.  

Nodes with high NSC values are characterized by both strong individual influence (high Lobby index and/or \(k\)-shell index) and connections to other influential nodes, rendering them especially important for spreading processes and the structural integrity of the network.

\section{Network global structure-based centrality (NGSC)}

The \emph{network global structure-based centrality}\index{network global structure-based centrality (NGSC)} (NGSC) is a hybrid method designed to identify the most effective spreaders in a network by combining the degree and \(k\)-shell index of a node and its neighbors \cite{Namtirtha2021}. For a node \(i \in \mathcal{N}\), the NGSC score \(c_{NGSC}(i)\) is defined as
\begin{align*}
c_{NGSC}(i) 
&= \sum_{j \in \mathcal{N}(i)} \Bigl[ (w_1 k_s(i) + w_2 d_i) + (w_1 k_s(j) + w_2 d_j) \Bigr] \nonumber \\
&= d_i \,(w_1 k_s(i) + w_2 d_i) + \sum_{j \in \mathcal{N}(i)} \left( w_1 k_s(j) + w_2 d_j \right),
\end{align*}
where \(d_i\) and \(k_s(i)\) denote the degree and \(k\)-shell index of node \(i\), respectively, \(\mathcal{N}(i)\) is the set of neighbors of \(i\), and \(w_1\) and \(w_2\) are tunable parameters weighting the contributions of the \(k\)-shell and degree.  Experimentally, Namtirtha \textit{et al.}~\cite{Namtirtha2021} suggest parameter ranges \(w_1 \in [0.2, 0.4], w_2 = 0.9\) or \(w_1 = 0.9, w_2 \in [0.2, 0.4]\), depending on the network’s density and percolation threshold.

\section{New Evidential Centrality (NEC)}

\emph{New evidential centrality}\index{new evidential centrality (NEC)} (NEC) is a hybrid measure based on Dempster-Shafer evidence theory that combines a node's degree with the global network structure, as quantified by shortest path distances \cite{BianDeng2017}. 
NEC addresses limitations of the existing evidential centrality (EVC), which was originally designed for weighted networks \cite{Wei2013}. 
The concept of evidential centrality is analogous to multi-attribute decision making (MADM), in which multiple factors are combined to obtain a final ranking of nodes. 

For each node \(i\), NEC computes two basic probability assignments (BPAs): one based on degree and one based on shortest paths. 
The degree-based BPA is defined as
\begin{equation*}
   M_k(i) = \bigl(m_{ki}(h), m_{ki}(l), m_{ki}(\theta)\bigr),
\end{equation*}
where
\[
m_{ki}(h) = \frac{d_i - d_{\min}}{d_{\max} - d_{\min} + \mu}, \quad
m_{ki}(l) = \frac{d_{\max} - d_i}{d_{\max} - d_{\min} + \mu}, \quad
m_{ki}(\theta) = 1 - m_{ki}(h) - m_{ki}(l),
\]
with \(d_i\) the degree of node \(i\), \(d_{\min}\) and \(d_{\max}\) the minimum and maximum degrees in the network, \(\mu \in (0,1)\) a small constant, and \(\theta = \{h,l\}\) the frame of discernment. Here, \(m_{ki}(h)\) and \(m_{ki}(l)\) represent the degrees of belief that node \(i\) has high or low influence based on its degree, while \(m_{ki}(\theta)\) captures the remaining uncertainty.

The shortest path-based BPA is defined as
\begin{equation*}
   M_d(i) = \frac{1}{N} \sum_{j=1}^{N} 
   \bigl(m^j_{di}(h), m^j_{di}(l), m^j_{di}(\theta)\bigr),
\end{equation*}
where the mass functions for each target node \(j\) are
\begin{align*}
   m^j_{di}(h) &= \frac{d_{ij} - \min_k(d_{ik})}
                        {\max_k(d_{ik}) - \min_k(d_{ik}) + \epsilon},\\
   m^j_{di}(l) &= \frac{\max_k(d_{ik}) - d_{ij}}
                        {\max_k(d_{ik}) - \min_k(d_{ik}) + \epsilon},\\
   m^j_{di}(\theta) &= 1 - m^j_{di}(h) - m^j_{di}(l),
\end{align*}
with \(d_{ij}\) denoting the shortest path distance from node \(i\) to node \(j\), 
\(\epsilon \in (0,1)\), and \(k\) ranging over all nodes in the network. 
Wei \textit{et al.}\ \cite{Wei2013} use \(\mu = \epsilon = 0.5\).

The combined influence of node \(i\) is obtained by merging the degree- and shortest path-based BPAs using a modified Dempster's rule of combination \(\oplus\):
\begin{equation*}
    M(i) = M_k(i) \oplus M_d(i) = \bigl(m_i(h), m_i(l), m_i(\theta)\bigr),
\end{equation*}
where \(m_i(h)\), \(m_i(l)\), and \(m_i(\theta)\) denote the resulting masses assigned to hypotheses \(h\), \(l\), and \(\theta\), respectively. This fusion of \(M_k(i)\) and \(M_d(i)\) produces a single BPA for node \(i\), integrating information from both local connectivity (degree) and global position (shortest paths). 
Finally, the NEC centrality of node \(i\) is defined as
\begin{equation*}
    c_{\mathrm{NEC}}(i) = m_i(h) - m_i(l).
\end{equation*}

An extension of NEC, called \emph{Multi-Evidence Centrality}\index{multi-evidence centrality (MeC)} (MeC), was proposed by Mo and Deng in \cite{Mo2019}. 
MeC integrates four centrality measures: degree, betweenness, harmonic, and correlation within the evidential framework. 
Each measure contributes a basic probability assignment (BPA) reflecting different aspects of node importance, and the BPAs are fused using Dempster's rule of combination to compute a single, comprehensive centrality score for each node.

\section{Nieminen's closeness centrality}

\emph{Nieminen's closeness centrality}\index{closeness centrality!Nieminen's}, originally designed for weakly connected directed graphs, measures a node's centrality by combining the total distance from node \(i\) to all other reachable nodes with the ability to reach a large number of nodes \cite{Nieminen1973}. Let \(RP(i)\) denote the set of nodes reachable from \(i\) in the network \(G\). By definition, \(i \in RP(i)\). Then, the Nieminen's closeness centrality of node \(i\) is defined as

\begin{equation*}
    c_{Nieminen}(i) =
    \begin{cases}
      \sum_{j \in RP(i)} \left(|RP(i)| - d_{ij} \right), & \text{if } |RP(i)| \geq 2,\\
      0, & \text{otherwise}.
    \end{cases}
\end{equation*}
where \(d_{ij}\) denotes the shortest-path distance between nodes \(i\) and \(j\).

For unweighted and strongly connected networks, the Nieminen's closeness centrality can be expressed as
\[
c_{Nieminen}(i) = N^2 - \sum_{j=1}^N d_{ij},
\]
which is directly related to the sum of shortest-path distances from node \(i\) to all other nodes. In this case, the ranking of nodes by Nieminen's closeness is identical to the ranking obtained from the closeness centrality.

\section{Node and neighbor layer information (NINL) centrality}

The \emph{node and neighbor layer information} (NINL)\index{node and neighbor layer information (NINL)} centrality identifies influential nodes by combining degree information of a node and its neighbors up to \(r\) hops \cite{Zhu2021}. 

First, a radius \(r\) is defined based on the average path length of the network, capturing the influence of the surrounding environment. The 0-order \textsc{NINL} score of node \(i\) is defined as
\[
\textsc{NINL}_0(i) = d_i + \sum_{j \in \mathcal{N}^{(\leq r)}(i)} d_j,
\]
where \(d_i\) is the degree of node \(i\) and \(\mathcal{N}^{(\leq r)}(i)\) denotes all neighbors of $i$ within \(r\) hops. 

To incorporate higher-order neighbor influence, the \(p\)-order \textsc{NINL} score is recursively defined as
\[
\textsc{NINL}_p(i) = \sum_{j \in \mathcal{N}(i)} \textsc{NINL}_{p-1}(j).
\]

Zhu and Wang \cite{Zhu2021} experimentally set \(p = 3\). This iterative aggregation captures both the local and slightly broader network environment around each node, providing a more comprehensive measure of influence. We note that as the order \(p\) increases, the \textsc{NINL} scores converge to the eigenvector centrality, since higher-order iterations progressively incorporate the influence of more distant neighbors throughout the network.

\section{NL centrality}

\emph{NL centrality}\index{NL centrality} is a semi-local measure that extends the DIL centrality \cite{JLiu2016} by accounting for the contributions of second-degree neighbors and the structural importance of edges \cite{Shao2019}. The NL centrality of node \(i\) is defined as
\begin{equation*}
    c_{NL}(i) = \sum_{j \in \mathcal{N}(i)} \left[ \phi(j) + 
    \left( \frac{(d_i - \Delta_{ij} - 1)(d_j - \Delta_{ij} - 1)}{\Delta_{ij}/2 + 1} \right)
    \left( \frac{d_i - 1}{d_i + d_j - 2} \right) \right],
\end{equation*}
where \(\mathcal{N}(i)\) is the set of neighbors of node \(i\), \(d_i\) is the degree of node \(i\), \(\Delta_{ij}\) denotes the number of triangles containing the edge \((i,j)\), and \(\phi(j)=|\mathcal{N}^{(\leq 2)}(j)|\) counts the number of nodes within two steps of node \(j\). The NL centrality considers both edge-level clustering and the connectivity of second-degree neighbors, capturing information beyond immediate neighbors.

\section{Node contraction (IMC) centrality}

The \emph{IMC method}\index{node contraction (IMC) centrality} is a centrality measure based on node contraction \cite{Tan2006}. In node contraction, a node and its neighboring nodes are merged into a single new node. If a node is central, contracting it will result in a more compact network structure. The IMC centrality of node \(i\) is defined as
\begin{equation*}
    c_{\mathrm{IMC}}(i) = 1 - \frac{\partial(G)}{\partial(G_{i})},
\end{equation*}
where \(\partial(G) = \frac{N-1}{\sum_{i \neq j} d_{ij}}\) is the agglomeration degree of the graph \(G\), \(d_{ij}\) is the shortest-path distance between nodes \(i\) and \(j\) and \(G_i\) denotes the graph obtained by removing node \(i\). A graph has a high agglomeration degree if its nodes are well connected, such that the average distance between nodes is small. Thus, nodes whose removal significantly reduces \(\partial(G)\) are considered more central according to the IMC measure.

\section{Node importance contribution correlation matrix (NICCM) method}

The \emph{node importance contribution correlation matrix}\index{node importance!contribution correlation matrix (NICCM)} (NICCM) method extends node centrality analysis by considering that a node contributes unevenly to the importance of both adjacent and non-adjacent nodes within a limited radius \cite{Hu2015}. This measure incorporates the influence of a node on others based on their shortest-path distance and the relative change in their centrality upon node removal.

The NICCM centrality of node $i$ is defined as
\[
c_{\mathrm{NICCM}}(i) = c_{h}(i) \cdot \left( \sum_{j \in \mathcal{N}^{(\leq r)}(i)} \frac{c_{h}(j)\,\delta_{ij}}{d_{ij}} \right),
\]
where $c_{h}(i)$ is the harmonic centrality of node $i$, $d_{ij}$ is the shortest-path distance between nodes $i$ and $j$, and $ \mathcal{N}^{(\leq r)}(i)$ denotes the set of nodes within radius $r$ from node $i$. The contribution probability $\delta_{ij}$ quantifies the extent to which node $i$ affects node $j$ and is given by
\[
\delta_{ij} = \frac{\Delta I_j}{\sum_{k \in  \mathcal{N}^{(\leq r)}(i)} \Delta I_k},
\]
where $\Delta I_j$ represents the change in the harmonic centrality of node $j$ after the removal of node $i$ from the network $G$. Following Hu \textit{et al.}~\cite{Hu2015}, the radius parameter is typically set to $r = 2$.

Nodes with high NICCM values exert strong influence on the importance of both nearby and moderately distant nodes, reflecting their broader structural impact on network connectivity.

\section{Node importance contribution matrix (NICM) method}

The \emph{node importance contribution matrix}\index{node importance!contribution matrix (NICM)} (NICM) method evaluates node influence by combining a node’s betweenness centrality with the contributions of its neighbors \cite{Zhao2009}. 

For a node \(i\), the \textsc{NICM} centrality \(c_{\textsc{NICM}}(i)\) is defined as
\[
c_{\textsc{NICM}}(i) = c_b(i) + \sum_{j \in \mathcal{N}(i)} \frac{c_b(j)}{d_j},
\]
where \(c_b(i)\) is the betweenness centrality of node \(i\), \(d_j\) is the degree of neighbor node \(j\), and \(\mathcal{N}(i)\) denotes the set of neighbors of node \(i\).  

The NICM measure captures both the node’s own centrality and the distributed influence of its neighbors, reflecting the combined effect of local connectivity and network position.

\section{Node importance evaluation matrix (NIEM) method}

The \emph{node importance evaluation matrix}\index{node importance!evaluation matrix (NIEM)} (NIEM) method assesses the relative importance of nodes in a network by integrating both degree and harmonic centrality measures \cite{Hu2015}. This approach captures not only a node's local connectivity but also its accessibility to other nodes within the network.

Formally, the NIEM centrality of node $i$, denoted as $c_{\mathrm{NIEM}}(i)$, is defined as
\[
c_{\mathrm{NIEM}}(i) = c_{h}(i) \cdot \left( \sum_{j \in \mathcal{N}(i)} \frac{c_{h}(j) \, d_j}{\langle d \rangle} \right),
\]
where $c_{h}(i)$ is the harmonic centrality of node $i$, $d_i$ denotes its degree, $\mathcal{N}(i)$ represents the set of its neighboring nodes, and $\langle d \rangle$ is the average degree of the network. The first term $c_h(i)$ reflects the global influence of node $i$ based on shortest-path distances, while the summation term accounts for the contributions of its neighbors weighted by their degree and harmonic centrality. Nodes with high NIEM values are efficiently reachable and well connected to other central nodes.

\section{Node information dimension (NID)}

The \emph{node information dimension}\index{node information dimension (NID)} (NID) is a centrality measure for identifying influential nodes based on the local dimension framework \cite{Bian2018}. Let \(d_{\max}(i) = \max_j d_{ij}\) denote the maximal shortest-path distance between node \(i\) and all other nodes in the network. Similar to the local dimension (LD) in \cite{Silva2012}, the local dimension coefficient \(d_j(r)\) is computed for each topological distance scale \(r = 1, \dots, d_{\max}(i)\) as
\[
d_j(r) = j \frac{n_r(j)}{B_r(j)}, \quad \forall j = 1, \dots, S_i(r),
\]
where \(B_r(j)\) is the number of nodes within distance \(j\) from node \(i\) with respect to topological distance scale $r$, \(n_r(j)\) is the number of nodes at exact distance \(j\) from node \(i\), and \(S_i(r) = \lceil d_{\max}(i)/r \rceil\). 

The information entropy of node \(i\) at distance \(r\) is
\[
I_i(r) = - \sum_{j=1}^{S_i(r)} \frac{d_j(r)}{\sum_{k=1}^{S_i(r)} d_k(r)} \ln \frac{d_j(r)}{\sum_{k=1}^{S_i(r)} d_k(r)}.
\]

The node information dimension of node \(i\) is then defined as
\[
c_{\textsc{NID}}(i) = -\lim_{r \to 0} \frac{I_i(r)}{\ln r}.
\]

The NID centrality of node $i$ is estimated numerically as the slope of the linear regression of \(I_i(r)\) against \(\ln r\).

\section{Node local centrality (NLC)}

The \emph{Node local centrality}\index{node local centrality (NLC)} (NLC) is a centrality measure for identifying influential spreaders by combining network embedding (NE) with local network information \cite{YangX2021}. 

For a node \(i\), the NLC centrality \(c_{\textsc{NLC}}(i)\) is defined as
\[
c_{\textsc{NLC}}(i) = \sum_{j \in \mathcal{N}(i)} k_s(i) \, e^{-\|x_i - x_j\|^2},
\]
where \(k_s(i)\) is the \(k\)-shell index of node \(i\) and \(x_i \in \mathbb{R}^{r \times 1}\) is the vector embedding of node \(i\) obtained via the DeepWalk network representation method \cite{Perozzi2014}. 

Yang \textit{et al.} \cite{YangX2021} set the embedding dimension to \(r = N/2\). The measure captures both the hierarchical position of a node in the network (via \(k\)-shell) and its proximity in the embedded low-dimensional space, reflecting local structural similarity.

\section{Non-backtracking centrality}

\emph{Non-backtracking (NB) centrality}\index{non-backtracking (NB) centrality} is a spectral measure designed to mitigate localization effects, where a hub with high centrality artificially inflates the centrality of its neighbors, which in turn feed back and further exaggerate the hub's centrality \cite{Martin2014}. 

The NB centrality is defined using the \(L \times L\) non-symmetric \emph{non-backtracking matrix}\index{matrix!non-backtracking} \(B\), where each row and column corresponds to a directed edge \((i,j)\), with elements
\[
B_{(k,l),(i,j)} = \delta_{jk} (1 - \delta_{il}),
\]
and \(\delta_{jk}\) is the Kronecker delta. The element \(v_{(j,i)}\) of the leading eigenvector of \(B\) represents the centrality of node \(j\) ignoring contributions from node \(i\). The full non-backtracking centrality of node \(i\) is then
\begin{equation*}
    c_{NB}(i) = \sum_{j \in \mathcal{N}(i)} a_{ji} \, v_{(j,i)}.
\end{equation*}

The non-backtracking centrality can be efficiently computed as the first \(N\) elements of the leading eigenvector of the \(2N \times 2N\) matrix
\[
M = \begin{bmatrix} A & I - D \\ I & 0 \end{bmatrix},
\]
where \(A\) is the adjacency matrix, \(I\) is the \(N \times N\) identity matrix, and \(D\) is the diagonal matrix of node degrees.

\section{Normalized local centrality (NLC)}
\emph{Normalized local centrality}\index{normalized local centrality (NLC)} (NLC) considers the topology of the local network around a node as well as the influence feedback of the node’s nearest neighbor nodes \cite{Zhao2018}. The centrality \(c_{NLC}(i)\) of node \(i\) is given by
\begin{equation*}
    c_{NLC}(i) = \sum_{j \in \mathcal{N}(i)}{Q(j)|\mathcal{N}^{(\leq 2)}(j)|},
\end{equation*}
where $\mathcal{N}^{(\leq 2)}(j)$ denotes the set of nearest and next nearest neighbors of node \(j\) and \(Q(j)\) is the influence feedback of nearest neighbor node $j$ with  
\begin{equation*}
    Q(j) = \sum_{l \in \mathcal{N}^{(\leq 2)}(j)}{\left(\frac{d_l}{\sqrt{\sum_{u \in \mathcal{N}^{(\leq 2)}(j)}{d_u^2}}} + \frac{c_l}{\sqrt{\sum_{u \in \mathcal{N}^{(\leq 2)}(j)}{c_u^2}}} \right)},
\end{equation*}
where $d_l$ and $c_l$ are the degree and the clustering coefficient of node $l$. Hence, $Q(j)$ is the normalized sum of the number of nodes in the local network and the local clustering coefficient that denotes the tightness of node topology connections, which represents the local structural attribute of the network.

\section{Normalized wide network ranking (NWRank)}

The \emph{normalized wide network ranking}\index{normalized wide network ranking (NWRank)}\index{NWRank} (NWRank) is a variation of WRank that incorporates the mutual reinforcement feature of HITS and the weight normalization feature of PageRank \cite{Wang2015}. Unlike WRank, which distributes node scores evenly among incident links, NWRank assigns link weights proportional to the neighboring nodes’ degree and the betweenness centrality of the link. 

Let \(Z\) be the \(L \times N\) link-node matrix with elements
\[
z_{li} =
\begin{cases}
\alpha \frac{d_l}{\sum_{j \in \mathcal{N}(i)}{d_{(i,j)}}} + (1-\alpha) \dfrac{bc_l}{\sum_{j \in \mathcal{N}(i)} bc_{(i,j)}}, & i \in l, \\
0, & \text{otherwise},
\end{cases}
\]
where \(\mathcal{N}(i)\) the set of neighbors of node \(i\), \(d_l\) denotes the degree of the node at the other end of link \(l\), \(bc_l\) is the edge betweenness centrality of link \(l\), and \(\alpha = 0.5\). 

Let \(W\) be the \(N \times L\) binary node-link incidence matrix with elements
\[
w_{il} =
\begin{cases}
1, & \text{if node } i \text{ is incident to link } l, \\[0.8em]
0, & \text{otherwise}.
\end{cases}
\]

The principal eigenvector of \(WZ\) defines the NWRank centrality of the nodes. The NWRank algorithm preserves the mutual reinforcement between nodes and links, analogous to HITS, while normalizing the contributions of each link in a manner similar to PageRank. Links that connect highly central nodes or have high betweenness receive greater weight in determining node centrality.

\section{Odd subgraph centrality}

\emph{Odd subgraph centrality}\index{subgraph centrality!odd} is a variant of subgraph centrality that counts the number of closed walks of \emph{odd} length in a network \cite{Estrada2007}. Focusing on odd-length walks highlights genuine cycles, since even-length walks can arise from trivial back-and-forth movements in acyclic subgraphs. The odd subgraph centrality of node $i$, denoted $c_{odd}(i)$, is defined as
\[
c_{odd}(i) = \sum_{k=0}^{\infty} \frac{(A^{2k+1})_{ii}}{(2k+1)!} 
= \sum_{j=1}^{N} \left( v_j(i) \right)^2 \sinh(\lambda_j),
\]
where $A$ is the adjacency matrix of the network, $v_j(i)$ is the $i$-th component of the eigenvector $v_j$ corresponding to eigenvalue $\lambda_j$, and $N$ is the number of nodes. Odd subgraph centrality has been applied to empirical food web networks to identify keystone species involved in cyclic trophic interactions.

\section{Onion decomposition (OD)}

The \emph{onion decomposition}\index{onion decomposition} (OD) extends the \textit{k}-core decomposition by assigning to each node not only a \emph{core index} but also a \emph{layer index} that records the iteration at which the node is removed during the peeling process \cite{HebertDufresne2016}. The OD is obtained through the following steps:

\begin{enumerate}
    \item Initialize the core index $k = 1$ and the layer index $\ell = 1$. Compute the degree $d_i$ of each node $i$ in the network $G$.
    \item Identify all nodes with degree $d_i \le k$. Assign to each such node a \emph{core index} $c_i = k$ and a \emph{layer index} $\ell_i = \ell$, then remove them from the network and update the degrees of their neighbors.
    \item Increment $\ell$ by 1. Repeat step 2 until no nodes with degree $d_i \le k$ remain in the network.
    \item Update $k$ to the minimal degree among the remaining nodes and repeat steps 2-3 until all nodes have been assigned both a core index $c_i$ and a layer index $\ell_i$.
\end{enumerate}

Each node $i$ in the onion decomposition is characterized by the pair $(c_i, \ell_i)$, where $c_i$ is the coreness and $\ell_i$ the removal iteration within its core. The pair $(c_i, \ell_i)$ captures both the core hierarchy and intra-core connectivity, with higher values associated with nodes in denser regions of the network.

The onion decomposition has been applied in a variety of contexts \cite{Thibault2024}, including detecting atypical structures within $k$-cores of empirical networks \cite{HebertDufresne2016}, ranking nodes according to their structural position in the network \cite{Young2019,Mimar2022,Lu2024}, and designing heuristics for NP-hard optimization problems \cite{GarciaPerez2019}. Other applications include using the OD to parameterize dynamical models of water distribution \cite{Zhou2023} and organizational networks \cite{HebertDufresne2023}, as well as for retrieving the underlying filamentary structure of the cosmic web \cite{Bonnaire2020}.

\section{Outward accessibility}

The \emph{outward accessibility}\index{accessibility!outward} quantifies the ability of a node to reach other nodes in a network after a fixed number of steps along self-avoiding walks \cite{Travençolo2008}. Let \( N \) be the total number of nodes in the network, and let \( P_h(i,j) \) denote the transition probability that an agent starting from node \( i \) reaches node \( j \) in exactly \( h \) steps along a self-avoiding walk (i.e., a simple path without revisiting nodes). The outward accessibility of node \( i \) after \( h \) steps is defined as
\begin{equation*}
    c_{OA_h}(i)
    = \frac{1}{N - 1}
    e^{\left(
    -\sum_{j:\, P_h(i,j) \neq 0} 
    P_h(i,j) \log P_h(i,j)
    \right)}.
\end{equation*}

The expression inside the exponential represents the Shannon entropy of the transition probability distribution for node \( i \). A higher entropy indicates that node \( i \) can reach many other nodes through distinct paths with similar probabilities, reflecting greater accessibility. The normalization factor \( 1/(N - 1) \) ensures that the measure is comparable across networks of different sizes.

\section{\textit{p}-means centrality}

The \emph{\textit{p}-means centrality}\index{\textit{p}-means centrality} is a distance-based measure designed to identify the most influential nodes in a network by generalizing several classical centrality metrics \cite{Andrade2019}. The centrality is parameterized by \(p\), which governs the aggregation of distances from node \(i\) to all other nodes in the network. For a node \(i \in \mathcal{N}\), the \textit{p}-means centrality is then defined as
\[
c_{p\text{-means}}(i) =
\begin{cases} 
\displaystyle \left( \frac{\sum_{j \neq i} d_{ji}^p}{N-1} \right)^{-\frac{1}{p}}, & \text{if } p \neq 0,\\[2mm]
\displaystyle \left( \prod_{j \neq i} d_{ji} \right)^{-\frac{1}{N-1}}, & \text{if } p = 0,
\end{cases}
\]
where \(d_{ji}\) is the shortest-path distance from node \(j\) to node \(i\), and \(N\) is the total number of nodes.  

By varying \(p\), p-means centrality interpolates between several classical measures:  
\begin{itemize}
    \item \(p = 1\) corresponds to closeness centrality;
    \item \(p = -1\) corresponds to harmonic centrality;
    \item \(p \to \infty\) approaches eccentricity centrality;
    \item \(p \to -\infty\) yields the same node ranking as degree centrality.
\end{itemize}

The performance of \textit{p}-means centrality has been evaluated using the susceptible-infected-recovered (SIR) model. Additionally, Andrade and Rêgo \cite{Andrade2019} investigated the values of \(p\) for which p-means centrality satisfies the size, density, and score monotonicity axioms.

\section{PageRank}
The \emph{PageRank}\index{PageRank} algorithm evaluates the relative importance of nodes based on the concept of a random walk with restart~\cite{Newman2018,BrinPage1998}. According to PageRank, the importance \(c_{PR}(i)\) of node \(i\) depends on the probability that it be visited by a random walker, that is,
\begin{equation*}
    c_{PR}(i) = \alpha \sum_{j=1}^{N} a_{ji} \frac{c_{PR}(j)}{\sum_{k=1}^{N} a_{jk}} + \frac{1 - \alpha}{N},
\end{equation*}
where \(a_{ji}\) denotes the element of the adjacency matrix \(A\) indicating a directed link from node \(j\) to node \(i\). The parameter \(\alpha \in [0,1]\) is the damping factor representing the probability that the random walker follows an outgoing link from the current node rather than teleporting to a randomly chosen node. In practice, \(\alpha\) is typically set to \(0.85\), balancing exploration through the network structure and random teleportation to ensure ergodicity and convergence of the ranking vector. When \(\alpha = 1\), the walker never teleports, and the process reduces to a pure random walk on the network, with the PageRank vector given by the principal eigenvector of the column-normalized adjacency matrix \(P\), where \(P_{ij} = a_{ji} / \sum_{k=1}^{N} a_{jk}\). The PageRank score $c_\mathrm{PR}(i)$ of node $i$ represents the stationary probability that a random walker occupies node $i$ in the long-term limit, and is given by the principal eigenvector of the stochastic matrix
\[
P_\mathrm{PR} = \alpha D^{-1} A + (1-\alpha) \frac{1}{N} \mathbf{1}_N \mathbf{1}_N^\top,
\]
where $D$ is the $N \times N$ diagonal out-degree matrix, $\mathbf{1}_N$ denotes the $N \times 1$ vector of ones.

\section{Pairwise disconnectivity centrality}

\emph{Pairwise disconnectivity centrality}\index{pairwise disconnectivity centrality} quantifies the topological importance of a node by comparing the number of ordered node pairs that are reachable before and after its removal \cite{Potapov2008}. A node is more central if its removal disconnects a larger fraction of node pairs. Formally, the centrality of node \(i\) is defined as
\begin{equation*}
c_{\mathrm{pd}}(i) = \frac{n_G - n_{G_i}}{n_G},
\end{equation*}
where \(n_G\) is the total number of ordered pairs of nodes connected by at least one directed path in the graph \(G\), and \(G_i\) denotes the graph obtained by removing node \(i\).

\section{Participation coefficient}

The \emph{participation coefficient}\index{participation coefficient} is a community-based measure that quantifies how evenly a node's links are distributed among different communities (modules) \cite{Guimerà2005}. Consider a graph \(G\) with a community structure consisting of \(K\) communities \(C_1, \dots, C_K\). For example, \cite{Oldham2019} applied the participation coefficient to networks whose communities were detected using the Louvain algorithm. The participation coefficient quantifies how a node's links are distributed within its own module compared to other modules:
\begin{equation*}
   c_{\mathrm{Part.coeff}}(i) = 1 - \sum_{s=1}^{K} \left( \frac{d_{is}}{d_i} \right)^2 
   = 1 - \sum_{s=1}^{K} \left( \frac{\sum_{j \in C_s \setminus \{i\}} a_{ij}}{d_i} \right)^2,
\end{equation*}
where \(d_{is}\) is the number of links from node \(i\) to nodes in module \(C_s\), and \(d_i\) is the degree of node \(i\). 

The participation coefficient \(c_{\mathrm{Part.coeff}}(i)\) approaches 1 when a node's links are uniformly distributed across all modules, and is close to 0 when all links are confined within its own module. While the participation coefficient was originally intended to describe the roles of nodes within community-structured networks, it has also proven useful for identifying highly connected hub nodes in real-world networks \cite{Power2013}.

\section{Partition-based spreaders identification (PBSI) method}

The \emph{partition-based spreaders identification (PBSI) method}\index{partition-based spreaders identification (PBSI)} is a community-aware algorithm for identifying influential nodes in a network \cite{Yanez-Sierra2021}. It assumes that the network $G$ has a community structure consisting of $K$ communities, denoted by $C_1, \dots, C_K$.  The PBSI procedure consists of two main steps:

\begin{enumerate}
    \item \textit{Community detection and node ranking:} the network is partitioned into communities using a community detection algorithm, specifically the Louvain method. Within each community, nodes are ranked according to their gravity centrality \cite{Ma2015}.
    \item \textit{Spreader selection:} a total of $m$ spreaders are selected from different communities. If $m \leq K$, one node with the highest gravity centrality is chosen from each of the $m$ largest communities. If $m > K$, a number of nodes $n_i$ is selected from each community $C_i$, proportional to its size, such that $\sum_{i=1}^K n_i = m$. In all cases, the nodes with the highest gravity centrality within each community are selected.
\end{enumerate}

\section{Path-transfer centrality}

\emph{Path-transfer centrality}\index{path-transfer centrality}, also called entropy path centrality (EPTC), is an entropy-based measure that quantifies node importance based on the way traffic flows through a network \cite{Tutzauer2007}.The key idea is that flows originating from highly central nodes spread broadly and evenly through the network, whereas flows from less central nodes are concentrated along fewer paths and reach fewer nodes.

Consider a flow originating at node $i$. At each step, the flow may either stop at the current node $j$ or transfer to an unvisited neighbor $k$. The overall probability $p_{ij}$ that a flow starting at $i$ ends at $j$ is the sum of probabilities over all simple paths from $i$ to $j$:
\begin{equation*}
    p_{ij} = \sum_{l=1}^{K(i,j)} \sigma_l(j) \prod_{t=0}^{|P_l(i,j)|-1} \tau_l(u_t),
\end{equation*}
where $K(i,j)$ is the number of simple paths from $i$ to $j$, $P_l(i,j)$ denotes the $l$th path, $\sigma_l(j)$ is the stopping probability at $j$ along path $P_l(i,j)$, and $\tau_l(u_t)$ is the transition probability from node $u_t$ to the next node in that path.  

The path-transfer centrality of node $i$ is then given by the Shannon entropy of the flow distribution:
\begin{equation*}
    c_{\text{PT}}(i) = -\sum_{j=1}^N p_{ij} \log p_{ij}.
\end{equation*}

Nodes with high path-transfer centrality are those from which flows can reach many other nodes with relatively uniform probability, highlighting their structural importance in facilitating traffic or information spread.

\section{PathRank}

\emph{PathRank}\index{PathRank} is a centrality measure that quantifies a node's importance based on all simple paths terminating at it \cite{Mussone2022}. Let \(P^{(1)}_{ij}, P^{(2)}_{ij}, \dots, P^{(s_{ij})}_{ij}\) denote the set of simple paths from node \(i\) to node \(j\) of length at most \(K\). Each path contributes to the centrality of node \(j\) with a weight inversely proportional to the square of its length:
\[
b_{ij} = \sum_{s=1}^{s_{ij}} \frac{1}{|P^{(s)}_{ij}|^2},
\]
where \(|P^{(s)}_{ij}|\) is the length of the \(s\)-th path. This weighting emphasizes shorter paths, reflecting their greater significance in connectivity.  The PathRank centrality of node \(j\) is obtained by summing these contributions over all source nodes:
\[
c_{PathRank}(j) = \sum_{i=1}^{N} b_{ij}.
\]

Thus, PathRank accumulates the influence of all paths terminating at a node, giving higher weight to shorter paths.  In practice, a maximum path length of \(K = 5\) is typically used \cite{Mussone2022}.

\section{Percolation centrality}

\emph{Percolation centrality}\index{percolation centrality} (PC) quantifies the importance of nodes in facilitating percolation through a network, such as in epidemic spreading or information diffusion \cite{Piraveenan2013}. In this model, each node $i$ is assigned a percolation state $x_i$ with $0 \leq x_i \leq 1$, representing the extent to which the node is percolated.  

The percolation centrality $c_{PC}(i)$ of node $i$ is defined as the fraction of ``percolated paths'', which are shortest paths whose source nodes are percolated, that pass through $i$, i.e.,
\begin{equation*}
c_{PC}(i) = \frac{1}{N-2} \sum_{j \neq i \neq k} \frac{\sigma_{jk}(i)}{\sigma_{jk}} \cdot \frac{x_j}{\sum_{s=1}^{N} x_s - x_i},
\end{equation*}
where $\sigma_{jk}$ denotes the total number of shortest paths from node $j$ to node $k$, and $\sigma_{jk}(i)$ is the number of those paths that pass through node $i$.  Piraveenan \textit{et al.}~\cite{Piraveenan2013} demonstrated that percolation centrality reduces to standard \emph{betweenness centrality} when all nodes have the same percolation state.

\section{Physarum centrality}
\emph{Physarum centrality}\index{Physarum centrality} is a bio-inspired measure based on the behavior of the slime mold \textit{Physarum polycephalum}~\cite{Zhang2012}. The organism can be modeled as an undirected weighted network, where each link represents a tube and each node represents a junction between tubes. The weight $w_{ij}$ denotes the length of link $(i,j)$. 

The core idea is that, during exploration for optimal paths, long and narrow tubes tend to weaken, while short and wide tubes strengthen due to positive feedback from flux. The Physarum dynamically adjusts flux through its network to identify efficient paths connecting two specified nodes. 

For each source node $s$ and target node $t$ (representing food sources), the flux $Q_{ij}^{st}$ through link $(i,j)$ is determined according to Poiseuille flow:
\[
Q_{ij}^{st} = \frac{D_{ij}}{w_{ij}}(p_i - p_j),
\]
where $p_i$ is the pressure at node $i$ and $D_{ij}$ is the conductivity of link $(i,j)$. Assuming flow conservation at each node, the network satisfies the Poisson equation:
\[
\sum_{j=1}^N \frac{D_{ij}}{w_{ij}}(p_i - p_j) =
\begin{cases}
-I_0 & \text{for } i = s,\\
I_0 & \text{for } i = t,\\
0 & \text{otherwise,}
\end{cases}
\]
where $I_0$ is the total flux from the source (typically $I_0=1$). The conductivities are initialized as $D_{ij} = 0.5$ and updated over time according to
\[
\frac{d}{dt} D_{ij} = \frac{(1+a)(Q_{ij}^{st})^\mu}{1 + a (Q_{ij}^{st})^\mu} - \alpha D_{ij},
\]
where $\alpha$ is the decay rate of the tube (e.g.  $\alpha=0.05$), $\mu = 27$ and $a = 2$. The flux $Q_{ij}^{st}$ is iteratively updated, and the model is terminated after $4 \log N$ iterations.

Finally, the Physarum centrality of node $i$ is defined as
\[
c_{\mathrm{Physarum}}(i) = \sum_{j \in \mathcal{N}(i)} c_{ij}, \quad
c_{ij} = \sum_{s \neq t} Q_{ij}^{st},
\]
where $c_{ij}$ represents the criticality of link $(i,j)$, computed as the total flux through the link across all source-target pairs.

\section{PhysarumSpreader}

\emph{PhysarumSpreader}\index{PhysarumSpreader} is a centrality measure that combines \textit{LeaderRank} with a positive feedback mechanism inspired by the amoeboid organism \textit{Physarum polycephalum} \cite{HWang2015}. Similarly to LeaderRank, PhysarumSpreader introduces a ground node $g=N+1$ connected bidirectionally to every node. The weight $w_{ig}$ of the link from node $i$ to the ground node $g$ is defined as
\[
w_{ig} =
\begin{cases}
\frac{\sum_{j=1}^{N} w_{ij}}{d_i}, & \text{if } d_i > 0,\\[2mm]
1, & \text{if } d_i = 0,
\end{cases}
\]
where $d_i$ is the degree of node $i$. The reverse weight $w_{gi}$ is set equal to $w_{ig}$ to allow bidirectional flux between node $i$ and the ground node.

PhysarumSpreader iteratively proceeds as follows:

\begin{enumerate}
    \item Initialize all nodes (except the ground node) with a unit of resource: $I_i(0) = 1$, $\forall i \neq g$, and set the ground node’s score to zero: $I_g(0) = 0$.
    \item At each discrete time step $k$, distribute each node’s flux to its neighbors along outgoing edges according to their weights $Q_{ij}(k)$:
    \[
        Q_{ij}(k) = \frac{w_{ij} D_{ij}(k)}{\sum_{l=1}^{N+1} w_{il} D_{il}(k)},
    \]
    where $D_{ij}(k)$ is the conductivity of edge $(i,j)$ at time $k$. Initially, $D_{ij}(0) = 1$ for all edges.
    \item Update the conductivity matrix based on the flux through each edge $(i,j)$ ($i,j \neq g$):
    \[
        D_{ij}(k+1) = \frac{D_{ij}(k) + Q_{ij}(k)}{2}.
    \]
    \item Update the resource of each node according to the incoming flux:
    \[
        I_i(k+1) = \sum_{j=1}^{N+1} Q_{ji}(k) I_j(k).
    \]
    \item Repeat steps 2--4 until a steady state $t_c$ is reached. The centrality of node $i$ is then
    \[
        c_{PS}(i) = I_i(t_c) + \frac{I_g(t_c)}{N}.
    \]
\end{enumerate}

PhysarumSpreader is applicable to both directed and undirected networks, and can handle weighted or unweighted edges. Nodes with high PhysarumSpreader centrality correspond to those that accumulate the most resource flux in the network, reflecting their structural importance and influence in propagating flow throughout the system.

\section{Pivotal index}

The \textit{pivotal index}\index{pivotal index} is a power index that quantifies both the individual and collective influence of nodes within a network \cite{YAleskerov2020}. It is based on the assumption that each node \(i\) possesses an individual \textit{influence threshold} \(q_i\), which specifies the level of accumulated influence required for the node to become affected (e.g., \(q_i = 3\)).  

A subset of nodes \(\Omega(i) \subset \mathcal{N}\) is called \textit{critical} for node \(i\) if the total influence exerted by the members of \(\Omega(i)\) on \(i\) meets or exceeds its threshold value:
\begin{equation*}
    \sum_{k \in \Omega(i)} w_{ki} \geq q_i,
\end{equation*}
where \(w_{ki}\) denotes the influence weight of node \(k\) on node \(i\).  

A node \(k\) is said to be \textit{pivotal} for the group \(\Omega(i)\) if its exclusion from this group makes the group non-critical. The set of all pivotal members within \(\Omega(i)\) is denoted by \(\Omega^{p}(i)\). The \textit{pivotal influence index} \(c_{PI}(i)\) of node \(i\) is defined as the total number of pivotal groups for that node:
\[
c_{PI}(i) = \sum_{\Omega(i) \subseteq \mathcal{N}(i)} |\Omega(i)| \cdot |\Omega^{p}(i)|.
\]

Since the number of possible critical groups can grow exponentially with network size, resulting in high computational complexity, a practical variant of the pivotal index considers only subsets of size up to \(k\). Aleskerov and Yakuba \cite{YAleskerov2020} also proposed an \textit{order-\(k\) extension} of the pivotal index that accounts for indirect influences. In this extension, influence is evaluated over \((k{+}1)\)-hop neighborhoods, where link strength is defined as the maximum bottleneck capacity among all \((k{+}1)\)-length paths.  

For unweighted networks, the pivotal influence index \(c_{PI}(i)\) of node \(i\) simplifies to
\[
c_{PI}(i) = \bigl(\lceil q_i \rceil\bigr)^2 \binom{d_i}{\lceil q_i \rceil},
\]
where \(d_i\) denotes the degree of node \(i\), and \(\lceil \cdot \rceil\) denotes the ceiling function.

The pivotal index has been applied in various contexts, including the identification of influential countries in global food trade networks \cite{Aleskerov2022} and oil trade networks \cite{Aleskerov2023}, analysis of trade relations among economic sectors across countries \cite{Aleskerov2024a}, and bibliometric studies of publications on Parkinson’s disease \cite{Aleskerov2024b}.

\section{Principal component centrality (PCC)}

\emph{Principal component centrality}\index{principal component centrality (PCC)} (PCC) is a spectral measure of node influence that generalizes eigenvector centrality by incorporating multiple leading eigenvectors of the adjacency matrix, rather than relying solely on the dominant one \cite{Ilyas2011}. Each node is represented in the eigenspace spanned by the top $p$ eigenvectors of the adjacency matrix, and its importance is quantified as the weighted Euclidean distance of its coordinates from the origin, where the weights correspond to the associated eigenvalues.

Formally, let $A$ be the $N \times N$ adjacency matrix of graph $G$ and $X_p$ be the $N \times p$ matrix of the top $p$ eigenvectors $v_1, \dots, v_p$ corresponding to eigenvalues $\lambda_1, \dots, \lambda_p$ with $|\lambda_1| \ge |\lambda_2| \ge \cdots \ge |\lambda_p|$. Then, the PCC vector is defined as
\[
c_{PCC} = \sqrt{ (A X_p \circ A X_p) u },
\]
where $\circ$ denotes the Hadamard (element-wise) product and $u$ is a $p \times 1$ vector of ones. Equivalently, the PCC score of node $i$ is
\[
c_{PCC}(i) = \sqrt{\sum_{k=1}^p \big((AX_p)_{ik}\big)^2} = \sqrt{\sum_{k=1}^p \big(\lambda_k (v_i)_k\big)^2}.
\]

When $p=1$, PCC reduces to a scaled version of eigenvector centrality. Nodes farther from the origin in this weighted eigenspace are considered more central, reflecting stronger contributions along the most significant structural directions of the network.

\section{Probabilistic-jumping random walk (PJRW) centrality}

The \emph{probabilistic-jumping random walk} (\textsc{PJRW})\index{probabilistic-jumping random walk centrality} centrality simulates the propagation of messages in a network by randomly spreading them from nodes to their neighbors \cite{Yu2018}. Inspired by the PageRank algorithm, \textsc{PJRW} starts from a randomly selected node and, at each step, either jumps to a random node in the graph \(G\) with probability \(p_c\) or moves to one of the current node's neighbors with probability \(1 - p_c\). The total number of visits to each node \(i\) during this process defines its centrality.

Unlike PageRank, which relies on the stationary distribution of a random walk, Yu \textit{et al.} \cite{Yu2018} impose a maximum of \(5 \sqrt{N}\) messages and a fixed number of \(50 \sqrt{L}\) steps for each message, where \(N\) is the number of nodes and \(L\) is the number of edges. Consequently, multiple iterations of PJRW may produce different centrality values. The probability of selecting a random node is set as
\[
p_c = \frac{\langle d^2 \rangle - \langle d \rangle}{\langle d \rangle} \frac{\log_{10} N}{10 \sqrt{N}},
\]
where \(\langle d \rangle\) and \(\langle d^2 \rangle\) are the average degree and the second-order average degree of the network, respectively.

\vfill

\section{ProfitLeader}

\emph{ProfitLeader}\index{ProfitLeader} (PL) centrality evaluates node importance based on the capacity to provide resources to neighbors \cite{Yu2019}. For each neighbor \(j \in \mathcal{N}(i)\), the available resource from \(i\) to \(j\) is denoted \(AR(i,j)\), and the sharing probability based on similarity is \(SP(i,j)\). The centrality of \(i\) is defined as
\[
c_{PL}(i) = \sum_{j \in \mathcal{N}(i)} AR(i,j) \cdot SP(i,j) 
= \sum_{j \in \mathcal{N}(i)} \left( |\mathcal{N}(i)| {+} \sum_{k \in \mathcal{N}(i) \setminus \mathcal{N}(j)} |\mathcal{N}(k)| \right) \cdot \frac{|\mathcal{N}(i) {\cap} \mathcal{N}(j)|}{|\mathcal{N}(i) {\cup} \mathcal{N}(j)|},
\]
where \(\mathcal{N}(i)\) denotes the set of neighbors of \(i\). Nodes achieve high centrality scores if they provide substantial resources to many similar neighbors, reflecting both connectivity and neighborhood overlap.

\section{Proximal betweenness centrality}

\emph{Proximal betweenness centrality}\index{betweenness centrality!proximal} is a variant of traditional betweenness centrality that quantifies how frequently a node appears in a \emph{penultimate} position along a shortest path \cite{Borgatti2002,Brandes2008}. The proximal betweenness centrality of a node \(i\) is defined as
\begin{equation*}
c_{\mathrm{pr.betw}}(i) = \sum_{j=1}^{N} \sum_{k=1}^{N} \frac{b_{jk}(i)}{\sigma_{jk}},
\end{equation*}
where \(\sigma_{jk}\) denotes the total number of shortest paths from node \(j\) to node \(k\), and \(b_{jk}(i)\) is the number of those paths in which node \(i\) occupies the \emph{penultimate} position. In this context, a node \(i\) is considered penultimate if it lies either directly before the destination node \(k\) (proximal to the target) or directly after the source node \(j\) (proximal to the source) on a shortest path. Following Borgatti \textit{\textit{et al.}}~\cite{Borgatti2002}, two variants of proximal betweenness centrality can be distinguished:
\begin{itemize}
    \item \emph{Proximal source betweenness:} counts node \(i\) when it occurs immediately \emph{before} the destination node \(k\) on shortest paths originating from node \(j\);
    \item \emph{Proximal target betweenness:} counts node \(i\) when it occurs immediately \emph{after} the source node \(j\) on shortest paths terminating at node \(k\).
\end{itemize}

\section{Quantum Jensen-Shannon Divergence (QJSD) centrality}

\emph{Quantum Jensen-Shannon Divergence (QJSD) centrality}\index{quantum Jensen-Shannon divergence centrality} is a quantum-inspired measure of node importance based on the evolution of quantum walks on a graph \cite{Rossi2014}. Rossi \textit{et al.} define two quantum walks where node \(i\) is initially set to be in phase and in antiphase with respect to the other nodes. The QJSD centrality \(c_{QJSD}(i)\) quantifies how the initial phase of node \(i\) influences the evolution of the quantum walks by computing the \emph{Quantum Jensen-Shannon divergence} (QJSD) \(D_{JS}\) between the density operators \(\rho_i\) and \(\sigma_i\) representing the corresponding quantum states:
\[
c_{QJSD}(i) = D_{JS}(\rho_i, \sigma_i).
\]

Rossi \textit{et al.} \cite{Rossi2014} show that when the quantum walk is defined using the normalized Laplacian as the generator, the QJSD centrality can be expressed analytically as
\[
c_{QJSD}(i) = 1 - \frac{1}{2}\log_2(\mu_0+1) + \frac{\mu_0}{2} \log_2 \frac{\mu_0}{\mu_0 + 1},
\]
where
\[
\mu_0 = \left( 1 - \frac{d_i}{L} \right)^2,
\]
with \(d_i\) the degree of node \(i\) and \(L\) the total number of links in the graph. This expression quantifies how much the initial phase of node \(i\) affects the evolution of the quantum walk compared to its anti-phase counterpart.

\section{Quasi-Laplacian centrality (QC)}

The \emph{quasi-Laplacian centrality}\index{Laplacian centrality!quasi- (QC)} (QC) is an eigenvalue-based method for identifying influential nodes in complex networks, derived from the concept of graph energy \cite{Ma2019}. The main idea behind QC is that the importance (centrality) of a node \(i\) is reflected by the variation in the quasi-Laplacian energy resulting from the removal of node \(i\) from the network.  

The quasi-Laplacian centrality \(c_{QC}(i)\) of node \(i\) is defined as
\[
c_{QC}(i) = E_Q(G) - E_Q(G_i),
\]
where \(G_i\) denotes the subgraph of \(G\) obtained by removing node \(i\), and \(E_Q(G)\) is the quasi-Laplacian energy of the network \(G\), given by
\[
E_Q(G) = \sum_{j=1}^{N} \mu_j^2,
\]
where \(\mu_1, \dots, \mu_N\) are the eigenvalues of the \emph{quasi-Laplacian}\index{matrix!quasi-Laplacian} matrix \(Q = D + A\), with \(D\) being the diagonal degree matrix and \(A\) the adjacency matrix.  

The quasi-Laplacian centrality \(c_{QC}(i)\) can be further simplified as
\[
c_{QC}(i) = d_i^2 + d_i + \sum_{j \in \mathcal{N}(i)} d_j,
\]
where \(d_i\) denotes the degree of node \(i\) and \(\mathcal{N}(i)\) is the set of its neighboring nodes.  

The quasi-Laplacian centrality has been tested on Zachary’s karate club network and several terrorist networks, and its effectiveness has been validated using the susceptible-infected-recovered (SIR) epidemic model.

\section{Radiality centrality}

\emph{Radiality centrality}\index{radiality centrality} measures how efficiently a node’s ties reach other nodes, emphasizing outward paths \cite{Guimaraes1972,Valente1998}. The radiality centrality of node \(i\) is defined as
\begin{equation*}
c_{\mathrm{Radiality}}(i) = \frac{\sum_{j \neq i} \left( d_G + 1 - d_{ij} \right)}{N-1},
\end{equation*}
where \(d_G\) is the diameter of \(G\) and \(d_{ij}\) is the length of the shortest path from node \(i\) to node \(j\). High radiality indicates that a node is, on average, close to other nodes relative to the network diameter, while low radiality indicates a peripheral position. For undirected networks, radiality centrality coincides with integration centrality.

\vfill

\section{Random walk accessibility (RWA)}

\emph{Random-walk accessibility}\index{random-walk accessibility (RWA)} (RWA) quantifies the diversity of nodes that can be reached via random walks from a given node~\cite{Arruda2014}. The RWA score of node \(i\) is defined as the exponential of the Shannon entropy of transition probabilities:
\begin{equation*}
    c_{RWA}(i) = \exp \Bigg( -\sum_{j} M_{ij} \log M_{ij} \Bigg),
\end{equation*}
where 
\begin{equation*}
    M = \frac{1}{e} \sum_{k=1}^{\infty} \frac{P^k}{k!} = \frac{e^P}{e}.
\end{equation*}
Here, \(M\) incorporates walks of all lengths, weighted by the inverse factorial of their lengths, and \(P\) is the row-normalized adjacency matrix of the network (\(P_{ij} = a_{ij} / \sum_{k=1}^N a_{ik}\)).

Random-walk accessibility reflects both the number of nodes that can be reached from \(i\) and the diversity of paths leading to them, giving higher scores to nodes with more evenly distributed access across the network.

\section{Random walk decay (RWD) centrality}

\emph{Random Walk Decay}\index{random walk decay (RWD) centrality} (RWD) centrality \cite{Wąs2019} is a probabilistic extension of the classical \emph{decay centrality} measure \cite{Jackson2008}, reformulated within the framework of random walks. The centrality value \( c_{RWD}(i) \) for a node \( i \) in a graph \( G \) is defined as
\begin{equation*}
    c_{RWD}(i) = \beta(G) \sum_{t=0}^{\infty} a^{t} P_i(t),
\end{equation*}
where \( a \in (0,1) \) is a decay parameter that controls the influence of longer walks, \( P_i(t) \) denotes the probability that node \( i \) is visited for the first time at step \( t \) of a random walk starting from a uniformly selected node, and \( \beta(G) \) is a normalization constant determined by the graph structure; it is typically defined as the total sum of node weights,
\[
\beta(G) = \sum_{v \in \mathcal{N}} b(v),
\]
where \( b(v) \ge 0 \) denotes the weight of node \( v \). In the special case of an unweighted graph, this reduces to \( \beta(G) = N\), the number of nodes in \( G \).

The RWD centrality can be interpreted as a weighted expectation of first-visit probabilities, where nodes reached earlier in the random walk contribute more strongly due to the exponential decay factor \( a^{t} \). W\k{a}s et~al.~\cite{Wąs2019} showed that random walk decay centrality behaves similarly to PageRank, but differs in that it accounts only for the \emph{first} visit of the random walk to each node, rather than all subsequent visits. This property can make it more suitable in settings where early reachability or first-time discovery of nodes is of primary interest.

\section{Random walk centrality (RWC)}

The \emph{random walk centrality}\index{random walk centrality} (RWC) measures how quickly a node can be reached by information diffusing randomly through a network, accounting for all paths rather than only shortest paths \cite{Noh2004}.  

Let \(A\) be the adjacency matrix of a network and \(D = \mathrm{diag}(d_1, \dots, d_N)\) the degree matrix. The transition probability matrix of a classical random walk is
\[
M = D^{-1} A,
\]
so that a walker at node \(i\) moves to a randomly chosen neighbor at each step.  
Let \(P_{ij}(t)\) denote the probability that a walker starting at \(i\) is at node \(j\) after \(t\) steps, and let \(P_i^\infty = d_i / \sum_j d_j\) be the stationary probability. The random-walk centrality of node \(i\) is defined as
\[
c_{\rm RWC}(i) = \frac{P_i^\infty}{\sum_{t=0}^{\infty} \bigl(P_{ii}(t) - P_i^\infty\bigr)} = \frac{P_i^\infty}{\tau_i},
\]
where \(\tau_i\) is the \emph{characteristic relaxation time} of node \(i\), i.e., the expected time for a random walker starting at \(i\) to approach its stationary distribution.  

Nodes with larger RWC are, on average, reached more quickly by diffusive processes and thus serve as more effective recipients in network communication.

\section{Random walk-based gravity (RWG) centrality}

\emph{Random walk-based gravity}\index{random walk-based gravity (RWG)} (RWG), also referred to as DFS-Gravity, is a variant of the gravity model that incorporates random walks instead of shortest-path distances \cite{Zhao2022}. In this approach, the “distance” between nodes is measured by the number of steps it takes to reach a node along the random walk starting from the source node, rather than by the shortest-path distance. The centrality of node $i$ is defined as
\[
c_{RWG}(i) = \frac{1}{\gamma (l-1)} \sum_{m=1}^{\gamma} \sum_{k=2}^{l+1} \frac{d_i d_{j_{m(k)}}}{(k-1)^2},
\]
where $d_i$ is the degree of node $i$, $j_{m(k)}$ is the node at position $k$ in the $m$-th random walk starting from $i$, $\gamma$ is the number of random walks per node, and $l$ is the length of each random walk.  

Zhao \textit{et al.} \cite{Zhao2022} consider $l \in [5,20]$ and set $\gamma$ to ensure convergence of the centrality values. They compare different random walk strategies, with a particular focus on depth-first search (DFS) random walks, in which the transition probabilities depend on edge weights: $w = 1/p$ to return to the previous node $j_{m(k-1)}$ with $p=10$, $w = 1$ to move to a neighbor of $j_{m(k-1)}$, and $w = 1/q$ to move to other neighbors of $j_{m(k)}$ with $q=0.01$. Their experiments show that the choice of random walk strategy does not significantly affect the performance of the centrality measure.

\section{Randomized shortest paths (RSP) betweenness centrality}

\emph{Randomized shortest paths (RSP) betweenness centrality}\index{betweenness centrality!randomized shortest paths (RSP)} is a variant of betweenness centrality that refines traditional measures based solely on either shortest paths or random walks \cite{Kivimäki2016}. The RSP model defines a Boltzmann probability distribution over all possible paths between node pairs, giving higher probability to short (near-optimal) paths while still assigning nonzero probability to longer alternatives. The balance between optimality and randomness is governed by an inverse temperature parameter~$\beta$.

Formally, the simple RSP betweenness of node~$i$, denoted by~$c_{\mathrm{RSP}}(i)$, is defined as
\begin{equation}
    c_{\mathrm{RSP}}(i) = \sum_{s=1}^{N}\sum_{t=1}^{N} \bar{n}_i(s,t),
\end{equation}
where $\bar{n}_i(s,t)$ represents the total flow transiting from node~$s$ to node~$t$ through node~$i$, computed as
\begin{equation*}
    \bar{n}_i(s,t) = \sum_{(i,j) \in \mathcal{L}} \bar{\eta}_{ij}(s,t)
    = \left( \frac{z_{si}}{z_{st}} - \frac{z_{ti}}{z_{tt}} \right) z_{it}.
\end{equation*}
Here, $\bar{\eta}_{ij}(s,t)$ denotes the expected number of passages through edge~$(i,j)$ over all $s$-$t$ walks, and $z_{st}$ is the $(s,t)$-element of the fundamental matrix~$Z$ of non-absorbing paths, given by
\[
Z = (I - W)^{-1} = \left( I - D^{-1}A \odot e^{-\beta C} \right)^{-1},
\]
where $A$ is the adjacency matrix, $C$ is the cost matrix, and $D$ is the diagonal matrix of row sums of $A$. The elements of matrix $C$ represents the traversal cost, or distance, between a pair of adjacent nodes.

For large values of~$\beta$, the path distribution concentrates on the shortest paths, while for small~$\beta$, longer and more random paths receive higher weight. In the limiting case $\beta \rightarrow 0$, the RSP betweenness converges to the stationary distribution of a random walk on the network~\cite{Kivimäki2016}.

\section{Randomized shortest paths (RSP) net betweenness centrality}

\emph{Randomized shortest paths (RSP) net betweenness centrality}\index{betweenness centrality!randomized shortest paths (RSP) net} is a variation of the simple RSP betweenness centrality specifically designed for directed networks \cite{Kivimäki2016}. Unlike standard RSP betweenness, which sums the total flow of random walkers along edges, the net variant considers the \emph{net flow}, so that opposing flows along the same edge partially cancel each other. This can provide a more meaningful measure of node importance in directed networks where bidirectional flows exist.

The RSP net betweenness of node~$i$, denoted $c_{\mathrm{RSPnet}}(i)$, is defined as
\begin{equation*}
    c_{\mathrm{RSPnet}}(i) = \sum_{s=1}^{N}\sum_{t=1}^{N} \sum_{(i,j) \in \mathcal{L}} \left| \bar{\eta}_{ij}(s,t) - \bar{\eta}_{ji}(s,t) \right|
    = \sum_{(i,j) \in \mathcal{L}} \bar{\eta}_{ij}^{\mathrm{net}},
\end{equation*}
where
\[
\bar{\eta}_{ij}^{\mathrm{net}} = e^T \big| N^{ij} - N^{ji} \big| e, 
\quad
N^{ij} = w_{ij} \left( \frac{z_{si} z_{jt}}{z_{st}} - \frac{z_{ti} z_{jt}}{z_{tt}} \right),
\]
and $z_{st}$ are the elements of the fundamental matrix
\[
Z = (I - W)^{-1} = \left( I - D^{-1}A \odot e^{-\beta C} \right)^{-1}.
\]
Here, $A$ is the adjacency matrix, $D$ is the diagonal matrix of row sums of $A$, and $C$ is the cost matrix, whose elements represent the traversal cost, or distance, between adjacent nodes.

In the limit $\beta \to 0$, the RSP net betweenness converges to the \emph{current flow betweenness} centrality, highlighting the equivalence between random-walk and electrical-flow interpretations of betweenness in this regime \cite{Kivimäki2016}.

\section{Rank centrality}

\emph{Rank centrality}\index{rank centrality} is a spectral ranking algorithm for directed graphs that aggregates pairwise comparisons to derive a global ranking of nodes \cite{Negahban2017}. The method constructs a random walk on the comparison graph, where at each step the transition probability from node \( i \) to node \( j \) is proportional to the empirical probability that \( i \) beats \( j \) in pairwise comparisons. The transition probabilities are normalized by the maximum out-degree to ensure that the resulting matrix is stochastic. Formally, the random walk is described by an \( N \times N \) time-independent transition matrix \( P \) with elements
\begin{equation*} 
   p_{ij} =
   \begin{cases}
       \dfrac{A_{ij}}{d_{\max}}, & \text{if } i \neq j, \\[6pt]
       1 - \dfrac{1}{d_{\max}} \sum_{k \neq i} A_{ik}, & \text{if } i = j,
   \end{cases}
\end{equation*}
where \( A_{ij} = \dfrac{a_{ij}}{a_{ij} + a_{ji}} \) denotes the empirical probability that node \( i \) is preferred over node \( j \), and \( d_{\max} \) is the maximum out-degree among all nodes. Intuitively, the random walk tends to remain longer on nodes that perform well across comparisons. The rank centrality of a node corresponds to the stationary distribution \( \boldsymbol{\pi} \) satisfying \( \boldsymbol{\pi} = \boldsymbol{\pi} P \), which represents the relative importance of each node in the network as determined by the random walk.

\section{Ranking-betweenness centrality}

\emph{Ranking-betweenness centrality}\index{betweenness centrality!ranking-} evaluates the importance of nodes in urban networks by combining random-walk betweenness with an adapted PageRank algorithm (APA) \cite{Agryzkov2014}. Unlike classical random-walk betweenness, which assumes equal probability of starting a random walk from any node, ranking-betweenness centrality weights these random walks according to the APA-derived importance of each starting node.  

Formally, the ranking-betweenness centrality of node $i$, denoted $c_{rb}(i)$, is defined as
\[
c_{rb}(i) = \sum_{s \neq t} \pi_s \, \sigma_{st}(i),
\]
where $\sigma_{st}(i)$ is the expected fraction of random-walk paths from node $s$ to node $t$ that pass through node $i$, and $\pi_s$ is the APA-based starting probability of node $s$.  The APA probability scores are computed using an \emph{adapted PageRank algorithm}\index{PageRank!adapted}, where the starting probabilities can be set according to external information about node relevance. If no such information is available, a uniform starting probability is used for all nodes, so that the random walks are unbiased with respect to their starting points.

\section{Rapid identifying method (RIM)}

The \emph{Rapid identifying method}\index{rapid identifying method (RIM)} (RIM) is designed to efficiently identify a fraction of highly influential nodes based on node degree and network structure \cite{Song2015}. The core idea of RIM is to start from randomly selected seed nodes and iteratively move toward neighbors with higher degrees, under the assumption that high-degree nodes are more likely to play key roles in spreading dynamics. By exploring multiple such paths and selecting the most connected nodes discovered, RIM balances randomness with structural importance.  

The method proceeds as follows:

\begin{enumerate}
    \item \textit{Seed Selection:} randomly select \(m\) seed nodes \(s_i\), where \(i = 1, 2, \dots, m\).
    
    \item \textit{Greedy Expansion:} for each seed node \(s_i\), iteratively select its highest-degree neighbor (excluding already selected nodes) for \(j\) steps:
    \[
        s_i^{m(1)},\, s_i^{m(2)},\, \dots,\, s_i^{m(j)}.
    \]
    If multiple neighbors share the same highest degree, one is chosen at random. This process generates a total of \(m \times j\) candidate nodes.

    \item \textit{Target Selection:} rank the \(m \times j\) candidate nodes by degree and select the top \(j\) nodes as the final target set \(T_j\).

    \item Repeat steps 1-3 for \(n\) independent iterations.
\end{enumerate}

The final RIM score of each node is determined by counting the number of times it appears in the top \(j\) sets across all \(n\) iterations, with nodes appearing more frequently considered more influential.

Song \textit{et al.} \cite{Song2015} consider the parameters \(m = 1\), \(n = 500\), and \(j \in \{5, 10\}\). The performance of RIM is typically evaluated using the susceptible-infected-recovered (SIR) model and compared against four classical centrality measures.

\section{Redundancy measure}
The \emph{redundancy measure}\index{redundancy measure}, originally proposed in \cite{Borgatti1997}, is a simplified version of Burt's redundancy \cite{Burt1992}. The redundancy \(c_r(i)\) of a node \(i\) quantifies the average number of connections that a neighbor of \(i\) has to other neighbors of \(i\), and is defined as
\begin{equation*} 
   c_r(i) = 
   \begin{cases} 
      \dfrac{\sum_{j \in \mathcal{N}(i)} \sum_{\substack{k \in \mathcal{N}(i) , \ k \neq j}} a_{jk}}{d_i}, & \text{if } d_i > 1, \\
      0, & \text{otherwise,}
   \end{cases}
\end{equation*}
where \(\mathcal{N}(i)\) denotes the set of neighbors of node \(i\), and \(d_i = |\mathcal{N}(i)|\) is the degree of node \(i\).  As shown by Newman \cite{Newman2018}, the redundancy measure is related to the clustering coefficient \(c_{cl}(i)\) via
\begin{equation*} 
   c_r(i) = c_{cl}(i) \, (d_i - 1).
\end{equation*}

The redundancy measure was later independently introduced as the \emph{local average connectivity (LAC)}\index{local average connectivity (LAC)} in \cite{Li2011}.

\section{Relative entropy}

\emph{Relative entropy}\index{relative entropy} is a hybrid measure that integrates multiple centrality indices to evaluate node importance using linear programming and Kullback-Leibler divergence \cite{Chen2016}. Suppose there are $m$ centrality measures, each represented as a discrete distribution $u_j = (u_{j1}, \dots, u_{jN})$ for $j=1,\dots,m$, where $u_{ji} \ge 0$ and $\sum_{i=1}^N u_{ji} = 1$.

The relative entropy of the nodes, denoted by $w = (w_1, \dots, w_N)$, is obtained by solving the linear programming problem
\begin{equation*}
    \min_{w_1, \dots, w_N} \sum_{j=1}^m \sum_{i=1}^N w_i \log_2 \frac{w_i}{u_{ji}}, \quad
    \text{subject to } \sum_{i=1}^N w_i = 1, \quad w_i > 0 \ \forall i.
\end{equation*}

The solution for node $i$ is given by the normalized geometric mean of its centrality values:
\begin{equation*}
    w_i = \frac{\prod_{j=1}^m (u_{ji})^{1/m}}{\sum_{k=1}^N \prod_{j=1}^m (u_{jk})^{1/m}}.
\end{equation*}

Chen \textit{et al.} \cite{Chen2016} apply relative entropy using $m=4$ centrality measures: degree, closeness, betweenness, and Burt's constraint coefficient. By computing the normalized geometric mean of these measures for each node, the approach produces a single composite score $w_i$ that reflects the node’s overall importance across multiple centrality perspectives.

\section{Relative local-global importance (RLGI) measure}

The \emph{relative local-global importance}\index{relative local-global importance (RLGI) measure} (RLGI) measure is a hybrid method for identifying the top-$k$ influential nodes in complex networks, combining node degree and $k$-core decomposition \cite{Gupta2021}. 

First, the \emph{normalized global importance} (\textsc{NGI}) of node \(i\) is defined as
\[
\textsc{NGI}(i) = \frac{d_i \, k_s(i) \, \delta(i)}{N},
\]
where \(d_i\) is the degree, \(k_s(i)\) is the $k$-shell score of node \(i\) and \(\delta(i)\) is the normalized iteration number (\textsc{NIM}) given by
\[
\delta(i) = 1 + \frac{n(i)}{m(i)},
\]
with \(n(i)\) being the iteration at which node \(i\) is removed, corresponding to the layer of node $i$ in the onion decomposition, and \(m(i)\) the total number of iterations in that step. 

The \textsc{RLGI} score of node \(i\) is then computed as
\[
c_{\textsc{RLGI}}(i) = \frac{\textsc{NGI}(i)\, d_i}{\sum_{j \in \mathcal{N}(i)} \textsc{NGI}(j)}.
\]

Finally, \textsc{RLGI} scores are normalized by dividing each score by the maximal RLGI value in the network. The RLGI measure integrates both local connectivity and the global hierarchical position of nodes to identify influential spreaders.

\section{Renewed coreness centrality}

\emph{Renewed coreness}\index{renewed coreness centrality} centrality is a variant of the $k$-shell centrality that accounts for redundant links in a network \cite{YLiu2015}. Based on the analysis of core-like structures in real-world networks, Liu \textit{et al.} \cite{YLiu2015} argue that core nodes should not only connect to other core nodes but also maintain links to nodes outside the core. 

In this approach, weights $w_{ij}$ are assigned to each link $(i,j)$ in the graph $G$ based on the connection patterns of its endpoints:
\[
w_{ij} = \frac{n_{i \rightarrow j} + n_{j \rightarrow i}}{2},
\]
where $n_{i \rightarrow j}$ is the number of links of node $i$ connecting outside the immediate neighborhood of node $j$. In other words, $n_{i \rightarrow j}$ counts the number of $i$'s neighbors that are not neighbors of $j$, indicating how much $i$ extends beyond $j$'s local network:
\[
n_{i \rightarrow j} = |\mathcal{N}(i) \setminus \mathcal{N}(j)|.
\]

Links with low diffusion importance ($w_{ij} < 2$) are considered redundant and removed, producing a residual network $G'$. The $k$-shell decomposition is then applied to $G'$ to obtain the renewed coreness for each node.

Nodes with high renewed coreness are those that maintain strategic connections both within the core and to peripheral nodes, reflecting their importance in facilitating diffusion and maintaining network cohesion.

\section{Residual closeness centrality}
\emph{Residual closeness centrality}\index{closeness centrality!residual}, also known as Dangalchev closeness centrality\index{closeness centrality!Dangalchev}, evaluates node closeness after the removal of nodes or edges, with shortest-path lengths weighted by an exponential decay \cite{Dangalchev2006}. Let \(G_i\) denote the subgraph of \(G\) obtained by removing node \(i\). Then the residual closeness centrality of node \(i\) equals
\begin{equation*}
    c_{residual}(i) = \sum_{j \neq k \neq i}\frac{1}{2^{d_{jk}(G_i)}},
\end{equation*}
where \(d_{jk}(G_i)\) denote the shortest-path distance between nodes \(j\) and \(k\) in \(G_i\). Thus, residual closeness reflects the distance-based importance of a node by quantifying the effect of its removal on the overall connectivity of the network.

\section{Resilience centrality}

\emph{Resilience centrality}\index{resilience centrality} quantifies the ability of a single node to affect the overall resilience of a networked system \cite{Zhang2020}. Specifically, the resilience centrality \(c_{R}(i)\) of node \(i\) measures the importance of the node to system resilience by evaluating the relative change in resilience after its removal:
\[
c_{R}(i) = \frac{\beta_{\text{eff}}(G) - \beta_{\text{eff}}(G_i)}{\beta_{\text{eff}}(G)},
\]
where \(G_i\) denotes the subgraph obtained by removing node \(i\), and \(\beta_{\text{eff}}(G)\) captures the influence of the network structure on system resilience, defined as
\[
\beta_{\text{eff}}(G) = \frac{\langle d^2 \rangle}{\langle d \rangle},
\]
with \(\langle d \rangle = \frac{1}{N} \sum_{i=1}^{N} d_i\) being the average degree of the network, and \(\langle k^2 \rangle = \frac{1}{N} \sum_{i=1}^{N} d_i^2\) being the average squared degree.

For \textit{directed networks}, resilience centrality is expressed as
\[
c_{R}(i) = \frac{\sum_{j=1}^N (a_{ij} d_i^{\text{in}} + a_{ji} d_i^{\text{out}}) + d_i^{\text{in}} d_i^{\text{out}}}{\sum_{j=1}^N d_j^{\text{in}} d_j^{\text{out}}} - \frac{d_i^{\text{in}} + d_i^{\text{out}}}{\sum_{j=1}^N d_j^{\text{in}}},
\]
where \(d_i^{\text{in}}\) and \(d_i^{\text{out}}\) denote the in-degree and out-degree of node \(i\), respectively.

For \textit{undirected networks}, the expression simplifies to
\[
c_{R}(i) = \frac{2 \sum_{j=1}^N a_{ij} d_j + d_i^2}{\sum_{j=1}^N d_j^2} - \frac{2 d_i}{\sum_{j=1}^N d_j}.
\]

Resilience centrality has been applied to epidemic spreading dynamics modeled by the SIS process and validated by comparing its predictions with node importance rankings derived from system state changes and simulated collapse scenarios \cite{Zhang2020}.

\section{Resistance curvature}

\emph{Resistance curvature}\index{resistance curvature} is a discrete curvature measure based on the effective resistance between nodes in a network \cite{Devriendt2022}. The resistance curvature of node $i$, denoted $c_{RC}(i)$, is defined as
\[
c_{RC}(i) = 1 - \frac{1}{2} \sum_{j \in \mathcal{N}(i)} \omega_{ij} w_{ij},
\]
where $w_{ij}$ is the weight of edge $(i,j)$, and $\omega_{ij}$ is the effective resistance between nodes $i$ and $j$, computed using the pseudoinverse Laplacian $Q^{\dagger}$:
\[
\omega_{ij} = (e_i - e_j)^T Q^{\dagger} (e_i - e_j),
\]
with $e_i$ denoting the $i$th unit vector.

Intuitively, the effective resistance $\omega_{ij}$ measures the voltage difference between nodes $i$ and $j$ in the network. A redundant link, which connects nodes within a densely interconnected cluster, has low relative resistance, as its removal has little effect on overall connectivity. Conversely, a link with high relative resistance is critical for maintaining connectivity between its endpoints. Hence, if the neighborhood of node $i$ is tree-like with few short cycles, local relative resistances are high and the curvature $c_{RC}(i)$ is small. In contrast, in densely connected regions with many short cycles, resistances are lower and $c_{RC}(i)$ is larger, consistent with the notion of curvature.

\section{Resolvent betweenness (RB) centrality}

\emph{Resolvent betweenness (RB) centrality}\index{betweenness centrality!resolvent (RB)}, also known as $f$-betweenness, is a variant of betweenness centrality that quantifies the intermediate role of a node based on the matrix resolvent \cite{Estrada2010b}.  

Given the adjacency matrix $A$, the matrix resolvent is defined as
\[
f(A) = \sum_{k=0}^{\infty}s^k A^k = (I - sA)^{-1},
\]
where $s$ is a penalty factor satisfying $s \in \left(0, \frac{1}{\lambda_{\max}}\right)$, with $\lambda_{\max}$ being the largest eigenvalue of $A$. The parameter $s$ downweights longer walks, reflecting the intuition that short paths contribute more to communicability. Estrada and Higham \cite{Estrada2010b} suggest that a reasonable choice is $s = \frac{1}{N-1}$, based on comparison with the complete graph $K_N$.  

The entry $f(A)_{jl}$ measures the communicability between nodes $j$ and $l$. The resolvent betweenness of node $i$, denoted $c_{RB}(i)$, quantifies the overall relative change in communicability between all pairs of nodes when node $i$ is removed, i.e.,
\[
c_{RB}(i) = \frac{1}{(N-1)(N-2)} 
\sum_{j \neq i} \sum_{\substack{l \neq i \\ l \neq j}} 
\frac{f(A)_{jl} - f(A - E(i))_{jl}}{f(A)_{jl}},
\]
where $E(i)$ is the $N \times N$ matrix with nonzero entries only in row and column $i$, matching the positions of nonzero entries in $A$. In other words, $A - E(i)$ corresponds to the adjacency matrix obtained by removing all edges incident to node $i$.  

Thus, resolvent betweenness (RB) centrality captures the importance of node $i$ in maintaining communicability across the network.

\section{Resolvent centrality}

\emph{Resolvent centrality}\index{resolvent centrality}\index{subgraph centrality!resolvent}, also known as $f$-centrality, is a variant of subgraph centrality that measures node importance by counting the number of closed walks in a graph with a non-factorial scaling function \cite{Estrada2010b}. Given the adjacency matrix \(A\), the matrix resolvent is defined as
\[
f(A) = \sum_{k=0}^{\infty} s^k A^k = (I - sA)^{-1},
\]
where \(s\) is a penalty factor chosen such that \(s \in (0, 1/\lambda_{\max})\), with \(\lambda_{\max}\) being the largest eigenvalue of \(A\). Estrada and Higham \cite{Estrada2010b} suggest that a reasonable choice for \(s\) is \(s = 1/(N-1)\), motivated by comparing closed walk counts to those in the complete graph \(K_N\).

The resolvent centrality \(c_{\mathrm{res}}(i)\) of node \(i\) is then given by the diagonal entry
\[
c_{\mathrm{res}}(i) = f(A)_{ii} = \sum_{k=1}^{N} \frac{N-1}{N-1 - \lambda_k} \, v_k^2(i),
\]
where \(v_k(i)\) is the \(i\)-th component of the eigenvector \(v_k\) corresponding to the eigenvalue \(\lambda_k\) of \(A\). Thus, resolvent centrality captures the participation of node \(i\) in walks of all lengths, with longer walks penalized according to \(s\).

\section{Return Random Walk Gravity (RRWG) centrality}

\emph{Return Random Walk Gravity}\index{return random walk gravity (RRWG) centrality} (RRWG) centrality combines the concepts of return random walks, effective distance, and the gravity model to assess node importance in networks \cite{Curado2023}. For node \(i\), it is defined as
\[
c_{RRWG}(i) = \sum_{j \neq i} \frac{d_i d_j}{\left(D_{j|i}\right)^2},
\]
where \(d_i\) and \(d_j\) are the degrees of nodes \(i\) and \(j\), respectively, and \(D_{j|i}\) is the effective distance from node \(j\) to node \(i\), given by
\[
D_{j|i} = 1 - \log_2 \left( \max_{t \neq k} \left( p_{itj} \, p_{jki} \right) \right),
\]
with \(p_{itj}\) representing the probability of reaching node \(j\) from node \(i\) via a transition node \(t\), and \(p_{jki}\) representing the probability of returning from node \(j\) to node \(i\) via another transition node \(k\).

The RRWG centrality integrates three aspects: the gravity model captures the attractive power of a node based on its connectivity, the effective distance encodes both static and dynamic structural information of the network, and return random walks quantify a node's importance by accounting for the strength of indirect interactions with other nodes.

\section{RMD-weighted degree (WD) centrality}

The \emph{RMD-weighted degree}\index{RMD-weighted degree centrality} (WD) centrality, originally called weighted degree, is a centrality measure that evaluates node influence based on the \emph{remaining minimum degree (RMD) decomposition}\index{remaining minimum degree (RMD) decomposition} \cite{Yang2018}. The RMD decomposition iteratively removes the node with the minimum degree, capturing the structural importance of nodes in both local and global contexts.

Yang \textit{et al.} \cite{Yang2018} proposed that a node's influence largely depends on the importance of its neighbors. Accordingly, the RMD-weighted degree of node $i$ is defined as
\[
c_{RMD}(i) = \sum_{j \in \mathcal{N}(i)} \frac{Iter(j)}{MaxIter},
\]
where $Iter(j)$ is the iteration at which node $j$ is removed during the RMD decomposition, and $MaxIter$ is the total number of iterations. 

This formulation distinguishes the contributions of each neighbor and simultaneously accounts for both local connectivity and the global network structure. Nodes with high RMD-weighted degree are those connected to neighbors that are removed late in the RMD process, indicating structurally important and influential positions in the network.

\section{Routing betweenness centrality (RBC)}

\emph{Routing betweenness centrality}\index{betweenness centrality!routing} (RBC) is a generalization of traditional betweenness, load and flow betweenness centralities \cite{Dolev2010}. RBC quantifies the extent to which nodes are exposed to network traffic under any loop-free routing strategy. The RBC score of a node \(i\), denoted \(c_{\mathrm{RBC}}(i)\), is defined as
\[
    c_{\mathrm{RBC}}(i) = \sum_{j,k \in \mathcal{N}} \delta_{j,k}(i) \, T(j,k),
\]
where \(\delta_{j,k}(i)\) is the probability (or expected fraction of traffic) that a packet sent from node \(j\) to node \(k\) passes through node \(i\). The value of \(\delta_{j,k}(i)\) depends on the routing strategy: for deterministic shortest-path routing, \(\delta_{j,k}(i) = 1\) if node \(i\) lies on the shortest path between \(j\) and \(k\), and \(0\) otherwise; for probabilistic or load-balanced routing, \(\delta_{j,k}(i)\) represents the fraction of traffic routed through \(i\) according to the chosen routing algorithm.  The term \(T(j,k)\) denotes the number of packets sent from source node \(j\) to target node \(k\) and is usually defined by the traffic model. For uniform traffic scenarios, \(T(j,k) = 1\) for all node pairs.  

Compared to standard betweenness centrality, RBC accounts for both the routing probabilities and traffic volume, and it includes contributions from communications originating from or destined to the node under consideration.

 \section{Rumor centrality}

\emph{Rumor centrality}\index{rumor centrality} quantifies the likelihood that a node is the source of information spread under the susceptible-infected (SI) model \cite{Shah2010,Shah2011}. 
For a tree-structured infected subgraph \(G_N\) of \(N\) nodes, the rumor centrality of a candidate source node \(v\) is defined as
\[
R(v, G_N) = \frac{N!}{\prod_{u \in G_N} T^v_u},
\]
where \(T^v_u\) is the number of nodes in the subtree rooted at \(u\) when the tree is rooted at \(v\). The rumor centrality
\(R(v, G_N)\) counts the number of infection sequences consistent with the SI model if \(v\) were the source.

In regular trees, maximizing \(R(v, G_N)\) yields the exact maximum-likelihood rumor source (the \emph{rumor center}). 
For irregular trees, a randomized estimator weighted by \(R(v, G_N)\) is used, and for general graphs, \(R(v, G_N)\) is approximated on the breadth-first search (BFS) tree rooted at \(v\). 
Nodes with higher rumor centrality are more plausible spread origins and tend to be more influential in diffusion processes.

\section{\textit{s}-shell index}

The \emph{\textit{s}-shell index}\index{\textit{s}-shell index} is an extension of the traditional $k$-shell decomposition, designed to identify influential spreaders in weighted networks \cite{Liu2017}. Unlike the $k$-shell method, which removes nodes based on degree, the $s$-shell decomposition relies on node strength, incorporating asymmetric edge weights that reflect the potential of each link to facilitate spreading.

The strength \( s_i \) of node \( i \) is defined as the sum of the asymmetric weights of its outgoing links:
\[
s_i = \sum_{j \in \mathcal{N}(i)} w_{ij} = \sum_{j \in \mathcal{N}(i)} \left[ 1 + (d_i d_j^{out})^{a} \right],
\]
where \( w_{ij} \) quantifies the influence of edge \( (i,j) \) in the spreading process, \( d_i \) is the degree of node \( i \), \( d_j^{out}\) denotes the number of neighbors of node \( j \) not shared with \( i \), i.e.,
\[
d_j^{out} = \left| \{\, l \in \mathcal{N}(j) : l \notin \mathcal{N}(i) \cup \{i\} \,\} \right|,
\]
and \( a \) is a tunable parameter controlling the weight contribution (\( a = 0.5 \) in \cite{Liu2017}).

The formulation of \textit{s}-shell index reflects the idea that edges leading to new regions of the network contribute more to spreading than those confined within a node's local neighborhood. The resulting edge weights are asymmetric ($w_{ij} \neq w_{ji}$), capturing directional spreading potential. When $a = 0$, the $s$-shell index reduces to the standard $k$-shell index.

\section{Second-order centrality}

The \emph{second-order centrality}\index{second-order centrality} measures a node's importance as the standard deviation of return times of an unbiased random walk that starts and returns to the node \cite{Kermarrec2011}. For a node \(i\), the centrality \(c_{\mathrm{second}}(i)\) is defined as
\begin{equation*}
c_{\mathrm{second}}(i) = \sqrt{\frac{1}{K-1} \sum_{k=1}^{K} \Xi_i(k)^2 - \left[ \frac{1}{K-1} \sum_{k=1}^{K} \Xi_i(k) \right]^2},
\end{equation*}
where \(\Xi_i(k)\) denotes the \(k\)-th return time of a random walk starting and returning to node \(i\), and \(K\) is the total number of recorded return times. The random walk is unbiased, meaning that at each step, the walker chooses among all neighbors with equal probability, independent of node degree. This ensures that the standard deviation of return times reflects the node's position in the network rather than its degree. Nodes with lower second-order centrality values are considered more central in the network.

\vfill

\section{Seeley’s index}

\emph{Seeley’s index}\index{Seeley’s index} \cite{Seeley1949} is a counterpart to eigenvector centrality that modifies how a node distributes its influence. While eigenvector centrality assigns a node's importance recursively based on the sum of the centralities of its neighbors, effectively contributing fully to each neighbor, Seeley’s index assumes that a node divides its influence equally among its successors.  

Formally, each row of the adjacency matrix \(A\) is normalized by the node's out-degree, producing a row-stochastic matrix \(S = D^{-1} A\), where \(D\) is the diagonal matrix of out-degrees. The Seeley index is the principal eigenvector of \(S\), corresponding to the stationary distribution of the Markov chain defined by \(S\).  

If the graph is undirected, the stationary distribution of the Markov chain defined by \(S\) is proportional to node degrees. Therefore, Seeley’s index coincides with degree centrality in this case.

\section{Semi-global triangular centrality}

The \emph{semi-global triangular centrality}\index{semi-global triangular centrality} evaluates node importance based on the number of triangles associated with a node and its extended neighborhood, including nodes up to two hops away \cite{Namtirtha2022}. For a node \(i \in \mathcal{N}\), the centrality \(c_{st}(i)\) is defined as
\[
c_{st}(i) = \sum_{j \in \mathcal{N}^{(\leq r)}(i) \cup \{i\}} \frac{\Delta(j)}{2^{d_{ij}}},
\]
where \(\Delta(j)\) is the number of triangles that include node \(j\), \(d_{ij}\) is the shortest-path distance between nodes \(i\) and \(j\), and \(\mathcal{N}^{(\leq r)}(i)\) denotes the set of nodes within distance \(r=2\) from node \(i\).  

The semi-global triangular centrality was validated using the susceptible-infected-recovered (SIR) epidemic model on nine real-world networks and outperformed ten classical centrality measures in identifying effective spreaders.

\section{Semi-local degree and clustering coefficient (SLDCC)}

The \emph{semi-local degree and clustering coefficient}\index{semi-local!degree and clustering coefficient (SLDCC)} (SLDCC) is a hybrid centrality measure that evaluates the influence of a node based on its degree, clustering coefficient and the clustering coefficients of its second-level neighbors \cite{Berahmand2018}. Berahmand \textit{et al.} \cite{Berahmand2018} argue that a node with high degree but low clustering coefficient can be considered a structural hole in the network, bridging otherwise disconnected regions. Furthermore, if the sum of the clustering coefficients of a node's second-level neighbors is high, it indicates that these neighbors reside in a densely connected part of the network.

The centrality of node $i$ is defined as
\[
c_{SLDCC}(i) = \frac{d_i}{c_i + \frac{1}{d_i}} + \sum_{j \in \mathcal{N}^{(2)}(i)} c_j,
\]
where $d_i$ and $c_i$ are the degree and clustering coefficient of node $i$, and $\mathcal{N}^{(2)}(i)$ denotes the set of second-level neighbors of $i$. Thus, the SLDCC measure combines three aspects: the degree of the node, the negative effect of its own clustering coefficient, and the positive effect of the clustering coefficients of its second-level neighbors.

\section{Semi-local iterative algorithm (semi-IA)}

The \emph{semi-local iterative algorithm}\index{semi-local!iterative algorithm (semi-IA)} (semi-IA) is an iterative centrality measure in which the importance of a node depends on the importance of other nodes, weighted by both the shortest-path distance and the number of shortest paths (NSPs) connecting them \cite{Luan2021}. 

For a node \(i\), the update rule at iteration \(t+1\) is defined as
\[
X_i[t+1] = \sum_{j \in \mathcal{N}^{(\leq l)}(i)} \frac{n_{ij}^{\gamma}}{d_{ij}} X_j'[t], \quad X_i[0] = 1, \quad i = 1, \dots, N,
\]
where \(\mathcal{N}^{(\leq l)}(i)\) is the set of nodes within distance \(l\) from node \(i\), \(n_{ij}\) is the number of shortest paths between \(i\) and \(j\), \(0 \leq \gamma \leq 1\) is the NSP weighting factor, and \(X_j'[t]\) is the normalized influence of node \(j\) at iteration \(t\):
\[
X_j'[t] = \frac{X_j[t]}{\| X[t] \|}.
\]

Luan \textit{et al.} \cite{Luan2021} suggest setting the truncated radius \(l = 3\) for networks with fewer than 1000 nodes, and \(l = 5\) for larger networks, with \(\gamma = 0.2\). The semi-IA centrality of node \(i\) is the value \(\tilde{X}_i=\lim_{t \rightarrow \infty}X_i[t]\) at the steady state of the iteration.

\section{Semi-local ranking centrality (SLC)}

The \emph{semi-local ranking centrality}\index{semi-local! ranking centrality (SLC)} (SLC) quantifies node influence based on random walks within the \(l\)-neighborhood of each node \cite{Dong2018}. A node is considered influential if the random walk frequently encounters other influential nodes in its neighborhood. The SLC measure applies a PageRank-like process to capture this local and semi-local influence. 

At each step, the random walk either adds an immediate neighbor \(j\) of node \(i\) to the path with probability \(p_{ij}\), or stops. The probability of adding node $j$ to the random walk starting from node $i$ is defined as
\begin{equation*}
   p_{ij} = a \Biggl[ \lambda \frac{d_i}{d_{\max}} + (1-\lambda) HP(i,j) \Biggr],
\end{equation*}
where $d_i$ is the degree of node $i$, $d_{\max}$ is the maximum degree in the network, and 
\[
HP(i,j) = \frac{|\mathcal{N}(i) \cap \mathcal{N}(j)|}{\min(d_i, d_j)}
\] 
measures the similarity between nodes $i$ and $j$, with $\mathcal{N}(i)$ denoting the set of immediate neighbors of node $i$. 
The parameter $\lambda \in (0,1)$ balances the contributions of node degree and neighborhood similarity, and $a \in (0,1)$ is a decay factor controlling the probability of continuing the random walk.

Let $S_t(i)$ denote the set of nodes visited by the $t$-th random walker starting from node $i$. 
The semi-local ranking centrality of node $i$ is defined as
\begin{equation*}
   c_{\mathrm{SLC}}(i) = \sum_{t=1}^{T} |S_t(i)|,
\end{equation*}
where $T$ is the total number of random walkers. 
Dong \textit{et al.} \cite{Dong2018} suggest using $T=100$, $a=0.9$, and $\lambda=0.85$, with the length of each random walk limited to the network diameter.

\section{Shapley value}

A class of game-theoretic network centrality measures based on the \emph{Shapley value}\index{Shapley value} from the cooperative game theory is discussed in \cite{Michalak2013}. The Shapley value \(SV(i)\) of node \(i\) is defined as the average marginal contribution of \(i\) to all possible coalitions of nodes \(C \subseteq \mathcal{N} \setminus \{i\}\):
\begin{equation*} 
   SV(i) = \sum_{C \subseteq \mathcal{N} \setminus \{i\}} \frac{(|C|-1)!(|\mathcal{N}|-|C|)!}{|\mathcal{N}|!} \left[ v(C \cup \{ i \}) - v(C) \right],
\end{equation*}
where \(v\) is a \emph{characteristic function} that assigns to every coalition \(C\) a real number representing the value or performance of the coalition \cite{Shapley1953}.

Michalak \textit{et al.} \cite{Michalak2013} proposed efficient linear-time algorithms to compute Shapley value-based centralities for specific definitions of the characteristic function \(v\):

\begin{enumerate}
    \item \textit{Game 1:} \(v(C)\) is the number of nodes reachable from the coalition \(C\) in at most one hop. In that case, the Shapley value \(SV_1(i)\) of node \(i\) reduces to 
    \begin{equation*} 
        SV_1(i) = \frac{1}{1+d_i} + \sum_{j \in \mathcal{N}(i)} \frac{1}{1+d_j},
    \end{equation*}
    where \(d_i\) is the degree of node \(i\).

    \item \textit{Game 2:} \(v(C)\) is the number of nodes that are either in \(C\) or adjacent to at least \(k\) neighbors in \(C\).  In that case, the Shapley value \(SV_2(i)\) of node \(i\) reduces to 
    \begin{equation*} 
        SV_2(i) = \min \left(1, \frac{k}{1+d_i}\right) + \sum_{j \in \mathcal{N}(i)} \max \left(0, \frac{d_j - k + 1}{d_j(1+d_j)}\right).
    \end{equation*}
    For \(k = 1\), this game is equivalent to Game 1.

    \item \textit{Game 3:} \(v(C)\) is the number of nodes that are at most \(d_{\max}\) hops away from \(C\).  In that case, the Shapley value \(SV_3(i)\) of node \(i\) reduces to 
    \begin{equation*} 
        SV_3(i) = \frac{1}{1+|s_i(d_{\max})|} + \sum_{j \in s_i(d_{\max})} \frac{1}{1+|s_j(d_{\max})|},
    \end{equation*}
    where \(s_i(d_{\max})\) denotes the set of nodes within distance \(d_{\max}\) from node \(i\).
\end{enumerate}

\section{Shapley value based information delimiters (SVID)}

\emph{Shapley Value based Information Delimiters}\index{Shapley value!information delimiters (SVID)} (SVID) is a group-based, game-theoretic centrality measure that uses the Shapley value to quantify each node's contribution to network connectivity and information flow \cite{Saxena2018}. The method targets nodes whose removal would either increase shortest-path distances among remaining nodes or reduce the number of alternative paths connecting them. Intuitively, nodes with fewer common neighbors are more critical, as they limit the availability of alternative paths in the network.  

Saxena \textit{et al.} \cite{Saxena2018} propose an efficient algorithm to rank nodes by their marginal contributions across all possible coalitions. The marginal contribution of a link $(i,j)$ to the Shapley value of nodes $i$ and $j$ is defined as
\begin{equation*}
    \text{MC}(i,j) = \frac{1}{(K+1)(K+2)},
\end{equation*}
where $K$ is the number of common neighbors shared by nodes $i$ and $j$. At each iteration, the algorithm selects the node with the highest Shapley value, removes it from the graph $G$, and reduces the Shapley values of its neighbors according to their marginal contributions. The final ranking reflects each node’s positional power and functional influence in maintaining network connectivity.

\section{Shell clustering coefficient (SCC)}

The \emph{shell clustering coefficient}\index{shell clustering coefficient (SCC)} (SCC) quantifies node influence by considering the hierarchical similarity between a node and its neighbors \cite{Zareie2020}. Zareie \textit{et al.} introduce the \emph{shell vector}\index{shell vector} of node \(i\) as
\[
sv(i) = \left(|N_{ks}^{(1)}(i)|,...,|N_{ks}^{(f)}(i)|\right),
\]
where \(|N_{ks}^{(k)}(i)|\) is the number of neighbors of node \(i\) belonging to hierarchy \(k\) with respect to the $k$-shell centrality, and \(f\) is the maximum hierarchy in the network.

The \emph{shell clustering coefficient} \(SCC(i)\) of node \(i\) is defined as
\begin{equation*}
   SCC(i) = \sum_{j \in \mathcal{N}(i)} \left[ 2 - \mathrm{corr}[sv(i), sv(j)] + \left(\frac{2 d_j}{\max_l d_l} + 1 \right) \right],
\end{equation*}
where \(\mathrm{corr}[sv(i), sv(j)]\) is the Pearson correlation between the shell vectors of nodes \(i\) and \(j\), and \(d_j\) is the degree of neighbor \(j\). 
Intuitively, a high correlation between node \(i\) and its neighbors, combined with neighbors of low degree, negatively affects the spreading ability of node \(i\).

\section{Short-Range Interaction Centrality (SRIC)}
The \emph{Short-Range Interaction Centrality (SRIC)}\index{short-range interaction centrality (SRIC)} index, originally called the key-borrower index (KBI), is a power index based on the concept of individual and group influence of nodes in a network \cite{Aleskerov2014,Aleskerov2020,AleskerovBook}. Each node \(i\) is assumed to have an individual threshold of influence \(q_i\), which represents the level at which this node becomes affected. This threshold can be specified externally based on domain knowledge or determined from the network structure, for instance, as a function of the degree of each node. A group of nodes \(\Omega(i) \subset \mathcal{N}\) is called \emph{critical} for node \(i\) if their collective influence exceeds the threshold \(q_i\), i.e.,
\begin{equation*}
    \sum_{k \in \Omega(i)} a_{ki} \geq q_i.
\end{equation*}
A node \(k\) is termed \emph{pivotal} for the group \(\Omega(i)\) if its removal renders the group non-critical. The set of pivotal members of \(\Omega(i)\) is denoted by \(\Omega^{p}(i)\).

The SRIC index considers only direct and indirect influence through one intermediate node, i.e., short-range influence. Formally, the initial adjacency matrix \(A\) is transformed into a matrix of direct influence \(P\), where \(D\) is the out-degree matrix. The indirect influence \(p_{ihj}\) of node \(i\) on node \(j\) via an intermediate node \(h\) is defined as
\begin{equation*}
    p_{ihj} = 
    \begin{cases}
        \frac{\min(a_{ih}, a_{hj})}{D_{kk}}, & \text{if } a_{ih} > 0, \ a_{hj} > 0, \ i \neq j \neq h, \\
        0, & \text{otherwise}.
    \end{cases}
\end{equation*}

The SRIC centrality of node \(i\) is defined as the average of the normalized short-range influence \(\chi_i(j)\) of node \(i\) on all nodes \(j \in \mathcal{N}\). This is evaluated by considering all critical groups \(\Omega_k(j)\) in which node \(i\) is pivotal:
\begin{equation*}
    c_{SRIC}(i) = \frac{1}{N} \sum_{j \in \mathcal{N}}  \frac{\chi_i(j)}{\sum_{h \in \mathcal{N}} \chi_h(j)},
\end{equation*}
where 
\begin{equation*}
    \chi_i(j) = \sum_{k: i \in \Omega^{p}_k(j)} \frac{p_{ij} + \sum_{h \in \Omega_k(j)} p_{ihj}}{|\Omega_k(j)|}.
\end{equation*}

The SRIC index has been applied in diverse domains, including the identification of influential countries in global food trade networks \cite{Aleskerov2017b}, analysis of financial \cite{Aleskerov2020}, global arms transfer \cite{Shvydun2019}, international conflict \cite{Aleskerov2016}, and international migration networks \cite{Aleskerov2016b}, as well as the detection of key actors in terrorist networks \cite{Shvydun2019} and citation networks of economic journals \cite{Aleskerov2016c}.

\section{Shortest cycle closeness (SCC) centrality}

\emph{Shortest cycle closeness}\index{closeness centrality!shortest cycle} (SCC) centrality is a variation of closeness centrality that accounts for the lengths of the shortest cycles involving a node \cite{Zhou2018}. Zhou \textit{et al.} introduce the concept of the \emph{shortest cycle}\index{shortest cycle} containing two nodes, which provides an alternative to the shortest path by effectively forming two independent paths between the nodes.  

For node \(i\), the SCC centrality \(c_{SCC}(i)\) is defined as
\[
c_{SCC}(i) = \frac{1}{\sum_{j=1}^{N} l_{ij}},
\]
where \(l_{ij}\) is the length of the shortest cycle that contains both nodes \(i\) and \(j\). If no cycle exists between \(i\) and \(j\), then \(l_{ij} = d_{ij} + N\), where \(d_{ij}\) is the shortest path distance between \(i\) and \(j\), and \(N\) is the total number of nodes in the network.

The SCC centrality generalizes traditional closeness centrality by emphasizing nodes that participate in short cycles, thereby capturing robustness in networks where direct shortest paths may be unavailable.

\section{Silent node rank (SNR)}

\emph{Silent node rank}\index{silent node rank (SNR)} (SNR), also known as LurkerRank (LR) \cite{Tagarelli2013}, is a spectral centrality measure designed to identify nodes that play a passive role in information networks, which are the nodes that consume more information than they produce and often remain unnoticed \cite{Interdonato2015}. Such silent nodes typically have low connectivity and are common in many systems, including leechers in P2P networks or lurkers in online social networks.

While the in-degree/out-degree ratio provides a simple measure of silence, it ignores the influence of neighbors and often produces many nodes with identical scores. SNR improves upon this by incorporating neighbor behavior: a node is considered more silent if its in-neighbors are active and its out-neighbors are themselves silent. This network-aware approach yields a more accurate and diversified ranking of silent nodes.

Formally, the SNR score \(r_i\) of node \(i\) in a weighted, directed network is given by
\begin{equation*}
\begin{split}
r_i &= \alpha \Bigg(
\frac{1}{d_i^{\mathrm{out}} + 1} 
\sum_{j \in \mathcal{N}^{\mathrm{in}}(i)} w_{ji} \frac{d_j^{\mathrm{out}} {+} 1}{d_j^{\mathrm{in}} {+} 1} r_j 
\Bigg)
\Bigg(
1 {+} \frac{d_i^{\mathrm{in}} {+} 1}{\sum_{j \in \mathcal{N}^{\mathrm{out}}(i)} (d_j^{\mathrm{in}} {+} 1)} 
\sum_{j \in \mathcal{N}^{\mathrm{out}}(i)} w_{ij} \frac{d_j^{\mathrm{in}} {+} 1}{d_j^{\mathrm{out}} {+} 1} r_j
\Bigg) \\
&\quad + \frac{1{-}\alpha}{N},
\end{split}
\end{equation*}
where \(w_{ij}\) is the edge weight, \(\alpha\) is the damping factor, \(\mathcal{N}^{\mathrm{in}}(i)\) and \(\mathcal{N}^{\mathrm{out}}(i)\) are the in- and out-neighbors of \(i\), and \(d_j^{\mathrm{in}}\) and \(d_j^{\mathrm{out}}\) are the in- and out-degrees of node \(j\). The first term downweights the influence of node \(i\)’s in-neighbors based on its out-degree, while the second term boosts the score according to how silent its out-neighbors are, adjusted by \(i\)’s in-degree.

For undirected and unweighted networks, the SNR score simplifies to
\begin{equation*}
r_i = \alpha \left( \frac{1}{d_i + 1} \sum_{j \in \mathcal{N}(i)} r_j \right)
\left( 1 + \frac{d_i + 1}{d_i + \sum_{j \in \mathcal{N}(i)} d_j} \sum_{j \in \mathcal{N}(i)} r_j \right)
+ \frac{1-\alpha}{N}.
\end{equation*}

\section{Similarity-based PageRank}

The \emph{similarity-based PageRank}\index{PageRank!similarity-based} is a variant of the PageRank algorithm designed to identify influential nodes by guiding random walks according to the structural similarity between nodes \cite{Zhang2018,Zhao2021}. 

Let \(d_{\max}\) denote the largest degree in the graph \(G\). For each node \(i\), a \((d_{\max}+1)\)-dimensional vector is constructed containing the degree of node \(i\) and the degrees of its \(|\mathcal{N}(i)|\) neighbors; any remaining entries are set to a small constant \(10^{-8}\). The vector entries are \emph{sorted} to ensure a consistent alignment for KL-divergence computation. This vector is then normalized to form a probability vector \(p(i)\). 

The dissimilarity between nodes \(i\) and \(j\) is quantified using the symmetric Kullback-Leibler (KL) divergence:
\[
r_{ij} = D_{\mathrm{KL}}(p(i)\,\|\,p(j)) + D_{\mathrm{KL}}(p(j)\,\|\,p(i)),
\]
where larger values of \(r_{ij}\) indicate greater structural differences between nodes \(i\) and \(j\).

The structural similarity between nodes \(i\) and \(j\) is defined as
\[
s_{ij} = 1 - \frac{r_{ij}}{r_{\max}+1},
\]
where \(r_{\max} = \max_{(i,j)} r_{ij}\). Higher structural similarity \(s_{ij}\) corresponds to a higher probability of transition between adjacent nodes. The resulting similarity matrix \(S = [s_{ij}]\) is then used to generate the transition probability matrix \(P\) for the PageRank algorithm, which in turn determines the importance of each node.

\section{Simulations-based LRIC (LRIC-sim) index}
\emph{Simulations-based LRIC}\index{long-range interaction centrality (LRIC)!simulations-based} (LRIC-sim) quantifies the influence of nodes through chain reactions in the network, also referred to as domino or contagion effects \cite{MS2018}. Similar to LRIC, each node \(i\) has a threshold \(q_i\) that represents the level of influence required from a subset of its neighbors to become affected.  This threshold can be specified externally based on domain knowledge or determined from the network structure, for instance, as a function of the degree of each node. A group of nodes \(\Omega(i) \subset \mathcal{N}\) is called \emph{critical} for node \(i\) if their collective influence exceeds the threshold \(q_i\), i.e.,
\begin{equation*}
    \sum_{k \in \Omega(i)} a_{ki} \geq q_i.
\end{equation*}
A node \(k\) is termed \emph{pivotal} for the group \(\Omega(i)\) if its removal renders the group non-critical.

In each simulation step \(t\), \(k\) nodes are randomly selected to malfunction, which may trigger a cascade of failures represented by the sequence \(S_t\). The influence of node \(i\) on node \(j\) is quantified as the fraction of simulation runs in which the failure of \(i\) causes the failure of \(j\), considering only cases where \(i\) is pivotal, meaning that if \(i\) had not failed, \(j\) would not have appeared in the sequence \(S_t\). The LRIC-sim centrality of node \(i\) is then obtained by aggregating its indirect influence across all other nodes in the network, capturing the overall potential of \(i\) to trigger cascading failures.

\section{SingleDiscount}
\emph{SingleDiscount}\index{SingleDiscount} is a degree-discount heuristic for identifying influential nodes in a network~\cite{Chen2009}. The method proceeds iteratively as follows:

\begin{enumerate}
    \item Identify the node with the highest degree in the current network.
    \item Select this node as an influential node and remove it from the network.
    \item Update the degrees of all remaining nodes to account for the removal.
    \item Repeat the process until the desired number of influential nodes is selected.
\end{enumerate}

Hence, SingleDiscount considers the diminishing influence of neighboring nodes once a high-degree node has been selected.

\section{Spanning tree centrality (STC)}

\emph{Spanning tree centrality}\index{spanning tree centrality (STC)} (STC) is a centrality measure based on the enumeration of spanning trees in a graph \cite{Qi2015}. In a spanning tree, a node \(i\) can play two roles: either as a leaf, whose removal does not disconnect the tree, or as a cut-vertex, whose removal disconnects the tree. 

The STC of node \(i\) is defined as
\[
c_{STC}(i) = t_G - d_i \, t_{G_i},
\]
where \(d_i\) is the degree of node \(i\), \(t_G\) is the total number of spanning trees in \(G\), and \(G_i\) is the subgraph obtained by removing node \(i\) from \(G\). The term \(d_i \, t_{G_i}\) counts the number of spanning trees in which node \(i\) is a leaf. Therefore, \(c_{STC}(i)\) measures the number of spanning trees in which node \(i\) acts as a cut-vertex, capturing its importance in maintaining the connectivity of the network.

\section{SpectralRank (SR)}

\emph{SpectralRank}\index{SpectralRank (SR)} (SR) is a parameter-free extension of LeaderRank designed to evaluate node propagation capability in networks \cite{Xu2019}. Similar to LeaderRank, SR introduces a ground node \(N+1\) that connects bidirectionally to all nodes in the network \(G\). Each node \(i\) is assigned a score \(s_i[t]\) at discrete time \(t\), representing its propagation potential.  

The initial scores are
\[
s_{N+1}[0] = 0 \quad \text{for the ground node,} \qquad s_i[0] = 1 \quad \text{for all other nodes } i \in \mathcal{N}.
\]

At each time step, the score of node \(i\) is updated based on the scores of its neighbors:
\begin{align*}
\tilde{s}_i[t+1] &= c \sum_{j=1}^{N+1} a_{ij} \, s_j[t], \\
s_i[t+1] &= \frac{\tilde{s}_i[t+1]}{\max_j \tilde{s}_j[t+1]},
\end{align*}
where \(c = 1 / \lambda_{\max}\), and \(\lambda_{\max}\) is the leading eigenvalue of the augmented adjacency matrix
\[
\tilde{A} =
\begin{bmatrix}
A & \mathbf{1} \\
\mathbf{1}^T & 0
\end{bmatrix},
\]
which includes the ground node. Here, \(A\) is the original \(N \times N\) adjacency matrix, and \(\mathbf{1}\) is a column vector of ones. The SpectralRank of node \(i\) is defined as the steady-state score
\[
\Tilde{s}_i = \lim_{t \to \infty} s_i[t],
\]
which quantifies the node’s long-term propagation influence in the network.

\section{Spreading probability (SP) centrality}

The \emph{spreading probability}\index{spreading probability (SP) centrality} (SP) centrality is a hybrid measure that incorporates shortest distances, the number of shortest paths, and the transmission rate to quantify node influence \cite{Bao2017}. The SP centrality of node \(i\) is defined as
\begin{equation*}
    c_{SP}(i) = \sum_{j \in \mathcal{N}^{(\leq l)}(i)} \sigma_{ij} \left( \frac{1}{\langle d \rangle} \right)^{d_{ij}},
\end{equation*}
where \(\mathcal{N}^{(\leq l)}(i)\) is the set of nodes within \(l\) hops from node \(i\), \(\sigma_{ij}\) is the number of shortest paths between nodes \(i\) and \(j\), \(\langle d \rangle\) is the average degree of the network, and \(d_{ij}\) is the shortest distance from node \(i\) to node \(j\). Bao \textit{et al.} \cite{Bao2017} consider \(l=3\) and use \(\frac{1}{\langle d \rangle}\) to approximate the transmission rate \(\beta\). Hence, the term \(\sigma_{ij} \left( \frac{1}{\langle d \rangle} \right)^{d_{ij}}\) approximates the probability that node \(j\) is infected by node \(i\). SP centrality assigns higher values to nodes that can reach many others through multiple short paths, reflecting both local connectivity and spreading potential within the network.

\section{Spreading strength}

\emph{Spreading strength}\index{spreading strength} is a topological measure that quantifies a node's influence in spreading processes, taking into account its indirect propagation through the neighborhood \cite{Yu2019}. The spreading strength $c_{ss}(i)$ of node $i$ is defined as
\[
c_{ss}(i) = \sum_{j \in \mathcal{N}(i)} \left( 1 + d_j^{out} \left( 1 + \frac{|D_{ij,2}|}{4} \right)^{\alpha} \right),
\]
where $\mathcal{N}(i)$ is the set of neighbors of node $i$, $d_j^{out} = |\{ l \in \mathcal{N}(j) : l \notin \mathcal{N}(i) \cup \{i\} \}|$ is the number of neighbors of $j$ outside $i$'s neighborhood, $|D_{ij,2}|$ is the number of paths of length 2 connecting $i$ and $j$, and $\alpha$ is a tunable parameter (e.g., $\alpha = 0.5$). This formulation captures both the local connectivity of neighbors and the potential for spreading beyond the immediate neighborhood.

\vfill

\section{Stochastic Approach for Link Structure Analysis (SALSA)}

The \emph{Stochastic Approach for Link Structure Analysis}\index{SALSA} (SALSA) \cite{Lempel2001} is a variant of HITS centrality designed to mitigate the Tightly Knit Community (TKC) effect, in which rankings are biased toward small, highly interconnected communities. Unlike HITS, which directly uses the adjacency matrix \(A\), SALSA normalizes \(A\) based on node out-degrees.

Formally, let \(D = \mathrm{diag}(d_1, \dots, d_N)\) be the diagonal matrix of node out-degrees, where \(d_i = \sum_{j=1}^N a_{ij}\). The row-normalized adjacency matrix $\bar{A} = D^{-1} A$ is a row-stochastic matrix, where each row represents a probability distribution over the outgoing edges of the corresponding node. Thus, \(\bar{A}\) serves as the transition matrix of a Markov chain corresponding to a random walk on the network.

SALSA computes hub and authority scores of nodes via
\begin{equation*}
\begin{cases}
h = a \cdot \bar{A}^T, \\
a = h \cdot \bar{A},
\end{cases}
\end{equation*}
where \(h\) and \(a\) are the hub and authority score vectors, respectively.

SALSA is based on random walks on the bipartite graph \(\hat{G} = (\mathcal{N}_{hub}, \mathcal{N}_{authority}, E)\), where
\[
\mathcal{N}_{hub} = \{v_h \mid \sum_{j=1}^N a_{v_h j} > 0\}, \quad
\mathcal{N}_{authority} = \{v_a \mid \sum_{j=1}^N a_{j v_a} > 0\}, \quad
E = \{(v_h, v_a) \mid a_{v_h v_a} = 1\}.
\]
With \(\bar{A} = D^{-1}A\) the row-normalized adjacency matrix, the hub and authority scores are given by the stationary distributions of the Markov chains
\[
\bar{A} \bar{A}^T \quad (\text{hub-hub chain}), \qquad
\bar{A}^T \bar{A} \quad (\text{authority-authority chain}).
\]

Boldi and Vigna \cite{Boldi} observe that SALSA does not require iterative computation. One first identifies the connected components of the symmetric graphs \(A^T A\) (for authorities) and \(A A^T\) (for hubs). A node’s SALSA score is then calculated as the product of its in-degree fraction within its component and the component’s size relative to the total number of nodes \(N\) in \(G\).

\section{Stress centrality}

\emph{Stress centrality}\index{stress centrality} quantifies the amount of "stress" a node experiences due to the network's activity, quantified by how often the node lies on the shortest paths connecting pairs of other nodes \cite{Shimbel1953}. For a node \(i\), the stress centrality, denoted by \(c_{Stress}(i)\), is defined as the total number of shortest paths between all pairs of nodes that pass through \(i\):
\begin{equation*}
    c_{Stress}(i) = \sum_{j \neq k \neq i} \sigma_{jk}(i),
\end{equation*}
where \(\sigma_{jk}(i)\) denotes the number of shortest paths from node \(j\) to node \(k\) that pass through node \(i\).

\section{Synthesize centrality (SC)}

\emph{Synthesize centrality}\index{synthesize centrality (SC)} (SC) is a composite measure proposed to identify opinion leaders in social networks \cite{HLiu2013}. It combines multiple aspects of centrality, including local connectivity, intermediary influence, and global reach, into a single index. The synthesized centrality of a node \(i\) is defined as
\begin{equation*}
    c_{\mathrm{SC}}(i) = \frac{\overline{c}_d(i) + \overline{c}_b^2(i)}{\overline{c}_c^2(i)},
\end{equation*}
where \(\overline{c}_d(i)\), \(\overline{c}_b(i)\), and \(\overline{c}_c(i)\) denote the normalized degree, betweenness, and closeness centrality scores of node \(i\), respectively.

Nodes with high SC values tend to have both strong local and intermediary influence (via degree and betweenness) while maintaining relatively short average distances to other nodes (low closeness denominator), making them effective opinion leaders within the network.

\section{Subgraph centrality}
\emph{Subgraph centrality}\index{subgraph centrality}, also known as \emph{communicability centrality}, quantifies the extent to which a node participates in all subgraphs of a network. It is computed based on the number of closed walks of different lengths, where a walk is considered \emph{closed}\index{walk!closed} if its starting and ending nodes coincide~\cite{Estrada}. The centrality \(c_s(i)\) of node \(i\) is the sum of weighted closed walks of all lengths starting and ending at node \(i\)
\begin{equation*}
    c_{s}(i) = \sum_{k=0}^{\infty}{\frac{(A^k)_{ii}}{k!}}=[e^A]_{ii}=\sum_{j=1}^{N}{\left( v_j(i) \right)^2e^{\lambda_j}},
\end{equation*}
where \(A\) denotes the adjacency matrix of the graph $G$, \(v_j(i)\) is the \(i\)-th component of the eigenvector \(v_j\) associated with the eigenvalue \(\lambda_j\) of \(A\).

\section{Super mediator degree (SMD)}

\emph{Super mediator degree}\index{super mediator degree (SMD)} (SMD) is a vitality-based centrality measure that quantifies the importance of a node by evaluating the reduction in global influence spread caused by its removal \cite{Saito2016}. The influence of each node is approximated using repeated bond percolation simulations. In each simulation \(m = 1, \dots, M\), a percolated graph \(G_m\) is generated by independently retaining each edge of the original graph \(G\) with probability \(\mu\). A node is considered a super-mediator if its removal substantially decreases the average influence degree under the underlying diffusion model.

The SMD score of node \(i\) is defined as
\begin{equation}
c_{\mathrm{SMD}}(i) =
\frac{1}{M} \sum_{j \in \mathcal{N}} \sum_{m=1}^{M} |R(j, G_m)| \, \kappa(j)
- \frac{1}{M} \sum_{j \in \mathcal{N} \setminus \{ i \}} \sum_{m=1}^{M} |R(j, G_m \setminus \{i\})| \, \kappa(j),
\end{equation}
where \(R(j, G_m)\) denotes the set of nodes reachable from \(j\) in the percolated graph \(G_m\), and \(\kappa(j)\) is the probability that node \(j\) becomes an initial active node. This formulation captures the expected decrease in overall influence due to the removal of node \(i\), averaged across \(M\) independent percolation instances.

For general graphs, Saito \textit{et al.} \cite{Saito2016} set the initial activation probability uniformly as \(\kappa(j) = 1/N\) and define the edge retention probability as \(\mu = r / \overline{d}\), where \(r \in \{0.25, 0.5, 1\}\) and \(\overline{d}\) is the average out-degree of \(G\). Nodes with high SMD scores tend to appear frequently in longer diffusion paths but less often in short ones, reflecting their role as critical bridges that facilitate sustained, deep information propagation across the network.

\section{Support}

\emph{Support}\index{support} measures the ability of an individual to exchange favors and safely transact with others, based on a combination of network position and repeated interactions \cite{Jackson2020}. A relationship between nodes \(i\) and \(j\) is considered \emph{supported} if they have at least one friend in common.  

The support of node \(i\), denoted \(c_{\mathrm{sup}}(i)\), is defined as the number of neighbors that share at least one common neighbor with \(i\):
\[
    c_{\mathrm{sup}}(i) = \left| \{ j \in \mathcal{N}(i) : (A^2)_{ij} > 0 \} \right|,
\]
where \(A\) is the adjacency matrix of the network and \(\mathcal{N}(i)\) is the set of neighbors of node \(i\). The support measure captures the extent to which a node’s relationships are reinforced through shared connections, reflecting trust and reliability in the network.

\section{\(\boldsymbol{\theta}\)-Centrality}

The \emph{\(\theta\)-centrality}\index{\(\theta\)-centrality}, also known as the improved method \(\theta\), the KS-\(k\) method, the \(k\)-shell distance method, or the distance-to-network-core (DNC) method, is a \(k\)-shell decomposition-based approach designed to differentiate the spreading influence of nodes that share the same \(k\)-core value~\cite{Liu2013}. Denote by \(k_s(i)\) the \(k\)-shell index of node \(i\). The \(\theta\)-centrality of node \(i\), denoted by \(c_{\theta}(i)\), depends both on its \(k\)-core value and on its distance from the network core:
\begin{equation*}
c_{\theta}(i) = \bigl(\max_{l} k_s(l) - k_s(i) + 1\bigr) \sum_{j \in J} d_{ij},
\end{equation*}
where \(d_{ij}\) is the shortest-path distance between nodes \(i\) and \(j\), and \(J\) denotes the set of nodes with the maximum \(k\)-shell index, i.e., the innermost core of the network:
\[
J = \{\, j \in \mathcal{N} \mid k_s(j) = \max_{l} k_s(l) \,\}.
\]

Nodes with lower \(\theta\)-centrality values are considered more influential, as they are closer to the network core and occupy structurally central positions. If the network is disconnected, \(\theta\)-centrality is undefined because shortest-path distances cannot be computed between all node pairs.

\section{Third Laplacian energy centrality (LC)}

The \emph{third Laplacian energy centrality}\index{third Laplacian energy centrality} (LC) is an eigenvalue-based method for identifying influential nodes in complex networks, derived from the concept of Laplacian energy \cite{Zhao2023}. It represents a special case of the \(k\)-th Laplacian energy centrality with \(k = 3\). The centrality \(c_{LC}(i)\) of node \(i\) is defined as
\[
c_{LC}(i) = E^k_L(G) - E^k_L(G_i),
\]
where \(G_i\) denotes the subgraph of \(G\) obtained by removing node \(i\), and \(E^k_L(G)\) is the \(k\)-th Laplacian energy of the network \(G\), given by
\[
E^k_L(G) = \sum_{j=1}^{N} \mu_j^k,
\]
where \(\mu_1, \dots, \mu_N\) are the eigenvalues of the Laplacian matrix of \(G\).  

Zhao and Sun~\cite{Zhao2023} demonstrated that the case \(k = 3\) yields superior performance in identifying influential nodes compared with other values of \(k\). The third Laplacian energy centrality is particularly effective for detecting influential spreaders in complex networks and is typically evaluated using the susceptible-infected-recovered (SIR) epidemic model.

\section{Top candidate (TC) method}

The \emph{Top Candidate} (TC) method \index{top candidate (TC) method} is an iterative voting-based algorithm originally designed to detect experts in a community and later applied to identify innovators and early adopters in social networks \cite{Sziklai2018,Sziklai2022}. The TC method focuses on identifying a \emph{stable set} of mutually reinforcing expert candidates whose nominations support one another. The approach relies on the assumption that experts tend to nominate other experts more reliably than non-experts. The algorithm proceeds through the following steps:

\begin{enumerate}
    \item \textit{Initialization:} all nodes are initially considered experts.

    \item \textit{Nomination:} each node selects an \( \alpha \)-fraction of its most popular neighbors as nominees, with popularity defined as the (weighted) in-degree and \( \alpha \in [0, 1] \).

    \item \textit{Elimination:} nodes that receive no nominations are removed from the expert set. All nominations issued by these removed nodes are discarded as well.

    \item \textit{Update:} the elimination in Step 3 may cause additional nodes to lose all incoming nominations. These nodes are likewise removed, and the process is repeated until no further removals occur.
\end{enumerate}

The resulting set of nodes constitutes the \emph{stable expert set}. A set \( S \subseteq \mathcal{N} \) is stable if (i) every node in \( S \) is nominated by at least one other node in \( S \) and (ii) all nominees of any node in \( S \) are also contained in \( S \). The parameter \( \alpha \) controls the exclusiveness of the selection: smaller values of \( \alpha \) lead to more restrictive nomination sets and thus smaller expert groups, whereas larger values of \( \alpha \) generally produce more inclusive and larger stable sets.

\section{Topological centrality (TC)}

\emph{Topological centrality}\index{topological centrality} (TC) is a network centrality measure that iteratively captures the relative importance of both nodes and edges, accounting for their mutual influence \cite{Zhuge2010}. Initially, all nodes are assigned the same centrality value:
\[
c_{\mathrm{TC}}(i, 0) = 1, \quad \forall i.
\]

At each iteration step \(t+1\), the centrality of a node \(i\) is updated based on the influence of its neighbors and the weights of the edges connecting them:
\[
c_{\mathrm{TC}}(i, t+1) = c_{\mathrm{TC}}(i, t) + \sum_{j \in \mathcal{N}(i)} w_{ij}(t) \, c_{\mathrm{TC}}(j, t),
\]
where \(w_{ij}(t)\) denotes the weight of the edge between nodes \(i\) and \(j\) at iteration \(t\), with \(w_{ij}(0) = w_{ij}\).

The edge weights are updated at each step based on the current centrality values of their incident nodes:
\[
w_{ij}(t+1) = c_{\mathrm{TC}}(i, t+1) + c_{\mathrm{TC}}(j, t+1).
\]

To ensure stability and comparability, node centralities and edge weights are normalized at each iteration by their respective maximum values. The process is repeated until the node centralities converge, yielding the final topological centrality scores.

\section{Topological coefficient}

The \emph{topological coefficient}\index{topological coefficient} quantifies the extent to which a node shares its neighbors with other nodes in the network \cite{Stelzl2005}. For a given node $i$, the topological coefficient is defined as
\begin{equation*}
c_{\mathrm{top}}(i) = \frac{\sum_{j=1}^N |\mathcal{N}(i) \cap \mathcal{N}(j)|}{|\mathcal{N}(i)| \cdot |\{v: \mathcal{N}(i) \cap \mathcal{N}(v) \neq \emptyset\}|},
\end{equation*}
where \(\mathcal{N}(i)\) is the set of neighbors of \(i\).  

The topological coefficient ranges from 0 to 1. A value of \(c_{\mathrm{top}}(i) = 0\) indicates that node \(i\) does not share any neighbors with other nodes, 
while higher values indicate that a larger fraction of \(i\)'s neighbors are shared with other nodes. The coefficient equals 1 if and only if every neighbor of \(i\) is shared with every node that shares at least one neighbor with \(i\). Intuitively, nodes with a high topological coefficient are embedded in tightly interconnected neighborhoods, whereas nodes with low values are more topologically isolated.

\section{TOPSIS centrality}
\emph{TOPSIS centrality}\index{TOPSIS centrality} is a hybrid centrality measure that ranks nodes based on their similarity to an ideal solution using the TOPSIS (Technique for Order Preference by Similarity to Ideal Solution) approach~\cite{Du2014}. Let $R$ denote the normalized $N \times m$ decision matrix, where each entry characterizes the normalized influence of a node with respect to $m$ centrality metrics. Du \textit{et al.}~\cite{Du2014} consider $m=3$ (degree, closeness and betweenness/eigenvector centralities). 

The positive ideal solution $A^{+}$ and negative ideal solution $A^{-}$ are defined as
\[
A^{+} = [\max_{i} w_1 r_{i1}, \dots, \max_{i} w_m r_{im}], \quad
A^{-} = [\min_{i} w_1 r_{i1}, \dots, \min_{i} w_m r_{im}],
\]
where $w_j$ is the weight of the $j$-th centrality metric (e.g., $w_j = 1/m$ for equal weighting). 

The TOPSIS centrality of node $i$ is then given by its relative closeness to the ideal solution:
\[
c_{\mathrm{TOPSIS}}(i) = \frac{S_i^{-}}{S_i^{-} + S_i^{+}},
\]
where $S_i^{+}$ and $S_i^{-}$ are the Euclidean distances from node $i$ to the positive and negative ideal solutions, respectively:
\[
S_i^{+} = \sqrt{\sum_{j=1}^m (A_j^{+} - w_j r_{ij})^2}, \quad
S_i^{-} = \sqrt{\sum_{j=1}^m (A_j^{-} - w_j r_{ij})^2}.
\]

Nodes with higher TOPSIS centrality scores are simultaneously closer to the positive ideal solution and farther from the negative ideal solution, reflecting high overall importance across the selected centrality metrics.

\section{TOPSIS-RE centrality}

\emph{TOPSIS-RE}\index{TOPSIS-RE centrality} centrality is a hybrid method that combines relative entropy and the Technique for Order of Preference by Similarity to Ideal Solution (TOPSIS) to evaluate node influence in a network \cite{ZLiu2015}. Let $R$ denote the normalized $N \times m$ decision matrix, where each row corresponds to a node and each column represents a normalized centrality measure. Liu \textit{et al.} \cite{ZLiu2015} consider $m=4$ centralities: degree, closeness, betweenness and IKSD.  

The \emph{positive ideal solution} $A^{+}$ and \emph{negative ideal solution} $A^{-}$ are defined as
\[
A^{+} = [\max_{i}{w_{1} r_{i1}}, \dots, \max_{i}{w_{m} r_{im}}], \quad
A^{-} = [\min_{i}{w_{1} r_{i1}}, \dots, \min_{i}{w_{m} r_{im}}],
\]
where $w_j$ is the weight assigned to centrality $j$. Liu \textit{et al.} \cite{ZLiu2015} set
\[
[w_1, w_2, w_3, w_4] = [0.0625, 0.1875, 0.1875, 0.5625]
\]
using the Analytic Hierarchy Process (AHP). The centrality $c_{TOPSIS-RE}(i)$ of node $i$ is calculated as the \emph{relative closeness} to the ideal solution:
\begin{equation*}
c_{TOPSIS-RE}(i) = \frac{S_i^{-}}{S_i^{-} + S_i^{+}},
\end{equation*}
where $S_i^{+}$ and $S_i^{-}$ measure the relative entropy between node $i$ and the positive and negative ideal solutions, respectively:
\[
S_i^{+} = \sqrt{\sum_{j=1}^m \left( A_j^{+} \log \frac{A_j^{+}}{w_j r_{ij}} + (1-A_j^{+}) \log \frac{1-A_j^{+}}{1-w_j r_{ij}} \right)},
\]
\[
S_i^{-} = \sqrt{\sum_{j=1}^m \left( A_j^{-} \log \frac{A_j^{-}}{w_j r_{ij}} + (1-A_j^{-}) \log \frac{1-A_j^{-}}{1-w_j r_{ij}} \right)}.
\]

Nodes with higher TOPSIS-RE scores are considered more influential, being closer to the positive ideal solution and farther from the negative ideal solution.

\section{Total centrality}
\emph{Total centrality}\index{total centrality} is an induced (vitality) centrality measure in which the sum of centrality scores is a graph invariant \cite{Everett2010}. Let \(G_i\) denote the subgraph obtained by removing node \(i\) from \(G\). The total centrality of node \(i\) is defined as
\begin{equation*}
c_{T}(i) = \sum_{j \in \mathcal{N}} c(j,G) - \sum_{j \in \mathcal{N} \setminus \{i\}} c(j,G_i) 
= c(i,G) + \sum_{j \in \mathcal{N} \setminus \{i\}} \bigl( c(j,G) - c(j,G_i) \bigr),
\end{equation*}
where \(c(i,G)\) is the endogenous centrality of node \(i\) in \(G\), that is, its standard centrality measure (e.g., degree, closeness, or betweenness) computed on the original graph. For example, if the underlying centrality \(c(i,G)\) is the degree \(d_i\), then the total degree centrality equals \(2 d_i\). 

Total centrality accounts for both the node’s own centrality and its contribution to the centrality of other nodes.

\section{Total communicability centrality (TCC)}

The \emph{total communicability centrality}\index{total communicability centrality} (TCC) quantifies how effectively each node communicates with all other nodes in a network \cite{Benzi2013}. It extends the concept of subgraph centrality, which evaluates all possible closed walks of different lengths. The centrality of node \(i\), \(c_{\mathrm{TTC}}(i)\), is defined as the sum of weighted walks of all lengths starting at node \(i\) and ending at any node \(j\):
\begin{equation*}
    c_{\mathrm{TTC}}(i) = \sum_{k=0}^{\infty} \sum_{j=1}^{N} \frac{(A^k)_{ij}}{k!} 
    = \sum_{j=1}^{N} [e^A]_{ij} 
    = \sum_{j=1}^{N} \sum_{l=1}^{N} v_j(i) v_l(i) e^{\lambda_j},
\end{equation*}
where \(A\) is the adjacency matrix, \(v_j(i)\) is the \(i\)-th component of the eigenvector \(v_j\) corresponding to eigenvalue \(\lambda_j\) of \(A\). Benzi and Klymko \cite{Benzi2013} show that for networks with a large spectral gap (\(\lambda_1 \gg \lambda_2\)), the TTC and subgraph centralities converge to the eigenvector centrality, and they illustrate graph types for which TTC and subgraph centrality yield identical node rankings.

\section{Total effects centrality (TEC)}

\emph{Total effects centrality}\index{effects centrality!total (TEC)} (TEC) quantifies the cumulative influence of a node across the network, accounting for the number and length of all paths connecting nodes~\cite{Friedkin1991}. The \(N \times N\) matrix of interpersonal effects is computed as
\[
W^{\infty} = \lim_{\alpha \to 1} \big[(I + \alpha W + \alpha^2 W^2 + \dots)(1-\alpha)\big] = \lim_{\alpha \to 1} (I - \alpha W)^{-1} (1-\alpha),
\]
where \(W\) is the \(N \times N\) row-normalized adjacency matrix with self-loops.

The total effect of node \(i\) on other nodes corresponds to column \(i\) of \(W^{\infty}\), and the total effects centrality of node \(i\) is the average of these effects:
\begin{equation*}
    c_{TEC}(i) = \frac{\sum_{j \neq i} (W^{\infty})_{ji}}{N-1}.
\end{equation*}

Total effects centrality is closely related to Katz centrality~\cite{Friedkin1991}, reflecting both direct and indirect influences of a node.

\section{Transportation centrality (TC)}

\emph{Transportation centrality}\index{transportation centrality (TC)} (TC) is a variant of betweenness centrality that incorporates traffic flow dynamics in transportation networks \cite{Piraveenan2023}. Let \(P_{s,t}\) denote the set of all simple paths between nodes \(s\) and \(t\), and \(P^i_{s,t}\) the subset of those paths that pass through node \(i\). The transportation centrality of node \(i\) is defined as
\begin{equation*}
    c_{\mathrm{TC}}(i) = \frac{1}{(N-1)(N-2)} 
    \sum_{s \neq i \neq t} 
    \frac{\sum_{p \in P^i_{s,t}} e^{-\beta C^p_{s,t}}}
         {\sum_{p \in P_{s,t}} e^{-\beta C^p_{s,t}}},
\end{equation*}
where \(C^p_{s,t}\) denotes the total cost of path \(p\) from node \(s\) to node \(t\), which may include travel distance, travel time, or other generalized measures of effort. The parameter \(\beta \ge 0\) regulates the sensitivity of the model to path costs, with larger values of \(\beta\) exponentially reducing the contribution of higher-cost paths.

The transportation centrality assumes that the probability of selecting a path \(p\) decreases exponentially with its cost \(C^p_{s,t}\). In the limiting case \(\beta \to \infty\), transportation centrality converges to classical betweenness centrality, as only the lowest-cost paths dominate. When \(\beta = 0\), all paths are equally likely, and the measure reduces to the \textit{all-path betweenness centrality (ABC)}\index{betweenness centrality!all-path (ABC)} proposed by Piraveenan and Saripada \cite{Piraveenan2023}.

\section{Trophic level centrality}

\emph{Trophic level centrality}\index{trophic level centrality}, also known as the flow-based trophic level, measures the average trophic function of a node, that is, the expected length of the path over which the node obtains energy from the source \cite{Adams1983,Williams2004}. For a node \(i\), the trophic level centrality \(c_{\mathrm{trophic}}(i)\) is defined as
\begin{equation*}
    c_{\mathrm{trophic}}(i) =
    \begin{cases}
        1 + \frac{1}{d_i^{\mathrm{in}}} \sum_{j=1}^{N} a_{ij} \, c_{\mathrm{trophic}}(j), & \text{if } d_i^{\mathrm{in}} \neq 0, \\
        0, & \text{if } d_i^{\mathrm{in}} = 0,
    \end{cases}
\end{equation*}
where \(d_i^{\mathrm{in}}\) is the in-degree of node \(i\) and \(a_{ij}\) are the adjacency matrix elements.  

This measure requires the network to include at least one basal node, a node with no incoming edges that serves as the primary source of energy. The trophic level of each node represents the average number of steps separating it from the basal nodes and reflects its position within the network’s energy flow. Trophic level centrality is therefore useful in ecological studies for identifying key species and evaluating the efficiency and stability of food webs.

\section{Truncated curvature index}

The \emph{truncated curvature index}\index{curvature index!truncated} is a simplified version of the curvature index that limits the computation to simplices of dimension $d \leq 2$, i.e., cliques of size up to three \cite{Wu2015}. For a node $i$, it is defined as
\[
K_{\mathrm{trunc}}(i) = \sum_{k=0}^{2} (-1)^k \frac{V_{k-1}(i)}{k+1} =1 - \frac{d_i}{2} + \frac{t_i}{3},
\]
where $V_k(i)$ denotes the number of $(k{+}1)$-cliques incident to node $i$, $d_i$ is the degree of node $i$ and $t_i$ is the number of triangles passing through node $i$.  

The truncated curvature index is particularly useful for analyzing large-scale networks where computing higher-dimensional cliques is computationally expensive, yet the essential topological and curvature properties of nodes are still captured.

\section{Trust-PageRank}

\emph{Trust-PageRank}\index{Trust-PageRank} combines the traditional PageRank algorithm with a trust-value that reflects the reliability of information transmission between nodes \cite{Sheng2020}. The trust-value \(T_{ij}\) from node \(i\) to its adjacent node \(j\) is defined as a weighted combination of a similarity ratio and a degree ratio:
\begin{equation*}
    T_{ij} = (1-\beta) R_{s_{ij}} + \beta R_{d_{ij}},
\end{equation*}
where the degree ratio is
\begin{equation*}
    R_{d_{ij}} = \frac{d_i}{\sum_{l \in \mathcal{N}(j)} d_l},
\end{equation*}
and the similarity ratio is
\begin{equation*}
    R_{s_{ij}} = \frac{s_{ij}}{\sum_{l \in \mathcal{N}(j)} s_{jl}}.
\end{equation*}

The similarity \(s_{ij}\) between nodes \(i\) and \(j\) is computed using the SimRank algorithm \cite{Jeh2002}:
\begin{equation*}
    s_{ij} =
    \begin{cases}
        1, & i = j, \\
        \frac{C}{d_i d_j} \sum_{a \in \mathcal{N}(i)} \sum_{b \in \mathcal{N}(j)} s_{ab}, & i \neq j,
    \end{cases}
\end{equation*}
where \(C\) is an attenuation factor.

Analogous to PageRank, the Trust-PageRank influence of node \(i\) at time \(t\) is defined as
\begin{equation*}
    TPR(i,t) = \frac{1-\alpha}{N} + \alpha \sum_{j \in \mathcal{N}(i)} T_{ij} \, TPR(j,t-1),
\end{equation*}
where \(\alpha\) is the damping (jump) probability. The Trust-PageRank centrality of node \(i\) is given by its influence \(TPR(i,t^*)\) when the network reaches a steady state (\(t^* \to \infty\)) or after a fixed number of iterations (\(t^* = t^{\mathrm{max}}\)).

Hence, Trust-PageRank integrates structural connectivity and node-level trust, assigning higher centrality to nodes that are both well-connected and linked to trustworthy neighbors, reflecting their importance and reliability in information propagation.

\section{Two-step framework (IF) centrality}

The \emph{two-step framework (IF) centrality}\index{two-step framework (IF) centrality}, also known as the global diversity and local feature (GDLF) method, quantifies node influence using both global and local network information \cite{Fu2015,Fu2015b}. 
Global information is derived from the \(k\)-shell decomposition, with entropy used to assess the distribution of a node’s neighbors across shells. 
Local information is captured by the degree of neighboring nodes. 
The centrality \( c_{\mathrm{IF}}(i) \) of node \( i \) is defined as
\begin{equation*}
   c_{\mathrm{IF}}(i) = \left(- \sum_{k=1}^{ks_{\max}} p_i(k) \log_2 p_i(k) \right) 
                        \left( \log_2 \sum_{j \in \mathcal{N}(i)} d_j \right),
\end{equation*}
where 
\begin{equation*}
   p_i(k) = \frac{x_k(i)}{\sum_{l=1}^{ks_{\max}} x_l(i)}
\end{equation*}
is the fraction of node \(i\)'s neighbors in the \(k\)-core layer, 
\( x_k(i) \) is the number of neighbors in the \(k\)-core layer \(k\), 
\( d_j \) is the degree of neighbor \( j \), and 
\( \mathcal{N}(i) \) is the set of neighbors of node \( i \).

\section{Two-way random walk betweenness (2RW) centrality}

\emph{Two-way random walk betweenness}\index{random walk betweenness centrality!two-way (2RW)} (2RW) centrality is a variant of betweenness centrality based on two-way random walks \cite{Curado2022}. For a pair of nodes \(i\) and \(j\), the probability of reaching \(j\) from \(i\) through a transition node \(t\) is defined as
\[
p_{itj} = \frac{a_{it} a_{tj}}{d_i d_t},
\]
where \(a_{ij}\) is the adjacency matrix entry and \(d_i\) is the degree of node \(i\).

The most likely two-way random walk between nodes \(i\) and \(j\) passes through a pair of transition nodes \((t^*, k^*)\) that maximizes the probability
\[
(t^*, k^*) = \arg\max_{t,k} \, p_{itj} \, p_{jki}.
\]

The 2RW betweenness centrality of a node counts how often it appears as one of these optimal transition nodes \((t^*\) or \(k^*)\) for all pairs of nodes \((i,j)\) in the network. Nodes that frequently serve as high-probability intermediate steps are considered more central.

\section{Vertex-disjoint \textit{k}-path}

The \emph{vertex-disjoint \textit{k}-path centrality}\index{\textit{k}-path centrality!vertex-disjoint} is a variant of the \(k\)-path centrality \cite{Borgatti2006}, which counts the number of \emph{vertex-disjoint paths} of length at most \(k\) that originate or terminate at a given node. A vertex-disjoint path is a simple path that shares no nodes with any other counted path, except for the two end nodes. By definition, the set of vertex-disjoint paths is always a subset of the set of edge-disjoint paths. Nodes with higher vertex-disjoint \(k\)-path centrality are therefore more robustly connected, as there are multiple independent paths linking them to other nodes in the network.

\section{Vertex Entanglement (VE)}

\emph{Vertex entanglement}\index{vertex entanglement (VE)} (VE) is an induced centrality measure designed to quantify the influence of individual nodes based on their impact on the network's functional diversity \cite{Huang2024}. To evaluate the influence of a node \(v \in \mathcal{N}\), the network is locally perturbed to form the \(v\)-control network \(G_v\), where \(v\) and its neighbors are merged into a super-vertex represented as a fully connected probabilistic graph. The original total link weights are evenly redistributed within the super-vertex, while the remainder of the network remains unchanged.  

The VE score \(c_{VE}(v)\) of node \(v\) is defined as the change in spectral entropy caused by this local perturbation:
\[
c_{VE}(v) = S(G) - S(G_v),
\]
where \(S(G)\) and \(S(G_v)\) denote the von Neumann entropies of the original and perturbed networks, respectively. The spectral entropy of \(G\) is computed from a density matrix \(\rho\) derived from the network Laplacian \(L(G)\):
\[
\rho = \frac{e^{-\tau L(G)}}{\mathrm{tr}(e^{-\tau L(G)})}, \qquad
S(G) = -\mathrm{tr}(\rho \log \rho),
\]
with \(\tau > 0\) being a diffusion time parameter that controls the scale of information propagation.  

Intuitively, VE quantifies how strongly a single node affects the global structure and information flow in the network. Nodes with high VE substantially influence network connectivity and functional diversity, highlighting their critical role. The VE method has been validated on various empirical networks, including social, biological, and infrastructure systems, demonstrating its effectiveness in identifying critical nodes for network dismantling and capturing functional diversity.

\section{ViralRank}

\emph{ViralRank}\index{ViralRank} ranks nodes based on the random-walk effective distance, which closely approximates the hitting time of a reaction-diffusion process on the network~\cite{Iannelli2018}. The ViralRank score of node \(i\) is the average effective distance from all sources to all targets:
\begin{equation*}
    c_{VR}(i) = \frac{1}{N} \sum_{j} \left[ D_{ij}^{RW}(\lambda) + D_{ji}^{RW}(\lambda) \right],
\end{equation*}
where \(\lambda\) is a parameter and \(D_{ji}^{RW}(\lambda)\) is the effective distance \cite{Iannelli2017}:
\[
D_{ji}^{RW}(\lambda) = -\ln \left[ \sum_{k \neq j} \left( I^{(j)} - e^{-\lambda} P^{(j)} \right)^{-1}_{ik} e^{-\lambda} p_k^{(j)} \right], \quad i \neq j,
\]
with \(D_{ii}^{RW}(\lambda) = 0\). Here, \(P^{(j)}\) and \(I^{(j)}\) are the \((N-1) \times (N-1)\) matrices obtained by removing the \(j\)th row and column from the Markov matrix \(P\), which is the row-normalized adjacency matrix \((P_{ij} = a_{ij} / \sum_k a_{ik})\), and from the identity matrix \(I\), respectively. The vector \(p^{(j)}\) is the \(j\)th column of \(P\) with the \(j\)th entry removed.

The logarithm term counts all random walks starting at \(i\) and terminating at \(j\). In the limit \(\lambda \to 0\), the ViralRank score reduces to the sum of mean first-passage times (MFPT): the average time for a random walk starting at \(i\) to reach other nodes, plus the average time for a walk starting at other nodes to reach \(i\)~\cite{Iannelli2018}.

\section{VMM algorithm}

The \emph{VMM algorithm}\index{VMM algorithm} is a multi-attribute voting method for identifying key nodes in a network \cite{Wang2023}. It is a variant of VoteRank, in which the voting ability $v_i$ and voting score $s_i$ of node $i$ are defined as
\[
v_i = \frac{d_i}{(1+c_i) \, \max_j d_j},
\quad
s_i = d_i + \sqrt{d_i \sum_{j \in \mathcal{N}(i)} v_j},
\]
where $d_i$ and $c_i$ are the degree and clustering coefficient of node $i$, respectively, and $\mathcal{N}(i)$ denotes its set of neighbors.

At each iteration, the node $k$ with the highest voting score $s_k$ is selected as a key node. Once selected, the voting ability and voting score of node $k$ are set to zero, and it no longer participates in subsequent voting rounds. The process repeats until the desired number of key nodes is identified.

\section{Volume centrality}

\emph{Volume centrality}\index{volume centrality}, also known as the \textit{Distributed Assessment of the Closeness Centrality Ranking} (DACCER), is a semi-local centrality measure based on the degrees of nodes within an $r$-hop neighborhood \cite{Wehmuth2012}. The centrality score $c_{V}(i)$ of a node $i$ is defined as
\[
c_{V}(i) = \sum_{j \in \mathcal{N}^{(\leq r)}(i)} d_j,
\]
where $\mathcal{N}^{(\leq r)}(i)$ denotes the set of nodes located within a topological distance $r$ from node $i$ (including $i$ itself). 

Wehmuth and Ziviani \cite{Wehmuth2012} empirically demonstrated that volume centrality achieves good performance for $r = 2$. This measure generalizes the \textit{sphere degree}\index{degree!sphere} introduced by da F. Costa \textit{et al.} \cite{Rio2009}, which corresponds to the special case of volume centrality with $r = 2$. For $r = 1$, Pei \textit{et al.} \cite{Pei2014} showed that volume centrality can outperform in-degree and PageRank centralities in certain empirical networks.

\section{VoteRank centrality}

The \emph{VoteRank}\index{VoteRank} algorithm is designed to identify a set of decentralized spreaders with the highest spreading capability through an iterative voting procedure \cite{Zhang2016}. Each node \( i \) in the network is characterized by a tuple \((s_i, v_i)\), where \(s_i\) denotes the \emph{voting score} and \(v_i\) the \emph{voting ability} of node \(i\). Initially, all nodes are assigned equal voting ability and zero score, i.e., \((s_i, v_i) = (0, 1)\) for all \( i \in \mathcal{N} \). The algorithm proceeds iteratively according to the following steps:

\begin{enumerate}
    \item \textit{Voting:} each node distributes votes to its neighbors in proportion to its current voting ability. The voting score of node \( i \) is computed as:
    \[
        s_i = \sum_{j=1}^{N} a_{ji} v_j,
    \]
    where \( a_{ji} \) denotes the element of the adjacency matrix \( A \).

    \item \textit{Selection:} the node \( k \) with the highest voting score \( s_k \) is selected as a spreader. Its voting ability is then set to zero, i.e., \( v_k = 0 \), ensuring that it does not participate in subsequent voting rounds.

    \item \textit{Update:} the voting abilities of all neighbors of the selected node \( k \) are reduced to reflect the decreased influence of nearby nodes. Specifically, for each neighbor \( i \in \mathcal{N}(k) \), the updated voting ability is given by:
    \[
        v_i = \max(0, v_i - f),
    \]
    where \( f \) is the attenuation factor, typically defined as the inverse of the average degree of the network, i.e. \( f = 1 / \langle k \rangle \).
\end{enumerate}

This procedure is repeated until a predefined number of spreaders are identified or until the desired network coverage is achieved. The VoteRank centrality thus effectively identifies multiple influential nodes while minimizing redundancy in their spreading domains.

\section{VoteRank$^{++}$ centrality}

The \emph{VoteRank$^{++}$}\index{VoteRank$^{++}$} method is an enhanced variant of the VoteRank algorithm designed to identify a set of influential nodes that are broadly distributed across a network \cite{Liu2021}. Each node \(i\) is characterized by a tuple \((s_i, v_i)\), where \(s_i\) denotes its voting score and \(v_i\) its voting ability. Initially, these values are assigned as
\[
(s_i, v_i) = \left( 0, \log\left(1 + \frac{d_i}{d_{\max}}\right) \right),
\]
where \(d_i\) is the degree of node \(i\) and \(d_{\max}=\max_j{d_j}\) is the maximum degree in the network. The algorithm proceeds iteratively through the following steps:

\begin{enumerate}
    \item \textit{Vote:} each node \(i\) casts votes to its neighbors according to
    \[
    s_i = \sqrt{d_i \sum_{j=1}^{N} \left( \frac{a_{ji} d_i}{\sum_{l=1}^{N} a_{jl} d_l} v_j \right)},
    \]
    where \(a_{ji}\) is the element of the adjacency matrix of the network.
    
    \item \textit{Select:} the node \(k\) with the highest voting score \(s_k\) is selected as an influential node. This node is then excluded from subsequent voting rounds by setting its voting ability \(v_k = 0\).
    
    \item \textit{Update:} the voting abilities of nodes that voted for node \(k\) are reduced to \(\lambda v_i\), where \(\lambda \in [0,1]\) is a suppressing factor. For nodes within two hops of \(k\), the voting ability is reduced to \(\sqrt{\lambda} v_i\). Following Liu \textit{et al.}~\cite{Liu2021}, the suppressing factor is typically set to \(\lambda = 0.1\).
\end{enumerate}

Nodes with high VoteRank$^{++}$ scores are thus identified as influential spreaders that are not only central but also spatially dispersed, ensuring broad network coverage.

\section{Weight degree centrality (WDC, Liu)}

The \emph{weight degree centrality}\index{degree centrality!weight (WDC, Liu)} (WDC) quantifies the influence of a node by considering both its own degree and the degrees of its neighbors \cite{Liu2016}. The centrality of node \(i\) is defined as
\begin{equation*}
    c_{Wdc}(i) = \left( \sum_{j \in \mathcal{N}(i)} d_j - d_i \right) d_i^{\alpha},
\end{equation*}
where \(\mathcal{N}(i)\) denotes the set of neighbors of node \(i\), \(d_i\) is the degree of node \(i\), while \(\alpha\) is a tunable parameter that regulates the contribution of node \(i\)'s degree relative to its neighbors. Liu \textit{et al.} \cite{Liu2016} suggest setting \(\alpha = |r|\), where \(r\) is the degree assortativity coefficient. This allows the centrality measure to adapt to the network type: in \emph{assortative} networks (\(r>0\)), high-degree nodes connecting to other high-degree nodes are emphasized; 
in \emph{disassortative} networks (\(r<0\)), high-degree nodes connecting to low-degree nodes are emphasized; 
and in \emph{neutral} networks (\(r \approx 0\)), node degrees have uniform influence.

\section{Weight neighborhood centrality}

The \emph{weight neighborhood centrality}\index{neighborhood centrality!weight (WNC)} quantifies the influence of a node by combining its own centrality with the weighted centralities of its neighbors \cite{Wang2017}. Let \( f \) denote a benchmark centrality measure. Then, the weight neighborhood centrality \( c_{wnc}(i) \) of node \( i \) is defined as
\begin{equation*}
   c_{wnc}(i) = f(i) + \sum_{j \in \mathcal{N}(i)} \frac{w_{ij}}{\overline{w}} f(j),
\end{equation*}
where \( w_{ij} = (d_i d_j)^{\alpha} \), \( d_i \) and \( d_j \) are the degrees of nodes \( i \) and \( j \), 
\( \alpha \) is a tunable parameter, and \( \overline{w} \) is the average importance of all links in the network. Wang \textit{et al.} \cite{Wang2017} used \( \alpha = 1 \) and considered degree or $k$-shell centrality as the benchmark \( f \).

\section{Weighted community betweenness (WCB) centrality}

\emph{Weighted community betweenness}\index{betweenness centrality!weighted community (WCB)} (WCB) centrality is a centrality measure that integrates a node’s betweenness centrality at both the global network level and within its local community \cite{Ghalmane2018}. Let the graph \(G\) have a community structure consisting of \(K\) non-overlapping communities \(C_1, \dots, C_K\). This measure highlights the importance of nodes based on their contributions to both global connectivity and local community structure.

The centrality \(c_{\textsc{WCB}}(i)\) of node \(i\) is defined as
\begin{equation*}
   c_{\textsc{WCB}}(i) = (1-\mu_{C_l})\, c_b(i,C_l) + \mu_{C_l}\, c_b(i,G),
\end{equation*}
where \(C_l\) is the community to which node \(i\) belongs, \(c_b(i,C_l)\) is the local betweenness centrality of node \(i\) within \(C_l\), and \(c_b(i,G)\) is the global betweenness centrality of node \(i\) in the entire graph \(G\). The weighting factor \(\mu_{C_l}\) quantifies the relative importance of global connectivity by reflecting the proportion of inter-community links, and is calculated as
\begin{equation*}
   \mu_{C_l} = \frac{\sum_{i \in C_l} \sum_{j \in C_l} a_{ij}}{\sum_{i=1}^N \sum_{j=1}^N a_{ij}},
\end{equation*}
where \(a_{ij}\) are the elements of the adjacency matrix of \(G\).

\section{Weighted formal concept analysis (WFCA)}

\emph{Weighted formal concept analysis} (WFCA) centrality\index{weighted formal concept analysis (WFCA)} applies the principles of formal concept analysis (FCA) to rank nodes in a network \cite{Sun2017}. Let \(\mathbb{K} = (O, K, I)\) denote a formal context, where \(O = \mathcal{N}\) is a set of objects, \(K\) is a set of attributes, and \(I \subseteq O \times K\) is a binary relation between objects and attributes. A pair \((T, P)\), with \(T \subseteq O\) and \(P \subseteq K\), forms a formal concept if every object \(t \in T\) possesses all attributes in \(P\), and every attribute \(p \in P\) is shared by all objects in \(T\).

In the case of a graph \(G\) without external attributes, the attribute set is taken as \(K = \mathcal{N}\), and a formal concept \((T, P)\) corresponds to a subset of nodes \(T \subseteq \mathcal{N}\) that share a common set of neighbors \(P\). The WFCA centrality of node \(i\) is then defined as
\begin{equation*}
    c_{WFCA}(i) = \sum_{\substack{(T,P): \\ i \in P}} \frac{|T|}{|P|}.
\end{equation*}

Hence, the WFCA centrality captures global structure by ranking nodes according to the number of nodes sharing their neighbor set (\(|T|\)) relative to the size of that set (\(|P|\)).

\section{Weighted gravity model (WGravity)}

The \emph{weighted gravity model}\index{gravity model!weighted (WGravity)} (WGravity) is a variant of the traditional gravity model that incorporates a truncation radius and the eigenvector centrality of nodes \cite{Liu2020}. The WGravity centrality of node $i$ is defined as
\[
c_{WGravity}(i) = e_i \sum_{j \in \mathcal{N}^{(\leq r)}(i)} \frac{d_i d_j}{d_{ij}^2},
\]
where $d_i$ and $d_j$ are the degrees of nodes $i$ and $j$, $d_{ij}$ is the shortest distance between nodes $i$ and $j$, $e_i$ is the eigenvector centrality of node $i$, and $\mathcal{N}^{(\leq r)}(i)$ denotes the set of nodes within $r$ hops of $i$. The parameter $r$ defines the radius of influence for each node. 

Liu \textit{et al.} \cite{Liu2020} suggest setting $r = 0.5 \langle d \rangle$, where $\langle d \rangle$ is the average shortest path length of the network $G$. This truncation balances local and semi-local information while incorporating the global influence captured by eigenvector centrality.

\section{Weighted \textit{h}-index}

The \emph{weighted h-index}\index{Hirsch’s h-index!weighted} is an extension of the classical h-index to weighted networks \cite{Gao2019}. First, the weight $w_{ij}$ of the edge $(i,j)$ is defined as
\begin{equation*}
    w_{ij} = d_i d_j,
\end{equation*}
where $d_i$ and $d_j$ are the degrees of nodes $i$ and $j$.  

Next, each neighbor $j$ of node $i$ is conceptually cloned $d_j$ times. Each cloned neighbor $j_c$ is assigned a virtual edge weight $w_{ij_c} = w_{ij}$, effectively repeating the original edge weight $d_j$ times.  

The weighted h-index $c_{wh}(i)$ of node $i$ is then calculated as
\begin{equation*}
    c_{wh}(i) = H(w_{ij_1,1}, \dots, w_{ij_1,d_{j_1}}, \dots, w_{ij_{d_i},1}, \dots, w_{ij_{d_i},d_{j_{d_i}}}),
\end{equation*}
where $H$ is the \textit{h}-index operator, which returns the largest integer $h$ such that there are at least $h$ elements in the set with value no less than $h$.  In other words, the weighted \textit{h}-index reflects both the degrees of a node’s neighbors and the multiplicity of their connections, capturing a more nuanced measure of local influence.

\section{Weighted \textit{k}-shell decomposition (Wks) centrality}

\emph{Weighted \textit{k}-shell decomposition}\index{\textit{k}-shell decomposition!weighted (Wks)} (Wks) 
extends the classical $k$-shell method by incorporating edge weights \cite{Wei2015}. The motivation is to identify central nodes not only by their position in the network core but also by the strength of their connections. For nodes $i$ and $j$, the potential edge weight is defined as
\[
w_{ij} = d_i + d_j,
\]
where $d_i$ denotes the degree of node $i$. The weighted degree of node $i$ is then
\[
k^w_i = \alpha d_i + (1 - \alpha) \sum_{j \in \mathcal{N}(i)} w_{ij} = \alpha d_i + (1-\alpha)d_i^2 +(1-\alpha) \sum_{j \in \mathcal{N}(i)} d_j,
\]
with $\alpha \in (0,1)$ as a tunable parameter (typically $\alpha = 0.5$) and $\mathcal{N}(i)$ as the set of neighbors of node \(i\). The Wks centrality is obtained by performing $k$-shell decomposition using the weighted degree $k^w$, allowing nodes with stronger connectivity to be placed more accurately within the core-periphery structure.

\section{Weighted \textit{k}-shell degree neighborhood (Wksd)}

\emph{Weighted \textit{k}-shell degree neighborhood}\index{\textit{k}-shell degree neighborhood!weighted (Wksd)} (Wksd) is a hybrid centrality measure that combines node degree and $k$-shell values \cite{Namtirtha2018a,Namtirtha2020}. The motivation is to capture nodes that are central both in terms of local connectivity and their position in the network’s hierarchical core. The edge weight between nodes $i$ and $j$ is defined as
\[
w_{ij} = (\alpha d_i + \mu k_s(i)) (\alpha d_j + \mu k_s(j)),
\]
where $d_i$ and $k_s(i)$ denote the degree and $k$-shell value of node $i$, and $\alpha$, $\mu$ are tunable parameters (typically $\alpha \in \{0.2, 0.4\}$ and $\mu = 0.9$). The Wksd centrality of node $i$ is given by
\[
c_{Wksd}(i) = \sum_{j \in \mathcal{N}(i)} w_{ij},
\]
where $\mathcal{N}(i)$ denotes the set of neighbors of node \(i\). Hence, Wksd reflects the cumulative weighted influence of its neighbors and emphasize nodes that combine high connectivity with strategic placement in the network core.

\section{Weighted \textit{k}-shell degree neighborhood (WKSDN)}

The \textit{weighted \textit{k}-shell degree neighborhood}\index{\textit{k}-shell degree neighborhood!weighted (WKSDN)} (WKSDN) is a parameter-free hybrid centrality measure that integrates both the degree and $k$-shell index of nodes in a network \cite{Maji2020}. Unlike the weighted $k$-shell degree (Wksd) measure proposed by Namtirtha \textit{et al.} \cite{Namtirtha2020}, Maji's formulation estimates the weight of each edge $(i,j)$ as
\[
w_{ij} = (k_s(i) + k_s(j)) + \lambda (d_i + d_j),
\]
where \(d_i\) and \(k_s(i)\) denote the degree and $k$-shell index of node \(i\), respectively. The parameter $\lambda$ serves as a normalization factor and is defined as
\[
\lambda = \frac{\sum_{i=1}^{N} k_s(i)}{\sum_{i=1}^{N} d_i}.
\]
The centrality of node \(i\) is then calculated as the sum of the weights of all its incident edges:
\[
c_{\mathrm{WKSDN}}(i) = \sum_{j \in \mathcal{N}(i)} w_{ij},
\]
where $\mathcal{N}(i)$ denotes the set of neighbors of node \(i\).

\section{Weighted \textit{k}-short node-disjoint paths (WKPaths) centrality}

The \emph{weighted \(k\)-short node-disjoint paths (WKPaths) centrality}\index{WKPaths centrality} was introduced by White and Smith \cite{White2003} as a variant of closeness centrality. Rather than considering all shortest paths between nodes, WKPaths considers the set of \(k\)-short paths, defined as all paths of length at most \(k\) with no shared intermediate nodes. The WKPaths centrality of a node \(i\), denoted \(c_{WKPaths}(i)\), is defined as
\begin{equation*}
c_{WKPaths}(i) = \frac{1}{N} \sum_{j \in \mathcal{N}} \sum_{P \in \mathcal{P}_k(j,i)} \delta^{-|P|},
\end{equation*}
where \(\mathcal{P}_k(j,i)\) is the set of \(k\)-short paths from node \(j\) to node \(i\), \(|P|\) is the length of path \(P\), and \(\delta\) is a scalar weighting factor with \(1 \leq \delta \leq \infty\) (e.g., \(\delta = 2\)). Shorter paths contribute more heavily to the centrality due to the exponential decay factor \(\delta^{-|P|}\).

\section{Weighted LeaderRank}

\emph{Weighted LeaderRank}\index{LeaderRank!weighted} is an extension of the LeaderRank algorithm that incorporates a weighted mechanism to account for node in-degrees \cite{Li2014}. Similar to LeaderRank, Weighted LeaderRank introduces a ground node \(g=N+1\) that connects bidirectionally to all nodes in the network \(G\). The weight of the link from the ground node to node \(i\), denoted \(w_{gi}\), is proportional to the in-degree of \(i\).  

Initially, each node (except the ground node) is assigned one unit of resource, which is then distributed to its neighbors according to the link weights. The resource dynamics at discrete time \(t+1\) are described by
\begin{equation*}
    s_i[t+1] = \sum_{j=1}^{N+1} \frac{w_{ji}}{\sum_{l=1}^{N+1} w_{jl}} \, s_j[t],
\end{equation*}
where the link weights \(w_{ji}\) are defined as
\[
w_{ji} =
\begin{cases}
a_{ji}, & \text{if } g \notin \{i,j\}, \\
1, & \text{if } g = i \neq j, \\
(d_i^{\mathrm{in}})^\alpha, & \text{if } g = j \neq i, \\
0, & \text{if } g = i = j,
\end{cases}
\]
with \(d_i^{\mathrm{in}}\) being the in-degree of node \(i\) and \(\alpha\) a tunable parameter (e.g., \(\alpha = 1\)).  

This formulation ensures that nodes with high in-degree receive larger contributions from the ground node. As with LeaderRank, the steady-state scores \(\Tilde{s}_i =\lim_{t \to \infty}s_i[t]\) quantify the influence of each node in the network.

\section{Weighted top candidate (WTC) method}

The \emph{weighted top candidate} (WTC) method\index{top candidate (TC) method! weighted (WTC)}\index{weighted Top Candidate (WTC)} is an extension of the top candidate (TC) method designed for ranking nodes in a weighted network, such as a citation network \cite{Sziklai2021}. Similar to the TC method \cite{Sziklai2018}, the WTC approach aims to identify a \emph{stable set} of mutually reinforcing candidates, but it leverages weighted interactions to better capture node reputation. WTC assumes that nodes with higher reputation tend to nominate stronger candidates than lower-reputation nodes. The algorithm proceeds through the following steps:

\begin{enumerate}
    \item \textit{Initialization:} all nodes are initially included in the candidate set. For each node \(i\), its reputation \(r_i\) is defined as the total weight of incoming links:
    \[
        r_i = \sum_{j \in \mathcal{N}} w_{ji},
    \]
    where \(w_{ji}\) is the weight of the link \((j,i)\).

    \item \textit{Weighted nomination:} for each node \(i\), a neighbor \(j \in \mathcal{N}_i^{out}\) is nominated if
    \[
        r_j \geq \omega_i (1 - \alpha),
    \]
    where \(\alpha \in [0,1]\) controls the selectivity of nominations and 
    \(\omega_i = \max_{k \in \mathcal{N}_i^{out}} r_k\). All neighbors satisfying this condition are considered in the next iteration of the candidate set.

    \item \textit{Elimination:} nodes that are not nominated by any other node are removed from the candidate set. All outgoing nominations from these removed nodes are discarded.

    \item \textit{Update:} the elimination in Step 3 may cause additional nodes to lose all incoming nominations. These nodes are likewise removed, and the process is repeated until no further removals occur.
\end{enumerate}

The remaining nodes form the \emph{stable candidate set}. A set \(S \subseteq \mathcal{N}\) is stable if (i) every node in \(S\) is nominated by at least one other node in \(S\) and (ii) all nominees of any node in \(S\) are also contained in \(S\). The parameter \(\alpha\) regulates the inclusiveness of the selection: smaller values of \(\alpha\) produce a more selective set of nodes, while larger values yield a more inclusive set.

\section{Weighted TOPSIS (w-TOPSIS) centrality}
\emph{Weighted TOPSIS (w-TOPSIS)}\index{TOPSIS centrality!weighted (w-TOPSIS)} is a hybrid centrality measure that extends the classical TOPSIS method by incorporating attribute weights~\cite{Hu2016}. Let $R$ be the normalized $N \times m$ decision matrix, where each entry $r_{ij}$ characterizes the normalized influence of node $i$ with respect to centrality metric $j$. Hu \textit{et al.}~\cite{Hu2016} consider $m=3$ metrics: degree, betweenness and closeness centralities.

The weights $w_j$ for each centrality metric are derived based on the node spreading capability in the SIR model. Let $F_i(t)$ denote the average number of infected and recovered nodes at time $t$ if node $i$ is initially infected, with spreading probability $\alpha=0.3$, recovery probability $\beta=1$, and $t=100$. The auxiliary variable $v_{ij}$ aligns the normalized centrality with spreading influence:
\[
v_{ij} = \left| \frac{r_{ij}}{\sum_{l=1}^N r_{lj}} - \frac{F_i(t)}{\sum_{l=1}^N F_l(t)} \right|^{-1}.
\]
The weight of metric $j$ is then defined as
\[
w_j = \frac{\sum_{i=1}^N v_{ij}}{\sum_{i=1}^N \sum_{j=1}^m v_{ij}}.
\]

The positive and negative ideal solutions are defined as
\[
A^{+} = \bigl[\max_i w_1 r_{i1}, \dots, \max_i w_m r_{im}\bigr], \quad
A^{-} = \bigl[\min_i w_1 r_{i1}, \dots, \min_i w_m r_{im}\bigr].
\]

The w-TOPSIS centrality of node $i$ is computed as its relative closeness to the ideal solution:
\[
c_{\mathrm{w-TOPSIS}}(i) = \frac{S_i^{-}}{S_i^{-} + S_i^{+}},
\]
where $S_i^{+}$ and $S_i^{-}$ are the Euclidean distances from node $i$ to the positive and negative ideal solutions, respectively:
\[
S_i^{+} = \sqrt{\sum_{j=1}^m (A_j^{+} - w_j r_{ij})^2}, \quad
S_i^{-} = \sqrt{\sum_{j=1}^m (A_j^{-} - w_j r_{ij})^2}.
\]

Nodes with higher $c_{\mathrm{w-TOPSIS}}(i)$ values are considered more influential, as they are simultaneously closer to the positive ideal and farther from the negative ideal solutions.

\section{Weighted volume centrality}

\emph{Weighted volume centrality}\index{volume centrality!weighted (WVC)} (WVC) is an extension of volume centrality that incorporates both the distance between nodes and their clustering coefficients \cite{Kim2012}. The centrality score $c_{WV}(i)$ of node $i$ is defined as
\[
c_{WV}(i) = \sum_{j \in \mathcal{N}^{(\leq r)}(i)} \frac{d_j \left[1 - c(j)\right]}{2^{d_{ij}}},
\]
where $\mathcal{N}^{(\leq r)}(i)$ denotes the set of nodes located within a topological distance $r$ from node $i$ (excluding $i$ itself), $d_j$ is the degree of node $j$, $c(j)$ is its clustering coefficient and $d_{ij}$ represents the shortest-path distance between nodes $i$ and $j$. 

The exponential term $2^{-d_{ij}}$ serves as a distance-decay factor, giving higher weight to nearby nodes. Kim and Yoneki \cite{Kim2012} demonstrated that, for $r \geq 2$, weighted volume centrality provides a good approximation of closeness centrality while requiring significantly lower computational cost.

\section{Wide ranking (WRank)}

The \emph{Wide Ranking}\index{wide ranking (WRank)} (WRank) algorithm simultaneously ranks the nodes and links of a network \cite{Wang2015}. The method is based on the principle that an important node is incident to many critical links, and a critical link connects important nodes. Let \(x\) be an \(N \times 1\) vector of node centralities and \(y\) an \(L \times 1\) vector of link centralities. The relationship between nodes and links is expressed as
\begin{equation*}
    \begin{cases}
        x = Wy, \\
        y = Zx,
    \end{cases}
\end{equation*}
where \(W\) is an \(N \times L\) incidence matrix with entries \(w_{il} = 1\) if node \(i\) is an endpoint of link \(l\) and 0 otherwise, and \(Z = W^T\). Substituting, we obtain \(x = WZ x\), so the principal eigenvector of \(WZ\) defines the centralities of the nodes, while the centralities of links follow from \(y = Zx\).

\vfill

\section{WVoteRank}

The \emph{WVoteRank}\index{WVoteRank} centrality is a modification of VoteRank that incorporates both the number of neighbors and the weight of each link \cite{Sun2019}. Each node \(i\) is represented by the tuple \((s_i, v_i)\), where \(s_i\) is the voting score and \(v_i\) is the voting ability, initialized as \((s_i, v_i) = (0, 1)\) for all \(i \in \mathcal{N}\). The voting procedure iteratively performs the following steps:

\begin{enumerate}
    \item \textit{Vote:} Each node votes for its neighbors based on its voting ability. The voting score of node \(i\) is updated as
    \begin{equation*}
        s_i = \sqrt{d_i \sum_{j=1}^{N} w_{ji} v_j},
    \end{equation*}
    where \(w_{ji}\) is the weight of the link from node \(j\) to node \(i\) and \(d_i\) is the degree of node \(i\).
    
    \item \textit{Select:} The node \(k\) with the highest voting score \(s_k\) is elected. Node \(k\) will not participate in subsequent voting turns, i.e., its voting ability is set to zero (\(v_k = 0\)).
    
    \item \textit{Update:} The voting ability of nodes that voted for \(k\) is reduced to account for influence spread. For each neighbor \(i \in \mathcal{N}(k)\), the updated voting ability is
    \begin{equation*}
        v_i \leftarrow \max(0, v_i - f),
    \end{equation*}
    where \(f\) is typically set to the inverse of the average degree of the network, i.e. \(f=1/\langle d \rangle\).
\end{enumerate}

WVoteRank prioritizes nodes that are both well-connected and linked through high-weight edges, ensuring that influential nodes are identified while accounting for the propagation of influence through their neighbors.

\section{\textit{X}-degree centrality}

\emph{X-degree centrality} quantifies a contribution of each node to non-backtracking paths in a network \cite{Torres2021}. A \emph{non-backtracking}\index{path!non-backtracking} path is a sequence of edges in which the path never immediately revisits the previous node: for a path \(v_1 \to v_2 \to \dots \to v_m\), we require \(v_{k+1} \neq v_{k-1}\) for all \(k = 2,\dots,m-1\).

Let \(G\) be a graph with adjacency matrix \(A\) and node degrees \(d_j = \sum_k a_{jk}\). For node \(i\), let \(F\) be the NB matrix of the star graph centered at \(i\), \(D\) the matrix with rows indexed by edges not incident to \(i\) and columns by edges incident to \(i\), and \(E\) the matrix with rows indexed by edges incident to \(i\) and columns by edges not incident to \(i\). The entries of \(D\) and \(E\) enforce the non-backtracking condition:
\[
D_{k \to l, i \to j} = a_{ik} a_{ij} (1 - \delta_{kj}), \qquad
E_{i \to j, k \to l} = a_{ik} a_{ij} (1 - \delta_{kj}),
\]
where \(\delta_{kj}\) is the Kronecker delta preventing immediate backtracking.

Define \(X = D F E\) and \(P = \begin{bmatrix} 0 & I \\ I & 0 \end{bmatrix}\). The \textit{X}-degree centrality of node \(i\) is
\[
c_{X\text{-degree}}(i) = \mathbf{1}^T P X \mathbf{1} 
= \Big( \sum_j a_{ij} (d_j - 1) \Big)^2 - \sum_j a_{ij} (d_j - 1)^2,
\]
where \(\mathbf{1}\) is the all-ones vector. This measure captures the node’s importance in supporting non-backtracking paths using only local degree information and serves as an upper bound for the \textit{X}-non-backtracking centrality, providing a computationally simpler approximation.

\section{\textit{X}-non-backtracking (X-NB) centrality}

\emph{X-non-backtracking centrality}\index{\textit{X}-non-backtracking centrality} quantifies the effect of a node on the largest eigenvalue of the non-backtracking (NB) matrix of a network \cite{Torres2021}. For node \(i\), let \(B'\) be the NB matrix after removing \(i\). The NB matrix of the original graph can be partitioned as
\[
B = \begin{bmatrix} B' & D \\ E & F \end{bmatrix},
\]
where \(F\) is the NB matrix of the star graph centered at \(i\), \(D\) has rows indexed by edges not incident to \(i\) and columns by edges incident to \(i\), and \(E\) has rows indexed by edges incident to \(i\) and columns by edges not incident to \(i\).

Define \(X = D F E\) with entries \(X_{k \rightarrow l, i \rightarrow j} = a_{ik} a_{ij} (1 - \delta_{kj})\) and \(P = \begin{bmatrix} 0 & I \\ I & 0 \end{bmatrix}\), where \(\delta_{kj}\) is the Kronecker delta. Let \(v_1\) be the right eigenvector of \(B'\) and \(v_1^j = \sum_k a_{jk} v_{k \rightarrow j}\). Then the \textit{X}-NB centrality $c_{X\text{-}NB}(i)$ of node \(i\) is
\[
c_{X\text{-}NB}(i) = v_1^T P X v_1 = \Big( \sum_j a_{ij} v_1^j \Big)^2 - \sum_j a_{ij} (v_1^j)^2,
\]
where \(a_{ij}\) are adjacency matrix entries. Thus, X-non-backtracking centrality captures the impact of \(i\) on non-backtracking paths and network connectivity.

\section{Zeta vector centrality}
\emph{Zeta vector centrality}\index{zeta vector centrality}, also referred to as node displacement \cite{Estrada2010} or topological centrality \cite{Ranjan2013}, identifies the most effective “spreader” node in a network \cite{PVM2017}. Inspired by electrical flows in a resistor network, Van Mieghem \textit{et al.} define the best conducting node \(i\) in a graph \(G\) as the node that minimizes the diagonal element \(Q^{\dagger}_{ii}\) of the pseudoinverse \(Q^{\dagger}\) of the weighted Laplacian matrix of \(G\):
\begin{equation*}
c_{\rm Zeta}(i) = Q^{\dagger}_{ii} = \frac{1}{N} \sum_{j=1}^{N} \omega_{ji} - \frac{\tilde{R}_G}{N^2},
\end{equation*}
where \(\omega_{ji}\) is the effective resistance between nodes \(j\) and \(i\), and \(\tilde{R}_G\) is the effective graph resistance. In other words, \(c_{\rm Zeta}(i)\) represents the average effective resistance from node \(i\) to all other nodes, minus the mean effective resistance of the graph.

The node with minimal \(c_{\rm Zeta}(i)\) is the best electrical spreader, having the lowest energy or potential and the strongest connectivity to the rest of the network. Alternative interpretations of \(Q^{\dagger}_{ii}\) include detour overheads in random walks or the average connectedness of nodes when the network splits, highlighting the node’s role in global communication \cite{Ranjan2013}.  The ranking of nodes based on \(Q^{\dagger}_{ii}\) is equivalent to node displacement \cite{Estrada2010}, computed as
\begin{equation*}
\Delta x_i = \sqrt{\frac{Q^{\dagger}_{ii}}{\beta k}},
\end{equation*}
where \(k\) is a common spring constant and \(\beta\) is the inverse temperature, linking network topology to physical interpretations of displacement.

\renewcommand{\bibname}{References}

\clearpage     
\phantomsection 
\addcontentsline{toc}{endsection}{\large Index}
\footnotesize
\printindex


\begin{thebibliography}{99}
\addcontentsline{toc}{endsection}{\large References}
\footnotesize
\setlength{\itemsep}{-0.42em}

\vspace{-0.8cm}

\bibitem{Achard2007} Achard, S., \& Bullmore, E. (2007). Efficiency and cost of economical brain functional networks. PLoS computational biology, 3(2), e17. \href{https://doi.org/10.1371/journal.pcbi.0030017}{doi: 10.1371/journal.pcbi.0030017.}


\bibitem{Adali2012} Adali, S., Lu, X., \& Magdon-Ismail, M. (2012). Attentive betweenness centrality (abc): considering options and bandwidth when measuring criticality. In 2012 International Conference on Privacy, Security, Risk and Trust and 2012 International Confernece on Social Computing (pp. 358-367). IEEE. \href{https://doi.org/10.1109/SocialCom-PASSAT.2012.53}{doi: 10.1109/SocialCom-PASSAT.2012.53.}


\bibitem{Adams1983} Adams, S. M., Kimmel, B. L., \& Ploskey, G. R. (1983). Sources of organic matter for reservoir fish production: a trophic-dynamics analysis. Canadian Journal of Fisheries and Aquatic Sciences, 40(9), 1480-1495. \href{https://doi.org/10.1139/f83-170}{doi: 10.1139/f83-170.}

\bibitem{Agneessens2017} Agneessens, F., Borgatti, S. P., \& Everett, M. G. (2017). Geodesic based centrality: Unifying the local and the global. Social Networks, 49, 12-26. \href{https://doi.org/10.1016/j.socnet.2016.09.005}{doi: 10.1016/j.socnet.2016.09.005.}


\bibitem{Agryzkov2014} Agryzkov, T., Oliver, J. L., Tortosa, L., \& Vicent, J. (2014). A new betweenness centrality measure based on an algorithm for ranking the nodes of a network. Applied Mathematics and Computation, 244, 467-478. \href{https://doi.org/10.1016/j.amc.2014.07.026}{doi: 10.1016/j.amc.2014.07.026.}




\bibitem{Agryzkov2019} Agryzkov, T., Tortosa, L., \& Vicent, J. F. (2019). A variant of the current flow betweenness centrality and its application in urban networks. Applied Mathematics and Computation, 347, 600-615. \href{https://doi.org/10.1016/j.amc.2018.11.032}{doi: 10.1016/j.amc.2018.11.032.}



\bibitem{Ahajjam2018} Ahajjam, S., \& Badir, H. (2018). Identification of influential spreaders in complex networks using HybridRank algorithm. Scientific reports, 8(1), 11932. \href{https://doi.org/10.1038/s41598-018-30310-2}{doi: 10.1038/s41598-018-30310-2.}


\bibitem{Ahn2010} Ahn, Y. Y., Bagrow, J. P., \& Lehmann, S. (2010). Link communities reveal multiscale complexity in networks. Nature, 466(7307), 761-764. \href{https://doi.org/10.1038/nature09182}{doi: 10.1038/nature09182.}



\bibitem{Ai2017} Ai, X. (2017). Node importance ranking of complex networks with entropy variation. Entropy, 19(7), 303. \href{https://doi.org/10.3390/e19070303}{doi: 10.3390/e19070303.}




\bibitem{Aleskerov2014} Aleskerov, F., Andrievskaya, I. and Permjakova, E. (2014), Key Borrowers Detected by the Intensities of Their Short-Range Interactions (2014). Higher School of Economics Research Paper No. WP BRP 33/FE/2014. \href{https://doi.org/10.2139/ssrn.2479272}{doi: 10.2139/ssrn.2479272.}



\bibitem{Aleskerov2016c} Aleskerov, F., Badgaeva, D., Pislyakov, V., Sterligov, I., \& Shvydun, S. (2016). An importance of Russian and international economic journals: A network approach. Journal of the New Economic Association, 30(2), 193-205. \href{https://doi.org/10.31737/2221-2264-2016-30-2-10}{doi: 10.31737/2221-2264-2016-30-2-10.}



\bibitem{Aleskerov2024a}Aleskerov, F., Cinar, Y., Deseatnicov, I., Sergeeva, E., Tkachev, D., \& Yakuba, V. (2024). Network analysis of economic sectors in the world economy. Procedia Computer Science, 242, 420-427. \href{https://doi.org/10.1016/j.procs.2024.08.165}{doi: 10.1016/j.procs.2024.08.165.}


\bibitem{Aleskerov2020c} Aleskerov, F., Gavrilenkova, I., Shvydun, S., \& Yakuba, V. (2020). Power Distribution in the Networks of Terrorist Groups: 2001–2018. Group decision and negotiation, 29(3), 399-424.


\bibitem{Aleskerov2022} Aleskerov, F., Dutta, S., Egorov, D., \& Tkachev, D. (2022). Networks under deep uncertainty. Procedia Computer Science, 214, 1285-1292. \href{https://doi.org/10.1016/j.procs.2022.11.307}{doi: 10.1016/j.procs.2022.11.307.}


\bibitem{Aleskerov2017}Aleskerov, F., Meshcheryakova, N. and Shvydun, S. (2017). Power in Network Structures. In: Springer Proceedings in Mathematics \& Statistics, vol 197. Springer, Cham. \href{https://doi.org/10.1007/978-3-319-56829-4_7}{doi: 10.1007/978-3-319-56829-4\_7.}



\bibitem{Aleskerov2020} Aleskerov, F. , Andrievskaya, I. , Nikitina A. and Shvydun, S. (2020). Key Borrowers Detected by the Intensities of Their Interactions. Handbook of Financial Econometrics, Mathematics, Statistics, and Machine Learning (In 4 Volumes), 355-389 World Scientific: Singapore Volume 1, Chapter 9. \href{https://doi.org/10.1142/9789811202391_0009}{doi: 10.1142/9789811202391\_0009.}




\bibitem{Aleskerov2024b} Aleskerov, F., Khutorskaya, O., Yakuba, V., Stepochkina, A., \& Zinovyeva, K. (2024). Affiliations based bibliometric analysis of publications on parkinson’s disease. Computational Management Science, 21(1), 13. \href{https://doi.org/10.1007/s10287-023-00495-7}{doi: 10.1007/s10287-023-00495-7.}


\bibitem{Aleskerov2016} Aleskerov, F. T., Kurapova, M., Meshcheryakova, N., Mironyuk, M., \& Shvydun, S. (2016). A network approach to analysis of international conflicts. Political science (RU), (4), 111-137. 


\bibitem{Aleskerov2016b} Aleskerov, F., Meshcheryakova, N., Rezyapova, A., \& Shvydun, S. (2016, May). Network analysis of international migration. In International Conference on Network Analysis (pp. 177-185). Cham: Springer International Publishing. \href{https://doi.org/10.1007/978-3-319-56829-4_13}{doi: 10.1007/978-3-319-56829-4\_13.}


\bibitem{Aleskerov2020b}Aleskerov, F., Meshcheryakova, N., \& Shvydun, S. (2020). Indirect influence assessment in the context of retail food network. In Network Algorithms, Data Mining, and Applications: NET, Moscow, Russia, May 2018 8 (pp. 143-160). Springer International Publishing. \href{https://doi.org/10.1007/978-3-030-37157-9_10}{doi: 10.1007/978-3-030-37157-9\_10.}



\bibitem{Aleskerov2023} Aleskerov, F., Seregin, M., \& Tkachev, D. (2023). The network analysis of oil trade under deep uncertainty. Procedia Computer Science, 221, 1021-1028. \href{https://doi.org/10.1016/j.procs.2023.08.083}{doi: 10.1016/j.procs.2023.08.083.}


\bibitem{Aleskerov2017b} Aleskerov, F., Sergeeva, Z., \& Shvydun, S. (2017). Assessment of exporting economies influence on the global food network. In Optimization Methods and Applications: In Honor of Ivan V. Sergienko's 80th Birthday (pp. 1-10). Cham: Springer International Publishing.

\bibitem{AleskerovBook} Aleskerov, F., Shvydun, S. and Meshcheryakova, N. (2021). New centrality measures in networks: how to take into account the parameters of the nodes and group influence of nodes to nodes (1st ed.). Chapman and Hall/CRC. \href{https://doi.org/10.1201/9781003203421}{doi: 10.1201/9781003203421.}



\bibitem{YAleskerov2020} Aleskerov, F., \& Yakuba, V. (2020). Matrix-vector approach to construct generalized centrality indices in networks. Available at SSRN 3597948. \href{https://doi.org/10.2139/ssrn.3597948}{doi: 10.2139/ssrn.3597948.}




\bibitem{Alvarez-Socorro2015} Alvarez-Socorro, A. J., Herrera-Almarza, G. C., \& González-Díaz, L. A. (2015). Eigencentrality based on dissimilarity measures reveals central nodes in complex networks. Scientific reports, 5(1), 17095. \href{https://doi.org/10.1038/srep17095}{doi: 10.1038/srep17095.}



\bibitem{Andrade2019} de Andrade, R. L., \& Rêgo, L. C. (2019). p-Means centrality. Communications in Nonlinear Science and Numerical Simulation, 68, 41-55. \href{https://doi.org/10.1016/j.cnsns.2018.08.002}{doi: 10.1016/j.cnsns.2018.08.002.}


\bibitem{Anthonisse1971} Anthonisse, J. M., (1971) The rush in a directed graph, Technical Report BN 9/71, Stichting Mathematisch Centrum, Amsterdam.


\bibitem{Arrow2010} Arrow, K. J., Sen, A., \& Suzumura, K. (Eds.). (2010). Handbook of social choice and welfare (Vol. 2). Elsevier.

\bibitem{Artime2024} Artime, O., Grassia, M., De Domenico, M., Gleeson, J. P., Makse, H. A., Mangioni, G., Perc, M. \& Radicchi, F. (2024). Robustness and resilience of complex networks. Nature Reviews Physics, 6(2), 114-131. \href{https://doi.org/10.1038/s42254-023-00676-y}{doi: 10.1038/s42254-023-00676-y.}




\bibitem{Babul2022} Babul, S. A., Devriendt, K., \& Lambiotte, R. (2022). Gromov centrality: A multiscale measure of network centrality using triangle inequality excess. Physical Review E, 106(3), 034312. \href{https://doi.org/10.1103/PhysRevE.106.034312}{doi: 10.1103/PhysRevE.106.034312.}

\bibitem{Bae2014} Bae, J., \& Kim, S. (2014). Identifying and ranking influential spreaders in complex networks by neighborhood coreness. Physica A: Statistical Mechanics and its Applications, 395, 549-559. \href{https://doi.org/10.1016/j.physa.2013.10.047}{doi: 10.1016/j.physa.2013.10.047.}


\bibitem{Bandyopadhyay2018} Bandyopadhyay, S., Narayanam, R., \& Murty, M. N. (2018). A generic axiomatic characterization for measuring influence in social networks. In 2018 24th international conference on pattern recognition (icpr) (pp. 2606-2611). IEEE. \href{https://doi.org/10.1109/ICPR.2018.8546109}{doi: 10.1109/ICPR.2018.8546109.}


\bibitem{Banerjee2013} Banerjee, A., Chandrasekhar, A. G., Duflo, E., \& Jackson, M. O. (2013). The diffusion of microfinance. Science, 341(6144), 1236498. \href{https://doi.org/10.1126/science.1236498}{doi: 10.1126/science.1236498.}


\bibitem{Banerjee2020} Banerjee, S., Jenamani, M., \& Pratihar, D. K. (2020). A survey on influence maximization in a social network. Knowledge and Information Systems, 62(9), 3417-3455. \href{https://doi.org/10.1007/s10115-020-01461-4}{doi: 10.1007/s10115-020-01461-4.}
 


\bibitem{Bao2017} Bao, Z. K., Ma, C., Xiang, B. B., \& Zhang, H. F. (2017). Identification of influential nodes in complex networks: Method from spreading probability viewpoint. Physica A: Statistical Mechanics and its Applications, 468, 391-397. \href{https://doi.org/10.1016/j.physa.2016.10.086}{doi: 10.1016/j.physa.2016.10.086.}


\bibitem{Barrat2004} Barrat, A., Barthelemy, M., Pastor-Satorras, R., \& Vespignani, A. (2004). The architecture of complex weighted networks. Proceedings of the national academy of sciences, 101(11), 3747-3752. \href{https://doi.org/10.1073/pnas.0400087101}{doi: 10.1073/pnas.0400087101.}


\bibitem{Katsaros2013} Basaras, P., Katsaros, D., \& Tassiulas, L. (2013). Detecting influential spreaders in complex, dynamic networks. Computer, 46(04), 24-29. \href{https://doi.org/10.1109/MC.2013.75}{doi: 10.1109/MC.2013.75.}


\bibitem{Batagelj2002} Batagelj V., Zaveršnik M. (2002) Generalized Cores, arXiv:0202039. \href{https://doi.org/10.48550/arXiv.cs/0202039}{doi: 10.48550/arXiv.cs/0202039.}


\bibitem{Bauer2012} Bauer, F., \& Lizier, J. T. (2012). Identifying influential spreaders and efficiently estimating infection numbers in epidemic models: A walk counting approach. Europhysics Letters, 99(6), 68007. \href{https://doi.org/10.1209/0295-5075/99/68007}{doi: 10.1209/0295-5075/99/68007.}


\bibitem{Bavelas1948} Bavelas, A. (1948). A mathematical model for group structures. Human organization, 7(3), 16-30. \href{https://doi.org/10.17730/humo.7.3.f4033344851gl053}{doi: 10.17730/humo.7.3.f4033344851gl053.}



\bibitem{Bavelas1950} Bavelas, A. (1950). Communication patterns in task-oriented groups. Journal of the acoustical society of America. \href{https://doi.org/10.1121/1.1906679}{doi: 10.1121/1.1906679.}


\bibitem{Beauchamp1965} Beauchamp, M. A. (1965). An improved index of centrality. Behavioral science, 10(2), 161-163. \href{https://doi.org/10.1002/bs.3830100205}{doi: 10.1002/bs.3830100205.}



\bibitem{Benzi2013} Benzi, M., \& Klymko, C. (2013). Total communicability as a centrality measure. Journal of Complex Networks, 1(2), 124-149. \href{https://doi.org/10.1093/comnet/cnt007}{doi: 10.1093/comnet/cnt007.}


\bibitem{Berahmand2018} Berahmand, K., Bouyer, A., \& Samadi, N. (2018). A new centrality measure based on the negative and positive effects of clustering coefficient for identifying influential spreaders in complex networks. Chaos, Solitons \& Fractals, 110, 41-54. \href{https://doi.org/10.1016/j.chaos.2018.03.014}{doi: 10.1016/j.chaos.2018.03.014.}


\bibitem{Berahmand2019} Berahmand, K., Bouyer, A., \& Samadi, N. (2019). A new local and multidimensional ranking measure to detect spreaders in social networks. Computing, 101, 1711-1733. \href{https://doi.org/10.1007/s00607-018-0684-8}{doi: 10.1007/s00607-018-0684-8.}


\bibitem{Berge1958} Berge, C. (1958). Théorie des graphes et ses applications. Dunod, Paris, France. \href{https://doi.org/10.1002/zamm.19600400516}{doi: 10.1002/zamm.19600400516.}



\bibitem{BianDeng2017} Bian, T., \& Deng, Y. (2017). A new evidential methodology of identifying influential nodes in complex networks. Chaos, Solitons \& Fractals, 103, 101-110. \href{https://doi.org/10.1016/j.chaos.2017.05.040}{doi: 10.1016/j.chaos.2017.05.040.}



\bibitem{Bian2018} Bian, T., \& Deng, Y. (2018). Identifying influential nodes in complex networks: A node information dimension approach. Chaos: An Interdisciplinary Journal of Nonlinear Science, 28(4). \href{https://doi.org/10.1063/1.5030894}{doi: 10.1063/1.5030894.}


\bibitem{Bian2017} Bian, T., Hu, J., \& Deng, Y. (2017). Identifying influential nodes in complex networks based on AHP. Physica A: Statistical Mechanics and its Applications, 479, 422-436. \href{https://doi.org/10.1016/j.physa.2017.02.085}{doi: 10.1016/j.physa.2017.02.085.}

\bibitem{Bloch2023} Bloch, F., Jackson, M. O., \& Tebaldi, P. (2023). Centrality measures in networks. Social Choice and Welfare, 61(2), 413-453. \href{https://doi.org/10.1007/s00355-023-01456-4}{doi: 10.1007/s00355-023-01456-4.}

\bibitem{Blöchl2011} Blöchl, F., Theis, F. J., Vega-Redondo, F., \& Fisher, E. O. N. (2011). Vertex centralities in input-output networks reveal the structure of modern economies. Physical Review E—Statistical, Nonlinear, and Soft Matter Physics, 83(4), 046127. \href{https://doi.org/10.1103/PhysRevE.83.046127}{doi: 10.1103/PhysRevE.83.046127.}


\bibitem{Blöcker2022} Blöcker, C., Nieves, J. C., \& Rosvall, M. (2022). Map equation centrality: community-aware centrality based on the map equation. Applied Network Science, 7(1), 56. \href{https://doi.org/10.1007/s41109-022-00477-9}{doi: 10.1007/s41109-022-00477-9.}

\bibitem{Boldi} Boldi, P., \& Vigna, S. (2014). Axioms for centrality. Internet Mathematics, 10(3-4), 222-262. \href{https://doi.org/10.1080/15427951.2013.865686}{doi: 10.1080/15427951.2013.865686.}





\bibitem{Bollobas1984}  Bollobás, B. (1984). Graph Theory and Combinatorics: Proceedings of the Cambridge
Combinatorial Conference in Honor of P. Erdös Vol. 35 (Academic, 1984). 



\bibitem{Bonacich1972} Bonacich, P. (1972). Factoring and weighting approaches to status scores and clique identification. The Journal of Mathematical Sociology, 2(1), 113-120. \href{https://doi.org/10.1080/0022250X.1972.9989806}{doi: 10.1080/0022250X.1972.9989806.}



\bibitem{Bonacich1972b} Bonacich, P. (1972). Technique for analyzing overlapping memberships. Sociological methodology, 4, 176-185. \href{https://doi.org/10.2307/270732}{doi: 10.2307/270732.}



\bibitem{Bonacich1987} Bonacich, P. (1987). Power and centrality: A family of measures. American journal of sociology, 92(5), 1170-1182. 

\bibitem{Bonacich2001} Bonacich, P., \& Lloyd, P. (2001). Eigenvector-like measures of centrality for asymmetric relations. Social networks, 23(3), 191-201. \href{https://doi.org/10.1016/S0378-8733(01)00038-7}{doi: 10.1016/S0378-8733(01)00038-7.}


\bibitem{Bonato2020} Bonato, A., Eikmeier, N., Gleich, D. F., \& Malik, R. (2019, November). Centrality in dynamic competition networks. In International Conference on Complex Networks and Their Applications (pp. 105-116). Cham: Springer International Publishing. \href{https://doi.org/10.1007/978-3-030-36683-4_9}{doi: 10.1007/978-3-030-36683-4\_9.}


\bibitem{Bonnaire2020} Bonnaire, T., Aghanim, N., Decelle, A., \& Douspis, M. (2020). T-ReX: a graph-based filament detection method. Astronomy \& Astrophysics, 637, A18. \href{https://doi.org/10.1051/0004-6361/201936859}{doi: 10.1051/0004-6361/201936859.}



\bibitem{Borgatti1997} Borgatti, S. P. (1997). Structural holes: Unpacking Burt’s redundancy measures. Connections, 20(1), 35-38. 


\bibitem{Borgatti2002} Borgatti, S. P., Everett, M. G., \& Freeman, L. C. (2002). Ucinet for Windows: Software for social network analysis. Harvard, MA: analytic technologies, 6, 12-15. 


\bibitem{Borgatti2003} Borgatti, S. P. (2003). The key player problem. Washington, D.C.: National Academy of Sciences Press. pp. 241-252. \href{https://doi.org/10.2139/ssrn.1149843}{doi: 10.2139/ssrn.1149843.}


\bibitem{Borgatti2005} Borgatti, S. P. (2005). Centrality and network flow. Social networks, 27(1), 55-71. \href{https://doi.org/10.1016/j.socnet.2004.11.008}{doi: 10.1016/j.socnet.2004.11.008.}


\bibitem{Borgatti2006} Borgatti, S. P., \& Everett, M. G. (2006). A graph-theoretic perspective on centrality. Social networks, 28(4), 466-484. \href{https://doi.org/10.1016/j.socnet.2005.11.005}{doi: 10.1016/j.socnet.2005.11.005.}


\bibitem{Borgatti2006b} Borgatti, S. P., Carley, K. M., \& Krackhardt, D. (2006). On the robustness of centrality measures under conditions of imperfect data. Social networks, 28(2), 124-136. \href{https://doi.org/10.1016/j.socnet.2005.05.001}{doi: 10.1016/j.socnet.2005.05.001.}


\bibitem{Brandes2005} Brandes, U. (2005). Network analysis: methodological foundations (Vol. 3418). Springer Science \& Business Media. \href{https://doi.org/10.1007/b106453}{doi: 10.1007/b106453.}


\bibitem{Brandes2008} Brandes, U. (2008). On variants of shortest-path betweenness centrality and their generic computation. Social networks, 30(2), 136-145. \href{https://doi.org/10.1016/j.socnet.2007.11.001}{doi: 10.1016/j.socnet.2007.11.001.}


\bibitem{Brandes2016} Brandes, U., Borgatti, S. P., \& Freeman, L. C. (2016). Maintaining the duality of closeness and betweenness centrality. Social networks, 44, 153-159. \href{https://doi.org/10.1016/j.socnet.2015.08.003}{doi: 10.1016/j.socnet.2015.08.003.}


\bibitem{Brandes2022} Brandes, U., Laußmann, C., \& Rothe, J. (2022). Voting for centrality. In Proceedings of the 21st International Conference on Autonomous Agents and Multiagent Systems (pp. 1554-1556). \href{https://doi.org/10.5555/3535850.3536032}{doi: 10.5555/3535850.3536032.}



\bibitem{Brockmann2013} Brockmann, D., \& Helbing, D. (2013). The hidden geometry of complex, network-driven contagion phenomena. science, 342(6164), 1337-1342. \href{https://doi.org/10.1126/science.1245200}{doi: 10.1126/science.1245200.}



\bibitem{BrinPage1998} Brin, S., \& Page, L. (1998). The anatomy of a large-scale hypertextual web search engine. Computer networks and ISDN systems, 30(1-7), 107-117. \href{https://doi.org/10.1016/S0169-7552(98)00110-X}{doi: 10.1016/S0169-7552(98)00110-X.}


\bibitem{Bringmann2019} Bringmann, L. F., Elmer, T., Epskamp, S., Krause, R. W., Schoch, D., Wichers, M., Wigman, J. T. W., \& Snippe, E. (2019). What do centrality measures measure in psychological networks?. Journal of abnormal psychology, 128(8), 892–903. \href{https://doi.org/10.1037/abn0000446}{doi: 10.1037/abn0000446.}



\bibitem{Burt1992} Burt, R.S. \& Holes, S. (1992). Structural Holes: The Social Structure of Competition. Harvard University Press, Cambridge, MA. 

\bibitem{Burt2004} Burt, R. S. (2004). Structural Holes and Good Ideas. American Journal of Sociology, 110(2), 349-399. \href{https://doi.org/10.1086/421787}{doi: 10.1086/421787.}




\bibitem{Caporossi2012} Caporossi, G., Paiva, M., Vukičević, D., \& Marcelo, S. (2012). Centrality and betweenness: vertex and edge decomposition of the Wiener index. MATCH: communications in mathematical and in computer chemistry, 68(1), 293-302.


\bibitem{Carpenter2002} Carpenter, T., Karakostas, G., \& Shallcross, D. (2002). Practical issues and algorithms for analyzing terrorist networks. In Proceedings of the western simulation multiconference. 


\bibitem{Chen2016} Chen, B., Wang, Z., \& Luo, C. (2016). Integrated evaluation approach for node importance of complex networks based on relative entropy. Journal of Systems Engineering and Electronics, 27(6), 1219-1226. \href{https://doi.org/10.21629/JSEE.2016.06.10}{doi: 10.21629/JSEE.2016.06.10.}


\bibitem{Chen2013} Chen, D. B., Gao, H., Lü, L., \& Zhou, T. (2013). Identifying influential nodes in large-scale directed networks: the role of clustering. PloS one, 8(10), e77455. \href{https://doi.org/10.1371/journal.pone.0077455}{doi: 10.1371/journal.pone.0077455.}




\bibitem{Chen2012} Chen, D., Lü, L., Shang, M. S., Zhang, Y. C., \& Zhou, T. (2012). Identifying influential nodes in complex networks. Physica a: Statistical mechanics and its applications, 391(4), 1777-1787. \href{https://doi.org/10.1016/j.physa.2011.09.017}{doi: 10.1016/j.physa.2011.09.017.}




\bibitem{Chen2014} Chen, D. B., Xiao, R., Zeng, A., \& Zhang, Y. C. (2014). Path diversity improves the identification of influential spreaders. Europhysics letters, 104(6), 68006. \href{https://doi.org/10.1209/0295-5075/104/68006}{doi: 10.1209/0295-5075/104/68006.}


\bibitem{Chen2003} Chen, Y., Hu, A. Q., Yip, K. W., Hu, J., \& Zhong, Z. G. (2003). Finding the most vital node with respect to the number of spanning trees. In International Conference on Neural Networks and Signal Processing, 2003. Proceedings of the 2003 (Vol. 2, pp. 1670-1673). IEEE. \href{https://doi.org/10.1109/ICNNSP.2003.1281204}{doi: 10.1109/ICNNSP.2003.1281204.}




\bibitem{Chen2009} Chen, W., Wang, Y., \& Yang, S. (2009). Efficient influence maximization in social networks. In Proceedings of the 15th ACM SIGKDD international conference on Knowledge discovery and data mining (pp. 199-208). \href{https://doi.org/10.1145/1557019.1557047}{doi: 10.1145/1557019.1557047.}


\bibitem{Chin2014} Chin, C. H., Chen, S. H., Wu, H. H., Ho, C. W., Ko, M. T., \& Lin, C. Y. (2014). cytoHubba: identifying hub objects and sub-networks from complex interactome. BMC systems biology, 8(Suppl 4), S11. \href{https://doi.org/10.1186/1752-0509-8-S4-S11}{doi: 10.1186/1752-0509-8-S4-S11.}



\bibitem{Chin2003} Chin, C. S., \& Samanta, M. P. (2003). Global snapshot of a protein interaction network—a percolation based approach. Bioinformatics, 19(18), 2413-2419.  \href{https://doi.org/10.1093/bioinformatics/btg339}{doi: 10.1093/bioinformatics/btg339.}



\bibitem{Christensen2018} Christensen, A. P., Kenett, Y. N., Aste, T., Silvia, P. J., \& Kwapil, T. R. (2018). Network structure of the Wisconsin Schizotypy Scales-Short Forms: Examining psychometric network filtering approaches. Behavior Research Methods, 50, 2531-2550. \href{https://doi.org/10.3758/s13428-018-1032-9}{doi: 10.3758/s13428-018-1032-9.}


\bibitem{Cohen2008} Cohen, J. (2008). Trusses: Cohesive subgraphs for social network analysis. National security agency technical report, 16(3.1), 1-29. 

\bibitem{Costenbader2003} Costenbader, E., \& Valente, T. W. (2003). The stability of centrality measures when networks are sampled. Social networks, 25(4), 283-307. \href{https://doi.org/10.1016/S0378-8733(03)00012-1}{doi: 10.1016/S0378-8733(03)00012-1.}



\bibitem{Csató2017} Csató, L. (2017). Measuring centrality by a generalization of degree. Central European Journal of Operations Research, 25(4), 771-790. \href{https://doi.org/10.1007/s10100-016-0439-6}{doi: 10.1007/s10100-016-0439-6.}


\bibitem{Curado2022} Curado, M., Rodriguez, R., Tortosa, L., \& Vicent, J. F. (2022). A new centrality measure in dense networks based on two-way random walk betweenness. Applied Mathematics and Computation, 412, 126560. \href{https://doi.org/10.1016/j.amc.2021.126560}{doi: 10.1016/j.amc.2021.126560.}



\bibitem{Curado2023} Curado, M., Tortosa, L., \& Vicent, J. F. (2023). A novel measure to identify influential nodes: return random walk gravity centrality. Information Sciences, 628, 177-195. \href{https://doi.org/10.1016/j.ins.2023.01.097}{doi: 10.1016/j.ins.2023.01.097.}


\bibitem{Dai2019} Dai, J., Wang, B., Sheng, J., Sun, Z., Khawaja, F. R., Ullah, A., Dejene, D. A. \& Duan, G. (2019). Identifying influential nodes in complex networks based on local neighbor contribution. IEEE Access, 7, 131719-131731. \href{https://doi.org/10.1109/ACCESS.2019.2939804}{doi: 10.1109/ACCESS.2019.2939804.}


\bibitem{DalCol2023} Dal Col, A., \& Petronetto, F. (2023). Graph regularization centrality. Physica A: Statistical Mechanics and its Applications, 628, 129188. \href{https://doi.org/10.1016/j.physa.2023.129188}{doi: 10.1016/j.physa.2023.129188.}

\bibitem{Dangalchev2006} Dangalchev, C. (2006). Residual closeness in networks. Physica A: Statistical Mechanics and its Applications, 365(2), 556-564. \href{https://doi.org/10.1016/j.physa.2005.12.020}{doi: 10.1016/j.physa.2005.12.020.}


\bibitem{Das2018} Das, K., Samanta, S., \& Pal, M. (2018). Study on centrality measures in social networks: a survey. Social network analysis and mining, 8(1), 13. \href{https://doi.org/10.1007/s13278-018-0493-2}{doi: 10.1007/s13278-018-0493-2.}


\bibitem{Arruda2014}  De Arruda, G. F., Barbieri, A. L., Rodriguez, P. M., Rodrigues, F. A., Moreno, Y., \& Costa, L. D. F. (2014). Role of centrality for the identification of influential spreaders in complex networks. Physical Review E, 90(3), 032812. \href{https://doi.org/10.1103/PhysRevE.90.032812}{doi: 10.1103/PhysRevE.90.032812.}

\bibitem{Rio2009} del Rio, G., Koschützki, D., \& Coello, G. (2009). How to identify essential genes from molecular networks?. BMC systems biology, 3(1), 102. \href{https://doi.org/10.1186/1752-0509-3-102}{doi: 10.1186/1752-0509-3-102.}


\bibitem{Debono2014} Debono, M., Lauri, J., \& Sciriha, I. (2014). Balanced centrality of networks. International Scholarly Research Notices, 2014(1), 871038. \href{https://doi.org/10.1155/2014/871038}{doi: 10.1155/2014/871038.}

\bibitem{DePaolis2022} DePaolis, F., Murphy, P., \& De Paolis Kaluza, M. C. (2022). Identifying key sectors in the regional economy: a network analysis approach using input-output data. Applied Network Science, 7(1), 86. \href{https://doi.org/10.1007/s41109-022-00519-2}{doi: 10.1007/s41109-022-00519-2.}




\bibitem{Devriendt2022}Devriendt, K., \& Lambiotte, R. (2022). Discrete curvature on graphs from the effective resistance. Journal of Physics: Complexity, 3(2), 025008. \href{https://doi.org/10.1088/2632-072X/ac730d}{doi: 10.1088/2632-072X/ac730d.}


\bibitem{Dolev2010} Dolev, S., Elovici, Y., \& Puzis, R. (2010). Routing betweenness centrality. Journal of the ACM (JACM), 57(4), 1-27. \href{https://doi.org/10.1145/1734213.1734219}{doi: 10.1145/1734213.1734219.}


\bibitem{Dong2018} Dong, J., Ye, F., Chen, W., \& Wu, J. (2018). Identifying influential nodes in complex networks via semi-local centrality. In 2018 IEEE International Symposium on Circuits and Systems (ISCAS) (pp. 1-5). IEEE. \href{https://doi.org/10.1109/ISCAS.2018.8351889}{doi: 10.1109/ISCAS.2018.8351889.}



\bibitem{Du2015} Du, Y., Gao, C., Chen, X., Hu, Y., Sadiq, R., \& Deng, Y. (2015). A new closeness centrality measure via effective distance in complex networks. Chaos: An Interdisciplinary Journal of Nonlinear Science, 25(3). \href{https://doi.org/10.1063/1.4916215}{doi: 10.1063/1.4916215.}

\bibitem{Du2014} Du, Y., Gao, C., Hu, Y., Mahadevan, S., \& Deng, Y. (2014). A new method of identifying influential nodes in complex networks based on TOPSIS. Physica A: Statistical Mechanics and its Applications, 399, 57-69. \href{https://doi.org/10.1016/j.physa.2013.12.031}{doi: 10.1016/j.physa.2013.12.031.}


\bibitem{Duron2020} Durón C (2020) Heatmap centrality: A new measure to identify super-spreader nodes in scale-free networks. PLOS ONE 15(7): e0235690. \href{https://doi.org/10.1371/journal.pone.0235690}{doi: 10.1371/journal.pone.0235690.}



\bibitem{Ellens2011} Ellens, W., Spieksma, F. M., Van Mieghem, P., Jamakovic, A., \& Kooij, R. E. (2011). Effective graph resistance. Linear algebra and its applications, 435(10), 2491-2506. \href{https://doi.org/10.1016/j.laa.2011.02.024}{doi: 10.1016/j.laa.2011.02.024.}




\bibitem{Engsig2024} Engsig, M., Tejedor, A., Moreno, Y., Foufoula-Georgiou, E., \& Kasmi, C. (2024). DomiRank Centrality reveals structural fragility of complex networks via node dominance. Nature communications, 15(1), 56. \href{https://doi.org/10.1038/s41467-023-44257-0}{doi: 10.1038/s41467-023-44257-0.}

\bibitem{Ravasz2012} Ercsey-Ravasz, M., Lichtenwalter, R. N., Chawla, N. V., \& Toroczkai, Z. (2012). Range-limited centrality measures in complex networks. Physical Review E—Statistical, Nonlinear, and Soft Matter Physics, 85(6), 066103. \href{https://doi.org/10.1103/PhysRevE.85.066103}{doi: 10.1103/PhysRevE.85.066103.}


\bibitem{Estrada2007} Estrada, E. (2007). Characterization of topological keystone species: Local, global and “meso-scale” centralities in food webs. Ecological Complexity, 4(1-2), 48-57. \href{https://doi.org/10.1016/j.ecocom.2007.02.018}{doi: 10.1016/j.ecocom.2007.02.018.}

\bibitem{Estrada2010c} Estrada, E. (2010). Generalized walks-based centrality measures for complex biological networks. Journal of theoretical biology, 263(4), 556-565. \href{https://doi.org/10.1016/j.jtbi.2010.01.014}{doi: 10.1016/j.jtbi.2010.01.014.}



\bibitem{Estrada2005} Estrada, E., \& Rodríguez-Velázquez, J. A. (2005). Spectral measures of bipartivity in complex networks. Physical Review E—Statistical, Nonlinear, and Soft Matter Physics, 72(4), 046105. \href{https://doi.org/10.1103/PhysRevE.72.046105}{doi: 10.1103/PhysRevE.72.046105.}


\bibitem{Estrada} Estrada, E., \& Rodríguez-Velázquez, J. A. (2005). Subgraph centrality in complex networks. Physical Review E, 71(5). \href{https://doi.org/10.1103/PhysRevE.71.056103}{doi: 10.1103/PhysRevE.71.056103.}



\bibitem{Estrada2010} Estrada, E., \& Hatano, N. (2010). A vibrational approach to node centrality and vulnerability in complex networks. Physica A: Statistical Mechanics and its Applications, 389(17), 3648-3660. \href{https://doi.org/10.1016/j.physa.2010.03.030}{doi: 10.1016/j.physa.2010.03.030.}




\bibitem{Estrada2010b} Estrada, E., \& Higham, D. J. (2010). Network properties revealed through matrix functions. SIAM review, 52(4), 696-714. \href{https://doi.org/10.1137/090761070}{doi: 10.1137/090761070.}


\bibitem{Estrada2009} Estrada, E., Higham, D. J., \& Hatano, N. (2009). Communicability betweenness in complex networks. Physica A: Statistical Mechanics and its Applications, 388(5), 764-774. \href{https://doi.org/10.1016/j.physa.2008.11.011}{doi: 10.1016/j.physa.2008.11.011.}




\bibitem{Everett2005} Everett, M., \& Borgatti, S. P. (2005). Ego network betweenness. Social networks, 27(1), 31-38. \href{https://doi.org/10.1016/j.socnet.2004.11.007}{doi: 10.1016/j.socnet.2004.11.007.}


\bibitem{Everett2010} Everett, M. G., \& Borgatti, S. P. (2010). Induced, endogenous and exogenous centrality. Social Networks, 32(4), 339-344. \href{https://doi.org/10.1016/j.socnet.2010.06.004}{doi: 10.1016/j.socnet.2010.06.004.}


\bibitem{Faghani2013} Faghani, M. R., \& Nguyen, U. T. (2013). A study of XSS worm propagation and detection mechanisms in online social networks. IEEE transactions on information forensics and security, 8(11), 1815-1826. \href{https://doi.org/10.1109/TIFS.2013.2280884}{doi: 10.1109/TIFS.2013.2280884.}




\bibitem{Fei2017} Fei, L., Mo, H., \& Deng, Y. (2017). A new method to identify influential nodes based on combining of existing centrality measures. Modern Physics Letters B, 31(26), 1750243. \href{https://doi.org/10.1142/S0217984917502438}{doi: 10.1142/S0217984917502438.}




\bibitem{Fei2018} Fei, L., Zhang, Q., \& Deng, Y. (2018). Identifying influential nodes in complex networks based on the inverse-square law. Physica A: Statistical Mechanics and its Applications, 512, 1044-1059. \href{https://doi.org/10.1016/j.physa.2018.08.135}{doi: 10.1016/j.physa.2018.08.135.}


\bibitem{Fiedler1973} Fiedler, M. (1973). Algebraic connectivity of graphs. Czechoslovak mathematical journal, 23(2), 298-305. 

\bibitem{Ford1962} Ford, L.R., \& Fulkerson, D.R. (1962). Flows in Networks. Princeton University Press, Princeton. 


\bibitem{Frantz2009} Frantz, T. L., Cataldo, M., \& Carley, K. M. (2009). Robustness of centrality measures under uncertainty: Examining the role of network topology. Computational and Mathematical Organization Theory, 15(4), 303-328. \href{https://doi.org/10.1007/s10588-009-9063-5}{doi: 10.1007/s10588-009-9063-5.}


\bibitem{Freeman1977} Freeman, L. C. (1977). A set of measures of centrality based on betweenness. Sociometry, 35-41. \href{https://doi.org/10.2307/3033543}{doi: 10.2307/3033543.}

\bibitem{Freeman1978} Freeman, L. C. (1978). Centrality in social networks conceptual clarification. Social networks, 1(3), 215-239. \href{https://doi.org/10.1016/0378-8733(78)90021-7}{doi: 10.1016/0378-8733(78)90021-7.}


\bibitem{Freeman1982}  Freeman, L. C. (1982).  Centered graphs and the construction of ego networks. Mathematical Social Sciences 3, 291-304. \href{https://doi.org/10.1016/0165-4896(82)90076-2}{doi: 10.1016/0165-4896(82)90076-2.}





\bibitem{Freeman1991} Freeman, L. C., Borgatti, S. P., \& White, D. R. (1991). Centrality in valued graphs: A measure of betweenness based on network flow. Social networks, 13(2), 141-154. \href{https://doi.org/10.1016/0378-8733(91)90017-N}{doi: 10.1016/0378-8733(91)90017-N.}




\bibitem{Friedkin1991} Friedkin, N. E. (1991). Theoretical Foundations for Centrality Measures. American Journal of Sociology, 96(6), 1478-1504. \href{https://doi.org/10.1086/229694}{doi: 10.1086/229694.}




\bibitem{Fu2010} Fu, L., Gao, L., \& Ma, X. (2010). A centrality measure based on spectral optimization of modularity density. Science China Information Sciences, 53, 1727-1737. \href{https://doi.org/10.1007/s11432-010-4043-4}{doi: 10.1007/s11432-010-4043-4.}


\bibitem{Fu2015} Fu, Y. H., Huang, C. Y., \& Sun, C. T. (2015). Identifying super‐spreader nodes in complex networks. Mathematical Problems in Engineering, 2015(1), 675713. \href{https://doi.org/10.1155/2015/675713}{doi: 10.1155/2015/675713.}

\bibitem{Fu2015b} Fu, Y. H., Huang, C. Y., \& Sun, C. T. (2015). Using global diversity and local topology features to identify influential network spreaders. Physica A: Statistical Mechanics and its Applications, 433, 344-355. \href{https://doi.org/10.1016/j.physa.2015.03.042}{doi: 10.1016/j.physa.2015.03.042.}


\bibitem{Gao2013} Gao, C., Lan, X., Zhang, X., \& Deng, Y. (2013). A bio-inspired methodology of identifying influential nodes in complex networks. PloS one, 8(6), e66732. \href{https://doi.org/10.1371/journal.pone.0066732}{doi: 10.1371/journal.pone.0066732.}


\bibitem{Gao2019} Gao, L., Yu, S., Li, M., Shen, Z., \& Gao, Z. (2019). Weighted H-index for identifying influential spreaders. Symmetry, 11(10), 1263. \href{https://doi.org/10.3390/sym11101263}{doi: 10.3390/sym11101263.}


\bibitem{Gao2014} Gao, S., Ma, J., Chen, Z., Wang, G., \& Xing, C. (2014). Ranking the spreading ability of nodes in complex networks based on local structure. Physica A: Statistical Mechanics and its Applications, 403, 130-147. \href{https://doi.org/10.1016/j.physa.2014.02.032}{doi: 10.1016/j.physa.2014.02.032.}

\bibitem{Gao2009} Gao, W., Li, Q., Zhao, B., \& Cao, G. (2009). Multicasting in delay tolerant networks: a social network perspective. In Proceedings of the tenth ACM international symposium on Mobile ad hoc networking and computing (pp. 299-308). \href{https://doi.org/10.1145/1530748.1530790}{doi: 10.1145/1530748.1530790.}



\bibitem{GarciaPerez2019} García-Pérez, G., Allard, A., Serrano, M. Á., \& Boguñá, M. (2019). Mercator: uncovering faithful hyperbolic embeddings of complex networks. New Journal of Physics, 21(12), 123033. \href{https://doi.org/10.1088/1367-2630/ab57d2}{doi: 10.1088/1367-2630/ab57d2.}


\bibitem{Geisberger2008} Geisberger, R., Sanders, P., \& Schultes, D. (2008). Better approximation of betweenness centrality. In 2008 Proceedings of the Tenth Workshop on Algorithm Engineering and Experiments (ALENEX) (pp. 90-100). Society for Industrial and Applied Mathematics. \href{https://doi.org/10.1137/1.9781611972887.9}{doi: 10.1137/1.9781611972887.9.}


\bibitem{Ghalmane2018} Ghalmane, Z., El Hassouni, M., \& Cherifi, H. (2018). Betweenness centrality for networks with non-overlapping community structure. In 2018 IEEE workshop on complexity in engineering (COMPENG) (pp. 1-5). IEEE. \href{https://doi.org/10.1109/CompEng.2018.8536229}{doi: 10.1109/CompEng.2018.8536229.}


\bibitem{Ghalmane2019} Ghalmane, Z., Hassouni, M. E., \& Cherifi, H. (2019). Immunization of networks with non-overlapping community structure. Social Network Analysis and Mining, 9, 1-22. \href{https://doi.org/10.1007/s13278-019-0591-9}{doi: 10.1007/s13278-019-0591-9.}


\bibitem{Ghavasieh2021} Ghavasieh, A., Stella, M., Biamonte, J., \& De Domenico, M. (2021). Unraveling the effects of multiscale network entanglement on empirical systems. Communications Physics, 4(1), 129. \href{https://doi.org/10.1038/s42005-021-00633-0}{doi: 10.1038/s42005-021-00633-0.}



\bibitem{Gialampoukidis2016} Gialampoukidis, I., Kalpakis, G., Tsikrika, T., Vrochidis, S., \& Kompatsiaris, I. (2016). Key player identification in terrorism-related social media networks using centrality measures. In 2016 European Intelligence and Security Informatics Conference (EISIC) (pp. 112-115). IEEE. \href{https://doi.org/10.1109/EISIC.2016.029}{doi: 10.1109/EISIC.2016.029.}



\bibitem{Goh2001} Goh, K. I., Kahng, B., \& Kim, D. (2001). Universal behavior of load distribution in scale-free networks. Physical review letters, 87(27), 278701. \href{https://doi.org/10.1103/PhysRevLett.87.278701}{doi: 10.1103/PhysRevLett.87.278701.}

\bibitem{Guimaraes1972} Guimaraes, L. L. (1973). Communication integration in modern and traditional social systems: a comparative analysis across twenty communities of Minas Gerais, Brazil. PhD Thesis, Michigan State Univ., East Lansing. \href{https://doi.org/10.25335/jcw2-6990}{doi: 10.25335/jcw2-6990.}


\bibitem{Guimerà2005} Guimera, R., \& Nunes Amaral, L. A. (2005). Functional cartography of complex metabolic networks. nature, 433(7028), 895-900. \href{https://doi.org/10.1038/nature03288}{doi: 10.1038/nature03288.}



\bibitem{Guo2020} Guo, C., Yang, L., Chen, X., Chen, D., Gao, H., \& Ma, J. (2020). Influential nodes identification in complex networks via information entropy. Entropy, 22(2), 242. \href{https://doi.org/10.3390/e22020242}{doi: 10.3390/e22020242.}

\bibitem{Guo2024} Guo, H., Wang, S., Yan, X., \& Zhang, K. (2024). Node importance evaluation method of complex network based on the fusion gravity model. Chaos, Solitons \& Fractals, 183, 114924. \href{https://doi.org/10.1016/j.chaos.2024.114924}{doi: 10.1016/j.chaos.2024.114924.}


\bibitem{Guo2016} Guo, L., Lin, J. H., Guo, Q., \& Liu, J. G. (2016). Identifying multiple influential spreaders in term of the distance-based coloring. Physics Letters A, 380(7-8), 837-842. \href{https://doi.org/10.1016/j.physleta.2015.12.031}{doi: 10.1016/j.physleta.2015.12.031.}



\bibitem{Gupta2016} Gupta, N., Singh, A., \& Cherifi, H. (2016). Centrality measures for networks with community structure. Physica A: Statistical Mechanics and its Applications, 452, 46-59. \href{https://doi.org/10.1016/j.physa.2016.01.066}{doi: 10.1016/j.physa.2016.01.066.}


\bibitem{Gupta2021} Gupta, M., \& Mishra, R. (2021). Spreading the information in complex networks: Identifying a set of top-N influential nodes using network structure. Decision Support Systems, 149, 113608. \href{https://doi.org/10.1016/j.dss.2021.113608}{doi: 10.1016/j.dss.2021.113608.}


\bibitem{Hage1995} Hage, P., \& Harary, F. (1995). Eccentricity and centrality in networks. Social networks, 17(1), 57-63. \href{https://doi.org/10.1016/0378-8733(94)00248-9}{doi: 10.1016/0378-8733(94)00248-9.}


\bibitem{Hajarathaiah2022} Hajarathaiah, K., Enduri, M. K., \& Anamalamudi, S. (2022). Efficient algorithm for finding the influential nodes using local relative change of average shortest path. Physica A: Statistical Mechanics and its Applications, 591, 126708. \href{https://doi.org/10.1016/j.physa.2021.126708}{doi: 10.1016/j.physa.2021.126708.}

\bibitem{Hanneman2005} Hanneman, R. A., \& Riddle, M. (2005). Introduction to social network methods.


\bibitem{Harris1954} Harris, C. D. (1954). The, Market as a Factor in the Localization of Industry in the United States. Annals of the association of American geographers, 44(4), 315-348. \href{https://doi.org/10.1080/00045605409352140}{doi: 10.1080/00045605409352140.}


\bibitem{HebertDufresne2016} Hébert-Dufresne, L., Grochow, J. A., \& Allard, A. (2016). Multi-scale structure and topological anomaly detection via a new network statistic: The onion decomposition. Scientific reports, 6(1), 31708. \href{https://doi.org/10.1038/srep31708}{doi: 10.1038/srep31708.}



\bibitem{HebertDufresne2023} Hébert-Dufresne, L., St-Onge, G., Meluso, J., Bagrow, J., \& Allard, A. (2023). Hierarchical team structure and multidimensional localization (or siloing) on networks. Journal of Physics: Complexity, 4(3), 035002. \href{https://doi.org/10.1088/2632-072X/ace602}{doi: 10.1088/2632-072X/ace602.}


\bibitem{Hernández2011} Hernández, J. M., \& Van Mieghem, P. (2011). Classification of graph metrics. Delft University of Technology. Tech Rep.


\bibitem{Higley1991} Higley, J., Hoffmann-Lange, U., Kadushin, C., \& Moore, G. (1991). Elite integration in stable democracies: a reconsideration. European Sociological Review, 7(1), 35-53. \href{https://doi.org/10.1093/oxfordjournals.esr.a036576}{doi: 10.1093/oxfordjournals.esr.a036576.}


\bibitem{Honey2007} Honey, C. J., Kötter, R., Breakspear, M., \& Sporns, O. (2007). Network structure of cerebral cortex shapes functional connectivity on multiple time scales. Proceedings of the National Academy of Sciences, 104(24), 10240-10245. \href{https://doi.org/10.1073/pnas.0701519104}{doi: 10.1073/pnas.0701519104.}

\bibitem{Hou2012} Hou, B., Yao, Y., \& Liao, D. (2012). Identifying all-around nodes for spreading dynamics in complex networks. Physica A: Statistical Mechanics and its Applications, 391(15), 4012-4017. \href{https://doi.org/10.1016/j.physa.2012.02.033}{doi: 10.1016/j.physa.2012.02.033.}


\bibitem{Hu2016} Hu, J., Du, Y., Mo, H., Wei, D., \& Deng, Y. (2016). A modified weighted TOPSIS to identify influential nodes in complex networks. Physica A: Statistical Mechanics and its Applications, 444, 73-85. \href{https://doi.org/10.1016/j.physa.2015.09.028}{doi: 10.1016/j.physa.2015.09.028.}


\bibitem{Hu2015} Hu, P., Fan, W., \& Mei, S. (2015). Identifying node importance in complex networks. Physica A: Statistical Mechanics and its Applications, 429, 169-176. \href{https://doi.org/10.1016/j.physa.2015.02.002}{doi: 10.1016/j.physa.2015.02.002.}

\bibitem{Hu2018} Hu, P., \& Mei, T. (2018). Ranking influential nodes in complex networks with structural holes. Physica A: Statistical Mechanics and its Applications, 490, 624-631. \href{https://doi.org/10.1016/j.physa.2017.08.049}{doi: 10.1016/j.physa.2017.08.049.}


\bibitem{Huang2020}Huang, X., Chen, D., Wang, D., \& Ren, T. (2020). Identifying influencers in social networks. Entropy, 22(4), 450. \href{https://doi.org/10.3390/e22040450}{doi: 10.3390/e22040450.}


\bibitem{Huang2024} Huang, Y., Wang, H., Ren, X. L., \& Lü, L. (2024). Identifying key players in complex networks via network entanglement. Communications physics, 7(1), 19. \href{https://doi.org/10.1038/s42005-023-01483-8}{doi: 10.1038/s42005-023-01483-8.}


\bibitem{Hubbell1965} Hubbell, C. H. (1965). An Input-Output Approach to Clique Identification. Sociometry, 28(4), 377-399. \href{https://doi.org/10.2307/2785990}{doi: 10.2307/2785990.}



\bibitem{Hwang2009} Hwang, Y. C., Lin, C. C., Chang, J. Y., Mori, H., Juan, H. F., \& Huang, H. C. (2009). Predicting essential genes based on network and sequence analysis. Molecular BioSystems, 5(12), 1672-1678. \href{https://doi.org/10.1039/B900611G}{doi: 10.1039/B900611G.}



\bibitem{Hwang2006} Hwang, W., Cho, Y. R., Zhang, A., \& Ramanathan, M. (2006). Bridging centrality: identifying bridging nodes in scale-free networks. In Proceedings of the 12th ACM SIGKDD international conference on Knowledge discovery and data mining (pp. 20-23). 


\bibitem{Iannelli2017} Iannelli, F., Koher, A., Brockmann, D., Hövel, P., \& Sokolov, I. M. (2017). Effective distances for epidemics spreading on complex networks. Physical Review E, 95(1), 012313. \href{https://doi.org/10.1103/PhysRevE.95.012313}{doi: 10.1103/PhysRevE.95.012313.}


\bibitem{Iannelli2018} Iannelli, F., Mariani, M. S., \& Sokolov, I. M. (2018). Influencers identification in complex networks through reaction-diffusion dynamics. Physical Review E, 98(6), 062302. \href{https://doi.org/10.1103/PhysRevE.98.062302}{doi: 10.1103/PhysRevE.98.062302.}


\bibitem{Ibnoulouafi2018} Ibnoulouafi, A., El Haziti, M., \& Cherifi, H. (2018). M-centrality: identifying key nodes based on global position and local degree variation. Journal of Statistical Mechanics: Theory and Experiment, 2018(7), 073407. \href{https://doi.org/10.1088/1742-5468/aace08}{doi: 10.1088/1742-5468/aace08.}


\bibitem{Ilyas2011} Ilyas, M. U., \& Radha, H. (2011). Identifying influential nodes in online social networks using principal component centrality. In 2011 IEEE International Conference on Communications (ICC) (pp. 1-5). IEEE. \href{https://doi.org/10.1109/icc.2011.5963147}{doi: 10.1109/icc.2011.5963147.}


\bibitem{Interdonato2015} Interdonato, R., \& Tagarelli, A. (2015). Ranking Silent Nodes in Information Networks: a quantitative approach and applications. Physics Procedia, 62, 36-41. \href{https://doi.org/10.1016/j.phpro.2015.02.008}{doi: 10.1016/j.phpro.2015.02.008.}



\bibitem{Jackson2008} Jackson, M. O. (2008). Social and economic networks (Vol. 3, p. 519). Princeton: Princeton university press. \href{https://doi.org/10.2307/j.ctvcm4gh1}{doi: 10.2307/j.ctvcm4gh1.}


\bibitem{Jackson2020} Jackson, M. O. (2020). A typology of social capital and associated network measures. Social choice and welfare, 54(2), 311-336. \href{https://doi.org/10.1007/s00355-019-01189-3}{doi: 10.1007/s00355-019-01189-3.}


\bibitem{Jalili2015} Jalili, M., Salehzadeh-Yazdi, A., Asgari, Y., Arab, S. S., Yaghmaie, M., Ghavamzadeh, A., \& Alimoghaddam, K. (2015). CentiServer: a comprehensive resource, web-based application and R package for centrality analysis. PloS one, 10(11), e0143111. \href{https://doi.org/10.1371/journal.pone.0143111}{doi: 10.1371/journal.pone.0143111.}


\bibitem{Jeh2002} Jeh, G., \& Widom, J. (2002). Simrank: a measure of structural-context similarity. In Proceedings of the eighth ACM SIGKDD international conference on Knowledge discovery and data mining (pp. 538-543). \href{https://doi.org/10.1145/775047.775126}{doi: 10.1145/775047.775126.}

\bibitem{Jia2011} Jia-sheng, W., Xiao-ping, W., Bo, Y., \& Jiang-wei, G. (2011). Improved method of node importance evaluation based on node contraction in complex networks. Procedia Engineering, 15, 1600-1604. \href{https://doi.org/10.1016/j.proeng.2011.08.298}{doi: 10.1016/j.proeng.2011.08.298.}



\bibitem{Joyce2010} Joyce, K. E., Laurienti, P. J., Burdette, J. H., \& Hayasaka, S. (2010). A new measure of centrality for brain networks. PloS one, 5(8), e12200. \href{https://doi.org/10.1371/journal.pone.0012200}{doi: 10.1371/journal.pone.0012200.}



\bibitem{Kalinka2011} Kalinka, A. T., \& Tomancak, P. (2011). linkcomm: an R package for the generation, visualization, and analysis of link communities in networks of arbitrary size and type. Bioinformatics, 27(14), 2011-2012. \href{https://doi.org/10.1093/bioinformatics/btr311}{doi: 10.1093/bioinformatics/btr311.}



\bibitem{Kamvar2003} Kamvar, S. D., Schlosser, M. T., \& Garcia-Molina, H. (2003). The eigentrust algorithm for reputation management in p2p networks. In Proceedings of the 12th international conference on World Wide Web (pp. 640-651). \href{https://doi.org/10.1145/775152.775242}{doi: 10.1145/775152.775242.}



\bibitem{Kang2011} Kang, U., Papadimitriou, S., Sun, J., \& Tong, H. (2011). Centralities in large networks: Algorithms and observations. In Proceedings of the 2011 SIAM international conference on data mining, pp. 119-130. Society for Industrial and Applied Mathematics. \href{https://doi.org/10.1137/1.9781611972818.11}{doi: 10.1137/1.9781611972818.11.}




\bibitem{Katz1953} Katz, L. (1953). A new status index derived from sociometric analysis. Psychometrika, 18(1), 39-43. \href{https://doi.org/10.1007/BF02289026}{doi: 10.1007/BF02289026.}

\bibitem{Kempe2005} Kempe, D., Kleinberg, J., \& Tardos, É. (2005, July). Influential nodes in a diffusion model for social networks. In international colloquium on automata, languages, and programming (pp. 1127-1138). Berlin, Heidelberg: Springer Berlin Heidelberg. \href{https://doi.org/10.1007/11523468_91}{doi: 10.1007/11523468\_91.}

\bibitem{Kendall1955} Kendall, M. G. (1955). Further contributions to the theory of paired comparisons. Biometrics, 11(1), 43-62. \href{https://doi.org/10.2307/3001479}{doi: 10.2307/3001479.}


\bibitem{Kermarrec2011} Kermarrec, A. M., Le Merrer, E., Sericola, B., \& Trédan, G. (2011). Second order centrality: Distributed assessment of nodes criticity in complex networks. Computer Communications, 34(5), 619-628. \href{https://doi.org/10.1016/j.comcom.2010.06.00}{doi: 10.1016/j.comcom.2010.06.00.}



\bibitem{Kim2012} Kim, H., \& Yoneki, E. (2012). Influential neighbours selection for information diffusion in online social networks. In 2012 21st international conference on computer communications and networks (ICCCN) (pp. 1-7). IEEE. \href{https://doi.org/10.1109/ICCCN.2012.6289230}{doi: 10.1109/ICCCN.2012.6289230.}



\bibitem{Kirkland2010} Kirkland, S. (2010). Algebraic connectivity for vertex-deleted subgraphs, and a notion of vertex centrality. Discrete Mathematics, 310(4), 911-921. \href{https://doi.org/10.1016/j.disc.2009.10.011}{doi: 10.1016/j.disc.2009.10.011.}


\bibitem{Kitsak2010} Kitsak, M., Gallos, L. K., Havlin, S., Liljeros, F., Muchnik, L., Stanley, H. E., \& Makse, H. A. (2010). Identification of influential spreaders in complex networks. Nature physics, 6(11), 888-893. \href{https://doi.org/10.1038/nphys1746}{doi: 10.1038/nphys1746.}



\bibitem{Kivimäki2016} Kivimäki, I., Lebichot, B., Saramäki, J., \& Saerens, M. (2016). Two betweenness centrality measures based on randomized shortest paths. Scientific reports, 6(1), 19668. \href{https://doi.org/10.1038/srep19668}{doi: 10.1038/srep19668.}


\bibitem{Kleinberg1999} Kleinberg, J. (1999). Authoritative sources in a hyperlinked environment. J. ACM 46, 604-632. \href{https://doi.org/10.1145/324133.324140}{doi: 10.1145/324133.324140.}

\bibitem{Klemm2012} Klemm, K., Serrano, M. Á., Eguíluz, V. M., \& Miguel, M. S. (2012). A measure of individual role in collective dynamics. Scientific reports, 2(1), 292. \href{https://doi.org/10.1038/srep00292}{doi: 10.1038/srep00292.}




\bibitem{Knill2011} Knill, O. (2011). A graph theoretical Gauss-Bonnet-Chern theorem. arXiv preprint arXiv:1111.5395. \href{https://doi.org/10.48550/arXiv.1111.5395}{doi: 10.48550/arXiv.1111.5395.}


\bibitem{Knill2012} Knill, O. (2012). On index expectation and curvature for networks. arXiv preprint arXiv:1202.4514. \href{https://doi.org/10.48550/arXiv.1202.4514}{doi: 10.48550/arXiv.1202.4514.}



\bibitem{Konstantinova2006} Konstantinova, E. V. (2006). On some applications of information indices in chemical graph theory. General Theory of Information Transfer and Combinatorics, 831-852. \href{https://doi.org/10.1016/j.endm.2005.07.071}{doi: 10.1016/j.endm.2005.07.071.}

\bibitem{Korn2009} Korn, A., Schubert, A., \& Telcs, A. (2009). Lobby index in networks. Physica A: Statistical Mechanics and its Applications, 388(11), 2221-2226. \href{https://doi.org/10.1016/j.physa.2009.02.013}{doi: 10.1016/j.physa.2009.02.013.}


\bibitem{Koschützki2005} Koschützki, D., Lehmann, K. A., Tenfelde-Podehl, D., \& Zlotowski, O. (2005). Advanced centrality concepts. In Network analysis: Methodological foundations (pp. 83-111). Berlin, Heidelberg: Springer Berlin Heidelberg. \href{https://doi.org/10.1007/978-3-540-31955-9_5}{doi: 10.1007/978-3-540-31955-9\_5.}

\bibitem{Kovács2010} Kovács, I. A., Palotai, R., Szalay, M. S., \& Csermely, P. (2010). Community landscapes: an integrative approach to determine overlapping network module hierarchy, identify key nodes and predict network dynamics. PloS one, 5(9), e12528. \href{https://doi.org/10.1371/journal.pone.0012528}{doi: 10.1371/journal.pone.0012528.}


\bibitem{Kumar2020} Kumar, S., \& Panda, B. S. (2020). Identifying influential nodes in Social Networks: Neighborhood Coreness based voting approach. Physica A: Statistical Mechanics and its Applications, 553, 124215. \href{https://doi.org/10.1016/j.physa.2020.124215}{doi: 10.1016/j.physa.2020.124215.}




\bibitem{Kumar2022} Kumar, S., \& Panda, A. (2022). Identifying influential nodes in weighted complex networks using an improved WVoteRank approach. Applied intelligence, 52(2), 1838-1852. \href{https://doi.org/10.1109/ACCESS.2017.2679038}{doi: 10.1109/ACCESS.2017.2679038.}



\bibitem{Kundu2011} Kundu, S., Murthy, C. A., \& Pal, S. K. (2011). A new centrality measure for influence maximization in social networks. In International conference on pattern recognition and machine intelligence (pp. 242-247). Berlin, Heidelberg: Springer Berlin Heidelberg. \href{https://doi.org/10.1007/978-3-642-21786-9_40}{doi: 10.1007/978-3-642-21786-9\_40.}




\bibitem{Landau1895} Landau, E. (1895). Zur relativen wertbemessung der turnierresultate. Deutsches Wochenschach, 11(366-369), 3.

\bibitem{Landherr2010} Landherr, A., Friedl, B., \& Heidemann, J. (2010). A critical review of centrality measures in social networks. Business \& Information Systems Engineering, 2(6), 371-385. \href{https://doi.org/10.1007/s12599-010-0127-3}{doi: 10.1007/s12599-010-0127-3.}

\bibitem{Latora2001} Latora, V., \& Marchiori, M. (2001). Efficient behavior of small-world networks. Physical review letters, 87(19), 198701. \href{https://doi.org/10.1103/PhysRevLett.87.198701}{doi: 10.1103/PhysRevLett.87.198701.}


\bibitem{Latora2007} Latora, V., \& Marchiori, M. (2007). A measure of centrality based on network efficiency. New Journal of Physics, 9(6), 188. \href{https://doi.org/10.1088/1367-2630/9/6/188}{doi: 10.1088/1367-2630/9/6/188.}


\bibitem{Leavitt1951} Leavitt, H. J. (1951). Some effects of certain communication patterns on group performance. The journal of abnormal and social psychology, 46(1), 38. \href{https://doi.org/10.1037/h0057189}{doi: 10.1037/h0057189.}


\bibitem{Lawyer2015} Lawyer, G. (2015). Understanding the influence of all nodes in a network. Scientific reports, 5(1), 8665. \href{https://doi.org/10.1038/srep08665}{doi: 10.1038/srep08665.}



\bibitem{Lempel2001} Lempel, R., \& Moran, S. (2001). SALSA: the stochastic approach for link-structure analysis. ACM Transactions on Information Systems (TOIS), 19(2), 131-160. \href{https://doi.org/10.1145/382979.383041}{doi: 10.1145/382979.383041.}





\bibitem{PVM2015} Li, C., Li, Q., Van Mieghem, P., Stanley, H. E., \& Wang, H. (2015). Correlation between centrality metrics and their application to the opinion model. The European Physical Journal B, 88(3), 65. \href{https://doi.org/10.1140/epjb/e2015-50671-y}{doi: 10.1140/epjb/e2015-50671-y.}




\bibitem{CLi2018} Li, C., Wang, L., Sun, S., \& Xia, C. (2018). Identification of influential spreaders based on classified neighbors in real-world complex networks. Applied Mathematics and Computation, 320, 512-523. \href{https://doi.org/10.1016/j.amc.2017.10.001}{doi: 10.1016/j.amc.2017.10.001.}


\bibitem{HLi2021b} Li, H., \& Deng, Y. (2021). Local volume dimension: A novel approach for important nodes identification in complex networks. International Journal of Modern Physics B, 35(05), 2150069. \href{https://doi.org/10.1142/S0217979221500697}{doi: 10.1142/S0217979221500697.}




\bibitem{HLi2021} Li, H., Shang, Q., \& Deng, Y. (2021). A generalized gravity model for influential spreaders identification in complex networks. Chaos, Solitons \& Fractals, 143, 110456. \href{https://doi.org/10.1016/j.chaos.2020.110456}{doi: 10.1016/j.chaos.2020.110456.}


\bibitem{Li2009} Li, J., \& Willett, P. (2009). ArticleRank: a PageRank‐based alternative to numbers of citations for analysing citation networks. In Aslib Proceedings (Vol. 61, No. 6, pp. 605-618). Emerald Group Publishing Limited. \href{https://doi.org/10.1108/00012530911005544}{doi: 10.1108/00012530911005544.}


\bibitem{Li2011} Li, M., Wang, J., Chen, X., Wang, H., \& Pan, Y. (2011). A local average connectivity-based method for identifying essential proteins from the network level. Computational biology and chemistry, 35(3), 143-150. \href{https://doi.org/10.1016/j.compbiolchem.2011.04.002}{doi: 10.1016/j.compbiolchem.2011.04.002.}


\bibitem{Li2018} Li, M., Zhang, R., Hu, R., Yang, F., Yao, Y., \& Yuan, Y. (2018). Identifying and ranking influential spreaders in complex networks by combining a local-degree sum and the clustering coefficient. International Journal of Modern Physics B, 32(06), 1850118. \href{https://doi.org/10.1142/S0217979218501187}{doi: 10.1142/S0217979218501187.}



\bibitem{Li2014} Li, Q., Zhou, T., Lü, L., \& Chen, D. (2014). Identifying influential spreaders by weighted LeaderRank. Physica A: Statistical Mechanics and its Applications, 404, 47-55. \href{https://doi.org/10.1016/j.physa.2014.02.041}{doi: 10.1016/j.physa.2014.02.041.}


\bibitem{SLi2021} Li, S., \& Xiao, F. (2021). The identification of crucial spreaders in complex networks by effective gravity model. Information Sciences, 578, 725-749. \href{https://doi.org/10.1016/j.ins.2021.08.026}{doi: 10.1016/j.ins.2021.08.026.}


\bibitem{YLi2019} Li, Y., Cai, W., Li, Y., \& Du, X. (2019). Key node ranking in complex networks: A novel entropy and mutual information-based approach. Entropy, 22(1), 52. \href{https://doi.org/10.3390/e22010052}{doi: 10.3390/e22010052.}


\bibitem{Li2021} Li, Z., \& Huang, X. (2021). Identifying influential spreaders in complex networks by an improved gravity model. Scientific reports, 11(1), 22194. \href{https://doi.org/10.1038/s41598-021-01218-1}{doi: 10.1038/s41598-021-01218-1.}


\bibitem{Li2022} Li, Z., \& Huang, X. (2022). Identifying influential spreaders by gravity model considering multi-characteristics of nodes. Scientific Reports, 12(1), 9879. \href{https://doi.org/10.1038/s41598-022-14005-3}{doi: 10.1038/s41598-022-14005-3.}


\bibitem{Li2019} Li, Z., Ren, T., Ma, X., Liu, S., Zhang, Y. \& Zhou, T. Identifying influential spreaders by gravity model. Sci Rep 9, 8387 (2019). \href{https://doi.org/10.1038/s41598-019-44930-9}{doi: 10.1038/s41598-019-44930-9.} 



\bibitem{LiYa2019} Li-Ya, H., Ping-Chuan, T., You-Liang, H., Yi, Z., \& Xie-Feng, C. (2019). Node importance based on the weighted K-order propagation number algorithm. Acta Physica Sinica, 68(12). \href{https://doi.org/10.3390/e22030364}{doi: 10.3390/e22030364.} 



\bibitem{Lin2008} Lin, C. Y., Chin, C. H., Wu, H. H., Chen, S. H., Ho, C. W., \& Ko, M. T. (2008). Hubba: hub objects analyzer—a framework of interactome hubs identification for network biology. Nucleic acids research, 36, W438-W443. \href{https://doi.org/10.1093/nar/gkn257}{doi: 10.1093/nar/gkn257.} 



\bibitem{Lin2014} Lin, J. H., Guo, Q., Dong, W. Z., Tang, L. Y., \& Liu, J. G. (2014). Identifying the node spreading influence with largest k-core values. Physics Letters A, 378(45), 3279-3284. \href{https://doi.org/10.1016/j.physleta.2014.09.054}{doi: 10.1016/j.physleta.2014.09.054.} 


\bibitem{Lin1976} Lin, N. (1976). Foundations of social research. McGraw-Hill, New York. \href{https://doi.org/10.2307/2066441}{doi: 10.2307/2066441.} 


\bibitem{Liu2020} Liu, F., Wang, Z., \& Deng, Y. (2020). GMM: A generalized mechanics model for identifying the importance of nodes in complex networks. Knowledge-Based Systems, 193, 105464. \href{https://doi.org/10.1016/j.knosys.2019.105464}{doi: 10.1016/j.knosys.2019.105464.} 


\bibitem{HLiu2013} Liu, H., Yu, X., \& Lu, J. (2013). Identifying TOP-N opinion leaders on local social network. In IET International Conference on Smart and Sustainable City 2013 (ICSSC 2013) (pp. 268-271). Stevenage UK: IET. \href{https://doi.org/10.1049/cp.2013.1970}{doi: 10.1049/cp.2013.1970.} 




\bibitem{Liu2013} Liu, J. G., Ren, Z. M., \& Guo, Q. (2013). Ranking the spreading influence in complex networks. Physica A: Statistical Mechanics and its Applications, 392(18), 4154-4159. \href{https://doi.org/10.1016/j.physa.2013.04.037}{doi: 10.1016/j.physa.2013.04.037.} 


\bibitem{JLiu2016} Liu, J., Xiong, Q., Shi, W., Shi, X., \& Wang, K. (2016). Evaluating the importance of nodes in complex networks. Physica A: Statistical Mechanics and its Applications, 452, 209-219. \href{https://doi.org/10.1016/j.physa.2016.02.049}{doi: 10.1016/j.physa.2016.02.049.} 


\bibitem{JGLiu2016} Liu, J. G., Lin, J. H., Guo, Q., \& Zhou, T. (2016). Locating influential nodes via dynamics-sensitive centrality. Scientific reports, 6(1), 21380. \href{https://doi.org/10.1038/srep21380}{doi: 10.1038/srep21380.} 


\bibitem{LiuJ2023} Liu, J., \& Zheng, J. (2023). Identifying important nodes in complex networks based on extended degree and E-shell hierarchy decomposition. Scientific Reports, 13(1), 3197. \href{https://doi.org/10.1038/s41598-023-30308-5}{doi: 10.1038/s41598-023-30308-5.} 


\bibitem{Liu2021} Liu, P., Li, L., Fang, S., \& Yao, Y. (2021). Identifying influential nodes in social networks: A voting approach. Chaos, Solitons \& Fractals, 152, 111309. \href{https://doi.org/10.1016/j.chaos.2021.111309}{doi: 10.1016/j.chaos.2021.111309.} 


\bibitem{Liu2018} Liu, Q., Zhu, Y. X., Jia, Y., Deng, L., Zhou, B., Zhu, J. X., \& Zou, P. (2018). Leveraging local h-index to identify and rank influential spreaders in networks. Physica A: Statistical Mechanics and its Applications, 512, 379-391. \href{https://doi.org/10.1016/j.physa.2018.08.053}{doi: 10.1016/j.physa.2018.08.053.} 




\bibitem{Liu2014} Liu, Y., Jin, J., Zhang, Y., \& Xu, C. (2014). A new clustering algorithm based on data field in complex networks. The Journal of Supercomputing, 67, 723-737. \href{https://doi.org/10.1007/s11227-013-0984-x}{doi: 10.1007/s11227-013-0984-x.} 

\bibitem{YLiu2015} Liu, Y., Tang, M., Yue, J., \& Gong, J. (2015). Identify influential spreaders in complex real-world networks. In 2015 IEEE 12th Intl Conf on Ubiquitous Intelligence and Computing and 2015 IEEE 12th Intl Conf on Autonomic and Trusted Computing and 2015 IEEE 15th Intl Conf on Scalable Computing and Communications and Its Associated Workshops (UIC-ATC-ScalCom) (pp. 1144-1148). IEEE. \href{https://doi.org/10.1109/UIC-ATC-ScalCom-CBDCom-IoP.2015.209}{doi: 10.1109/UIC-ATC-ScalCom-CBDCom-IoP.2015.209.} 




\bibitem{YLiuTang2015} Liu, Y., Tang, M., Zhou, T., \& Do, Y. (2015). Improving the accuracy of the k-shell method by removing redundant links: From a perspective of spreading dynamics. Scientific reports, 5(1), 13172. \href{https://doi.org/10.1038/srep13172}{doi: 10.1038/srep13172.} 





\bibitem{YLiu2016} Liu, Y., Tang, M., Zhou, T., \& Do, Y. (2016). Identify influential spreaders in complex networks, the role of neighborhood. Physica A: Statistical Mechanics and its Applications, 452, 289-298. \href{https://doi.org/10.1016/j.physa.2016.02.028}{doi: 10.1016/j.physa.2016.02.028.} 




\bibitem{Liu2017} Liu, Y., Tang, M., Do, Y., \& Hui, P. M. (2017). Accurate ranking of influential spreaders in networks based on dynamically asymmetric link weights. Physical Review E, 96(2), 022323. \href{https://doi.org/10.1103/PhysRevE.96.022323}{doi: 10.1103/PhysRevE.96.022323.} 


\bibitem{Liu2016} Liu, Y., Wei, B., Du, Y., Xiao, F., \& Deng, Y. (2016). Identifying influential spreaders by weight degree centrality in complex networks. Chaos, Solitons \& Fractals, 86, 1-7.  \href{https://doi.org/10.1016/j.chaos.2016.01.030}{doi: 10.1016/j.chaos.2016.01.030.} 


\bibitem{ZLiu2015} Liu, Z., Jiang, C., Wang, J., \& Yu, H. (2015). The node importance in actual complex networks based on a multi-attribute ranking method. Knowledge-Based Systems, 84, 56-66. \href{https://doi.org/10.1016/j.knosys.2015.03.026}{doi: 10.1016/j.knosys.2015.03.026.} 


\bibitem{Lu2024} Lu, Y., Huang, Y., Nie, J., Chen, Z., \& Xuan, Q. (2024). RK-CORE: An Established Methodology for Exploring the Hierarchical Structure within Datasets. In ICASSP 2024-2024 IEEE International Conference on Acoustics, Speech and Signal Processing (ICASSP) (pp. 3150-3154). IEEE.  \href{https://doi.org/10.1109/ICASSP48485.2024.10447791}{doi: 10.1109/ICASSP48485.2024.10447791.} 



\bibitem{Lü2016} Lü, L., Chen, D., Ren, X. L., Zhang, Q. M., Zhang, Y. C., \& Zhou, T. (2016). Vital nodes identification in complex networks. Physics reports, 650, 1-63. \href{https://doi.org/10.1016/j.physrep.2016.06.007}{doi: 10.1016/j.physrep.2016.06.007.} 


\bibitem{Lü2011} Lü, L., Zhang, Y. C., Yeung, C. H., \& Zhou, T. (2011). Leaders in social networks, the delicious case. PloS one, 6(6), e21202. \href{https://doi.org/10.1371/journal.pone.0021202}{doi: 10.1371/journal.pone.0021202.} 


\bibitem{Lu2016} Lü, L., Zhou, T., Zhang, Q. M., \& Stanley, H. E. (2016). The H-index of a network node and its relation to degree and coreness. Nature communications, 7(1), 10168. \href{https://doi.org/10.1038/ncomms10168}{doi: 10.1038/ncomms10168.} 


\bibitem{Luan2021}Luan, Y., Bao, Z., \& Zhang, H. (2021). Identifying influential spreaders in complex networks by considering the impact of the number of shortest paths. Journal of Systems Science and Complexity, 34(6), 2168-2181. \href{https://doi.org/10.1007/s11424-021-0111-7}{doi: 10.1007/s11424-021-0111-7.} 


\bibitem{Lv2019} Lv, Z., Zhao, N., Xiong, F., \& Chen, N. (2019). A novel measure of identifying influential nodes in complex networks. Physica A: Statistical Mechanics and Its Applications, 523, 488-497. \href{https://doi.org/10.1016/j.physa.2019.01.136}{doi: 10.1016/j.physa.2019.01.136.} 


\bibitem{Ma2015} Ma, L. L., Ma, C., Zhang, H. F. \& Wang, B. H. Identifying influential spreaders in complex networks based on gravity formula. Physica A 451, 205-212 (2015). \href{https://doi.org/10.1016/j.physa.2015.12.162}{doi: 10.1016/j.physa.2015.12.162.} 


\bibitem{Ma2017} Ma, Q., \& Ma, J. (2017). Identifying and ranking influential spreaders in complex networks with consideration of spreading probability. Physica A: Statistical Mechanics and its Applications, 465, 312-330. \href{https://doi.org/10.1016/j.physa.2016.08.041}{doi: 10.1016/j.physa.2016.08.041.} 


\bibitem{Ma2019} Ma, Y., Cao, Z., \& Qi, X. (2019). Quasi-Laplacian centrality: A new vertex centrality measurement based on Quasi-Laplacian energy of networks. Physica A: Statistical mechanics and its applications, 527, 121130. \href{https://doi.org/10.1016/j.physa.2019.121130}{doi: 10.1016/j.physa.2019.121130.} 


\bibitem{Macker2016} Macker, J. P. (2016). An improved local bridging centrality model for distributed network analytics. In MILCOM 2016-2016 IEEE Military Communications Conference (pp. 600-605). IEEE. doi: 10.1109/MILCOM.2016.7795393. \href{https://doi.org/10.1109/MILCOM.2016.7795393}{doi: 10.1109/MILCOM.2016.7795393.} 


\bibitem{Madotto2016} Madotto, A., \& Liu, J. (2016). Super-spreader identification using meta-centrality. Scientific reports, 6(1), 38994. \href{https://doi.org/10.1038/srep38994}{doi: 10.1038/srep38994.} 


\bibitem{Magelinski2021} Magelinski, T., Bartulovic, M., \& Carley, K. M. (2021). Measuring node contribution to community structure with modularity vitality. IEEE Transactions on Network Science and Engineering, 8(1), 707-723. \href{https://doi.org/10.1109/TNSE.2020.3049068}{doi: 10.1109/TNSE.2020.3049068.} 



\bibitem{Mahdi2024} Mahdi J. (2024). Centiserver: the most comprehensive centrality resource and web application for centrality measures calculation. Available from: \href{https://www.centiserver.ir/centrality/list/}{https://www.centiserver.ir/centrality/list/} (accessed on 1 May 2025).

\bibitem{Maji2020} Maji, G. (2020). Influential spreaders identification in complex networks with potential edge weight based k-shell degree neighborhood method. Journal of Computational Science, 39, 101055. \href{https://doi.org/10.1016/j.jocs.2019.101055}{doi: 10.1016/j.jocs.2019.101055.} 




\bibitem{Maji2021} Maji, G., Dutta, A., Malta, M. C., \& Sen, S. (2021). Identifying and ranking super spreaders in real world complex networks without influence overlap. Expert Systems with Applications, 179, 115061. \href{https://doi.org/10.1016/j.eswa.2021.115061}{doi: 10.1016/j.eswa.2021.115061.} 


\bibitem{Maji2020c} Maji, G., Mandal, S., \& Sen, S. (2020). A systematic survey on influential spreaders identification in complex networks with a focus on K-shell based techniques. Expert Systems with Applications, 161, 113681. \href{https://doi.org/10.1016/j.eswa.2020.113681}{doi: 10.1016/j.eswa.2020.113681.} 



\bibitem{Maji2020b} Maji, G., Namtirtha, A., Dutta, A., \& Malta, M. C. (2020). Influential spreaders identification in complex networks with improved k-shell hybrid method. Expert Systems with Applications, 144, 113092. \href{https://doi.org/10.1016/j.eswa.2019.113092}{doi: 10.1016/j.eswa.2019.113092.} 








\bibitem{Malliaros2016} Malliaros, F. D., Rossi, M. E. G., \& Vazirgiannis, M. (2016). Locating influential nodes in complex networks. Scientific reports, 6(1), 19307. \href{https://doi.org/10.1038/srep19307}{doi: 10.1038/srep19307.} 


\bibitem{Marsden2002} Marsden, P. V. (2002). Egocentric and sociocentric measures of network centrality. Social networks, 24(4), 407-422. \href{https://doi.org/10.1016/S0378-8733(02)00016-3}{doi: 10.1016/S0378-8733(02)00016-3.} 




\bibitem{Markovsky1988} Markovsky, B., Willer, D., \& Patton, T. (1988). Power relations in exchange networks. American sociological review, 220-236. \href{https://doi.org/10.2307/2095634}{doi: 10.2307/2095634.} 


\bibitem{Martin2014} Martin, T., Zhang, X., \& Newman, M. E. (2014). Localization and centrality in networks. Physical review E, 90(5), 052808. \href{https://doi.org/10.1103/PhysRevE.90.052808}{doi: 10.1103/PhysRevE.90.052808.} 


\bibitem{Maslov2002} Maslov, S., \& Sneppen, K. (2002). Specificity and stability in topology of protein networks. Science, 296(5569), 910-913. \href{https://doi.org/10.1126/science.1065103}{doi: 10.1126/science.1065103.} 



\bibitem{Mavroforakis2015} Mavroforakis, C., Mathioudakis, M., \& Gionis, A. (2015). Absorbing random-walk centrality: Theory and algorithms. In 2015 IEEE International Conference on Data Mining (pp. 901-906). IEEE. \href{https://doi.org/10.1109/ICDM.2015.103}{doi: 10.1109/ICDM.2015.103.} 





\bibitem{MS2018} Meshcheryakova, N. and Shvydun, S. (2018). Power in Network Structures Based on Simulations. In: Studies in Computational Intelligence, vol 689. Springer, Cham. \href{https://doi.org/10.1007/978-3-319-72150-7_83}{doi: 10.1007/978-3-319-72150-7\_83.}


\bibitem{MS2023} Meshcheryakova, N., \& Shvydun, S. (2023). Perturbation analysis of centrality measures. In Proceedings of the International Conference on Advances in Social Networks Analysis and Mining (pp. 407-414).


\bibitem{MS2024} Meshcheryakova, N., \& Shvydun, S. (2024). A comparative analysis of centrality measures in complex networks. Automation and Remote Control, 85(8), 685-695. \href{https://doi.org/10.1134/S0005117924700127}{doi: 10.1134/S0005117924700127.} 



\bibitem{Michalak2013} Michalak, T. P., Aadithya, K. V., Szczepanski, P. L., Ravindran, B., \& Jennings, N. R. (2013). Efficient computation of the Shapley value for game-theoretic network centrality. Journal of Artificial Intelligence Research, 46, 607-650. \href{https://doi.org/10.5555/2512538.2512553}{doi: 10.5555/2512538.2512553.} 



\bibitem{Milenković2011} Milenković, T., Memišević, V., Bonato, A., \& Pržulj, N. (2011). Dominating biological networks. PloS one, 6(8), e23016. \href{https://doi.org/10.1371/journal.pone.0023016}{doi: 10.1371/journal.pone.0023016.} 



\bibitem{Milenković2008} Milenković, T., \& Pržulj, N. (2008). Uncovering biological network function via graphlet degree signatures. Cancer informatics, 6, CIN-S680. \href{https://doi.org/10.4137/CIN.S680}{doi: 10.4137/CIN.S680.}


\bibitem{Mimar2022} Mimar, S., \& Ghoshal, G. (2022). A sampling-guided unsupervised learning method to capture percolation in complex networks. Scientific Reports, 12(1), 4147. \href{https://doi.org/10.1038/s41598-022-07921-x}{doi: 10.1038/s41598-022-07921-x.}


\bibitem{Min2018} Min, B. (2018). Identifying an influential spreader from a single seed in complex networks via a message-passing approach. The European Physical Journal B, 91(1), 18. \href{https://doi.org/10.1140/epjb/e2017-80597-1}{doi: 10.1140/epjb/e2017-80597-1.}





\bibitem{Mo2019} Mo, H., \& Deng, Y. (2019). Identifying node importance based on evidence theory in complex networks. Physica A: Statistical Mechanics and its Applications, 529, 121538. \href{https://doi.org/10.1016/j.physa.2019.121538}{doi: 10.1016/j.physa.2019.121538.}




\bibitem{Mones2012} Mones, E., Vicsek, L., \& Vicsek, T. (2012). Hierarchy measure for complex networks. PloS one, 7(3), e33799. \href{https://doi.org/10.1371/journal.pone.0033799}{doi: 10.1371/journal.pone.0033799.}



\bibitem{Morone2015} Morone, F., \& Makse, H. A. (2015). Influence maximization in complex networks through optimal percolation. Nature, 524(7563), 65-68. \href{https://doi.org/10.1038/nature14604}{doi: 10.1038/nature14604.}




\bibitem{Moxley1974} Moxley, R. L., \& Moxley, N. F. (1974). Determining Point-Centrality in Uncontrived Social Networks. Sociometry, 37(1), 122-130. \href{https://doi.org/10.2307/2786472}{doi: 10.2307/2786472.}




\bibitem{Mukhtar2023} Mukhtar, M. F., Abal Abas, Z., Baharuddin, A. S., Norizan, M. N., Fakhruddin, W. F. W. W., Minato, W., Rasib, A. H. A, Abidin, Z. Z., Rahman, A. F. N. A. \& Anuar, S. H. H. (2023). Integrating local and global information to identify influential nodes in complex networks. Scientific Reports, 13(1), 11411. \href{https://doi.org/10.1038/s41598-023-37570-7}{doi: 10.1038/s41598-023-37570-7.}




\bibitem{Mussone2022} Mussone, L., Viseh, H., \& Notari, R. (2022). Novel centrality measures and applications to underground networks. Physica A: Statistical Mechanics and its Applications, 589, 126595. \href{https://doi.org/10.1016/j.physa.2021.126595}{doi: 10.1016/j.physa.2021.126595.}

\bibitem{Myerson}
Myerson, R.B., Graphs and Cooperation in Games, \textit{Math. Oper. Res.}, 1977, vol. 2., pp. 225-229. \href{https://doi.org/10.1287/moor.2.3.225}{doi: 10.1287/moor.2.3.225.}


\bibitem{Myerson2}
Myerson, R.B., Conference Structures and Fair Allocation Rules,  \textit{Int. J. Game Theory}, 1980, vol. 9, pp. 169-182. \href{https://doi.org/10.1007/BF01781371}{doi: 10.1007/BF01781371.}




\bibitem{Namtirtha2018a} Namtirtha, A., Dutta, A., \& Dutta, B. (2018). Weighted kshell degree neighborhood method: An approach independent of completeness of global network structure for identifying the influential spreaders. In 2018 10th international conference on communication systems \& networks (COMSNETS) (pp. 81-88). IEEE. \href{https://doi.org/10.1109/COMSNETS.2018.8328183}{doi: 10.1109/COMSNETS.2018.8328183.}


\bibitem{Namtirtha2018} Namtirtha, A., Dutta, A., \& Dutta, B. (2018). Identifying influential spreaders in complex networks based on kshell hybrid method. Physica A: Statistical Mechanics and Its Applications, 499, 310-324. \href{https://doi.org/10.1016/j.physa.2018.02.016}{doi: 10.1016/j.physa.2018.02.016.}



\bibitem{Namtirtha2020} Namtirtha, A., Dutta, A., \& Dutta, B. (2020). Weighted kshell degree neighborhood: A new method for identifying the influential spreaders from a variety of complex network connectivity structures. Expert Systems with Applications, 139, 112859. \href{https://doi.org/10.1016/j.eswa.2019.112859}{doi: 10.1016/j.eswa.2019.112859.}



\bibitem{Namtirtha2021} Namtirtha, A., Dutta, A., Dutta, B., Sundararajan, A., \& Simmhan, Y. (2021). Best influential spreaders identification using network global structural properties. Scientific reports, 11(1), 2254. \href{https://doi.org/10.1038/s41598-021-81614-9}{doi: 10.1038/s41598-021-81614-9.}




\bibitem{Namtirtha2022} Namtirtha, A., Dutta, B., \& Dutta, A. (2022). Semi-global triangular centrality measure for identifying the influential spreaders from undirected complex networks. Expert Systems with Applications, 206, 117791. \href{https://doi.org/10.1016/j.eswa.2022.117791}{doi: 10.1016/j.eswa.2022.117791.}



\bibitem{Nanda2008} Nanda, S., \& Kotz, D. (2008). Localized bridging centrality for distributed network analysis. In 2008 proceedings of 17th international conference on computer communications and networks (pp. 1-6). IEEE. \href{https://doi.org/10.1109/ICCCN.2008.ECP.31}{doi: 10.1109/ICCCN.2008.ECP.31.}




\bibitem{Negahban2017} Negahban, S., Oh, S., \& Shah, D. (2017). Rank Centrality: Ranking from Pairwise Comparisons. Operations Research, 65(1), 266-287. \href{https://doi.org/10.1287/opre.2016.1534}{doi: 10.1287/opre.2016.1534.}



\bibitem{Newman2005} Newman, M. E. (2005). A measure of betweenness centrality based on random walks. Social networks, 27(1), 39-54. \href{https://doi.org/10.1016/j.socnet.2004.11.009}{doi: 10.1016/j.socnet.2004.11.009.}




\bibitem{Newman2018} Newman, M. (2018). Networks. Oxford university press. \href{https://doi.org/10.1093/oso/9780198805090.001.0001}{doi: 10.1093/oso/9780198805090.001.0001.}




\bibitem{Ni2020} Ni, C., Yang, J., \& Kong, D. (2020). Sequential seeding strategy for social influence diffusion with improved entropy-based centrality. Physica A: Statistical Mechanics and its Applications, 545, 123659. \href{https://doi.org/10.1016/j.physa.2019.123659}{doi: 10.1016/j.physa.2019.123659.}


\bibitem{Nie2016} Nie, T., Guo, Z., Zhao, K., \& Lu, Z. M. (2016). Using mapping entropy to identify node centrality in complex networks. Physica A: Statistical Mechanics and its Applications, 453, 290-297. \href{https://doi.org/10.1016/j.physa.2016.02.009}{doi: 10.1016/j.physa.2016.02.009.}






\bibitem{Nieminen1973} Nieminen, U. J. (1973). On the centrality in a directed graph. Social science research, 2(4), 371-378. \href{https://doi.org/10.1016/0049-089X(73)90010-0}{doi: 10.1016/0049-089X(73)90010-0.}





\bibitem{Nikolaev2015} Nikolaev, A. G., Razib, R., \& Kucheriya, A. (2015). On efficient use of entropy centrality for social network analysis and community detection. Social Networks, 40, 154-162. \href{https://doi.org/10.1016/j.socnet.2014.10.002}{doi: 10.1016/j.socnet.2014.10.002.}





\bibitem{Noh2004} Noh, J. D., \& Rieger, H. (2004). Random walks on complex networks. Physical review letters, 92(11), 118701. \href{https://doi.org/10.1103/PhysRevLett.92.118701}{doi: 10.1103/PhysRevLett.92.118701.}



\bibitem{Oldham2019} Oldham, S., Fulcher, B., Parkes, L., Arnatkevi\v{c}i\={u}t\.{e}, A., Suo, C., \& Fornito, A. (2019). Consistency and differences between centrality measures across distinct classes of networks. PloS one, 14(7), e0220061. \href{https://doi.org/10.1371/journal.pone.0220061}{doi: 10.1371/journal.pone.0220061.}


\bibitem{Omar2020} Omar, Y. M., \& Plapper, P. (2020). A survey of information entropy metrics for complex networks. Entropy, 22(12), 1417. \href{https://doi.org/10.3390/e22121417}{doi: 10.3390/e22121417.}




\bibitem{Opsahl2010} Opsahl, T., Agneessens, F., \& Skvoretz, J. (2010). Node centrality in weighted networks: Generalizing degree and shortest paths. Social networks, 32(3), 245-251. \href{https://doi.org/10.1016/j.socnet.2010.03.006}{doi: 10.1016/j.socnet.2010.03.006.}




\bibitem{OrtizArroyo2008} Ortiz-Arroyo, D., Hussain, D.M.A. (2008). An Information Theory Approach to Identify Sets of Key Players. In: Lecture Notes in Computer Science, vol 5376. Springer, Berlin, Heidelberg. \href{https://doi.org/10.1007/978-3-540-89900-6_5}{doi: 10.1007/978-3-540-89900-6\_5.}



\bibitem{Oztemiz2024} Öztemiz, F., \& Yakut, S. (2024). An Effective Method for Determining Node Dominance Values: Malatya Centrality Algorithm. In 2024 32nd Signal Processing and Communications Applications Conference (SIU) (pp. 1-4). IEEE. \href{https://doi.org/10.1109/SIU61531.2024.10601155}{doi: 10.1109/SIU61531.2024.10601155.}




\bibitem{Papadimitriou2009} Papadimitriou, A., Katsaros, D., \& Manolopoulos, Y. (2009). Social network analysis and its applications in wireless sensor and vehicular networks. In International Conference on e-Democracy (pp. 411-420). Berlin, Heidelberg: Springer Berlin Heidelberg. \href{https://doi.org/10.1007/978-3-642-11631-5_37}{doi: 10.1007/978-3-642-11631-5\_37.}



\bibitem{Pei2014} Pei, S., Muchnik, L., Andrade, Jr, J. S., Zheng, Z., \& Makse, H. A. (2014). Searching for superspreaders of information in real-world social media. Scientific reports, 4(1), 5547. \href{https://doi.org/10.1038/srep05547}{doi: 10.1038/srep05547.}


\bibitem{Peng2017} Peng, S., Yang, A., Cao, L., Yu, S.,\& Xie, D. (2017). Social influence modeling using information theory in mobile social networks. Information Sciences, 379, 146-159. \href{https://doi.org/10.1016/j.ins.2016.08.023}{doi: 10.1016/j.ins.2016.08.023.}


\bibitem{Perozzi2014} Perozzi, B., Al-Rfou, R., \& Skiena, S. (2014). Deepwalk: Online learning of social representations. In Proceedings of the 20th ACM SIGKDD international conference on Knowledge discovery and data mining (pp. 701-710). \href{https://doi.org/10.1145/2623330.2623732}{doi: 10.1145/2623330.2623732.}


\bibitem{Piraveenan2013} Piraveenan, M., Prokopenko, M., \& Hossain, L. (2013). Percolation centrality: Quantifying graph-theoretic impact of nodes during percolation in networks. PloS one, 8(1), e53095. \href{https://doi.org/10.1371/journal.pone.0053095}{doi: 10.1371/journal.pone.0053095.}



\bibitem{Piraveenan2023} Piraveenan, M., \& Saripada, N. B. (2023). Transportation centrality: quantifying the relative importance of nodes in transportation networks based on traffic modelling. IEEE Access, vol. 11, pp. 142214-142234. \href{https://doi.org/10.1109/ACCESS.2023.3339121}{doi: 10.1109/ACCESS.2023.3339121.}




\bibitem{Potapov2008} Potapov, A. P., Goemann, B., \& Wingender, E. (2008). The pairwise disconnectivity index as a new metric for the topological analysis of regulatory networks. BMC bioinformatics, 9(1), 227. \href{https://doi.org/10.1186/1471-2105-9-227}{doi: 10.1186/1471-2105-9-227.}



\bibitem{Poulin2000} Poulin, R., Boily, M. C., \& Mâsse, B. R. (2000). Dynamical systems to define centrality in social networks. Social networks, 22(3), 187-220. \href{https://doi.org/10.1016/S0378-8733(00)00020-4}{doi: 10.1016/S0378-8733(00)00020-4.}





\bibitem{Power2013} Power, J. D., Schlaggar, B. L., Lessov-Schlaggar, C. N., \& Petersen, S. E. (2013). Evidence for hubs in human functional brain networks. Neuron, 79(4), 798-813. \href{https://doi.org/10.1016/j.neuron.2013.07.035}{doi: 10.1016/j.neuron.2013.07.035.}




\bibitem{Pozzi2013} Pozzi, F., Di Matteo, T., \& Aste, T. (2013). Spread of risk across financial markets: better to invest in the peripheries. Scientific reports, 3(1), 1665. \href{https://doi.org/10.1038/srep01665}{doi: 10.1038/srep01665.}




\bibitem{Przulj2004} Pržulj, N., Wigle, D. A., \& Jurisica, I. (2004). Functional topology in a network of protein interactions. Bioinformatics, 20(3), 340-348. \href{https://doi.org/10.1093/bioinformatics/btg415}{doi: 10.1093/bioinformatics/btg415.}



\bibitem{Pu2014} Pu, J., Chen, X., Wei, D., Liu, Q., \& Deng, Y. (2014). Identifying influential nodes based on local dimension. Europhysics Letters, 107(1), 10010. \href{https://doi.org/10.1209/0295-5075/107/10010}{doi: 10.1209/0295-5075/107/10010.}



\bibitem{Qi2012} Qi, X., Fuller, E., Wu, Q., Wu, Y., \& Zhang, C. Q. (2012). Laplacian centrality: A new centrality measure for weighted networks. Information Sciences, 194, 240-253. \href{https://doi.org/10.1016/j.ins.2011.12.027}{doi: 10.1016/j.ins.2011.12.027.}



\bibitem{Qi2015} Qi, X., Fuller, E., Luo, R., \& Zhang, C. Q. (2015). A novel centrality method for weighted networks based on the Kirchhoff polynomial. Pattern Recognition Letters, 58, 51-60. \href{https://doi.org/10.1016/j.patrec.2015.02.007}{doi: 10.1016/j.patrec.2015.02.007.}





\bibitem{Qiao2017} Qiao, T., Shan, W., \& Zhou, C. (2017). How to identify the most powerful node in complex networks? A novel entropy centrality approach. Entropy, 19(11), 614. \href{https://doi.org/10.3390/e19110614}{doi: 10.3390/e19110614.}



\bibitem{Qiao2018} Qiao, T., Shan, W., Yu, G., \& Liu, C. (2018). A novel entropy-based centrality approach for identifying vital nodes in weighted networks. Entropy, 20(4), 261. \href{https://doi.org/10.3390/e20040261}{doi: 10.3390/e20040261.}




\bibitem{Ranjan2013} Ranjan, G., \& Zhang, Z. L. (2013). Geometry of complex networks and topological centrality. Physica A: Statistical Mechanics and its Applications, 392(17), 3833-3845. \href{https://doi.org/10.1016/j.physa.2013.04.013}{doi: 10.1016/j.physa.2013.04.013.}




\bibitem{Ren2019} Ren, X. L., Gleinig, N., Helbing, D., \& Antulov-Fantulin, N. (2019). Generalized network dismantling. Proceedings of the national academy of sciences, 116(14), 6554-6559. \href{https://doi.org/10.1073/pnas.1806108116}{doi: 10.1073/pnas.1806108116.}




\bibitem{Ren2014} Ren, Z. M., Zeng, A., Chen, D. B., Liao, H., \& Liu, J. G. (2014). Iterative resource allocation for ranking spreaders in complex networks. Europhysics Letters, 106(4), 48005. \href{https://doi.org/10.1209/0295-5075/106/48005}{doi: 10.1209/0295-5075/106/48005.}




\bibitem{Restrepo2006} Restrepo, J. G., Ott, E., \& Hunt, B. R. (2006). Characterizing the dynamical importance of network nodes and links. Physical review letters, 97(9), 094102. \href{https://doi.org/10.1103/PhysRevLett.97.094102}{doi: 10.1103/PhysRevLett.97.094102.}


\bibitem{Riquelme2018} Riquelme, F., Gonzalez-Cantergiani, P., Molinero, X., \& Serna, M. (2018). Centrality measure in social networks based on linear threshold model. Knowledge-Based Systems, 140, 92-102. \href{https://doi.org/10.1016/j.knosys.2017.10.029}{doi: 10.1016/j.knosys.2017.10.029.}


\bibitem{Riveros2020} Riveros, C., \& Salas, J. (2020). A family of centrality measures for graph data based on subgraphs. In 23rd international conference on database theory (icdt 2020) (pp. 23-1). Schloss Dagstuhl–Leibniz-Zentrum für Informatik. \href{https://doi.org/10.1145/3649134}{doi: 10.1145/3649134.}


\bibitem{Rocco2022} Rocco, C. M., \& Barker, K. (2022). Deriving a minimum set of indicators to assess network component importance. Decision Analytics Journal, 5, 100145. \href{https://doi.org/10.1016/j.dajour.2022.100145}{doi: 10.1016/j.dajour.2022.100145.}




\bibitem{Rochat2009} Rochat, Y. (2009). Closeness centrality extended to unconnected graphs: The harmonic centrality index (Tech. Rep.). 


\bibitem{Rodriguez2007} Rodríguez, J. A., Estrada, E., \& Gutiérrez, A. (2007). Functional centrality in graphs. Linear and Multilinear Algebra, 55(3), 293-302. \href{https://doi.org/10.1080/03081080601002221}{doi: 10.1080/03081080601002221.}


\bibitem{Rossi2014} Rossi, L., Torsello, A., \& Hancock, E. R. (2014). Node centrality for continuous-time quantum walks. In Joint IAPR International Workshops on Statistical Techniques in Pattern Recognition (SPR) and Structural and Syntactic Pattern Recognition (SSPR) (pp. 103-112). Berlin, Heidelberg: Springer Berlin Heidelberg. \href{https://doi.org/10.1007/978-3-662-44415-3_11}{doi: 10.1007/978-3-662-44415-3\_11.}




\bibitem{Rosvall2005} Rosvall, M., Trusina, A., Minnhagen, P., \& Sneppen, K. (2005). Networks and cities: An information perspective. Physical review letters, 94(2), 028701. \href{https://doi.org/10.1103/PhysRevLett.94.028701}{doi: 10.1103/PhysRevLett.94.028701.}



\bibitem{Rubinov2011} Rubinov, M., \& Sporns, O. (2011). Weight-conserving characterization of complex functional brain networks. Neuroimage, 56(4), 2068-2079. \href{https://doi.org/10.1016/j.neuroimage.2011.03.069}{doi: 10.1016/j.neuroimage.2011.03.069.}


\bibitem{Rusinowska2011} Rusinowska, A., Berghammer, R., De Swart, H., \& Grabisch, M. (2011). Social networks: prestige, centrality, and influence. In International Conference on Relational and Algebraic Methods in Computer Science (pp. 22-39). Berlin, Heidelberg: Springer Berlin Heidelberg. \href{https://doi.org/10.1007/978-3-642-21070-9_2}{doi: 10.1007/978-3-642-21070-9\_2.}




\bibitem{Vargas2014} Ruiz Vargas, E., \& Wahl, L. M. (2014). The gateway coefficient: a novel metric for identifying critical connections in modular networks. The European Physical Journal B, 87(7), 161. \href{https://doi.org/10.1140/epjb/e2014-40800-7}{doi: 10.1140/epjb/e2014-40800-7.}


\bibitem{Sabidussi1966} Sabidussi, G. (1966). The centrality index of a graph. Psychometrika, 31(4), 581-603. \href{https://doi.org/10.1007/BF02289527}{doi: 10.1007/BF02289527.}




\bibitem{Sade1989} Sade, D. S. (1989). Sociometrics of macaca mulatta III: N-path centrality in grooming networks. Social Networks, 11(3), 273-292. \href{https://doi.org/10.1016/0378-8733(89)90006-3}{doi: 10.1016/0378-8733(89)90006-3.} 




\bibitem{Saito2016} Saito, K., Kimura, M., Ohara, K., \& Motoda, H. (2016). Super mediator-A new centrality measure of node importance for information diffusion over social network. Information Sciences, 329, 985-1000. \href{https://doi.org/10.1016/j.ins.2015.03.034}{doi: 10.1016/j.ins.2015.03.034.} 




\bibitem{Salavati2018} Salavati, C., Abdollahpouri, A., \& Manbari, Z. (2018). BridgeRank: A novel fast centrality measure based on local structure of the network. Physica A: Statistical Mechanics and its Applications, 496, 635-653. \href{https://doi.org/10.1016/j.physa.2017.12.087}{doi: 10.1016/j.physa.2017.12.087.} 





\bibitem{Saxena2018} Saxena, C., Doja, M. N., \& Ahmad, T. (2018). Group based centrality for immunization of complex networks. Physica A: Statistical Mechanics and Its Applications, 508, 35-47. \href{https://doi.org/10.1016/j.physa.2018.05.107}{doi: 10.1016/j.physa.2018.05.107.} 


\bibitem{Saxena2020} Saxena, A., \& Iyengar, S. (2020). Centrality measures in complex networks: A survey. arXiv preprint arXiv:2011.07190. \href{https://doi.org/10.48550/arXiv.2011.07190}{doi: 10.48550/arXiv.2011.07190.} 


\bibitem{Saxena2020b} Saxena, C., Doja, M. N., \& Ahmad, T. (2020). Entropy based flow transfer for influence dissemination in networks. Physica A: Statistical Mechanics and its Applications, 555, 124630. \href{https://doi.org/10.1016/j.physa.2020.124630}{doi: 10.1016/j.physa.2020.124630.}


\bibitem{Scardoni2009} Scardoni, G., Petterlini, M., \& Laudanna, C. (2009). Analyzing biological network parameters with CentiScaPe. Bioinformatics, 25(21), 2857-2859. \href{https://doi.org/10.1093/bioinformatics/btp517}{doi: 10.1093/bioinformatics/btp517.}




\bibitem{Schoch2024} Schoch D. Periodic Table of Network Centrality. (2024). Available from: \href{http://schochastics.net/sna/periodic.html}{http://schochastics.net/sna/periodic.html} (accessed on 1 October 2025).


\bibitem{Schoch2016}Schoch, D., \& Brandes, U. (2016). Re-conceptualizing centrality in social networks. European Journal of Applied Mathematics, 27(6), 971-985. \href{https://doi.org/10.1017/S0956792516000401}{doi: 10.1017/S0956792516000401.}



\bibitem{Schoch2018}Schoch, D. (2018). Centrality without indices: Partial rankings and rank probabilities in networks. Social Networks, 54, 50-60. \href{https://doi.org/10.1016/j.socnet.2017.12.003}{doi: 10.1016/j.socnet.2017.12.003.}




\bibitem{Seidman1983} Seidman, S. B. (1983). Network structure and minimum degree. Social networks, 5(3), 269-287. \href{https://doi.org/10.1016/0378-8733(83)90028-X}{doi: 10.1016/0378-8733(83)90028-X.}


\bibitem{Seidman2015}Segarra, S., \& Ribeiro, A. (2015). Stability and continuity of centrality measures in weighted graphs. IEEE Transactions on Signal Processing, 64(3), 543-555. \href{https://doi.org/10.1109/ICASSP.2015. 7178599}{doi: 10.1109/ICASSP.2015. 7178599.}




\bibitem{Servedio2025} Servedio, V. D., Bellina, A., Calò, E., \& De Marzo, G. (2025). Fitness centrality: a non-linear centrality measure for complex networks. Journal of Physics: Complexity, 6(1), 015002. \href{https://doi.org/10.1088/2632-072X/ada845}{doi: 10.1088/2632-072X/ada845.}



\bibitem{Shah2010} Shah, D., \& Zaman, T. (2010). Detecting sources of computer viruses in networks: theory and experiment. In Proceedings of the ACM SIGMETRICS international conference on Measurement and modeling of computer systems (pp. 203-214). \href{https://doi.org/10.1145/1811039.1811063}{doi: 10.1145/1811039.1811063.}




\bibitem{Shah2011} Shah, D., \& Zaman, T. (2011). Rumors in a network: Who's the culprit?. IEEE Transactions on information theory, 57(8), 5163-5181. \href{https://doi.org/10.1109/TIT.2011.2158885}{doi: 10.1109/TIT.2011.2158885.}



\bibitem{Shang2021} Shang, Q., Deng, Y., \& Cheong, K. H. (2021). Identifying influential nodes in complex networks: Effective distance gravity model. Information Sciences, 577, 162-179. \href{https://doi.org/10.1016/j.ins.2021.01.053}{doi: 10.1016/j.ins.2021.01.053.}



\bibitem{Shao2019} Shao, Z., Liu, S., Zhao, Y., \& Liu, Y. (2019). Identifying influential nodes in complex networks based on Neighbours and edges. Peer-to-Peer Networking and Applications, 12, 1528-1537. \href{https://doi.org/10.1007/s12083-018-0681-x}{doi: 10.1007/s12083-018-0681-x.}



\bibitem{Shapley1953} Shapley, L. S. (1953). A value for n-person games. In Kuhn, H., \& Tucker, A. (Eds.), In Contributions to the Theory of Games, volume II, pp. 307-317. Princeton University Press. \href{https://doi.org/10.1515/9781400881970-018}{doi: 10.1515/9781400881970-018.}




\bibitem{Shaw1954} Shaw, M. E. (1954). Group structure and the behavior of individuals in small groups. The Journal of psychology, 38(1), 139-149. \href{https://doi.org/10.1080/00223980.1954.9712925}{doi: 10.1080/00223980.1954.9712925.}


\bibitem{Sheikhahmadi2015} Sheikhahmadi, A., Nematbakhsh, M. A., \& Shokrollahi, A. (2015). Improving detection of influential nodes in complex networks. Physica A: Statistical Mechanics and its Applications, 436, 833-845. \href{https://doi.org/10.1016/j.physa.2015.04.035}{doi: 10.1016/j.physa.2015.04.035.}


\bibitem{Sheikhahmadi2017} Sheikhahmadi, A., \& Nematbakhsh, M. A. (2017). Identification of multi-spreader users in social networks for viral marketing. Journal of Information Science, 43(3), 412-423. \href{https://doi.org/10.1177/0165551516644171}{doi: 10.1177/0165551516644171.}




\bibitem{Sheng2020} Sheng, J., Dai, J., Wang, B., Duan, G., Long, J., Zhang, J., Guan, K., Hu, S., Chen, L. \& Guan, W. (2020). Identifying influential nodes in complex networks based on global and local structure. Physica A: Statistical Mechanics and its Applications, 541, 123262. \href{https://doi.org/10.1016/j.physa.2019.123262}{doi: 10.1016/j.physa.2019.123262.}



\bibitem{Shimbel1953} Shimbel, A. (1953). Structural parameters of communication networks. The bulletin of mathematical biophysics, 15(4), 501-507. \href{https://doi.org/10.1007/BF02476438}{doi: 10.1007/BF02476438.}



\bibitem{Shvydun2016} Shvydun, S. (2016). Normative properties of multi-criteria choice procedures and their superpositions: I. arXiv preprint arXiv:1611.00524. \href{https://doi.org/10.48550/arXiv.1611.00524}{doi: 10.48550/arXiv.1611.00524.}


\bibitem{Shvydun2019} Shvydun, S. (2019). Influence of Countries in the Global Arms Transfers Network: 1950–2018. In International Conference on Complex Networks and Their Applications (pp. 736-748). Cham: Springer International Publishing. \href{https://doi.org/10.1007/978-3-030-36683-4_59}{doi: 10.1007/978-3-030-36683-4\_59.}


\bibitem{Shvydun2020} Shvydun, S. (2020). Power of Nodes Based on Their Interdependence. In Complex Networks XI: Proceedings of the 11th Conference on Complex Networks CompleNet 2020 (pp. 70-82). Cham: Springer International Publishing. \href{https://doi.org/10.1007/978-3-030-40943-2_7}{doi: 10.1007/978-3-030-40943-2\_7.}


\bibitem{Shvydun2020b} Shvydun, S. (2020). Dynamic analysis of the global financial network. In 2020 IEEE/ACM International Conference on Advances in Social Networks Analysis and Mining (ASONAM) (pp. 374-378). IEEE. \href{https://doi.org/10.1109/ASONAM49781.2020.9381345}{doi: 10.1109/ASONAM49781.2020.9381345.}




\bibitem{Shvydun2025} Shvydun, S. (2025). Centrality in complex networks under incomplete data. PLOS Complex Systems, 2(5), e0000042. \href{https://doi.org/10.1371/journal.pcsy.0000042}{doi: 10.1371/journal.pcsy.0000042.}



\bibitem{Sikic2013} Šikić, M., Lančić, A., Antulov-Fantulin, N., \& Štefančić, H. (2013). Epidemic centrality—is there an underestimated epidemic impact of network peripheral nodes?. The European Physical Journal B, 86(10), 440. \href{https://doi.org/10.1140/epjb/e2013-31025-5}{doi: 10.1140/epjb/e2013-31025-5.}



\bibitem{Silva2012} Silva, F. N., \& Costa, L. D. F. (2012). Local dimension of complex networks. arXiv preprint arXiv:1209.2476. \href{https://doi.org/10.48550/arXiv.1209.2476}{doi: 10.48550/arXiv.1209.2476.}




\bibitem{Simko2013} Simko, G. I., \& Csermely, P. (2013). Nodes having a major influence to break cooperation define a novel centrality measure: game centrality. PloS one, 8(6), e67159. \href{https://doi.org/10.1371/journal.pone.0067159}{doi: 10.1371/journal.pone.0067159.}


\bibitem{Sinclair2009} Sinclair, P. A. (2009). Network centralization with the Gil Schmidt power centrality index. Social Networks, 31(3), 214-219. \href{https://doi.org/10.1016/j.socnet.2009.04.004}{doi: 10.1016/j.socnet.2009.04.004.}



\bibitem{Singh2017} Singh, R., Chakraborty, A., \& Manoj, B. S. (2017). GFT centrality: A new node importance measure for complex networks. Physica A: Statistical Mechanics and its Applications, 487, 185-195. \href{https://doi.org/10.1016/j.physa.2017.06.018}{doi: 10.1016/j.physa.2017.06.018.}




\bibitem{Skibski}
Skibski, O., Michalak, T. P., \& Rahwan, T. (2018). Axiomatic characterization of game-theoretic centrality. Journal of Artificial Intelligence Research, 62, 33-68. \href{https://doi.org/10.1613/jair.1.11202}{doi: 10.1613/jair.1.11202.}



\bibitem{Skibski2}
Skibski, O., \& Sosnowska, J. (2018). Axioms for distance-based centralities. In Proceedings of the AAAI Conference on Artificial Intelligence (Vol. 32, No. 1). pp. 1218-1225. \href{https://doi.org/10.1609/aaai.v32i1.11441}{doi: 10.1609/aaai.v32i1.11441.}


\bibitem{Skibski3} Skibski, O. (2018) Axioms4Centralities. Available from: \href{http://centrality.mimuw.edu.pl/}{http://centrality.mimuw.edu.pl/} (accessed on 1 May 2025).





\bibitem{Seeley1949} Seeley, J. R. (1949). The net of reciprocal influence; a problem in treating sociometric data. Canadian Journal of Psychology / Revue canadienne de psychologie, 3(4), 234-240. \href{https://doi.org/10.1037/h0084096}{doi: 10.1037/h0084096.}




\bibitem{Song2015} Song, B., Jiang, G. P., Song, Y. R., \& Xia, L. L. (2015). Rapid identifying high-influence nodes in complex networks. Chinese Physics B, 24(10), 100101. \href{https://doi.org/10.1088/1674-1056/24/10/100101}{doi: 10.1088/1674-1056/24/10/100101.}


\bibitem{Stella2018} Stella, M., \& De Domenico, M. (2018). Distance entropy cartography characterises centrality in complex networks. Entropy, 20(4), 268. \href{https://doi.org/10.3390/e20040268}{doi: 10.3390/e20040268.}




\bibitem{Stelzl2005} Stelzl, U., Worm, U., Lalowski, M., Haenig, C., Brembeck, F. H., Goehler, H., Stroedicke M., Zenkner M., Schoenherr A., Koeppen S., Timm J., Mintzlaff S., Abraham C., Bock N., Kietzmann S., Goedde A., Toksöz E., Droege A., Krobitsch S., Korn B., Birchmeier W., Lehrach H. \& Wanker, E. E. (2005). A human protein-protein interaction network: a resource for annotating the proteome. Cell, 122(6), 957-968. \href{https://doi.org/10.1016/j.cell.2005.08.029}{doi: 10.1016/j.cell.2005.08.029.}



\bibitem{Stephenson1989} Stephenson, K., \& Zelen, M. (1989). Rethinking centrality: Methods and examples. Social networks, 11(1), 1-37. \href{https://doi.org/10.1016/0378-8733(89)90016-6}{doi: 10.1016/0378-8733(89)90016-6.}





\bibitem{Sun2019} Sun, H. L., Chen, D. B., He, J. L., \& Ch’ng, E. (2019). A voting approach to uncover multiple influential spreaders on weighted networks. Physica A: Statistical Mechanics and its Applications, 519, 303-312. \href{https://doi.org/10.1016/j.physa.2018.12.001}{doi: 10.1016/j.physa.2018.12.001.}




\bibitem{Sun2017} Sun, Z., Wang, B., Sheng, J., Hu, Y., Wang, Y., \& Shao, J. (2017). Identifying influential nodes in complex networks based on weighted formal concept analysis. IEEE access, 5, 3777-3789. \href{https://doi.org/10.1109/ACCESS.2017.2679038}{doi: 10.1109/ACCESS.2017.2679038.}


\bibitem{Sziklai2018} Sziklai, B. (2018). How to identify experts in a community?. International Journal of Game Theory, 47(1), 155-173. \href{https://doi.org/10.1007/s00182-017-0582-x}{doi: 10.1007/s00182-017-0582-x.}


\bibitem{Sziklai2021} Sziklai, B. R. (2021). Ranking institutions within a discipline: The steep mountain of academic excellence. Journal of Informetrics, 15(2), 101133. \href{https://doi.org/10.1016/j.joi.2021.101133}{doi: 10.1016/j.joi.2021.101133.}


\bibitem{Sziklai2022} Sziklai, B. R., \& Lengyel, B. (2022). Finding early adopters of innovation in social networks. Social Network Analysis and Mining, 13(1), 4. \href{https://doi.org/10.1007/s13278-022-01012-5}{doi: 10.1007/s13278-022-01012-5.}



\bibitem{Tagarelli2013} Tagarelli, A., \& Interdonato, R. (2013). "Who's out there?" identifying and ranking lurkers in social networks. In Proceedings of the 2013 IEEE/ACM international conference on advances in social networks analysis and mining (pp. 215-222). \href{https://doi.org/10.1145/2492517.2492542}{doi: 10.1145/2492517.2492542.}


\bibitem{Takes2011} Takes, F. W., \& Kosters, W. A. (2011). Identifying prominent actors in online social networks using biased random walks. In Proceedings of the 23rd benelux conference on artificial intelligence (BNAIC) (pp. 215-222). 


\bibitem{Tan2006} Tan, Y. J., Wu, J., \& Deng, H. Z. (2006). Evaluation method for node importance based on node contraction in complex networks. Systems Engineering-Theory \& Practice, 11(11), 79-83. 


\bibitem{Thibault2024} Thibault, F., Hébert-Dufresne, L., \& Allard, A. (2024). On the uniform sampling of the configuration model with centrality constraints. arXiv preprint arXiv:2409.20493. \href{https://doi.org/10.48550/arXiv.2409.20493}{doi: 10.48550/arXiv.2409.20493.}


\bibitem{Torres2021} Torres, L., Chan, K. S., Tong, H., \& Eliassi-Rad, T. (2021). Nonbacktracking eigenvalues under node removal: X-centrality and targeted immunization. SIAM Journal on Mathematics of Data Science, 3(2), 656-675. \href{https://doi.org/10.1137/20M1352132}{doi: 10.1137/20M1352132.}




\bibitem{Travençolo2008} Travençolo, B. A. N., \& Costa, L. D. F. (2008). Accessibility in complex networks. Physics Letters A, 373(1), 89-95. \href{https://doi.org/10.1016/j.physleta.2008.10.069}{doi: 10.1016/j.physleta.2008.10.069.}


\bibitem{Tseng2024} Tseng, C. C., \& Lee, S. L. (2024). Fractional graph Fourier transform centrality and its application to social network. In 2024 IEEE 4th International Conference on Electronic Communications, Internet of Things and Big Data (ICEIB) (pp. 105-109). IEEE. \href{https://doi.org/10.1109/ICEIB61477.2024.10602678}{doi: 10.1109/ICEIB61477.2024.10602678.}



\bibitem{Tulu2018} Tulu, M. M., Hou, R., \& Younas, T. (2018). Identifying influential nodes based on community structure to speed up the dissemination of information in complex network. IEEE access, 6, 7390-7401. \href{https://doi.org/10.1109/ACCESS.2018.2794324}{doi: 10.1109/ACCESS.2018.2794324.}

\bibitem{Tutzauer2007} Tutzauer, F. (2007). Entropy as a measure of centrality in networks characterized by path-transfer flow. Social networks, 29(2), 249-265. \href{https://doi.org/10.1016/j.socnet.2006.10.001}{doi: 10.1016/j.socnet.2006.10.001.}

\bibitem{Ugurlu2022} Ugurlu, O. (2022). Comparative analysis of centrality measures for identifying critical nodes in complex networks. Journal of Computational Science, 62, 101738. \href{https://doi.org/10.1016/j.jocs.2022.101738}{doi: 10.1016/j.jocs.2022.101738.}



\bibitem{Ullah2021} Ullah, A., Wang, B., Sheng, J., Long, J., Khan, N., \& Sun, Z. (2021). Identification of nodes influence based on global structure model in complex networks. Scientific Reports, 11(1), 6173. \href{https://doi.org/10.1038/s41598-021-84684-x}{doi: 10.1038/s41598-021-84684-x.}



\bibitem{Ullah2021b} Ullah, A., Wang, B., Sheng, J., Long, J., Khan, N., \& Sun, Z. (2021). Identifying vital nodes from local and global perspectives in complex networks. Expert Systems with Applications, 186, 115778. \href{https://doi.org/10.1016/j.eswa.2021.115778}{doi: 10.1016/j.eswa.2021.115778.}





\bibitem{Valente1998} Valente, T. W., \& Foreman, R. K. (1998). Integration and radiality: Measuring the extent of an individual's connectedness and reachability in a network. Social networks, 20(1), 89-105. \href{https://doi.org/10.1016/S0378-8733(97)00007-5}{doi: 10.1016/S0378-8733(97)00007-5.}




\bibitem{vdBrink1994} van den Brink, R., \& Gilles, R. P. (1994). A social power index for hierarchically structured populations of economic agents. In Imperfections and Behavior in Economic Organizations (pp. 279-318). Dordrecht: Springer Netherlands. \href{https://doi.org/10.1007/978-94-011-1370-0_12}{doi: 10.1007/978-94-011-1370-0\_12.}




\bibitem{VanDenBrink2008} Van Den Brink, R., Borm, P., Hendrickx, R., \& Owen, G. (2008). Characterizations of the $\beta$-and the degree network power measure. Theory and Decision, 64(4), 519-536. \href{https://doi.org/10.1007/s11238-007-9077-8}{doi: 10.1007/s11238-007-9077-8.}




\bibitem{PVM2014} Van Mieghem P. \emph{Performance Analysis of Complex Networks and Systems}. Cambridge University Press; 2014. \href{https://doi.org/10.1017/CBO9781107415874}{doi: 10.1017/CBO9781107415874.}




\bibitem{PVM2017} Van Mieghem, P., Devriendt, K., \& Cetinay, H. (2017). Pseudoinverse of the Laplacian and best spreader node in a network. Physical Review E, 96(3), 032311. \href{https://doi.org/10.1103/PhysRevE.96.032311}{doi: 10.1103/PhysRevE.96.032311.}


\bibitem{Viswanath2009} Viswanath M. Ontology-based automatic text summarization (Doctoral dissertation, uga). 2009. 

\bibitem{Wan2021} Wan, Z., Mahajan, Y., Kang, B. W., Moore, T. J., \& Cho, J. H. (2021). A survey on centrality metrics and their network resilience analysis. IEEE Access, 9, 104773-104819. \href{https://doi.org/10.1109/ACCESS.2021.3094196}{doi: 10.1109/ACCESS.2021.3094196.}


\bibitem{Wandelt2018}Wandelt, S., Sun, X., Feng, D., Zanin, M., \& Havlin, S. (2018). A comparative analysis of approaches to network-dismantling. Scientific reports, 8(1), 13513. \href{https://doi.org/10.1038/s41598-018-31902-8}{doi: 10.1038/s41598-018-31902-8.}

\bibitem{Wang2023}Wang, G., Parthasarathy, R., \& Li, Y. (2023). Key node identification voting method based on multi-attributes in social complex networks. In Proceedings of the 2023 3rd Guangdong-Hong Kong-Macao Greater Bay Area Artificial Intelligence and Big Data Forum (pp. 394-399). \href{https://doi.org/10.1145/3660395.3660462}{doi: 10.1145/3660395.3660462.}


\bibitem{GWang2008} Wang, G., Shen, Y., \& Luan, E. (2008). A measure of centrality based on modularity matrix. Progress in Natural Science, 18(8), 1043-1047. \href{https://doi.org/10.1016/j.pnsc.2008.03.015}{doi: 10.1016/j.pnsc.2008.03.015.}


\bibitem{Wang2010} Wang, H., Li, M., Wang, J., \& Pan, Y. (2011). A new method for identifying essential proteins based on edge clustering coefficient. In International Symposium on Bioinformatics Research and Applications (pp. 87-98). Berlin, Heidelberg: Springer Berlin Heidelberg. \href{https://doi.org/10.1007/978-3-642-21260-4_12}{doi: 10.1007/978-3-642-21260-4\_12.}





\bibitem{HWang2015} Wang, H., Zhang, Y., Zhang, Z., Mahadevan, S., \& Deng, Y. (2015). Physarum Spreader: A New Bio-Inspired Methodology for Identifying Influential Spreaders in Complex Networks. PloS one, 10(12), e0145028. \href{https://doi.org/10.1371/journal.pone.0145028}{doi: 10.1371/journal.pone.0145028.}


\bibitem{Wang2017} Wang, J., Hou, X., Li, K., \& Ding, Y. (2017). A novel weight neighborhood centrality algorithm for identifying influential spreaders in complex networks. Physica A: Statistical Mechanics and its Applications, 475, 88-105. \href{https://doi.org/10.1016/j.physa.2017.02.007}{doi: 10.1016/j.physa.2017.02.007.}


\bibitem{Wang2018} Wang, J., Li, C., \& Xia, C. (2018). Improved centrality indicators to characterize the nodal spreading capability in complex networks. Applied Mathematics and Computation, 334, 388-400. \href{https://doi.org/10.1016/j.amc.2018.04.028}{doi: 10.1016/j.amc.2018.04.028.}




\bibitem{Wang2012} Wang, J., Li, M., Wang, H., \& Pan, Y. (2011). Identification of essential proteins based on edge clustering coefficient. IEEE/ACM Transactions on Computational Biology and Bioinformatics, 9(4), 1070-1080. \href{https://doi.org/10.1109/TCBB.2011.147}{doi: 10.1109/TCBB.2011.147.}




\bibitem{Wang2020} Wang, M., Li, W., Guo, Y., Peng, X., \& Li, Y. (2020). Identifying influential spreaders in complex networks based on improved k-shell method. Physica A: Statistical Mechanics and Its Applications, 124229. \href{https://doi.org/10.1016/j.physa.2020.124229}{doi: 10.1016/j.physa.2020.124229.}



\bibitem{QWang2018} Wang, Q., Ren, J., Wang, Y., Zhang, B., Cheng, Y., \& Zhao, X. (2018). CDA: A clustering degree based influential spreader identification algorithm in weighted complex network. IEEE Access, 6, 19550-19559. \href{https://doi.org/10.1109/ACCESS.2018.2822844}{doi: 10.1109/ACCESS.2018.2822844.}


\bibitem{Wang2016b} Wang, X., Zhang, X., Zhao, C., \& Yi, D. (2016). Maximizing the spread of influence via generalized degree discount. PloS one, 11(10), e0164393. \href{https://doi.org/10.1371/journal.pone.0164393}{doi: 10.1371/journal.pone.0164393.}


\bibitem{SWang2017} Wang, S., Du, Y., \& Deng, Y. (2017). A new measure of identifying influential nodes: Efficiency centrality. Communications in Nonlinear Science and Numerical Simulation, 47, 151-163. \href{https://doi.org/10.1016/j.cnsns.2016.11.008}{doi: 10.1016/j.cnsns.2016.11.008.}


\bibitem{XWang2016} Wang, X., Su, Y., Zhao, C., \& Yi, D. (2016). Effective identification of multiple influential spreaders by DegreePunishment. Physica A: Statistical Mechanics and its Applications, 461, 238-247. \href{https://doi.org/10.1016/j.physa.2016.05.020}{doi: 10.1016/j.physa.2016.05.020.}




\bibitem{Wang2008} Wang, X., Tao, T., Sun, J. T., Shakery, A., \& Zhai, C. (2008). Dirichletrank: Solving the zero-one gap problem of pagerank. ACM Transactions on Information Systems (TOIS), 26(2), 1-29. \href{https://doi.org/10.1145/1344411.1344416}{doi: 10.1145/1344411.1344416.}


\bibitem{Wang2021} Wang, X., Yang, Q., Liu, M., \& Ma, X. (2021). Comprehensive influence of topological location and neighbor information on identifying influential nodes in complex networks. Plos one, 16(5), e0251208. \href{https://doi.org/10.1371/journal.pone.0251208}{doi: 10.1371/journal.pone.0251208.}


\bibitem{Wang2019} Wang, Y., Chen, B., Li, W., \& Zhang, D. (2019). Influential Node Identification in Command and Control Networks Based on Integral k‐Shell. Wireless Communications and Mobile Computing, 2019(1), 6528431. \href{https://doi.org/10.1155/2019/6528431}{doi: 10.1155/2019/6528431.}


\bibitem{ZWang2017} Wang, Z., Du, C., Fan, J., \& Xing, Y. (2017). Ranking influential nodes in social networks based on node position and neighborhood. Neurocomputing, 260, 466-477. \href{https://doi.org/10.1016/j.neucom.2017.04.064}{doi: 10.1016/j.neucom.2017.04.064.}


\bibitem{Wang2015} Wang, Z., Duenas-Osorio, L., \& Padgett, J. E. (2015). A new mutually reinforcing network node and link ranking algorithm. Scientific reports, 5(1), 15141. \href{https://doi.org/10.1038/srep15141}{doi: 10.1038/srep15141.}



\bibitem{Wang2024} Wang, Z., Huang, R., Yang, D., Peng, Y., Zhou, B., \& Chen, Z. (2024). Identifying influential nodes based on the disassortativity and community structure of complex network. Scientific Reports, 14(1), 8453. \href{https://doi.org/10.1038/s41598-024-59071-x}{doi: 10.1038/s41598-024-59071-x.}


\bibitem{Wang2016} Wang, Z., Zhao, Y., Xi, J., \& Du, C. (2016). Fast ranking influential nodes in complex networks using a k-shell iteration factor. Physica A: Statistical Mechanics and its Applications, 461, 171-181. \href{https://doi.org/10.1016/j.physa.2016.05.048}{doi: 10.1016/j.physa.2016.05.048.}



\bibitem{Wąs2019} W\k{a}s, T., Rahwan, T., \& Skibski, O. (2019). Random walk decay centrality. In Proceedings of the AAAI Conference on Artificial Intelligence (Vol. 33, No. 01, pp. 2197-2204). \href{https://doi.org/10.1609/aaai.v33i01.33012197}{doi: 10.1609/aaai.v33i01.33012197.}



\bibitem{WassermanFaust1994} Wasserman, S., \& Faust, K. (1994). Social network analysis: Methods and applications. \href{https://doi.org/10.1017/CBO9780511815478}{doi: 10.1017/CBO9780511815478.}


\bibitem{Wiener1947} Wiener, H. (1947) Structural Determination of Paraffin Boiling Points. Journal of the American Chemical Society, 69, 17-20. \href{https://doi.org/10.1021/ja01193a005}{doi: 10.1021/ja01193a005.}




\bibitem{Wehmuth2012} Wehmuth, K., \& Ziviani, A. (2012). Distributed assessment of the closeness centrality ranking in complex networks. In Proceedings of the Fourth Annual Workshop on Simplifying Complex Networks for Practitioners (pp. 43-48). \href{https://doi.org/10.1145/2184356.2184368}{doi: 10.1145/2184356.2184368.}




\bibitem{Wei2015} Wei, B., Liu, J., Wei, D., Gao, C., \& Deng, Y. (2015). Weighted k-shell decomposition for complex networks based on potential edge weights. Physica A: Statistical Mechanics and its Applications, 420, 277-283. \href{https://doi.org/10.1016/j.physa.2014.11.012}{doi: 10.1016/j.physa.2014.11.012.}


\bibitem{Wei2013} Wei, D., Deng, X., Zhang, X., Deng, Y., \& Mahadevan, S. (2013). Identifying influential nodes in weighted networks based on evidence theory. Physica A: Statistical Mechanics and its Applications, 392(10), 2564-2575. \href{https://doi.org/10.1016/j.physa.2013.01.054}{doi: 10.1016/j.physa.2013.01.054.}



\bibitem{Wei1952} Wei, T. H. (1952). The Algebraic Foundations of Ranking Theory. PhD thesis, University of Cambridge. 


\bibitem{Wen2019} Wen, T., \& Jiang, W. (2019). Identifying influential nodes based on fuzzy local dimension in complex networks. Chaos, Solitons \& Fractals, 119, 332-342. \href{https://doi.org/10.1016/j.chaos.2019.01.011}{doi: 10.1016/j.chaos.2019.01.011.}

\bibitem{Wen2020} Wen, T., \& Deng, Y. (2020). Identification of influencers in complex networks by local information dimensionality. Information Sciences, 512, 549-562. \href{https://doi.org/10.1016/j.ins.2019.10.003}{doi: 10.1016/j.ins.2019.10.003.}

\bibitem{Wen2020b} Wen, T., Pelusi, D., \& Deng, Y. (2020). Vital spreaders identification in complex networks with multi-local dimension. Knowledge-Based Systems, 195, 105717. \href{https://doi.org/10.1016/j.knosys.2020.105717}{doi: 10.1016/j.knosys.2020.105717.}



\bibitem{Wenli2013} Wenli, F., Zhigang, L., \& Ping, H. (2013). Identifying node importance based on information entropy in complex networks. Physica Scripta, 88(6), 065201. \href{https://doi.org/10.1088/0031-8949/88/06/065201}{doi: 10.1088/0031-8949/88/06/065201.}

\bibitem{White2003} White, S., \& Smyth, P. (2003). Algorithms for estimating relative importance in networks. In Proceedings of the ninth ACM SIGKDD international conference on Knowledge discovery and data mining, pp. 266-275. \href{https://doi.org/10.1145/956750.956782}{doi: 10.1145/956750.956782.}



\bibitem{Williams2004} Williams, R. J., \& Martinez, N. D. (2004). Limits to trophic levels and omnivory in complex food webs: theory and data. The American Naturalist, 163(3), 458-468. \href{https://doi.org/10.1086/381964}{doi: 10.1086/381964.}



\bibitem{Wu2015} Wu, Z., Menichetti, G., Rahmede, C., \& Bianconi, G. (2015). Emergent complex network geometry. Scientific reports, 5(1), 10073. \href{https://doi.org/10.1038/srep10073}{doi: 10.1038/srep10073.}


\bibitem{Xu2019}Xu, S., Wang, P., Zhang, C. X., \& Lü, J. J. (2018). Spectral learning algorithm reveals propagation capability of complex networks. IEEE transactions on cybernetics, 49(12), 4253-4261. \href{https://doi.org/10.1109/TCYB.2018.2861568}{doi: 10.1109/TCYB.2018.2861568.}




\bibitem{Xu2017} Xu, S., \& Wang, P. (2017). Identifying important nodes by adaptive LeaderRank. Physica A: Statistical Mechanics and its Applications, 469, 654-664. \href{https://doi.org/10.1016/j.physa.2016.11.034}{doi: 10.1016/j.physa.2016.11.034.}


\bibitem{Yan2020} Yan, X. L., Cui, Y. P., \& Ni, S. J. (2020). Identifying influential spreaders in complex networks based on entropy weight method and gravity law. Chinese physics B, 29(4), 048902. \href{https://doi.org/10.1088/1674-1056/ab77fe}{doi: 10.1088/1674-1056/ab77fe.}



\bibitem{Yan2013} Yan, X., Zhai, L., \& Fan, W. (2013). C-index: A weighted network node centrality measure for collaboration competence. Journal of Informetrics, 7(1), 223-239. \href{https://doi.org/10.1016/j.joi.2012.11.004}{doi: 10.1016/j.joi.2012.11.004.}




\bibitem{Yanez-Sierra2021} Yanez-Sierra, J., Diaz-Perez, A., \& Sosa-Sosa, V. (2021). An efficient partition-based approach to identify and scatter multiple relevant spreaders in complex networks. Entropy, 23(9), 1216. \href{https://doi.org/10.3390/e23091216}{doi: 10.3390/e23091216.}


\bibitem{Yang2018} Yang, F., Li, X., Xu, Y., Liu, X., Wang, J., Zhang, Y., Zhang, R. \& Yao, Y. (2018). Ranking the spreading influence of nodes in complex networks: An extended weighted degree centrality based on a remaining minimum degree decomposition. Physics Letters A, 382(34), 2361-2371. \href{https://doi.org/10.1016/j.physleta.2018.05.032}{doi: 10.1016/j.physleta.2018.05.032.}


\bibitem{Yang2020} Yang, H., \& An, S. (2020). Critical nodes identification in complex networks. Symmetry, 12(1), 123. \href{https://doi.org/10.3390/sym12010123}{doi: 10.3390/sym12010123.}


\bibitem{Yang2023} Yang, P., Meng, F., Zhao, L., \& Zhou, L. (2023). AOGC: An improved gravity centrality based on an adaptive truncation radius and omni-channel paths for identifying key nodes in complex networks. Chaos, Solitons \& Fractals, 166, 112974. \href{https://doi.org/10.1016/j.chaos.2022.112974}{doi: 10.1016/j.chaos.2022.112974.}


\bibitem{Yang2021} Yang, X., \& Xiao, F. (2021). An improved gravity model to identify influential nodes in complex networks based on k-shell method. Knowledge-Based Systems, 227, 107198. \href{https://doi.org/10.1016/j.knosys.2021.107198}{doi: 10.1016/j.knosys.2021.107198.}



\bibitem{YangX2021} Yang, X. H., Xiong, Z., Ma, F., Chen, X., Ruan, Z., Jiang, P., \& Xu, X. (2021). Identifying influential spreaders in complex networks based on network embedding and node local centrality. Physica A: Statistical Mechanics and its Applications, 573, 125971. \href{https://doi.org/10.1016/j.physa.2021.125971}{doi: 10.1016/j.physa.2021.125971.}




\bibitem{YangWang2020} Yang, Y., Wang, X., Chen, Y., Hu, M., \& Ruan, C. (2020). A novel centrality of influential nodes identification in complex networks. IEEE access, 8, 58742-58751. \href{https://doi.org/10.1109/ACCESS.2020.2983053}{doi: 10.1109/ACCESS.2020.2983053.}




\bibitem{YangY2020} Yang, Y. Z., Hu, M., \& Huang, T. Y. (2020). Influential nodes identification in complex networks based on global and local information. Chinese Physics B, 29(8), 088903. \href{https://doi.org/10.1088/1674-1056/ab969f}{doi: 10.1088/1674-1056/ab969f.}



\bibitem{Young2019}Young, J. G., St-Onge, G., Laurence, E., Murphy, C., Hébert-Dufresne, L., \& Desrosiers, P. (2019). Phase transition in the recoverability of network history. Physical Review X, 9(4), 041056. \href{https://doi.org/10.1103/PhysRevX.9.041056}{doi: 10.1103/PhysRevX.9.041056.}



\bibitem{Yu2018} Yu, H., Chen, L., Cao, X., Liu, Z., \& Li, Y. (2018). Identifying top-k important nodes based on probabilistic-jumping random walk in complex networks. In Complex Networks \& Their Applications VI: Proceedings of Complex Networks 2017 (The Sixth International Conference on Complex Networks and Their Applications) (pp. 326-338). Springer International Publishing. \href{https://doi.org/10.1007/978-3-319-72150-7_27}{doi: 10.1007/978-3-319-72150-7\_27.}


\bibitem{Yu2019} Yu, Z., Shao, J., Yang, Q., \& Sun, Z. (2019). Profitleader: Identifying leaders in networks with profit capacity. World Wide Web, 22(2), 533-553. \href{https://doi.org/10.1007/s11280-018-0537-6}{doi: 10.1007/s11280-018-0537-6.}



\bibitem{Zachary1977} Zachary, W. W. (1977). An information flow model for conflict and fission in small groups. Journal of anthropological research, 33(4), 452-473. \href{https://doi.org/10.1086/jar.33.4.3629752}{doi: 10.1086/jar.33.4.3629752.}

\bibitem{Zareie2017} Zareie, A., Sheikhahmadi, A., \& Fatemi, A. (2017). Influential nodes ranking in complex networks: An entropy-based approach. Chaos, Solitons \& Fractals, 104, 485-494. \href{https://doi.org/10.1016/j.chaos.2017.09.010}{doi: 10.1016/j.chaos.2017.09.010.}

\bibitem{Zareie2018} Zareie, A., \& Sheikhahmadi, A. (2018). A hierarchical approach for influential node ranking in complex social networks. Expert Systems with Applications, 93, 200-211. \href{https://doi.org/10.1016/j.eswa.2017.10.018}{doi: 10.1016/j.eswa.2017.10.018.}




\bibitem{Zareie2018b}Zareie, A., Sheikhahmadi, A., \& Khamforoosh, K. (2018). Influence maximization in social networks based on TOPSIS. Expert Systems with Applications, 108, 96-107. \href{https://doi.org/10.1016/j.eswa.2018.05.001}{doi: 10.1016/j.eswa.2018.05.001.}


\bibitem{Zareie2019b} Zareie, A., \& Sheikhahmadi, A. (2019). EHC: Extended H-index centrality measure for identification of users’ spreading influence in complex networks. Physica A: Statistical Mechanics and its Applications, 514, 141-155. \href{https://doi.org/10.1016/j.physa.2018.09.064}{doi: 10.1016/j.physa.2018.09.064.}


\bibitem{Zareie2019} Zareie, A., Sheikhahmadi, A., \& Jalili, M. (2019). Influential node ranking in social networks based on neighborhood diversity. Future Generation Computer Systems, 94, 120-129. \href{https://doi.org/10.1016/j.future.2018.11.023}{doi: 10.1016/j.future.2018.11.023.}

\bibitem{Zareie2020} Zareie, A., Sheikhahmadi, A., Jalili, M., \& Fasaei, M. S. K. (2020). Finding influential nodes in social networks based on neighborhood correlation coefficient. Knowledge-based systems, 194, 105580. \href{https://doi.org/10.1016/j.knosys.2020.105580}{doi: 10.1016/j.knosys.2020.105580.}


\bibitem{Zdeborová2016} Zdeborová, L., Zhang, P., \& Zhou, H. J. (2016). Fast and simple decycling and dismantling of networks. Scientific reports, 6(1), 37954. \href{https://doi.org/10.1038/srep37954}{doi: 10.1038/srep37954.}

\bibitem{Zeng2013} Zeng, A., \& Zhang, C. J. (2013). Ranking spreaders by decomposing complex networks. Physics letters A, 377(14), 1031-1035. \href{https://doi.org/10.1016/j.physleta.2013.02.039}{doi: 10.1016/j.physleta.2013.02.039.}

\bibitem{Zhai2013} Zhai, L., Yan, X., \& Zhang, G. (2013). A centrality measure for communication ability in weighted network. Physica A: Statistical Mechanics and its Applications, 392(23), 6107-6117. \href{https://doi.org/10.1016/j.physa.2013.07.056}{doi: 10.1016/j.physa.2013.07.056.}


\bibitem{Zhang2021} Zhang, H., Zhong, S., Deng, Y., \& Cheong, K. H. (2021). LFIC: Identifying influential nodes in complex networks by local fuzzy information centrality. IEEE Transactions on Fuzzy Systems, 30(8), 3284-3296. \href{https://doi.org/10.1109/TFUZZ.2021.3112226}{doi: 10.1109/TFUZZ.2021.3112226.}


\bibitem{Zhang2016} Zhang, J. X., Chen, D. B., Dong, Q., \& Zhao, Z. D. (2016). Identifying a set of influential spreaders in complex networks. Scientific reports, 6(1), 27823. \href{https://doi.org/10.1038/srep27823}{doi: 10.1038/srep27823.}


\bibitem{Zhang2019} Zhang, J., Wang, B., Sheng, J., Dai, J., Hu, J., \& Chen, L. (2019). Identifying influential nodes in complex networks based on local effective distance. Information, 10(10), 311. \href{https://doi.org/10.3390/info10100311}{doi: 10.3390/info10100311.}


\bibitem{Zhang2022} Zhang, J., Zhang, Q., Wu, L., \& Zhang, J. (2022). Identifying influential nodes in complex networks based on multiple local attributes and information entropy. Entropy, 24(2), 293. \href{https://doi.org/10.3390/e24020293}{doi: 10.3390/e24020293.}



\bibitem{Zhang2018} Zhang, Q., Li, M., \& Deng, Y. (2018). Measure the structure similarity of nodes in complex networks based on relative entropy. Physica A: Statistical Mechanics and its Applications, 491, 749-763.  \href{https://doi.org/10.1016/j.physa.2017.09.042}{doi: 10.1016/j.physa.2017.09.042.}


\bibitem{Zhang2014} Zhang, Q., Li, M., Du, Y., \& Deng, Y. (2014). Local structure entropy of complex networks. arXiv preprint arXiv:1412.3910. \href{https://doi.org/10.48550/arXiv.1412.3910}{doi: 10.48550/arXiv.1412.3910.}


\bibitem{ZhangQ2022} Zhang, Q., Shuai, B., \& Lü, M. (2022). A novel method to identify influential nodes in complex networks based on gravity centrality. Information Sciences, 618, 98-117. \href{https://doi.org/10.1016/j.ins.2022.10.070}{doi: 10.1016/j.ins.2022.10.070.}


\bibitem{Zhang2020} Zhang, Y., Shao, C., He, S., \& Gao, J. (2020). Resilience centrality in complex networks. Physical Review E, 101(2), 022304. \href{https://doi.org/10.1103/PhysRevE.101.022304}{doi: 10.1103/PhysRevE.101.022304.}

\bibitem{Zhang2012} Zhang, Y., Zhang, Z., Wei, D., \& Deng, Y. (2012). Centrality measure in weighted networks based on an amoeboid algorithm. Journal of Information and Computational Science, 9(2), 369-376. 

\bibitem{Zhao2021} Zhao, J., Song, Y., Liu, F., \& Deng, Y. (2021). The identification of influential nodes based on structure similarity. Connection science, 33(2), 201-218. \href{https://doi.org/10.1080/09540091.2020.1806203}{doi: 10.1080/09540091.2020.1806203.}

\bibitem{Zhao2020} Zhao, J., Wang, Y., \& Deng, Y. (2020). Identifying influential nodes in complex networks from global perspective. Chaos, Solitons \& Fractals, 133, 109637. \href{https://doi.org/10.1016/j.chaos.2020.109637}{doi: 10.1016/j.chaos.2020.109637.}


\bibitem{Zhao2022} Zhao, J., Wen, T., Jahanshahi, H., \& Cheong, K. H. (2022). The random walk-based gravity model to identify influential nodes in complex networks. Information Sciences, 609, 1706-1720. \href{https://doi.org/10.1016/j.ins.2022.07.084}{doi: 10.1016/j.ins.2022.07.084.}


\bibitem{Zhao2023} Zhao, S., \& Sun, S. (2023). Identification of node centrality based on Laplacian energy of networks. Physica A: Statistical Mechanics and its Applications, 609, 128353. \href{https://doi.org/10.1016/j.physa.2022.128353}{doi: 10.1016/j.physa.2022.128353.}

\bibitem{ZhaoX2015} Zhao, X. Y., Huang, B., Tang, M., Zhang, H. F., \& Chen, D. B. (2015). Identifying effective multiple spreaders by coloring complex networks. Europhysics Letters, 108(6), 68005. \href{https://doi.org/10.1209/0295-5075/108/68005}{doi: 10.1209/0295-5075/108/68005.}




\bibitem{Zhao2018} Zhao, X., Liu, F. A., Xing, S., \& Wang, Q. (2018). Identifying influential spreaders in social networks via normalized local structure attributes. IEEE Access, 6, 66095-66104. \href{https://doi.org/10.1109/ACCESS.2018.2879116}{doi: 10.1109/ACCESS.2018.2879116.}



\bibitem{Zhao2009} Zhao, Y., Wang, Z., Zheng, J., \& Guo, X. (2009). Finding most vital node by node importance contribution matrix in communication netwoks. Journal of Beijing University of Aeronautics and Astronautics, 35(9), 1076. 


\bibitem{Zhao2015} Zhao, Z., Wang, X., Zhang, W., \& Zhu, Z. (2015). A community-based approach to identifying influential spreaders. Entropy, 17(4), 2228-2252. \href{https://doi.org/10.3390/e17042228}{doi: 10.3390/e17042228.}



\bibitem{Zhong2015} Zhong, L. F., Liu, J. G., \& Shang, M. S. (2015). Iterative resource allocation based on propagation feature of node for identifying the influential nodes. Physics Letters A, 379(38), 2272-2276. \href{https://doi.org/10.1016/j.physleta.2015.05.021}{doi: 10.1016/j.physleta.2015.05.021.}



\bibitem{Zhong2022} Zhong, S., Zhang, H., \& Deng, Y. (2022). Identification of influential nodes in complex networks: A local degree dimension approach. Information Sciences, 610, 994-1009. \href{https://doi.org/10.1016/j.ins.2022.07.172}{doi: 10.1016/j.ins.2022.07.172.}


\bibitem{Zhou2023} Zhou, X., Duenas‐Osorio, L., Doss‐Gollin, J., Liu, L., Stadler, L., \& Li, Q. (2023). Mesoscale modeling of distributed water systems enables policy search. Water Resources Research, 59(5), e2022WR033758. \href{https://doi.org/10.1029/2022WR033758}{doi: 10.1029/2022WR033758.}



\bibitem{Zhou2018} Zhou, X., Liang, X., Zhao, J., \& Zhang, S. (2018). Cycle based network centrality. Scientific Reports, 8(1), 11749. \href{https://doi.org/10.1038/s41598-018-30249-4}{doi: 10.1038/s41598-018-30249-4.}



\bibitem{Zhou2012} Zhou, X., Zhang, F. M., Li, K. W., Hui, X. B., \& Wu, H. S. (2012). Finding vital node by node importance evaluation matrix in complex networks. Acta Phys. Sin., 61(5): 050201. \href{https://doi.org/10.7498/aps.61.050201}{doi: 10.7498/aps.61.050201.}


\bibitem{JZhou2018} Zhou, J., Yu, X., \& Lu, J. A. (2018). Node importance in controlled complex networks. IEEE Transactions on Circuits and Systems II: Express Briefs, 66(3), 437-441. \href{https://doi.org/10.1109/TCSII.2018.2845940}{doi: 10.1109/TCSII.2018.2845940.}


\bibitem{Zhu2017} Zhu, C., Wang, X., \& Zhu, L. (2017). A novel method of evaluating key nodes in complex networks. Chaos, Solitons \& Fractals, 96, 43-50. \href{https://doi.org/10.1016/j.chaos.2017.01.007}{doi: 10.1016/j.chaos.2017.01.007.}

\bibitem{Zhu2021} Zhu, J., \& Wang, L. (2021). Identifying influential nodes in complex networks based on node itself and neighbor layer information. Symmetry, 13(9), 1570. \href{https://doi.org/10.3390/sym13091570}{doi: 10.3390/sym13091570.}

\bibitem{Zhu2022} Zhu, J. C., \& Wang, L. W. (2022). An extended improved global structure model for influential node identification in complex networks. Chinese Physics B, 31(6), 068904. \href{https://doi.org/10.1088/1674-1056/ac380d}{doi: 10.1088/1674-1056/ac380d.}


\bibitem{Zhuge2010} Zhuge, H., \& Zhang, J. (2010). Topological centrality and its e‐Science applications. Journal of the American Society for Information Science and Technology, 61(9), 1824-1841. \href{https://doi.org/10.1002/asi.21353}{doi: 10.1002/asi.21353.}

\end{thebibliography}
\end{document}